\documentclass[12pt,preprint]{aastex}

\usepackage{graphicx}

\newcommand{\nosne}{50}
\newcommand{\nosnenir}{45}

\newcommand{\dm}{\Delta m_{15}}

\shorttitle{Photometry of CSP SNe~Ia}
\shortauthors{Stritzinger et al.}
\begin{document}

\title{The Carnegie Supernova Project:\\ Second Photometry Data Release of
 Low-Redshift Type~Ia Supernovae}

\author{Maximilian D. Stritzinger\altaffilmark{1,2,3}
M.~M.~Phillips\altaffilmark{3},
Luis Boldt S.\altaffilmark{4},
Chris Burns\altaffilmark{5},
Abdo Campillay\altaffilmark{3},
Carlos~Contreras\altaffilmark{6},
Sergio Gonzalez\altaffilmark{7},
Gast\'on~Folatelli\altaffilmark{8},
Nidia~Morrell\altaffilmark{3}, 
Wojtek~Krzeminski\altaffilmark{3},
Miguel~Roth\altaffilmark{3},
Francisco~Salgado\altaffilmark{9},
Darren L. Depoy\altaffilmark{10},
Mario~Hamuy\altaffilmark{11},
Wendy~L.~Freedman\altaffilmark{5},
Barry~F.~Madore\altaffilmark{5,12},
Jennifer~L.~Marshall\altaffilmark{10},
Sven~E.~Persson\altaffilmark{5},
Jean-Philippe Rheault\altaffilmark{10},
Nicholas~B.~Suntzeff\altaffilmark{10},
Steven Villanueva\altaffilmark{10},
Weidong Li\altaffilmark{13}, and
Alexei~V.~Filippenko\altaffilmark{13}
}

\altaffiltext{1}{
The Oskar Klein Centre, Department of Astronomy, Stockholm University, AlbaNova, 10691 Stockholm, Sweden; {max.stritzinger@astro.su.se}}
\altaffiltext{2}{Dark Cosmology Centre, Niels Bohr Institute, University
of Copenhagen, Juliane Maries Vej 30, 2100 Copenhagen \O, Denmark; {max@dark-cosmology.dk}}
\altaffiltext{3}{Carnegie Observatories, Las Campanas Observatory, 
  Casilla 601, La Serena, Chile; {mstritzinger@lco.cl}}
  \altaffiltext{4}{Argelander Institut f\"ur Astronomie, Universit\"at Bonn, Auf dem H\"ugel 71, D-53111 Bonn, Germany}
\altaffiltext{5}{Observatories of the Carnegie Institution for
 Science, 813 Santa Barbara St., Pasadena, CA 91101, USA}
  \altaffiltext{6}{Centre for Astrophysics \& Supercomputing, Swinburne University of Technology, P.O. Box 218, Victoria 3122, Australia}
\altaffiltext{7}{Atacama Large Millimeter/Submillimeter Array, European Southern Observatory, Chile}
\altaffiltext{8}{Institute for the Physics and Mathematics of the Universe (IPMU), University of Tokyo, 5-1-5 Kashiwanoha, Kashiwa, Chiba 277-8583, Japan}
\altaffiltext{9}{Leiden Observatory, Leiden University, PO Box 9513, NL-2300 RA Leiden, The Netherlands}
\altaffiltext{10}{George P. and Cynthia Woods Mitchell Institute for Fundamental Physics and Astronomy, Department of Physics and Astronomy, Texas A\&M University, College Station, TX 77843, USA}
\altaffiltext{11}{Departamento de Astronom\'{\i}a, Universidad de Chile,
  Casilla 36-D, Santiago, Chile}
\altaffiltext{12}{Infrared Processing and Analysis Center, Caltech/Jet
  Propulsion Laboratory, Pasadena, CA 91125, USA}
\altaffiltext{13}{Department of Astronomy, University of California,
  Berkeley, CA 94720-3411, USA}

\begin{abstract}
  \noindent 
The Carnegie Supernova Project (CSP) was a five-year observational survey conducted at 
Las Campanas Observatory that obtained, among other things,  
high-quality light curves of $\sim 100$ low-redshift Type~Ia supernovae (SNe~Ia).  
Presented here is the second data release of nearby SN~Ia 
photometry consisting of {\nosne} objects, with a subset of {\nosnenir} 
having near-infrared follow-up observations. 
Thirty-three objects have optical pre-maximum coverage with a subset of 15 beginning at least 5 days before maximum light. 
In the near-infrared,  27 objects have  coverage beginning before the epoch of $B$-band maximum, with 
a subset of 13 beginning at least 5 days before maximum.
In addition, we present results of a photometric calibration program to
measure the CSP optical ($uBgVri$)
bandpasses with an accuracy of  $\sim 1$\%. 
Finally, we report the discovery of a second SN~Ia, SN~2006ot, similar in its characteristics to the peculiar SN~2006bt.

\end{abstract}
\keywords{galaxies: distances and redshifts -- supernovae: general}

\section{INTRODUCTION}
\label{sec:intro}
The current cosmological paradigm ($\Lambda$-cold-dark-matter)
suggests that the Universe is composed of  $\sim$ 4.4\% baryonic matter,  $\sim$ 21.3\%  cold dark matter, and  $\sim$ 74.2\% dark energy \citep[e.g.,][]{dunkley09}. The luminosity distances of Type~Ia supernovae (SNe~Ia) provided the 
first strong evidence that the expansion rate of the Universe is currently accelerating \citep{riess98,perlmutter99}, and that dark energy having negative
pressure is the major contributor to the overall energy budget of the Universe.
These  findings have been further substantiated through the detection of the 
late-time integrated Sachs-Wolfe effect \citep{giannantonio08} and X-ray cluster distances
\citep[e.g.,][]{allen07}.

To date, much effort and resources have been directed toward characterizing 
the nature of dark energy, resulting in more than 1000 SNe~Ia being observed,
extending from the smooth Hubble flow out to a redshift ($z$) of 1.7. The current data 
have revealed that  the  equation-of-state parameter of the dark energy, $w = P / (\rho c^{2})$, is  consistent with a cosmological constant, $w = -1$
 \citep{astier06,wood07,riess07,freedman09,sullivan11}.
Advancements in our understanding of dark energy will come from 
future experiments that seek to measure the time derivative of $w$.
The success and credibility of these experiments will require percent-level accuracy in SN distances, something that has yet to be demonstrated.  To achieve this will require both a deeper theoretical understanding of the progenitors and explosion mechanism as well as, on the observational side, a significant sample of well-calibrated low-$z$ SNe Ia monitored over a wide wavelength range.

The Carnegie Supernova Project (CSP) is the umbrella designation of
a five-year (2004--2009) National Science Foundation (NSF) funded observational campaign consisting of  low- and high-$z$ components.
The  overriding goal of the low-$z$ component was to construct an
atlas of $\sim$ 100 high signal-to-noise ratio (S/N) SN~Ia light curves in a 
homogeneous and well-defined photometric system in the optical and near-infrared (IR).
These data will serve as a fundamental reference to anchor future  
Hubble diagrams that will contain thousands of high-$z$ SNe~Ia observed 
by future all-sky surveys. 
The objective of the high-$z$ portion of the CSP was to construct a rest-frame $i$-band Hubble diagram in order to minimize the effects of dust reddening.
At the end of the latter observing program, 71 SNe~Ia were observed at cosmological distances out to $z$ $\approx 0.7$.  A subset of these objects was combined with our initial low-$z$ sample to construct the Hubble diagram presented by   \citet{freedman09}. 

By the end of the fifth and final year (June 2009),
 the low-$z$ part of the CSP had obtained photometric follow-up observations of 
129 SNe~Ia. In the first of a series of papers, we presented optical ($uBgVri$)  and 
near-IR ($YJHK_s$) light curves of 35 SNe~Ia \citep{contreras10}, along with a  detailed 
analysis  \citep{folatelli10}. A highlight of  the  \citeauthor{folatelli10} study is the 
discrepancy found between the value of the reddening parameter, $R_{V}$,
as determined through the comparison of color excesses to the method of
leaving $R_{V}$ as a free parameter
and minimizing  the dispersion in the Hubble diagram. When excluding the two
 most highly reddened  objects from the first sample, a comparison of colors or 
color excesses suggests agreement with the canonical Milky Way value, 
 $R_{V} \approx 3.0$. However, when one minimizes the dispersion in the 
 Hubble diagram with $R_{V}$  treated as a free parameter,  
 lower values of $R_{V}$ $\approx$  $1 - 2$ are obtained.
This discrepancy implies that there is an 
intrinsic color variation  of SNe~Ia that correlates with luminosity, but is 
  independent of the light-curve decline-rate parameter 
  $\Delta m_{15}$.\footnote[14]{The decline-rate parameter is traditionally defined as the 
 difference in magnitude of the 
  $B$-band light curve from peak brightness to 15 days later.
 The value of $\Delta m_{15}$ is correlated with absolute peak magnitude 
 in such a way that low-luminosity SNe~Ia exhibit a faster decline rate
 \citep{phillips93}. } 
  To address this apparent inconsistency demands an expanded sample of
  well-observed multi-band SNe~Ia.
 
In this paper we present the second CSP release of SN~Ia light curves.
The sample consists of  $\nosne$ objects observed between 2005 and 2009, with a subset of
 $\nosnenir$ objects having accompanying near-IR observations.
The light curves presented here are a significant addition to a growing 
sample of low-$z$ SNe~Ia that have been observed in well-understood
and stable photometric systems \citep{wood08,hicken09,contreras10,mo10}.
A third and final paper containing optical and near-IR 
light curves of an additional $\sim$ 20 to 25
objects will complete the full data release of 
low-$z$ SNe~Ia observed by the CSP.

Additional aims of the CSP included the construction of substantial samples of 
Type~II SNe ($\sim$ 100) and  Type Ib/c  SNe ($\sim$ 25). 
By the end of our observational program, these goals were successfully achieved, and 
a majority of the data are currently being  analyzed.  
The CSP was not limited to only photometric follow-up observations; a notable commitment
was also made to obtain contemporaneous optical spectroscopy. 
In a companion paper currently in preparation an analysis of 
 spectrophotometry of $\sim$ 80 SNe~Ia monitored by the CSP will be presented. Additionally, a second analysis 
 paper combining the CSP's first 85 published low-$z$ 
SNe~Ia will be presented in the near future.

The organization of this paper is as follows. Section~\ref{sec:obs} briefly describes
the observations, \S~\ref{bandpasses} discusses results obtained from 
a calibration program  to measure the optical and near-IR transmission functions of the  
Henrietta Swope 1-m and Ir\'en\'ee du Pont 2.5-m telescopes equipped with CSP filters,
\S~\ref{lightcurves} presents the final light
curves, and \S~\ref{conclusion} contains our conclusions.

\section{Observations}
\label{sec:obs}

A complete overview of the CSP including the goals of the project, 
the facilities at Las Campanas Observatory (LCO) used to conduct photometric and spectroscopic 
data, and the observing procedures is given by \citet{hamuy06}.  A detailed and  up-to-date description of the data-reduction procedures, the host-galaxy subtraction techniques, and computation of definitive photometry in the natural system of the LCO facilities is given in  the first data release of SN~Ia photometry \citep{contreras10}. It is not our intention to repeat the material contained in these previous publications, so in this section we provide only a brief summary of our methodology.

The CSP was a follow-up program of southern and equatorial SNe discovered 
by various sources ranging from dedicated  backyard amateurs to professional searches.
Five observational campaigns, each of duration 9 months, were conducted between 2004 and 2009.
The primary criteria used to select an object for monitoring 
were (i) that it was caught young, preferably before maximum light, or within a few weeks after explosion in the case of 
Type II plateau objects,
 and  (ii) that the peak magnitude was $\sim$ 18 or brighter. 
 The workhorse of the CSP was the Henrietta Swope 1-m telescope equipped
with both a direct imaging CCD camera that we refer to as SITe3 and a near-IR imager called RetroCam.
Additional near-IR imaging and limited optical observations were obtained with the
du Pont 2.5-m telescope equipped with the  Wide Field IR Camera (WIRC) and
a direct optical imager referred to as Tek 5. 
Typically a year or more after the last observation 
of any given SN, deep ``template" images of its host galaxy were obtained with the du Pont telescope.
These high S/N templates  are  used to subtract away the host-galaxy light at the position of the SN in each science image. This procedure is essential for obtaining precise photometry,
particularly if the SN occurred  in the inner region of the
host galaxy or in a spiral arm.

Template subtraction has been performed on the science images of  all objects 
presented here except SN~2007if (near-IR) and SN~2009dc (optical and near-IR).
As these two  events  are super-Chandrasekhar candidates,  
we recognize the interests within the community to promptly make their photometry public.  
Given the position of both objects in their respective host galaxies, we suspect 
that the photometry suffers negligible background contamination.  
Nevertheless, in the near future  templates of these objects will be acquired,  and the subsequent template-subtracted photometry 
will be available on the CSP webpage.

Photometry of  each SN is computed differentially  with respect to a local sequence of stars.
Absolute photometry of each local sequence is determined relative to 
photometric standard stars which were observed over a minimum of three photometric nights.
The master list of optical photometric standard stars used to calibrate the local sequences
is composed of stars in common with the \citet{landolt92} and  
\citet{smith02} catalogs. In the near-IR, absolute photometry of each local sequence
is computed relative to \citet{persson98} standards. 

The CSP has elected to publish photometry on  the {\em natural system}
of the Swope and du Pont telescopes. 
 The reasons for this are highlighted by \citet{contreras10}, and boil down to 
 providing the photometry in a form which simplifies the process of combining it
 with datasets obtained by other groups. Armed with spectral templates and well-defined 
 system-response functions, CSP photometry can be transformed or 
 ``S-corrected" \citep{stritzinger02} to any desired user-defined photometric system in a 
 straightforward manner.

 The SNe contained in this data release were observed from mid-2005 to 
 early-2009. A mosaic consisting of  a $V$-band image of each SN is presented in 
 Figure ~\ref{fig:fcharts}.
 General properties related to each object are provided in Table~\ref{tab:snproperties}. 
 Specifically, this includes  (i) the coordinates, (ii) identification, morphology, and redshift of the host galaxy,
 (iii) a reference to the IAUC or CBET reporting the discovery, and 
 (iv) the name of the discovery individual(s) or group(s). 
 
Basic  photometric and spectroscopic information on  each object is compiled in  
 Table~\ref{tab:sne2},  consisting of
 (i) an estimate of the decline-rate parameter, $\Delta m_{15}$ 
 (see \S~\ref{lightcurves} for details), 
 (ii) the number of nights each SN was observed, 
 (iii) the number of nights that standard-star fields were observed under photometric conditions, and 
 (iv) the SN spectral subtype. 
  Values of $\Delta m_{15}$ were computed using the SNooPy Python package (SNooPy; Burns et al. 2011), while  the SuperNova IDentification code (SNID; Blondin \& Tonry 2007) provided aid in determining spectral subtypes.
Finally, the last column  of Table~\ref{tab:sne2} lists the epoch of the spectrum (relative to the time of $B$-band maximum) used to determine the spectral subtype.

Examination of the light-curve properties and optical spectra
reveals that 38 of the objects in this data release are normal SNe~Ia,
seven are of the low-luminosity SN~1986G \citep{phillips87} or  
 SN~1991bg \citep[][]{filippenko92a,leibundgut93} variety, two are of the slow-declining SN~1991T-like subtype \citep[][]{filippenko92b,phillips92}, two are considered as
possible super-Chandrasekhar SNe \citep[e.g., see][]{howell06},
and one object, SN~2006bt, is  
uniquely peculiar \citep{foley10}. 
 
Optical and near-IR photometry of  each local sequence in the {\em standard} 
Landolt ($BV$; Landolt 1992),  Smith ($u'g'r'i'$; Smith et al. 2002), 
and Persson ($JHK_s$; Persson et al. 1998) photometric systems is given in Table~\ref{tab:optir_stds}. 
The $Y$-band photometry is  calibrated relative to  the ($Y-K_s$) and ($J-K_s$) color relation
given by \citet[][see Appendix C, Equation C2]{hamuy06}. 
The quoted uncertainty accompanying the magnitudes of each star corresponds to the weighted average of the instrumental errors computed from several measurements  
obtained over the course of multiple photometric nights (see Columns 5 and 6 of 
Table~\ref{tab:sne2}). 
 
\section{CSP Bandpasses}
\label{bandpasses}
 In the past, photometry of local SNe~Ia has often been obtained with poorly quantified
 system (telescope $+$ instrument $+$ filters) response functions. 
 This problem makes it difficult to precisely combine low- and high-$z$ SN~Ia samples,
 and  contributes to the systematic error budget of cosmological parameters. 
To maximize the potential of our low-$z$ sample, we 
have devised a calibration program to measure with 1\% accuracy 
the response  of the CSP bandpasses on the Swope (optical $+$ near-IR) and du Pont (near-IR) telescopes. 
In this section we describe the photometric calibration program, 
and present the scanned optical response functions.  
The near-IR scanned response functions will be presented in a forthcoming
data paper.
A more complete discussion of the technical aspects of this experiment, 
including a schematic of the setup, 
is presented by \citet{rheault10}.

  \subsection{Experimental Setup}
  
  The experimental setup consists of a broad-band light source that illuminates a monochromator. Two light sources provide sufficient illumination
  in the optical and near-IR regimes. 
  In the optical, a 75 W Xenon lamp  placed in an ellipsoidal reflector gives a smooth continuum 
  from 3000 \AA\  to 8000 \AA, while redward from 8000 \AA\ to 24000 \AA\ a 100 W Quartz  
   Tungsten Halogen lamp is used.   
   An optimized setup of collimating optics directs the light 
   to the entrance slit of a monochromator. 
    The collimating optics  used for 
   projection consist of UV grade fused silica.
    The monochromator selects a narrow bandwidth of light  (between 20 to 50 \AA) 
that is fed into a fiber bundle  made of  10 individual 600 $\mu$m diameter fibers arranged as a vertical line at the  exit slit of the monochromator, while at the end of the fiber line
they are arranged 
as  a compact spot. 
The custom fiber bundle was manufactured by Fibertech Optical using 600 micron 
fibers from Polymicro (model FBP).  
The fibers are made of a special broad-band glass that has high transmission from 3500 \AA\ up
to 18000 \AA. Outside of this wavelength range, the fiber transmission begins to drop, but
there is still enough power to measure the near-IR throughput accurately.
The output of the fiber bundle is 
projected onto a highly reflective flat-field screen using an Engineered Diffuser from RPC Photonics (model EDC--40) that ensures a uniform distribution of  the light.
The  glass optics of the diffuser have a polymer coating to enhance the UV transmission.
One of the fibers is 
routed to a MS125 spectrometer from Newport, which measures in real time the central wavelength and the full width at half-maximum intensity (FWHM) of the illuminating source  to within an accuracy of 1 \AA.
 The flat-field screen is made of 
lightweight honeycomb aluminum panels covered by a highly diffusive coating from Labsphere with reflectivity above 95\% from
3000 \AA\ to 16000 \AA. From 16000 \AA\ to 22000 \AA\ the reflectivity hovers around 90\%, and
beyond 22000--25000 \AA\ the reflectivity drops to 75\%.

Placed behind the secondary mirror of the telescope are four
photodiodes with NIST traceable calibrations  that measure the power of  the light  shining 
off the flat-field screen. 
The photodiodes monitor the screen brightness and provide a signal proportional to the
amount of light that enters the telescope. 
Two types of photodiodes are employed depending 
on the wavelength range being calibrated.
Silicon photodiodes are used between 2500 \AA\ and 10000 \AA, and
Germanium 
photodiodes best serve the purpose from 8000 \AA\ to 16500 \AA.
The signal from the photodiodes is amplified by a high-gain low-noise amplifier and 
read by an analog-to-digital converter acquisition system.
The monochromator, lamps, and photodiodes  are controlled remotely from the 
telescope control room. This setup is placed on the platform at the base of the telescope mount below the flat-field screen for the duration of the calibration run.
 Finally, from our own in lab tests we calibrated both the Silicon 
and Germanium photodiodes using the same NIST traceable standards and found that 
they both agree  in their region of overlap.

\subsection{Calibration Procedure}

The calibration is made by comparing the number of photons measured by the photodiodes
aimed at the flat-field screen to the number of counts measured
during the same period by the science imager mounted on the telescope.
 It is a relative measurement. 
The acquisition system is automated; however, it is unable to
communicate with the instrument control software, so part of the procedure is
accomplished manually.
 
 The optical bandpasses are scanned using a  narrow bandwidth 
 (FWHM $=  26$ \AA)  with  a 20 \AA\ step between each measurement. 
 Near-IR scans are performed with a fine step interval of 50 \AA\ and a coarser 
 step interval of 200 \AA.  
 An image is  taken at each interval step with a  30~s exposure time.
  Due to the low transmission sensitivity in the blue of the SITe3 chip in the 
  direct-camera CCD,
  120~s exposure times are required to accurately scan the $u$ band.
  To evaluate the repeatability of the calibration, each filter is scanned at least twice during non-sequential nights over the course of a calibration run.

\subsection{Calibration Results}
 \label{calibrationresults}

Two calibration runs were conducted at LCO during 2010.
During the first of these runs (January 2010) 
the $uBgVri$ passbands of the SITe3 CCD imager on the Swope  telescope
were measured twice per filter, point-to-point by wavelength,
except for the $i$-band filter which was scanned only once.
In a second run (July 2010) a second scan was obtained of the Swope $i$ bandpass,
and the near-IR bandpasses 
of both RetroCam on the Swope telescope and WIRC on the du Pont were
scanned.   The data from the second run are currently being analyzed and will be presented in a future publication (Stritzinger et al., in prep.).
The repeatability of the $uBgVr$ measurements from the first calibration run
was $\leq$ 1\%; 
by this we mean that there is a $\leq$ 1\% point-to-point (by wavelength) 
uncertainty in the filter-response profiles. 
In other words, the ratios of the two scans of each individual filter 
fall  within an envelope $\pm$ 1\% wide.

 Plotted in Figure~\ref{fig:optfilters} are the $uBgVri$ bandpasses measured 
 from our calibration program multiplied by 
 a CTIO atmospheric transmission function, including telluric 
 absorption features.
  Also included  are the tracings previously published by \citet[][see their Figure 8]{contreras10}.  
In addition to measuring the bandpasses,  the response of the Swope telescope
 was measured with and without the  SITe3 Direct CCD camera. 
 The scanning without the CCD camera  was performed with a photodiode 
 placed at the  focal plane of the telescope. 
 These measurements yielded the optical throughput of the telescope which, when
 combined with the total throughput measurements, allowed the instrument response function (CCD + dewar window) to be deduced.

Under close inspection of the scanned bandpasses it was evident 
that in all cases the wings  did not 
reach zero transmission.  The cause of this
effect was determined to be white light that leaked into the 
calibration setup.
In general the offset produced by the white light 
was reasonably constant across the bandpasses; thus, to produce the 
final corrected curves the same constant value was subtracted from each bandpass trace.

 As a direct result of our measurements, we managed to
 resolve two issues  with 
 the bandpasses  published by \citet{contreras10}.
First, we now have an accurate measurement of the $u$ filter response function
which, as discussed by \citet{contreras10}, was a source of considerable uncertainty.
Second,  on 14 January 2006 UT the $V$ filter
(called ``LC-3014") was damaged and subsequently replaced with another $V$ filter
(``LC-3009"). After several nights of use, it was determined that the replacement 
filter  had a significantly different color term compared to the original.  
This filter was  therefore replaced with a third filter  (``LC-9844") on 25 January 2006 UT, which was used for the remainder of the CSP campaigns. 
During the first photometric calibration run a portion of the broken LC-3014 filter and 
the temporary replacement, LC-3009, were also scanned using the monochromator while
mounted on an optical bench. 
A comparison of these  tracings indicates that the 
 LC-9844 and  LC-3014 $V$-band filters, though similar, are not identical 
   (see Figure~\ref{fig:optfilters}). 
Details regarding the color terms for these three $V$-band filters are provided in 
the Appendix B. 

\section{Final Light Curves}
\label{lightcurves}

Final optical and near-IR light curves of $\nosne$ SNe~Ia in the {\em natural} system 
are displayed in Figure~\ref{fig:flcurves}.
The corresponding optical and near-IR photometry 
of each object is provided in Table~\ref{tab:opt_sne} and Table~\ref{tab:ir_sne}, 
respectively.\footnote[15]{Electronic files in {\tt ASCII} format are  available on the CSP webpage
{\tt  http://obs.carnegiescience.edu/CSP/data .}}$^{,}$\footnote[16]{The small amount of optical  photometry obtained with the du Pont telescope ($+$ Tek 5 CCD) is included in 
Table~~\ref{tab:opt_sne}. These data, which are indicated with superscript ``a" in the table, 
 were reduced assuming that they are in 
the  same natural system as the Swope photometry.
This is a reasonable assumption since the exact same filters were used
with both telescopes, and  with CCD detectors that had similar quantum efficiencies as a function of wavelength.} 
 The quoted uncertainties in these tables correspond to the  sum in quadrature 
of the instrumental error and the nightly zero-point error.
In general, the light curves provide excellent temporal coverage and 
exhibit photometric  precision at the 0.01 to 0.03 mag level. 

The photometry of the SNe with either ``normal" or ``SN 1991T-like" spectral subtypes 
(see Table 2) is overplotted in Figure~\ref{fig:flcurves} with
light-curve fits (solid red lines) computed by SNooPy \citep{burns11}
 using the  ``max model" \citep[see][Eq. 5]{stritzinger10}.  
From the SNooPy fits we obtain estimates of the peak magnitudes, the time of $B$-band maximum, $T(B_{\rm max}$),  the decline-rate parameter $\Delta m_{15}$  \citep{burns11}, and 
a covariance matrix of these parameters.
Currently, SNooPy is unable to 
satisfactorily 
fit  template light curves to peculiar 
 SNe~Ia like SN~2006bt \citep{foley10},  the  fast-declining low-luminosity
 SN~1986G/SN 1991bg-like objects, nor the possible super-Chandrasekhar mass SNe.
 In these cases we interpolated the observed photometry 
 with the use of  spline functions, from which the peak magnitudes 
 may be estimated, and the values of  $T(B_{\rm max}$) and  $\Delta m_{15}$
 were derived directly from the $B$-band spline fit. 
Overplotted on the observed photometry
in Figure~\ref{fig:flcurves} are  the resulting spline fits (red dashed lines). 
Due to insufficient photometric coverage, we
were unable to accurately determine estimates of  $\Delta m_{15}$
for the SN 1991bg-like objects SNe 2006bd and 2006hb, nor for the possible 
 super-Chandrasekhar SN~2007if.

The $\Delta m_{15}$ values derived from either the SNooPy template light-curve fits 
or the spline fits are listed in Table~\ref{tab:sne2},
 and are plotted in a histogram in Figure~\ref{fig:dm15}.
  We note that because a single value of  $\Delta m_{15}$ 
 is used in SNooPy to parameterize all the filters when performing the ``max model'' fit to  a SN~Ia, the values reported here can differ slightly from the value one would obtain by measuring it directly from the decline of just the $B$-band light curve.
 Also included in  Figure~\ref{fig:dm15} are the $\Delta m_{15}$ values for the 
 SNe~Ia presented in the first CSP data release of SNe~Ia \citep{contreras10}.

We find that for the objects presented in this paper, 33  have pre-maximum optical coverage, 
and of these, 15 have observations beginning at least 5 days before $T(B_{\rm max}$). 
Of the 45 SN~Ia with near-IR observations,  27  have
observations that commence  prior to $T(B_{\rm max}$), with a subset of 13   
beginning at least 5 days earlier than maximum $B$-band light. 

\section{SNe 2006bt and 2006ot}
\label{sne06bt_06ot}

In a recent paper, \citet{foley10} called attention to one of the objects 
in our sample, SN~2006bt, which showed photometric and spectroscopic 
characteristics unlike those of any other known SN~Ia.  While possessing
a broad, slow-declining ($\dm = 1.1$) $B$ light curve similar to a typical
luminous SN~Ia, SN~2006bt also displayed certain photometric and spectroscopic 
properties that are more similar to those of fast-declining, low-luminosity
events.  In particular, the $i$-band light curve showed only a weak 
secondary maximum, and the maximum and pre-maximum optical spectra displayed
strong \ion{Si}{2}~$\lambda$5972 absorption relative to 
\ion{Si}{2}~$\lambda$6355 and a depression at $\sim 4100$~\AA\ due to 
\ion{Ti}{2}, all of which are properties characteristic of low-luminosity 
SN~1991bg-like objects.  As mentioned earlier, our attempts to fit the
light curves of SN~2006bt with SNooPy met with failure because of the 
combination of slow-declining light curves with a weak secondary maximum in
the $i$ band.  The $i$ light curve of SN~2006bt is also peculiar in 
reaching a primary maximum at essentially the same epoch as $B$ maximum,
whereas for normal slow-declining SNe~Ia, $i$ maximum occurs several days
before $B$ maximum.  Nevertheless, \citet{foley10} showed that the luminosity 
of SN~2006bt was quite typical of normal, slow-declining SNe~Ia.

Interestingly, a second object in our sample, SN~2006ot, appears to be closely 
related to SN~2006bt.  This is illustrated in Figure~\ref{fig:06otphot} where
the $uBgVriYJH$ light curves of both SNe are overplotted.  Also included in 
this figure are SNooPy fits to the CSP photometry of the normal, 
slow-declining ($\dm = 1.0$) SN~2006af \citep{contreras10}.  Once again, in
the observations of SN~2006ot we find the combination of broad, slow-declining 
light curves with a weak secondary maximum in $i$.  The morphology of the 
$Y$ and $J$ light curves of SN~2006ot is also peculiarly flat and consistent 
with the more limited CSP coverage obtained for SN~2006bt.  
Figure~\ref{fig:06otspec} shows a comparison of the spectra of SNe~2006bt and 
2006ot near maximum light, and approximately 3--4~weeks after maximum.  Again, 
spectra of the normal SN~2006ax are included for comparison.  The peculiar
nature of the maximum-light spectrum of SN~2006bt compared to SN~2006ax is 
apparent in this figure, in particular the strength of the 
\ion{Si}{2}~$\lambda$5972 absorption compared to $\lambda$6355, and the 
broad \ion{Ti}{2} absorption at $\sim 4100$~\AA.  The maximum-light spectrum 
of SN~2006ot is different yet again, appearing in some ways intermediate 
between the spectra of SNe~2006bt and 2006ax (e.g., in the ratio of the
equivalent widths of the \ion{Si}{2}~$\lambda$5972 and $\lambda$6355 
features).  Most striking is the strength, width, and blueshift of the 
\ion{Si}{2}~$\lambda$6355 line, the minimum of which yields an expansion 
velocity of $\sim$14500~km~s$^{-1}$.  At 3--4~weeks past maximum, the
spectra of SNe~2006bt and 2006ot closely resemble each other, while the
spectrum of the normal SN~2006ax differs significantly at wavelengths
shortward of 6000~\AA.

According to the maps of \citet{schlegel98}, interstellar reddening of 
SNe~2006bt and 2006ot due to dust in the Milky Way was small (A$_V$ = 0.17~mag
and A$_V$ = 0.05~mag, respectively).  \citet{foley10} did not find evidence 
for \ion{Na}{1}~D absorption in the spectrum of SN~2006ot at the redshift of 
its host galaxy, nor is \ion{Na}{1}~D detectable in our maximum-light 
spectrum of SN~2006ot.  Hence, it is unlikely that the light of either SN 
suffered large dust extinction.  The difference in host galaxy recession
velocities corrected to the cosmic microwave background (CMB) reference frame
(v$_{CMB}$ = 9737  km s$^{-1}$ for SN~2006bt and v$_{CMB}$ = 15695 km s$^{-1}$
for SN~2006ot) implies a 
brightness difference of $\sim$1.0~mag if both objects were unreddened and
shared the same peak luminosity.  This is only 0.1~mag different from the 
shift of $\sim$0.9~mag required to register the light curves of both SNe (see 
Figure~\ref{fig:06otphot}).  Thus, the peak luminosities were likely quite 
similar.  Interestingly, the host of SN~2006bt was an S0/a galaxy and NED 
lists the morphology of the host of SN~2006ot as Sa, implying a relatively
old progenitor \citep{foley10}.

It is not at all obvious how SNe~2006bt and 2006ot fit into the overall 
scheme of SNe~Ia, but it is clear that they can be easily confused
with normal events unless high-quality photometry and/or spectroscopy is
obtained.  This is particularly true at high redshifts where rest system
$i$ or near-IR light curves (where the photometric differences are greatest) are
almost never obtained.  \citet{foley10} concluded that contamination of 
SNe~Ia samples by similar objects will both increase the scatter of the 
Hubble diagram and systematically bias measurements of cosmological 
parameters.  Fortunately, they do not appear to be very common in the 
local Universe as only two such events are among the 87 SNe~Ia with photometry 
published to date by the CSP.

\section{Conclusion, and Prospects for the Future}
\label{conclusion}

We have presented the second data release of SN~Ia photometry obtained 
by the CSP.  Combined with the first sample 
 \citep{contreras10}, the CSP dataset currently consists of 85 objects.
 The high quality and dense temporal coverage, extending over 9 filters
 from the optical to the near-IR,
 provides a sample that is well-suited for tackling the issues related
 to accurately estimating host-galaxy reddening \citep[e.g.,][]{folatelli10}.  
  
 With the addition of the 50 low-$z$ objects presented in this paper, 
 we are in a position to extend upon  our first analysis and, 
 among other things, to provide a detailed investigation of the fastest declining 
 SNe~Ia. The CSP sample now consists of  16 rapid decliners
 ($\Delta$$m_{15} > 1.5$). 
  This sample is nearly evenly split between the morphologies of the near-IR maximum of those objects that peak in the near-IR before 
  $T(B_{\rm max}$), and those that peak after. 
\citet{krisciunas09} suggest that objects whose near-IR light curves peak after 
   $T(B_{\rm max}$) appear to be 0.5 to 0.8 mag less luminous in the optical and near-IR than those events whose 
   near-IR light curves peak before  $T(B_{\rm max}$). It will be particularly interesting to
   re-examine this issue using the CSP sample,
    as well as to compare other properties of the light curves and spectra of this 
  SN~Ia subtype.

When completely reduced and analyzed, the CSP optical and near-IR light curves of SNe~Ia already in hand will constitute a fundamental reference for cosmological studies. However, at present, the full utility of the CSP sample as a reference for observations at high redshift is compromised by the fact that the median redshift of the sample is $\sim 0.02$. 
At this redshift, the root-mean-square (rms) error in distance modulus due to peculiar velocities is $\pm$0.13 mag, and dominates the observed Hubble diagram dispersion 
\citep{neill07,folatelli10}.
 This problem can be minimized by extending observations to SNe further out in the smooth Hubble flow. We have therefore proposed and been granted NSF funding to carry out a second stage of the CSP with the goal of producing optical and near-IR light curves of 100--200 SNe~Ia in the redshift range $0.03 < z < 0.08$. With such a sample 
 we expect the  rms error due to peculiar motions to drop to $\pm$ 3\% or less 
 in distance.

\acknowledgments 

The CSP extends special thanks to the mountain staff of the Las Campanas Observatory for 
their assistance throughout the duration of our observational program, and to Jim Hughes for his superb support of our network of computers.
This material is based upon work supported by NSF under 
grants AST--0306969, AST--0908886, AST--0607438, and AST--1008343. 
 The Dark Cosmology Centre is funded by the Danish National Research Foundation. 
M.H. acknowledges support by CONICYT through grants FONDECYT Regular
1060808, Centro de Astrofisica FONDAP 15010003, Centro BASAL CATA
 (PFB--06), and by the Millennium Center for Supernova Science (P06--045-F).
A.V.F. is grateful for the financial support of the NSF and the TABASGO Foundation.
This research has made use of the NASA/IPAC Extragalactic Database (NED) which is operated by the Jet Propulsion Laboratory, California Institute of Technology, under contract with the National Aeronautics and Space Administration. 

\appendix
\section{Photometric Zero Points}

With the scanned system response functions presented in 
$\S$~\ref{calibrationresults}, we proceed to compute photometric 
zero points following the prescription described by 
\citet[][see their Appendix C]{contreras10}. 

The photometric zero point for bandpass $X$, with a response function
denoted by $S_{X}(\lambda)$, can be expressed as 
\begin{equation}
\label{eqn:zp}
ZP_{X}=2.5\log\left[\int_0^{\infty}
  S_{X}\left(\lambda\right)\, F_{o}\left(\lambda\right) \frac{\lambda}{hc}\, d\lambda\right]+m_{0}.
\end{equation}
\noindent Here $F_{o}(\lambda)$ is the spectral energy distribution (SED)
of a fiducial source and $m_{o}$ is the
observed magnitude of this source through the bandpass $X$. 

To compute the $B$- and $V$-band zero points we adopt 
the SED of Vega ($\alpha$ Lyr)  as our fiducial source. 
In doing so, we use the  \citet{bohlin04} SED of 
Vega (version 5)  obtained from 
 CALSPEC\footnote[17]{ftp://ftp.stsci.edu/cdbs/current\_calspec/alpha\_lyr\_stis\_005.ascii .},
 and the  \citet{fukugita96} standard magnitudes $B_{\rm std}$ $= 0.03$ 
 and $V_{\rm std}$$= 0.03$.  
As we aim to compute zero points in the {\em natural} system 
 of the Swope telescope,  $B_{\rm std}$  and  $V_{\rm std}$ are  transformed to the Swope natural system using the following equations:

\begin{equation}
B_{\rm nat}  = B_{\rm std} - CT_{B} \times (B_{\rm std} - V_{\rm std}),
\end{equation}
\begin{equation}
\label{Vnat}
V_{\rm nat}  = V_{\rm std} - CT_{V} \times (V_{\rm std} - i_{\rm std}),
\end{equation}

\noindent 
where CT$_{X}$ are the color terms.  Color terms for the CSP bandpasses were given 
by \citet{hamuy06} and \citet{contreras10}; the former values correspond to
the average for the first of five yearly observing
campaigns, while the slightly different numbers given in the latter reference 
are based on the first four observing campaigns.
In Table~6 the final five-year average color terms are presented.
 The first column corresponds to the 
 unweighted average after applying a 5$\sigma$  clipping algorithm. 
 Listed in  the second column as a consistency check are the median values calculated 
 without the 5$\sigma$ clipping, with uncertainties corresponding to 1.49 multiplied by  the median absolute deviation.
 As is seen, the differences between these two methods of computation are minimal. 
 We therefore adopt the unweighted averages determined with the 5$\sigma$  clipping.
 
 In Equation~\ref{Vnat}  the value of 
 ``$i_{\rm std}$" was computed synthetically using the
 SED of Vega and the ``USNO 40" $i'$ bandpass. This yields $i'_{\rm std}$ $= 0.382$ mag. 
Executing  the transformations yields Vega magnitudes in the Swope {\em natural} system of $B_{\rm nat} = 0.03$ and $V_{\rm nat} = 0.009$.
  
Armed with $B_{\rm nat}$ and $V_{\rm nat}$,  and
the $B$ and $V$ (LC-9844) scanned bandpasses, zero points were computed directly 
using Equation~\ref{eqn:zp}. In addition, zero points for the temporary (LC-3009)  and replacement (LC-3014)  $V$-band filters discussed in $\S$~\ref{calibrationresults}
were also computed, and the results are  listed in  Table~7. 
 
We note that due  to the limited use of the temporary  LC-3009 $V$ filter, we 
 are unable to accurately determine the color term of this filter from observations of 
 standard-star fields. 
To remedy  this problem, the  color term was estimated  synthetically using 
 an atlas of \citet{landolt92} spectrophotometic standards \citep{stritzinger05},
 and the corresponding known $B_{\rm std}$  and  $V_{\rm std}$ magnitudes 
 (see Appendix~\ref{appcolor} for further details).
 
To compute the zero points of the  scanned $ugri$ bandpasses,  
BD$+$17$^{\circ}$4708 was chosen as the fiducial source.
Similar to the steps described above, the standard 
$u_{\rm std}=10.56$, $g_{\rm std}=9.64$, $r_{\rm std}=9.35$ and $i_{\rm std}=9.25$
magnitudes of   BD$+$17$^{\circ}$4708 \citep{fukugita96} were transformed to the 
Swope natural system using the color term in Table~6 and 
the following transformation equations:

\begin{equation}
u_{\rm nat}  = u_{\rm std} - CT_{u} \times (u_{\rm std} - g_{\rm std}),
\end{equation}
\begin{equation}
g_{\rm nat}  = g_{\rm std} - CT_{g} \times (g_{\rm std} - r_{\rm std}),
\end{equation}
\begin{equation}
r_{\rm nat}  = r_{\rm std} - CT_{r} \times (r_{\rm std} - i_{\rm std}),
\end{equation}
\begin{equation}
i_{\rm nat}  = i_{\rm std} - CT_{i} \times (r_{\rm std} - i_{\rm std}).
\end{equation}

\noindent These transformed magnitudes were then compared to synthetic magnitudes
computed with our scanned bandpasses and the 
\citet{bohlin04} SED of  BD$+$17$^{\circ}$4708 to compute the zero points via 
Equation~\ref{eqn:zp}; the results  are listed in Table~7. 


\section{$B$- and $V$-Band Color Terms}
\label{appcolor}
To test the accuracy of  the scanned $B$ and $V$ bandpasses, ``synthetic" color terms were computed by comparing synthetic photometry derived using the  
 \citet{stritzinger05} set of \citet{landolt92} spectrophotometric standards to the  
 published Landolt values. 
 If the scanned bandpasses are correct, the synthetic color terms derived from this 
 experiment should match the observed values reported in Table~6.

Plotted in the left panel of Figure~\ref{fig:CTs} are the differences between 
the synthetic $B$-band magnitudes computed with the scanned bandpass 
 and the published Landolt values vs.  synthetic  $B - V$ color.
 Applying a color cut to exclude  stars bluer than $B-V = 0$ or redder than 
 $B-V = 0.9$ mag yields the best-fit color term of $CT_{B} = +0.065$.
 This is reasonably consistent with the value of  $CT_{B} = +0.069$ given in Table~6.
 
 Presented in  the right panel of  Figure~\ref{fig:CTs} is a similar comparison 
for the three $V$ bandpasses  used by the CSP. 
The blue points correspond to values derived with the scan of the broken 
LC-3014 filter, while the red points are from the scan of the temporary 
 LC-3009 filter, and the green points are the values obtained with the scan of the replacement 
 filter  LC-9844. 
Adopting the previously mentioned color cut yields best-fit 
color terms  of  $CT_{V} =  -0.063$,   $CT_{V} =  -0.044$, and $CT_{V} =  -0.065$ for the  LC-3014, LC-3009, and LC-9844 bandpasses, respectively.  
These values are in excellent agreement with the color terms derived from the broad-band 
photometry of the Landolt standard-star observations (see Table~6). 
 
 The results of this experiment confirm the accuracy of the scanned $B$ and $V$ bandpasses.
Unfortunately, due to a lack of spectrophotometric measurements of the Smith et al. (2002) standard 
stars, it is not possible to accurately apply this test to the scanned $ugri$ bandpasses. 

\section{V-Band LC-3009 Photometry}

As the $V$-band LC-3009 color term has  a non-negligible difference 
 with respect to the two other $V$-band filters used,  
we re-examined the small amount of photometry obtained with LC-3009.
During the time the LC-3009 filter was used, 12 different SNe~Ia were observed 
during   1--3 epochs. Nine of these objects are included in \citet{contreras10}.
For completeness we returned to these nights in question and recomputed 
the photometry in the natural system of the Swope telescope.  We note that for these objects,
eliminating the observations taken with the  LC-3009 filter has little effect 
on the quality or coverage of the $V$-band light 
curves,  except for the case of SN~2006D whose  first three pre-maximum 
epochs were obtained with LC-3009.
Updated $V$-band photometry for these three epochs is  
available on the CSP webpage.

\clearpage

\newpage
\begin{deluxetable} {lcclllll}
\tabletypesize{\tiny}
\tablecolumns{15}
\tablewidth{0pt}
\tablecaption{General Properties of $\nosne$ Type Ia Supernovae \label{tab:snproperties}}
\tablehead{%
\colhead{} & 
\colhead{SN}   & 
\colhead{SN}   &
\colhead{Host} & 
\colhead{}   & 
\colhead{}   & 
\colhead{Discovery} &
\colhead{}   \\
\colhead{SN} &
\colhead{$\alpha (2000)$} & 
\colhead{$\delta (2000)$} & 
\colhead{Galaxy} &
\colhead{Morphology\tablenotemark{a}} & 
\colhead{$z_{\rm helio}$\tablenotemark{a}} & 
\colhead{Reference} &
\colhead{Discovery Individual/Group\tablenotemark{b}}
}
\startdata
2005hj & 01:26:48.27 &$-$01:14:16.8&  B012415.21$-$012950.3     & $\cdots$    & 0.0580$^{\dag}$ & CBET 266  & ROTSE-III \\ 
2005ku & 22:59:42.61 &$+$00:00:49.3&  2MASX J22594265$-$0000478 & $\cdots$    & 0.0454$^{\dag}$ & CBET 304  & SDSS-II \\
2005mc & 08:27:06.36 &$+$21:38:45.6&  UGC 04414                  & S0a        & 0.0252   & CBET 331  & THCA    \\
2006bd & 11:38:28.46 &$+$20:31:34.4&  UGC 6609                  & E           & 0.0257   & CBET 448  & Puckett \\
2006br & 13:30:01.80 &$+$13:24:56.8&  NGC 5185                  & Sb          & 0.0246   & CBET 482  & Puckett \\
2006bt & 15:56:30.53 &$+$20:02:45.3&  CGCG 108$-$013            & SA0/a       & 0.0322   & CBET 485  & LOSS    \\
2006ef & 02:04:19.51 &$-$08:43:42.2&  NGC 809                   & (R)S0       & 0.0179   & CBET 597  & LOSS    \\
2006ej & 00:38:59.77 &$-$09 00 56.6&  NGC 191A                  & S0 pec sp   & 0.0205   & CBET 603  & LOSS    \\
\enddata
\tablenotetext{a}{Host-galaxy morphology and heliocentric redshift are
  from the NASA/IPAC Extragalactic Database (NED) unless indicated.
Redshifts marked with a $\dag$ were computed via
host spectra obtained by the CSP, while $\ddagger$ and $\mp$ are 
from the SDSS-II SN survey and \citet{scalzo10}, respectively.}
  
\tablenotetext{b}{References/URLs: 
CHASE (http://www.das.uchile.cl/proyectoCHASE/);
CSS (http://www.lpl.arizona.edu/css/);
LOSS (see Filippenko et al. 2001, Filippenko 2005, and references therein); 
Perth \citep{williams97};
ROTSE-III \citep{quimby06}; Puckett (http://www.cometwatch.com);
Tenagra-II (http://www.tenagraobservatories.com/Discoveries.htm); 
SDSS-II \citep{frieman08}.}
 
\end{deluxetable}

\clearpage
\begin{deluxetable} {lccccccl}
\tabletypesize{\tiny}
\tablecolumns{15}
\tablewidth{0pt}
\tablecaption{Photometric/Spectroscopic Properties of $\nosne$ SN~Ia \label{tab:sne2}}
\tablehead{
\colhead{} & 
\colhead{$\Delta$m$_{15}$\tablenotemark{a}} & 
\colhead{Opt.} &  
\colhead{NIR.} &
\colhead{Opt. Std.} & 
\colhead{NIR. Std.} &
\colhead{Spectral}&
\colhead{$T(B)_{\rm max}$}\\
\colhead{SN} &
\colhead{(mag)} &
\colhead{Epochs} &
\colhead{Epochs} &
\colhead{Epochs} &
\colhead{Epochs} &
\colhead{Subtype}&
\colhead{Days\tablenotemark{b}}}
\startdata
2005hj & 0.787(003)& 17   & 12       & 4 & 5        & normal      & $\cdots$ \\ 
2005ku & 1.019(060)& 7    & 4        & 3 & 3        & normal      & $-1$     \\ 
2005mc & 1.533(005)& 17   & $\cdots$ & 3 & $\cdots$ & normal      & $\cdots$ \\ 
2006bd & $\cdots$  & 10   & 4        & 2 & 4        & 91bg-like   & $+7$     \\
2006br & 1.160(034)& 9    & 5        & 3   & 3        & normal      & 0  \\  
2006bt & 0.979(017)& 17   & 5        & 4 & 3        & peculiar    & $\cdots$ \\
2006ef & 1.326(015)& 10   & $\cdots$ & 3 & $\cdots$ & normal      & $+37$  \\ 
2006ej & 1.344(012)& 13   & 3        & 3 & 5        & normal      & $+30$  \\
\enddata
\tablenotetext{a}{Values of $\Delta m_{15}$ for the normal and SN~1991T-like objects were 
determined using the ``max" model of SNooPy \citep[see][Eq.~5]{stritzinger10}. 
In the case of the low luminosity SN~1991bg and SN~1986G-like objects, the possible 
super-Chandrasekhar mass object SN~2009dc, and the peculiar SN~2006bt,
estimates of $\Delta m_{15}$ were obtained from direct spline fits to the $B$-band light curve. 
Uncertainties are given in parentheses in millimagnitude.}
\tablenotetext{b}{Epoch of the spectrum (with respect to time of $B$-band maximum) used to determine spectral subtype. If no epoch is given, the spectral classification was taken from the 
literature: SN~2005hj \citep{quimby07}, SN~2005mc \citep{deng05}, SN~2006bt \citep{foley10},
SN2007hx \citep{bassett07a}, and in the case of SN~2007mm from a combination of 
\citet{bassett07b} and our own anaylsis of an SDSS-II spectrum.}

\end{deluxetable}

\clearpage
\begin{deluxetable} {lllcccccccccc}
\rotate
\tabletypesize{\tiny}
\tablecolumns{13}
\tablewidth{0pt}
\tablecaption{Optical and near-infrared photometry of the local sequences in the standard system\label{tab:optir_stds}}
\tablehead{
\colhead{} &
\colhead{}  &
\colhead{}  &
\colhead{$u'$}   &
\colhead{$g'$}   &
\colhead{$r'$}   &
\colhead{$i'$}   &
\colhead{$B$}   &
\colhead{$V$}  &
\colhead{$Y$}   &
\colhead{$J$}   &
\colhead{$H$}  &
\colhead{$K_s$}\\
\colhead{STAR} &
\colhead{$\alpha(2000)$}  &
\colhead{$\delta(2000)$}  &
\colhead{(mag)}   &
\colhead{(mag)}   &
\colhead{(mag)}   &
\colhead{(mag)}   &
\colhead{(mag)}   &
\colhead{(mag)}  &
\colhead{(mag)}   &
\colhead{(mag)}   &
\colhead{(mag)}  &
\colhead{(mag)}}
\startdata
\multicolumn{13}{c}{\bf SN 2005hj}\\
01 & 01:26:51.61 & $-$01:16:32.78 &  17.153(011)&  15.821(009)	&  15.345(009)	&  15.197(009)	&  16.184(008)	&  15.522(007)	&  14.540(007)	&  14.291(007)	&  13.958(008)	& $ \cdots $	\\
02 & 01:26:47.87 & $-$01:16:36.35 &  18.084(016)&  16.286(007)	&  15.659(007)	&  15.450(007)	&  16.727(008)	&  15.907(007)	&  14.700(007)	&  14.407(007)	&  13.996(006)	& $ \cdots $	\\
03 & 01:26:38.12 & $-$01:12:08.18 &  19.757(064)&  17.068(007)	&  15.730(009)	&  14.990(009)	&  17.752(012)	&  16.315(008)	& $ \cdots $	& $ \cdots $	& $ \cdots $	& $ \cdots $	\\
04 & 01:26:53.82 & $-$01:11:25.42 &  20.096(101)&  17.370(009)	&  15.987(011)	&  14.521(011)	&  18.211(015)	&  16.568(017)	&  12.906(008)	&  12.418(008)	&  11.870(008)	& $ \cdots $	\\
05 & 01:26:55.70 & $-$01:15:04.82 &  17.673(011)&  16.534(007)	&  16.140(007)	&  16.014(007)	&  16.855(008)	&  16.288(007)	&  15.391(007)	&  15.179(006)	&  14.875(009)	& $ \cdots $	\\
06 & 01:26:58.24 & $-$01:14:24.36 &  20.624(124)&  17.836(008)	&  16.699(007)	&  16.265(007)	&  18.446(013)	&  17.205(007)	&  15.319(006)	&  14.915(007)	&  14.314(006)	& $ \cdots $	\\
07 & 01:26:51.35 & $-$01:13:09.98 &  19.066(035)&  17.542(007)	&  16.919(007)	&  16.657(007)	&  17.974(009)	&  17.173(007)	&  15.880(008)	&  15.562(011)	&  15.092(016)	& $ \cdots $	\\
\enddata
\tablecomments{Uncertainties given in parentheses in thousandths of a
  magnitude correspond to an rms of the magnitudes obtained on
  photometric nights.}
\end{deluxetable}
\clearpage
\begin{deluxetable} {lcccccccccccc}
\setlength{\tabcolsep}{0.02in}
\tabletypesize{\scriptsize}
\tablecolumns{7}
\tablewidth{0pt}
\tablecaption{Optical Photometry of 50 SNe~Ia in the Swope Natural System\label{tab:opt_sne}}
\tablehead{
\colhead{JD} &
\colhead{$u$ }  &
\colhead{$g$ }   &
\colhead{$r$ }   &
\colhead{$i$ }   &
\colhead{$B$ }   &
\colhead{$V$ }  \\
\colhead{2453000+} &
\colhead{(mag)}  &
\colhead{(mag)}  &
\colhead{(mag)}  &
\colhead{(mag)}  &
\colhead{(mag)} &
\colhead{(mag)}   }
\startdata
\multicolumn{7}{c}{\bf SN 2005hj}\\ 
 671.66& 18.137(025)& 17.737(008)& 17.839(010)& 18.300(014)& 17.890(008)& 17.804(008)\\
 673.64& 18.157(025)& 17.696(008)& 17.743(012)& 18.351(014)& 17.849(010)& 17.764(008)\\
 675.62& 18.274(024)& 17.743(010)& 17.734(014)& 18.390(016)& 17.879(008)& 17.802(014)\\
 678.70$\tablenotemark{a}$& $ \cdots $ & $ \cdots $ & $ \cdots $ & $ \cdots $ & 17.920(008)& 17.820(008)\\ 
 682.67& 18.888(070)& 17.952(012)& 17.862(014)& 18.522(021)& 18.220(021)& 17.950(022)\\
 683.64& 19.104(043)& 18.014(014)& 17.928(014)& 18.521(021)& 18.227(012)& 18.000(012)\\
 690.67& 20.586(452)& 18.536(028)& 18.242(025)& 19.102(040)& 18.671(048)& 18.285(032)\\
\enddata
\tablenotetext{a}{Images obtained with the du Pont telescope.}
\tablecomments{Uncertainties given in parentheses in thousandths of a magnitude.}
\end{deluxetable}

\clearpage
\begin{deluxetable} {lccccc}
\tabletypesize{\scriptsize}
\tablecolumns{6}
\tablewidth{0pt}
\tablecaption{Near-infrared Photometry of 45 SNe~Ia \label{tab:ir_sne}}
\tablehead{
\colhead{JD}  &
\colhead{$Y$}   &
\colhead{$J$}   &
\colhead{$H$}   &
\colhead{$K_s$}   &
\colhead{}   \\
\colhead{2453000+}  &
\colhead{(mag)}   &
\colhead{(mag)}   &
\colhead{(mag)}   &
\colhead{(mag)}   &
\colhead{Telescope\tablenotemark{a}}   }
\startdata
\multicolumn{6}{c}{\bf SN 2005hj}\\ 
 672.60& 18.349(072)& 18.081(095)& 18.102(180)& $ \cdots $ &SWO \\
 674.65& 18.357(060)& 18.239(097)& 18.801(301)& $ \cdots $ &SWO \\
 676.60& 18.432(077)& 18.322(119)& 18.653(258)& $ \cdots $ &SWO \\
 677.66& 18.575(076)& 18.265(101)& 19.016(387)& $ \cdots $ &SWO \\
 680.69& 18.716(102)& 18.241(129)& 18.553(407)& $ \cdots $ &SWO \\
 681.70& 18.736(073)& 18.482(070)& 18.650(142)& $ \cdots $ &DUP \\ 
 686.73& 18.731(111)& 18.908(251)& 18.334(318)& $ \cdots $ &SWO \\
\enddata
\tablecomments{Uncertainties given in parentheses in thousandths of a
  magnitude.}
\tablenotetext{a}{DUP and SWO correspond to the du Pont and Swope telescopes, respectively.}
\end{deluxetable}

\clearpage
\begin{deluxetable} {ccc}
\tabletypesize{\normalsize}
\tablecolumns{3}
\tablewidth{0pt}
\tablecaption{Optical color terms for CSP bandpasses in the Landolt System\label{tab:cts}}
\tablehead{
\colhead{Filter} & 
\colhead{Method1} &
\colhead{Method2}}
\startdata
$u$            & $+$0.050(002) & $+$0.049(027) \\
$g$            & $-$0.014(001) & $-$0.015(013) \\
$r$            & $-$0.016(001) & $-$0.016(016) \\
$i$            & $+$0.000(001) & $+$0.000(015) \\
$B$            & $+$0.069(001) & $+$0.069(013) \\
$V$ LC$-$3014  & $-$0.059(002) & $-$0.060(015) \\
$V$ LC$-$3009  & $-$0.034(006) & $-$0.034(009) \\
$V$ LC$-$9844  & $-$0.063(001) & $-$0.065(010) \\
\enddata
\tablecomments{Uncertainties given in parentheses in thousandths of a
  magnitude.}
\end{deluxetable}

\clearpage
\begin{deluxetable} {cc}
\tabletypesize{\normalsize}
\tablecolumns{2}
\tablewidth{0pt}
\tablecaption{Photometric Zero Points of CSP Optical Bandpasses\label{tab:zpts}}
\tablehead{
\colhead{Filter} & 
\colhead{Zero Point}}
\startdata
$u$ & 13.044   \\
$g$ & 15.135    \\
$r$ & 14.915    \\
$i$ & 14.781    \\
$B$ & 14.344\\
$V$ LC$-$3014  & 14.540       \\
$V$ LC$-$3009  & 14.493      \\
$V$ LC$-$9844  & 14.450\\
\enddata
\end{deluxetable}

\clearpage

\figcaption[]{A mosaic of $V$-band CCD images of 50 SNe~Ia observed by the CSP.
 \label{fig:fcharts}}   

\figcaption[]{ Results obtained from the photometric calibration of the Swope telescope
equipped with CSP optical bandpasses. Top panel shows the throughput of the telescope 
with (dashed line) and without (solid line)  the CCD camera. 
Both tracings have been normalized so that they cross  at 6500 \AA. 
Below are the scanned (solid lines) $ugri$ and $BV$
bandpasses compared to those published by \citet{contreras10}.
Also included are the LC-3014  and LC-3009 $V$ bandpasses (see text for details).
 To facilitate comparison the throughput of each bandpass has been 
 normalized to unity. Note that all  bandpasses 
 have been multiplied by an atmospheric transmission function including 
 telluric absorption features.
  \label{fig:optfilters}}

\figcaption[]{Optical ($uBgVri$) and near-IR ($YJHK_s$) light curves of $\nosne$ SNe~Ia in the 
  natural system of the Swope and du Pont telescopes.  Uncertainties in the photometry are smaller
  than the points, unless shown.  The smooth red curves correspond to  the 
  best template fits derived with SNooPy, while the  dashed red curves  correspond to
   spline fits (see $\S$~\ref{lightcurves} for details). 
  \label{fig:flcurves}}

\figcaption[]{Comparison of light curves of  the normal Type~Ia SN 2006ax (solid line) to 
those of the peculiar SNe~2006bt (circles) and 2006ot (filled squares).
 \label{fig:06otphot}}   

\figcaption[]{Spectral comparisons of the normal Type~Ia SN 2006ax (red dashed line) to those of the peculiar SNe~2006ot (black solid line) and 2006bt (blue dotted line). The spectra of SN~2006bt are reproduced from \citet{foley10}, and the $-$1 day spectrum of SN~2006ot is from Silverman et al. 2011 (in preparation).  All other spectra were obtained by the CSP, and will be published in Folatelli et al. 2011 (in preparation).
 \label{fig:06otspec}}

\figcaption[]{Stacked bar chart of  the $\dm$ values for 47 of the 50  SNe~Ia presented in this paper (red)
and 34 of the 35 objects (blue) from the first CSP SN~Ia data release \citep{contreras10}.  
The  $\dm$ values were computed with the ``max model," or via spline fits in the case of the 11 non-standard objects
 (see Table~\ref{tab:sne2}). 
The current CSP sample covers the full range of decline rates, with the distribution peaking between
$\dm$ $=$ 1.1 and 1.2.
 \label{fig:dm15}}   

\figcaption[]{Derivations of $B$- and $V$-band color terms derived synthetically from 
the scanned bandpasses and the \citet{stritzinger05} atlas of Landolt spectrophotometric 
standards. The best linear fits are indicated with lines. Prior to determining the best linear fits, a color cut was applied to exclude those stars with $B - V$ values bluer than 0 and 
redder than 0.9 mag. This left a subsample of 44 standards in which the best fits were
determined. 
 \label{fig:CTs}}

\clearpage

\begin{figure}[t]
\epsscale{.54}
\plottwo{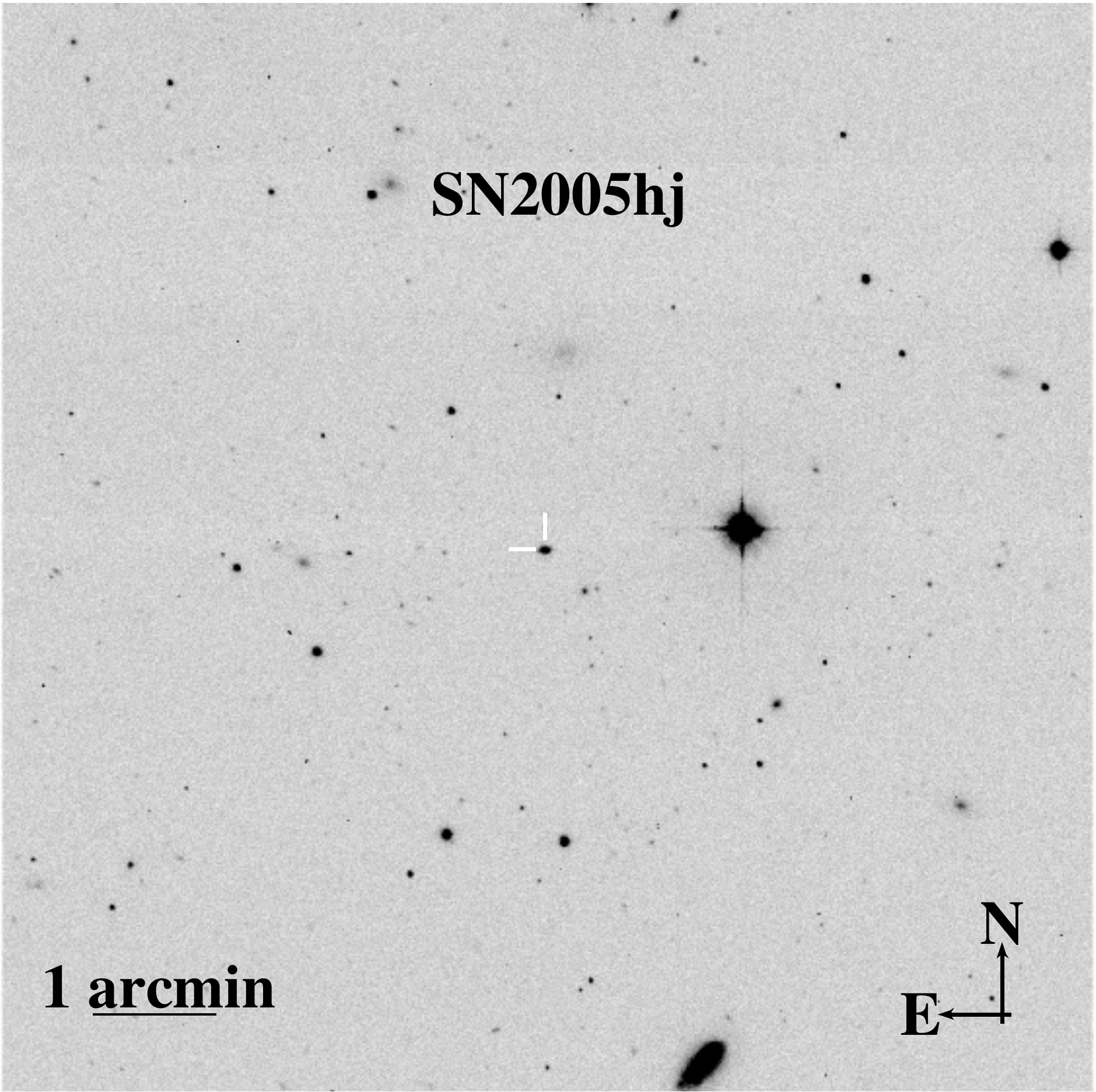}{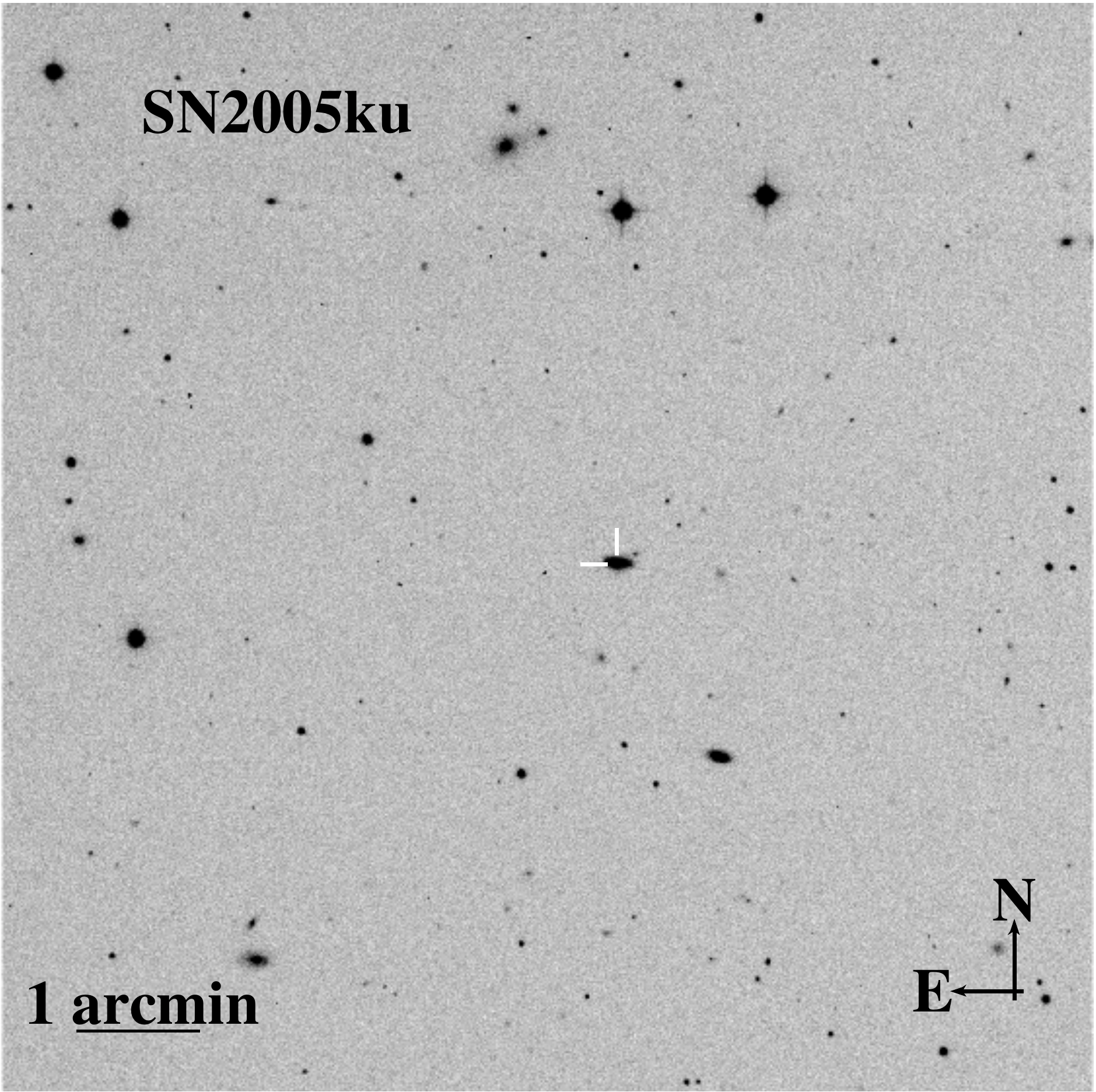}
\plottwo{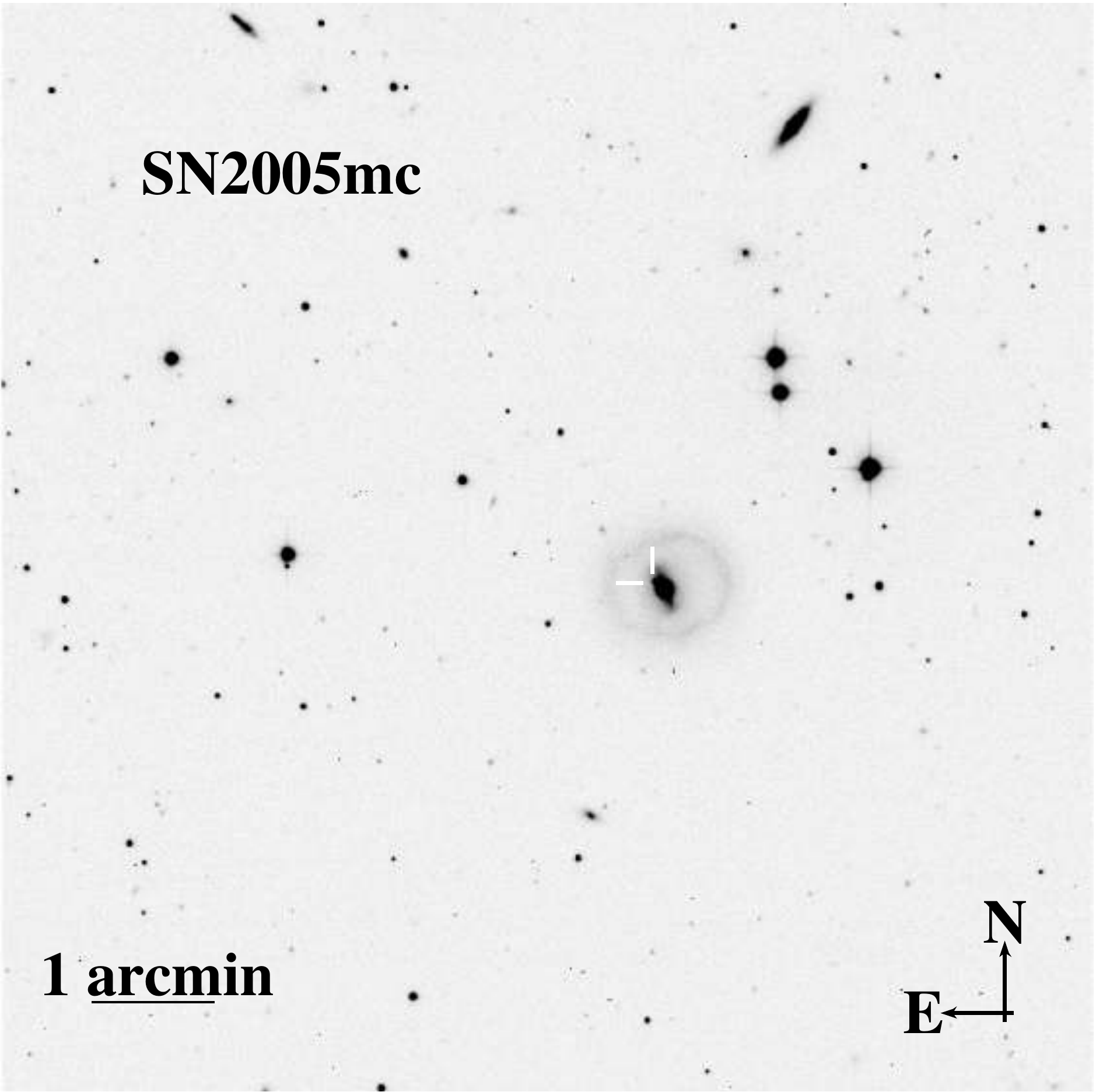}{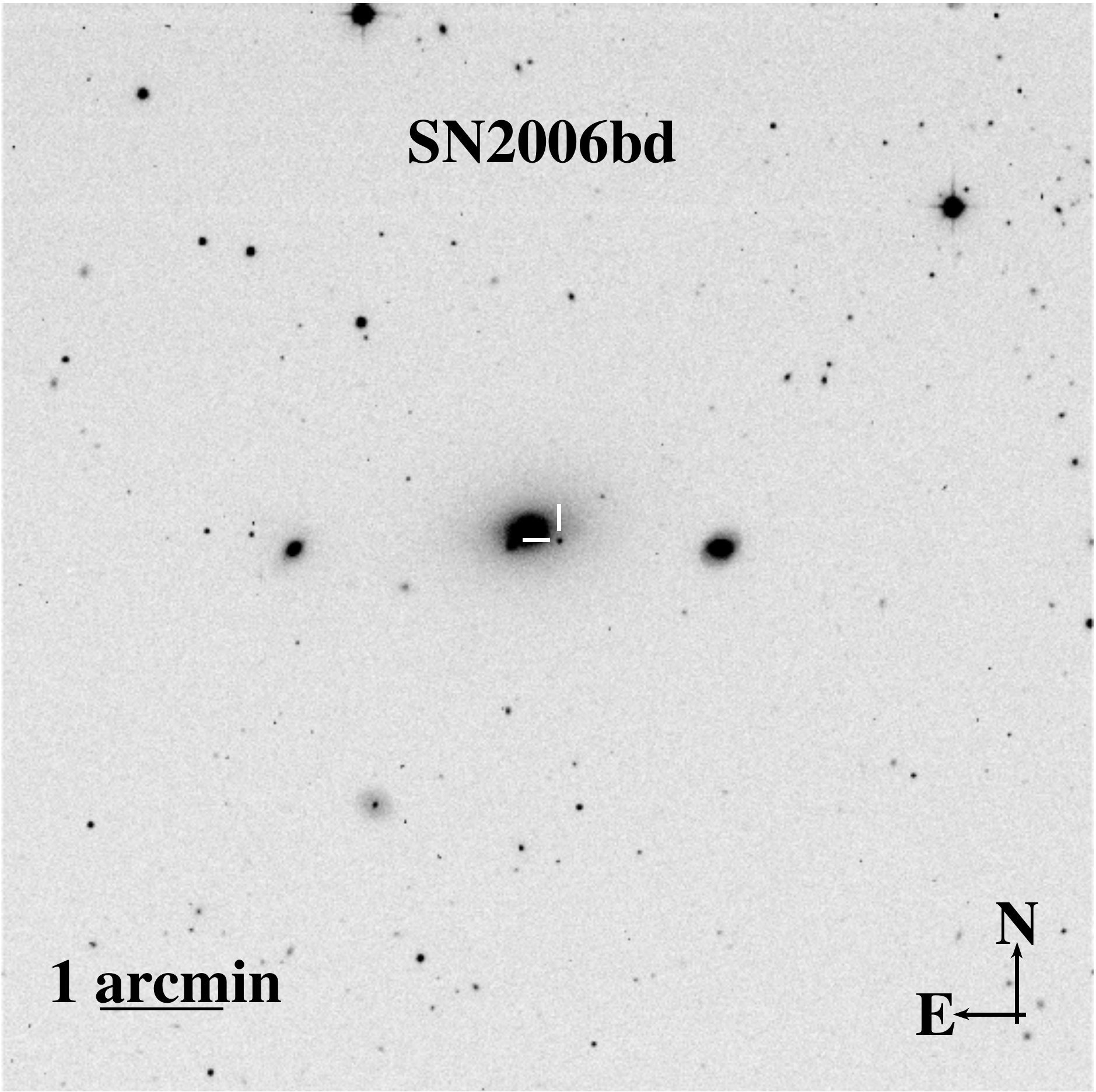}
\newline                                                                     
\plottwo{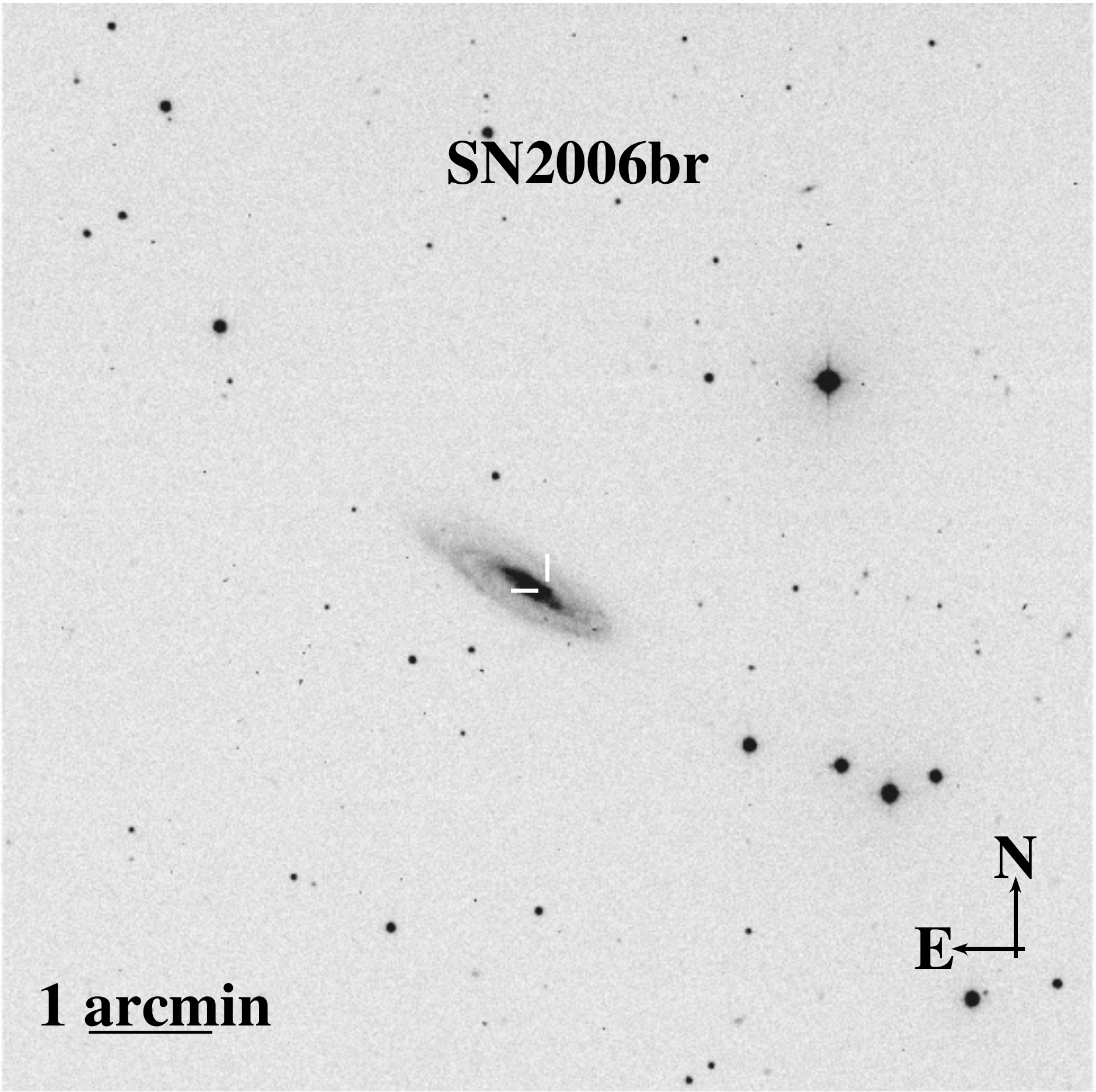}{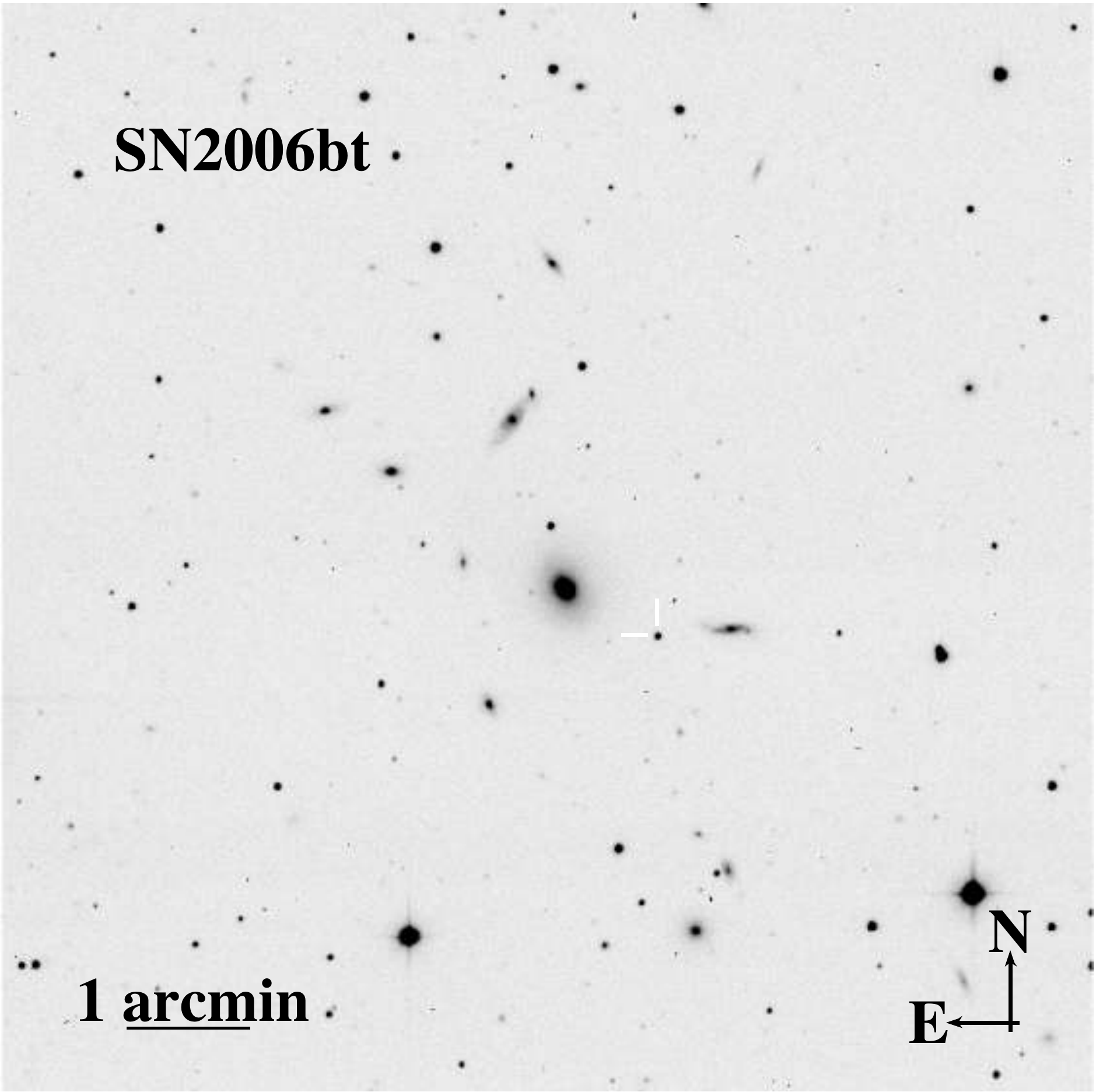}
\plottwo{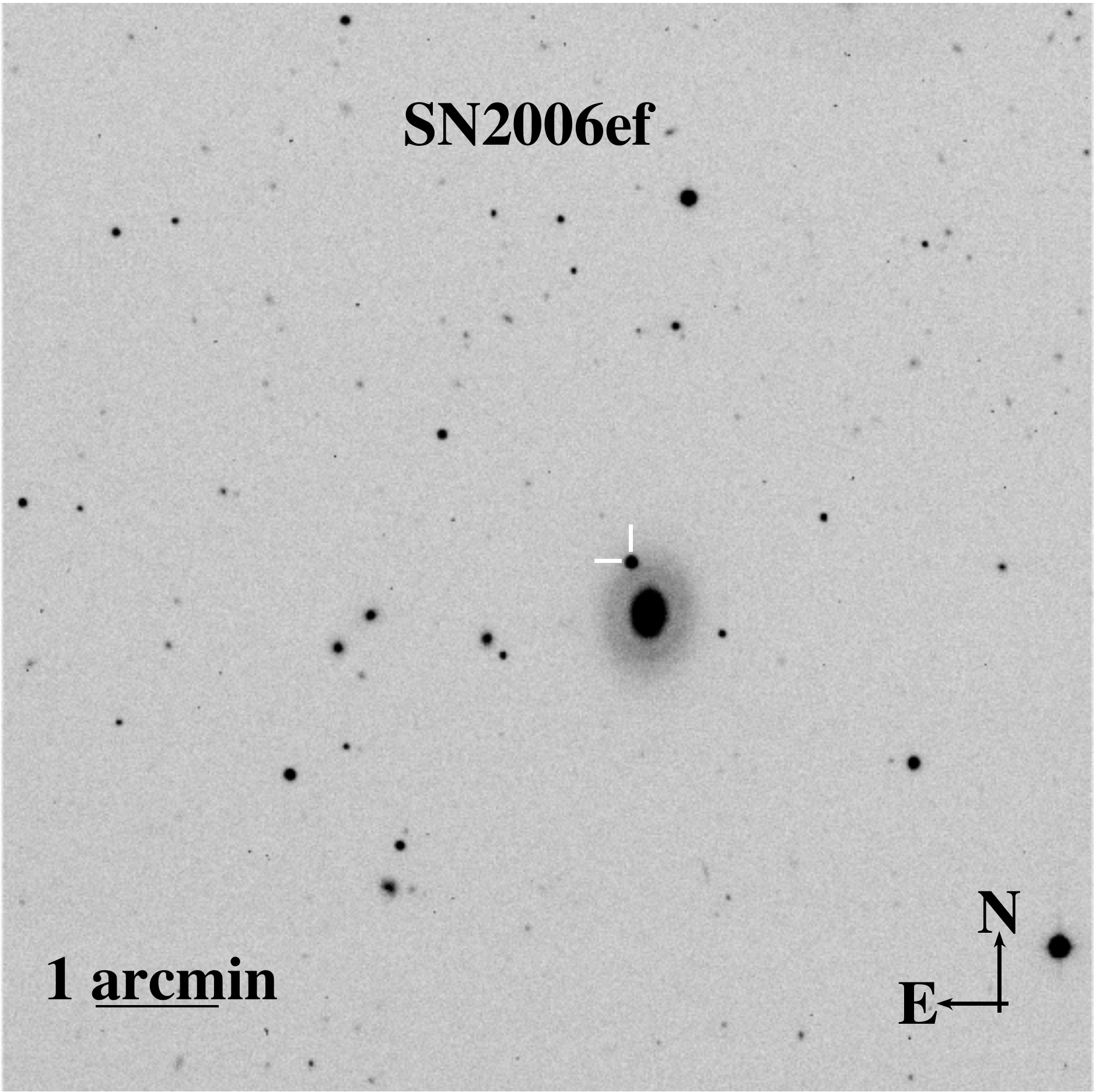}{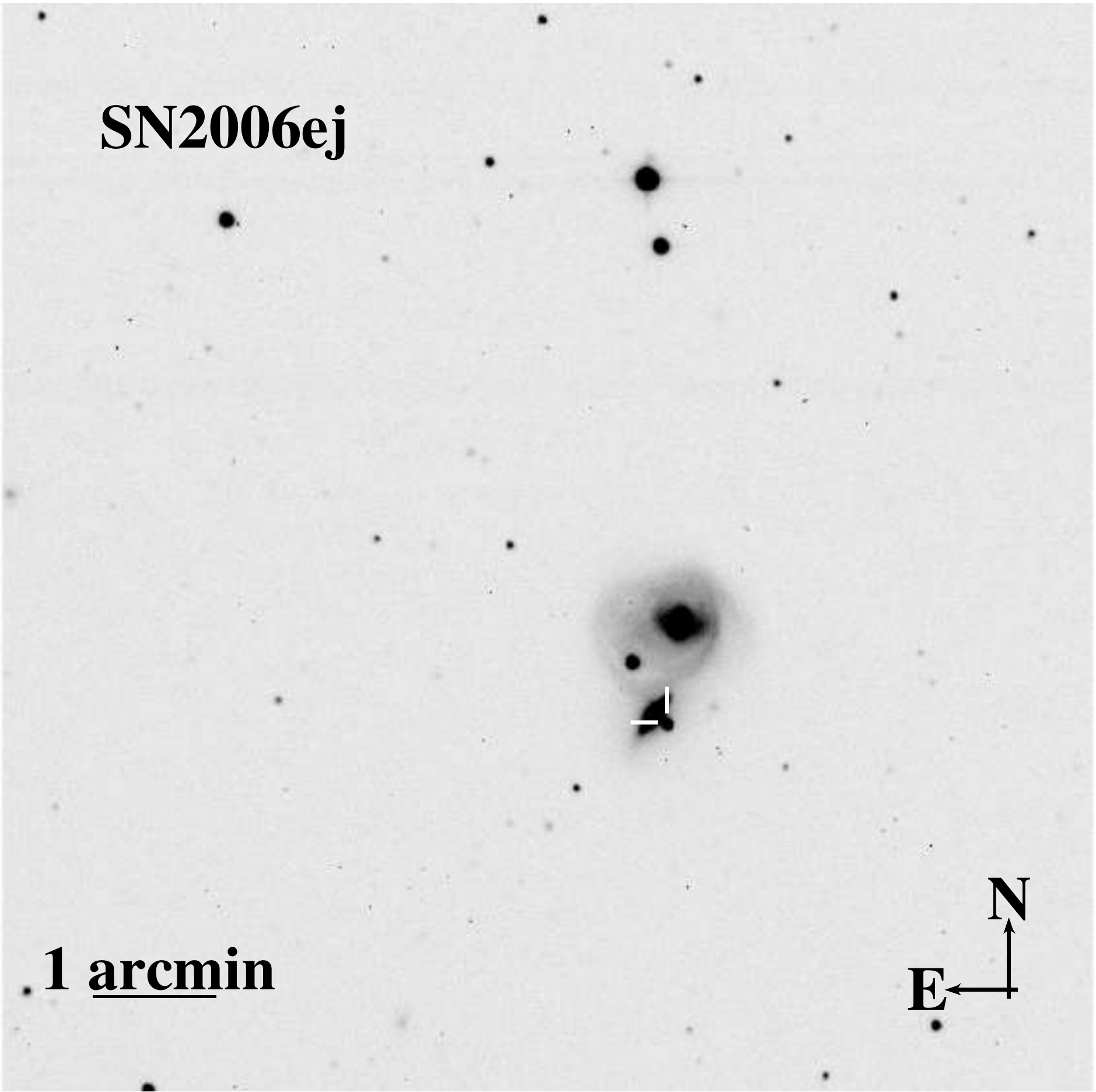}
\newline                                                                     
\plottwo{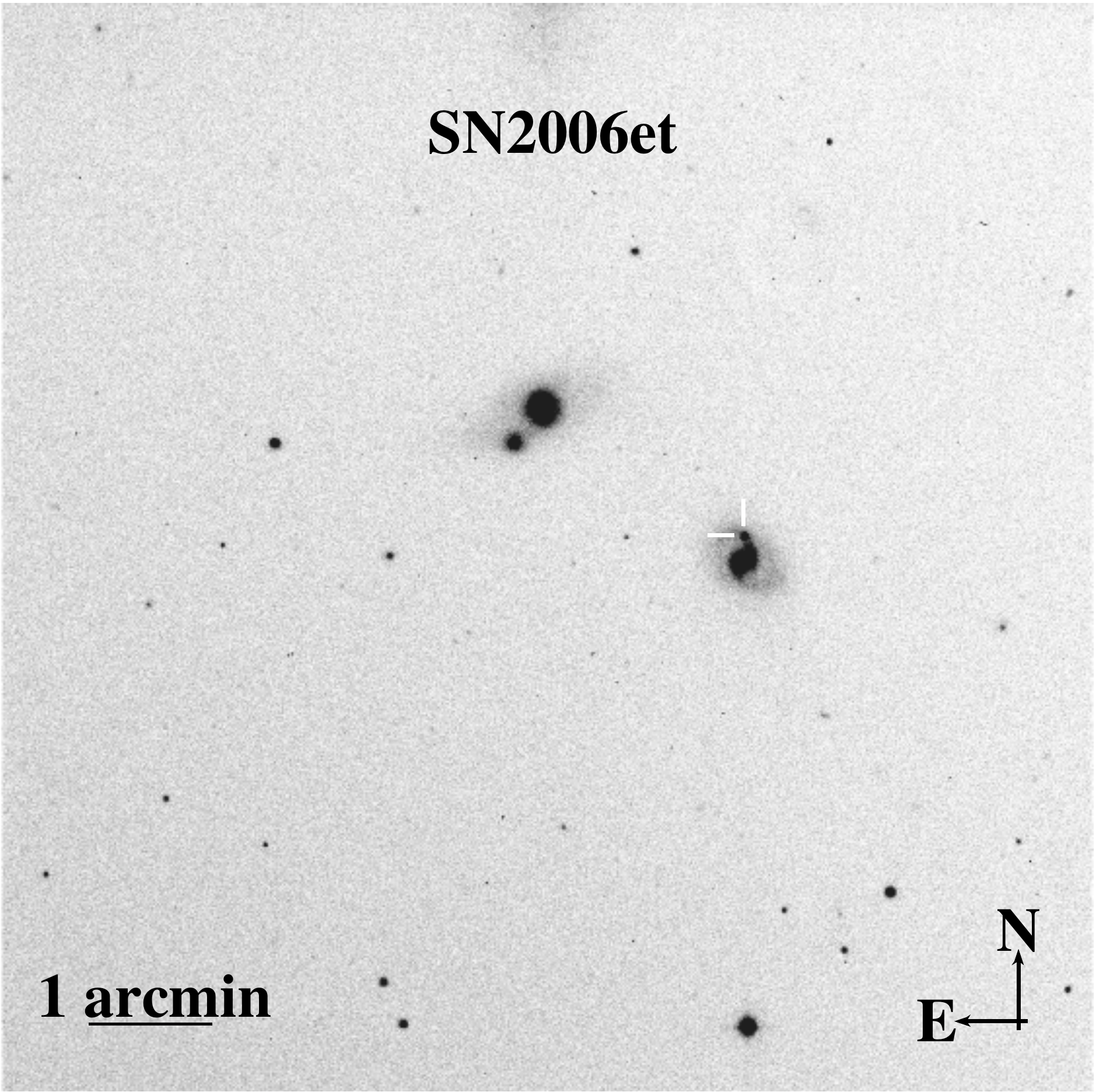}{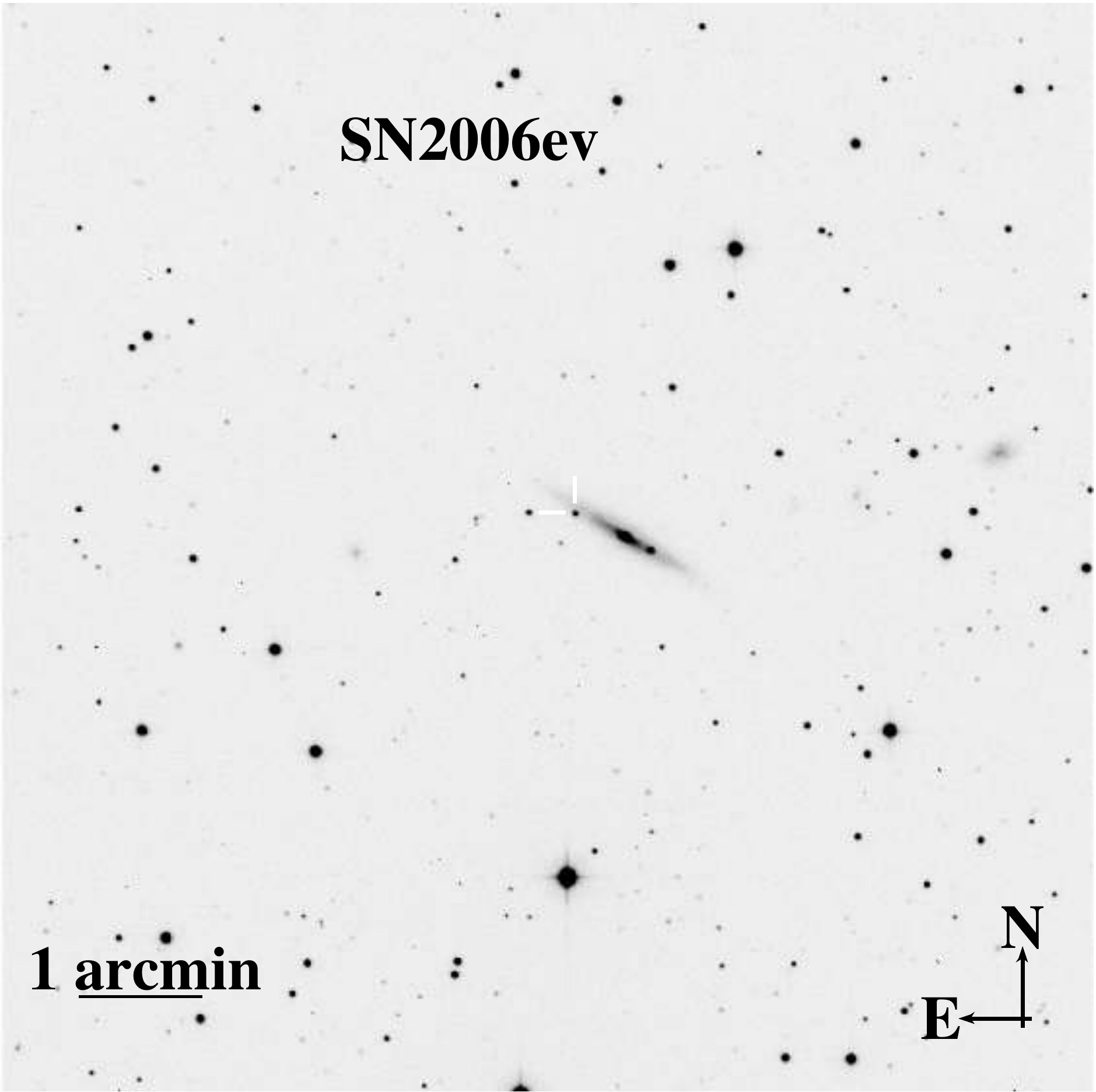}
\plottwo{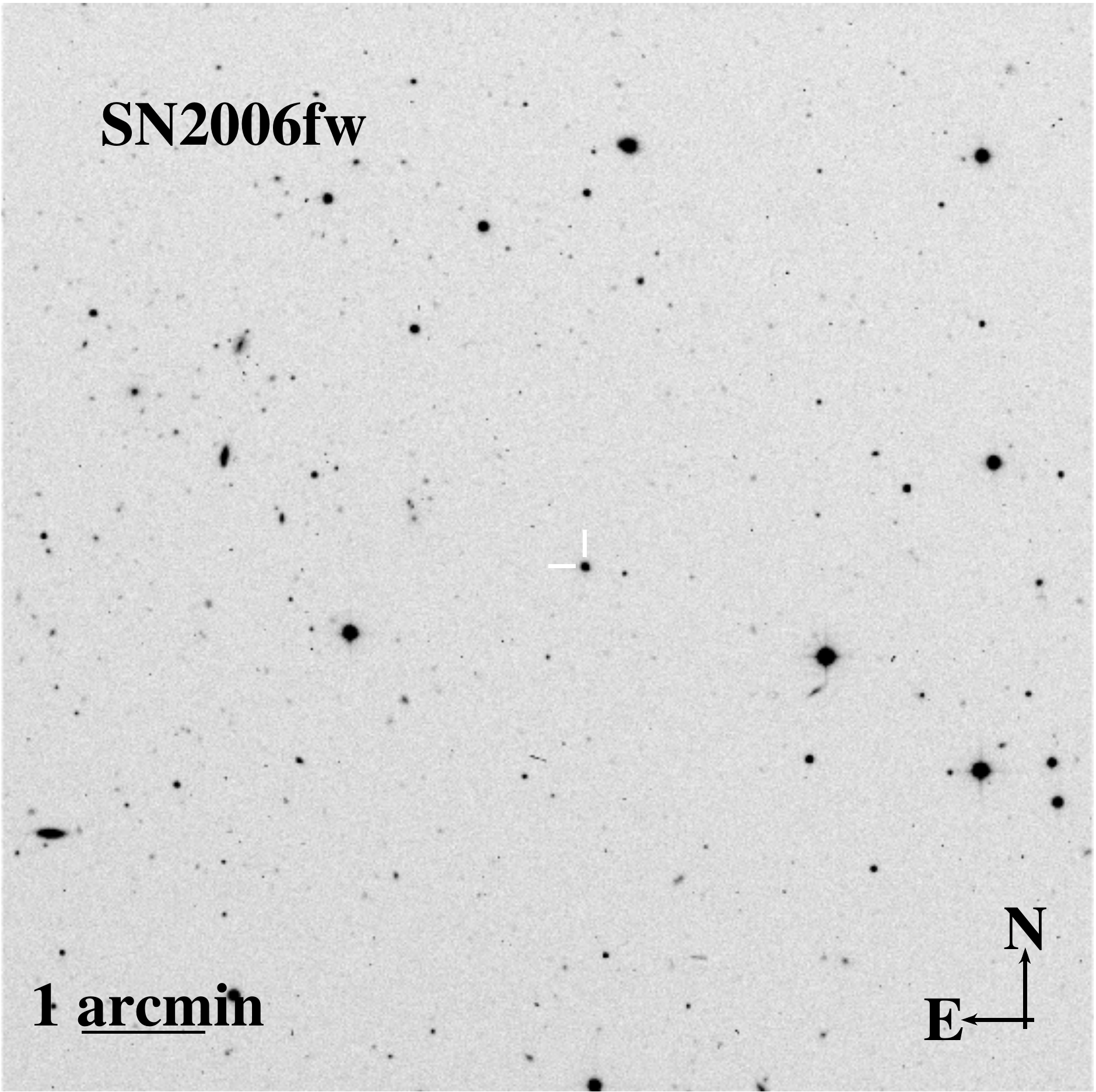}{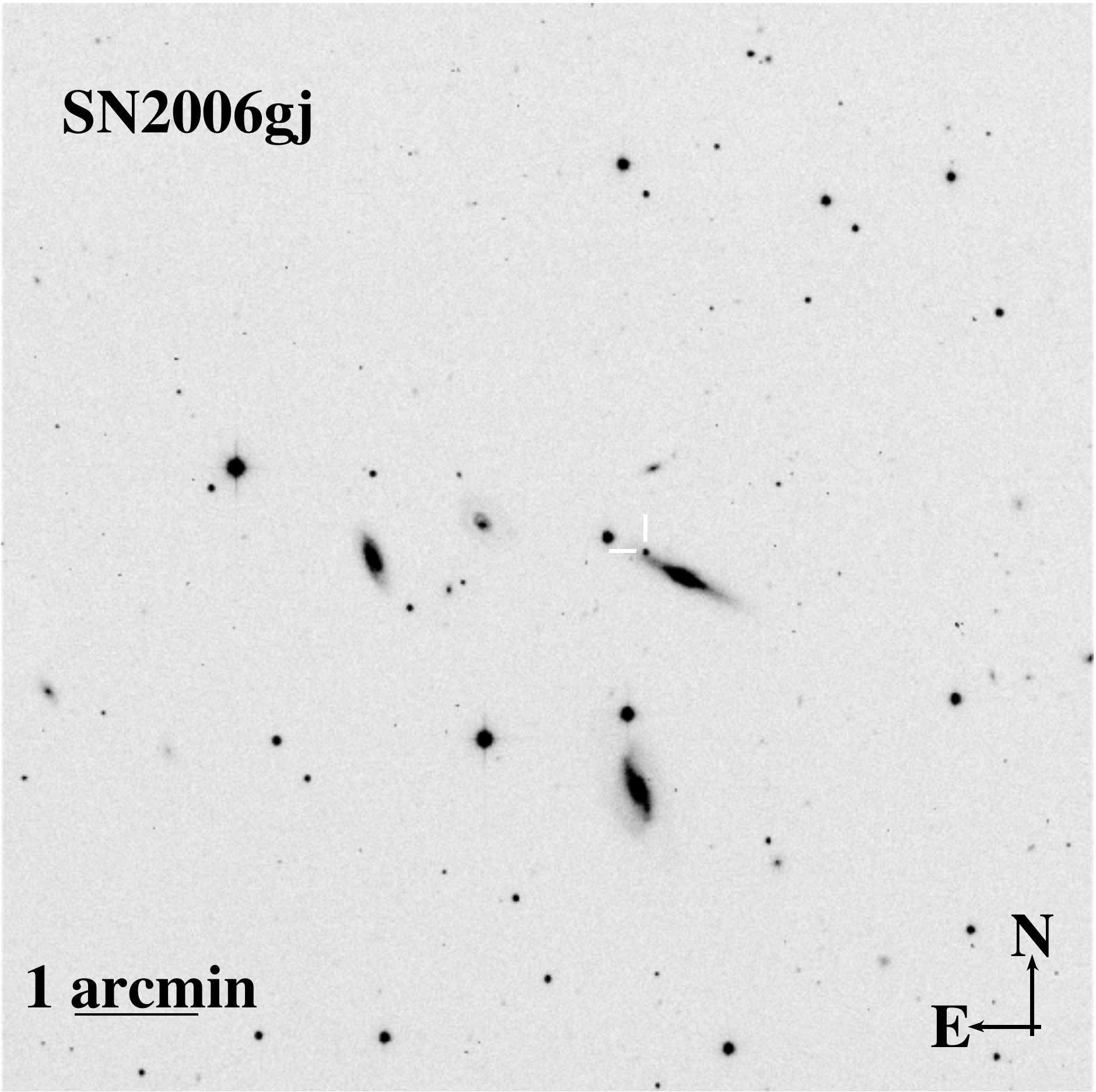}
\newline
\plottwo{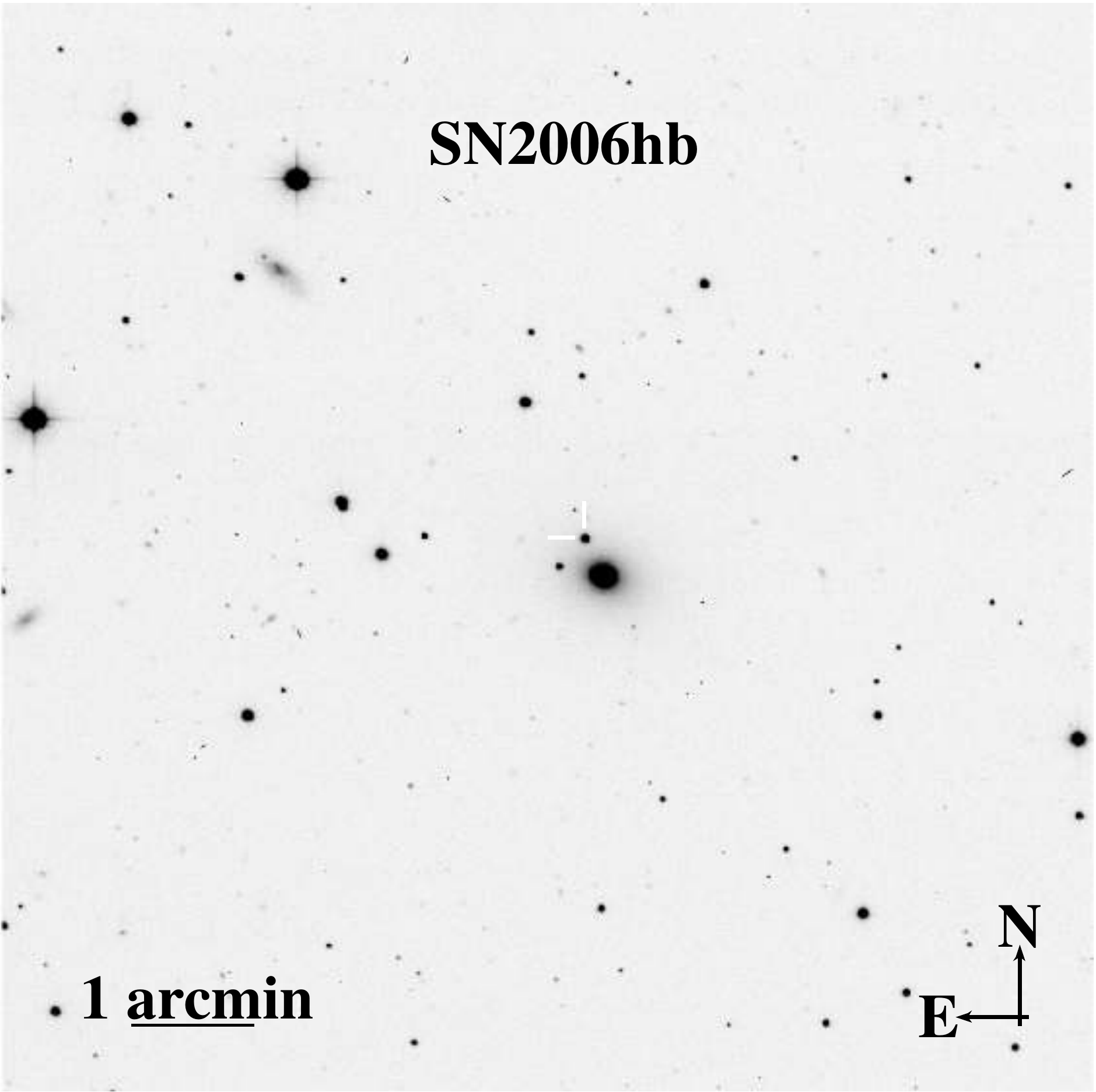}{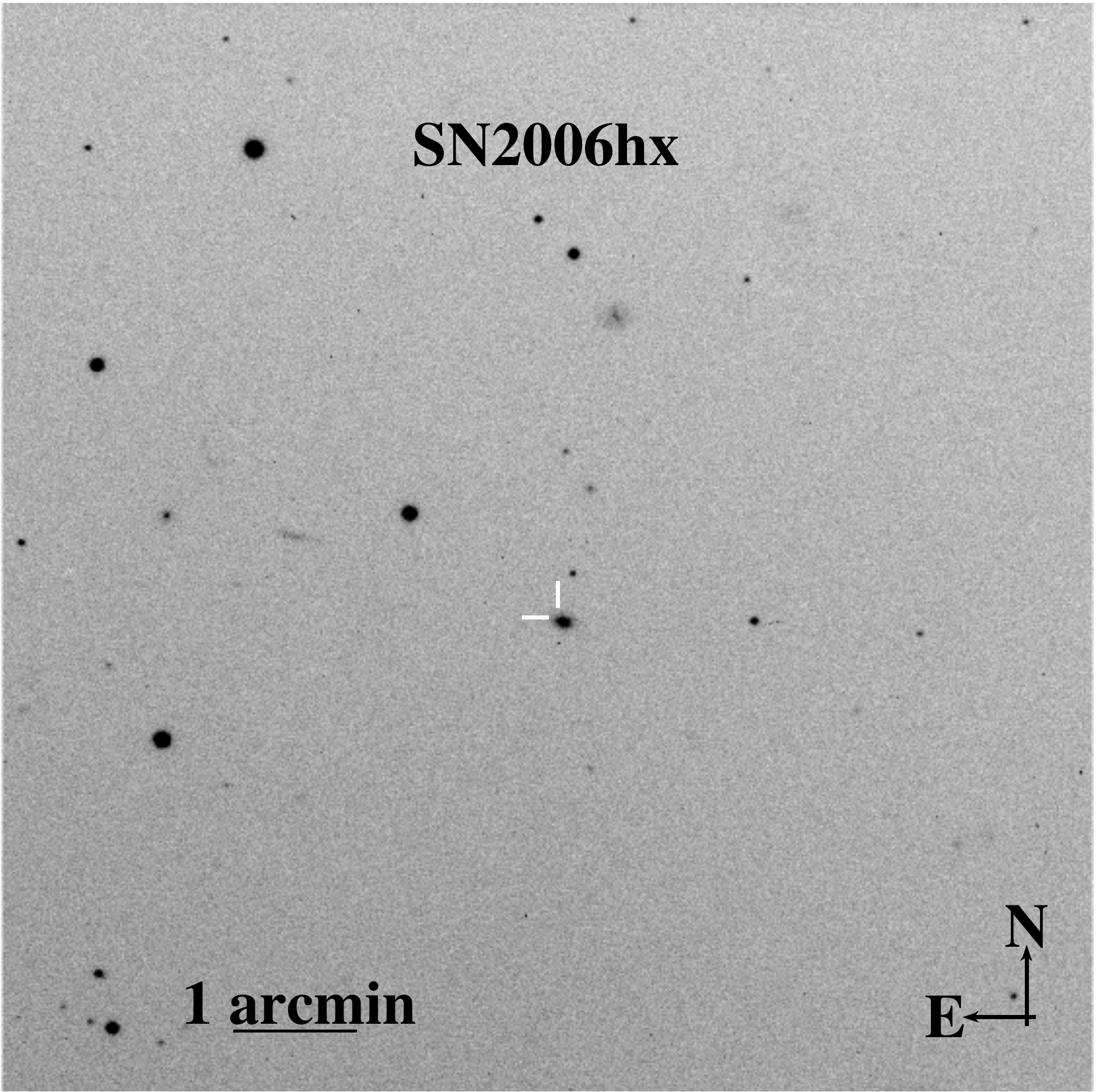}
\plottwo{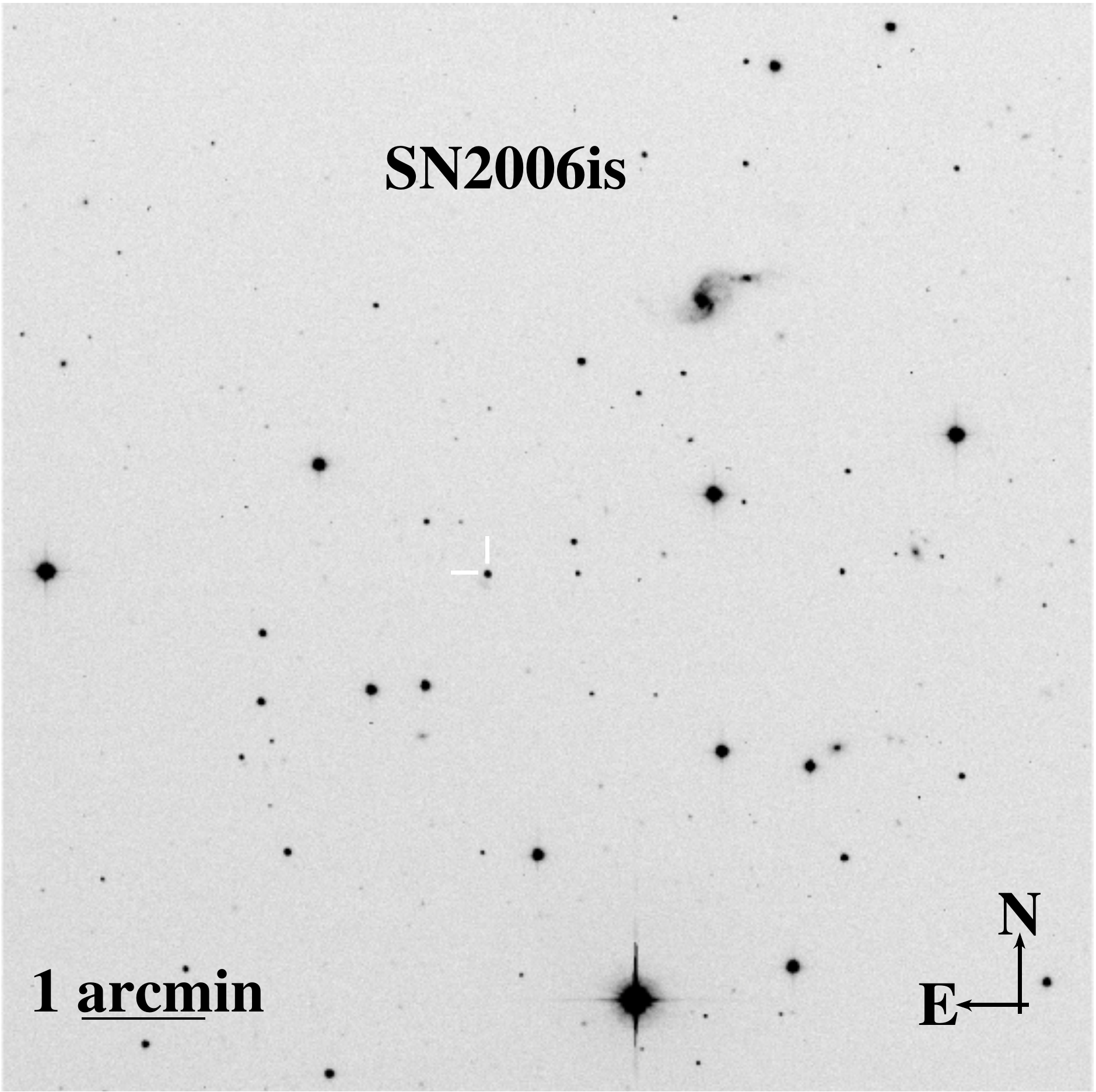}{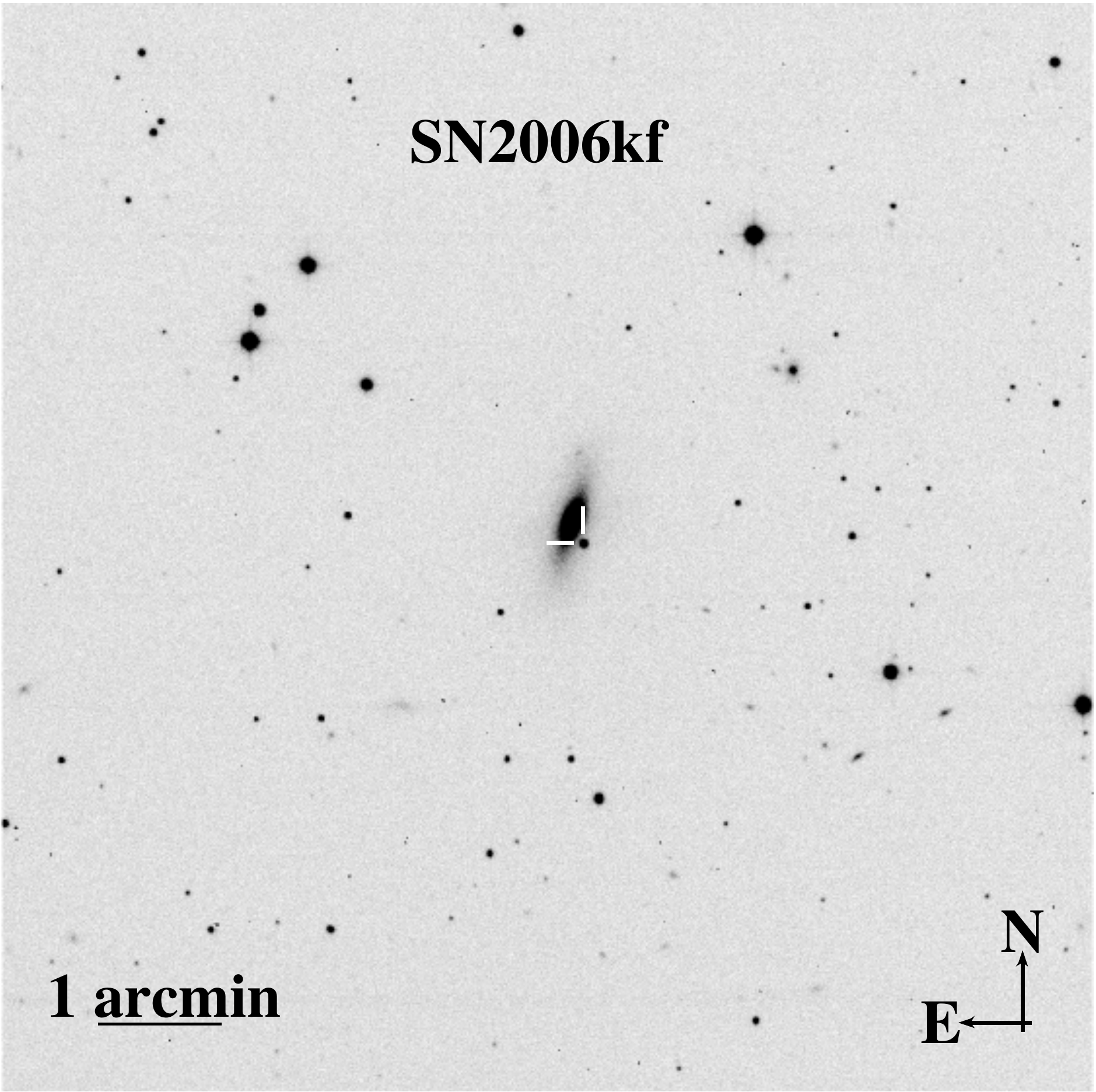}
\newline
\plottwo{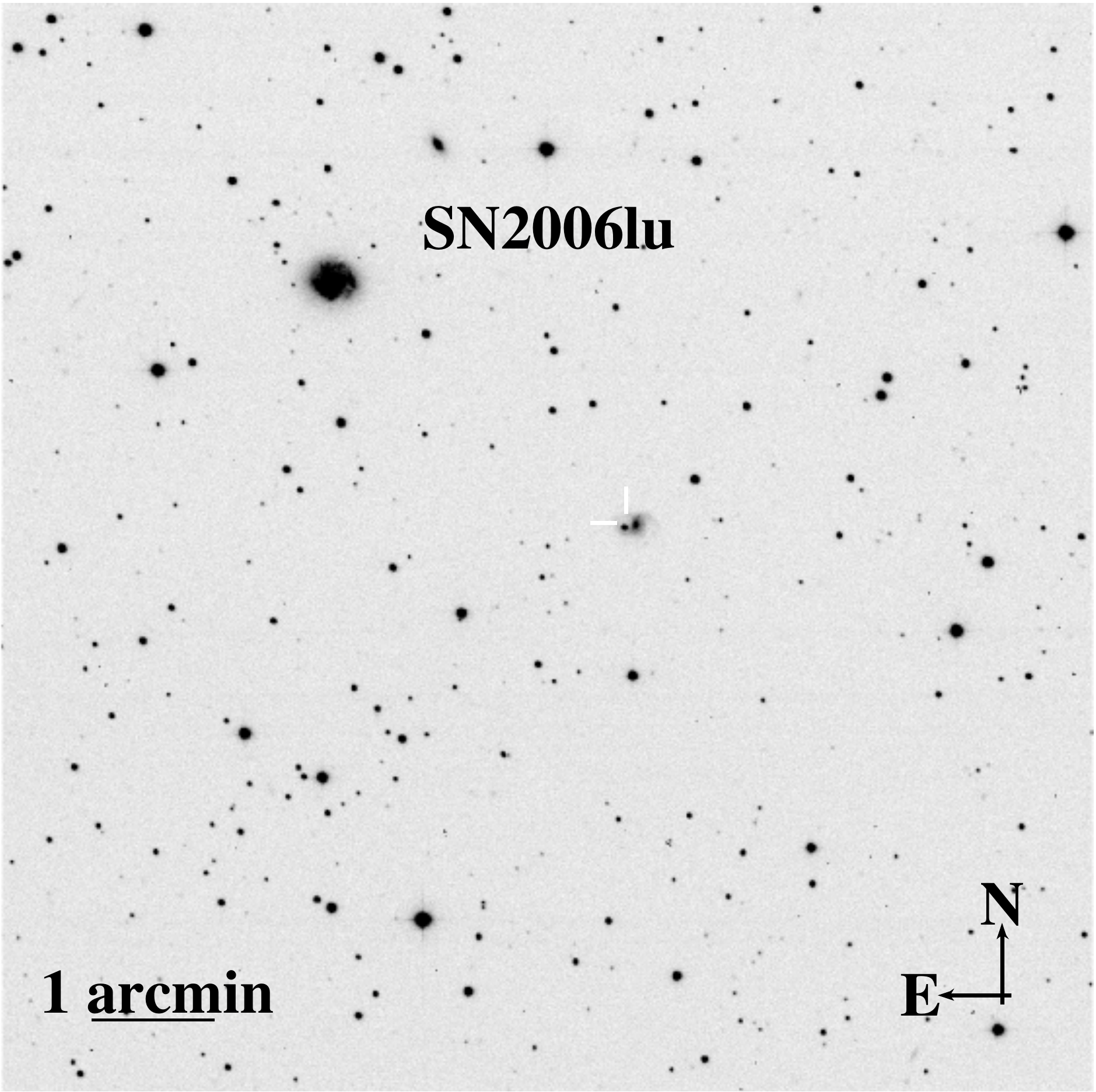}{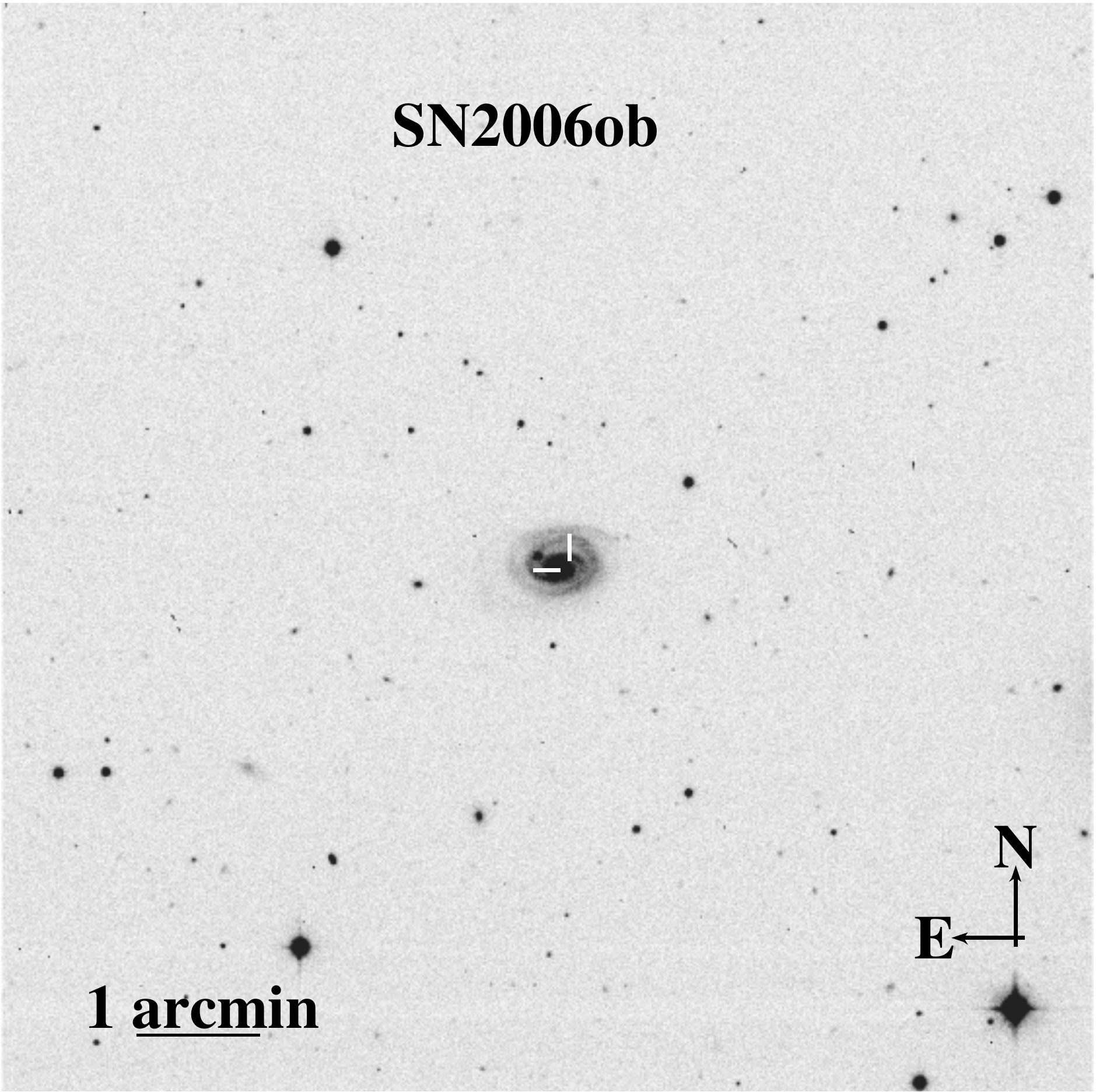}
\plottwo{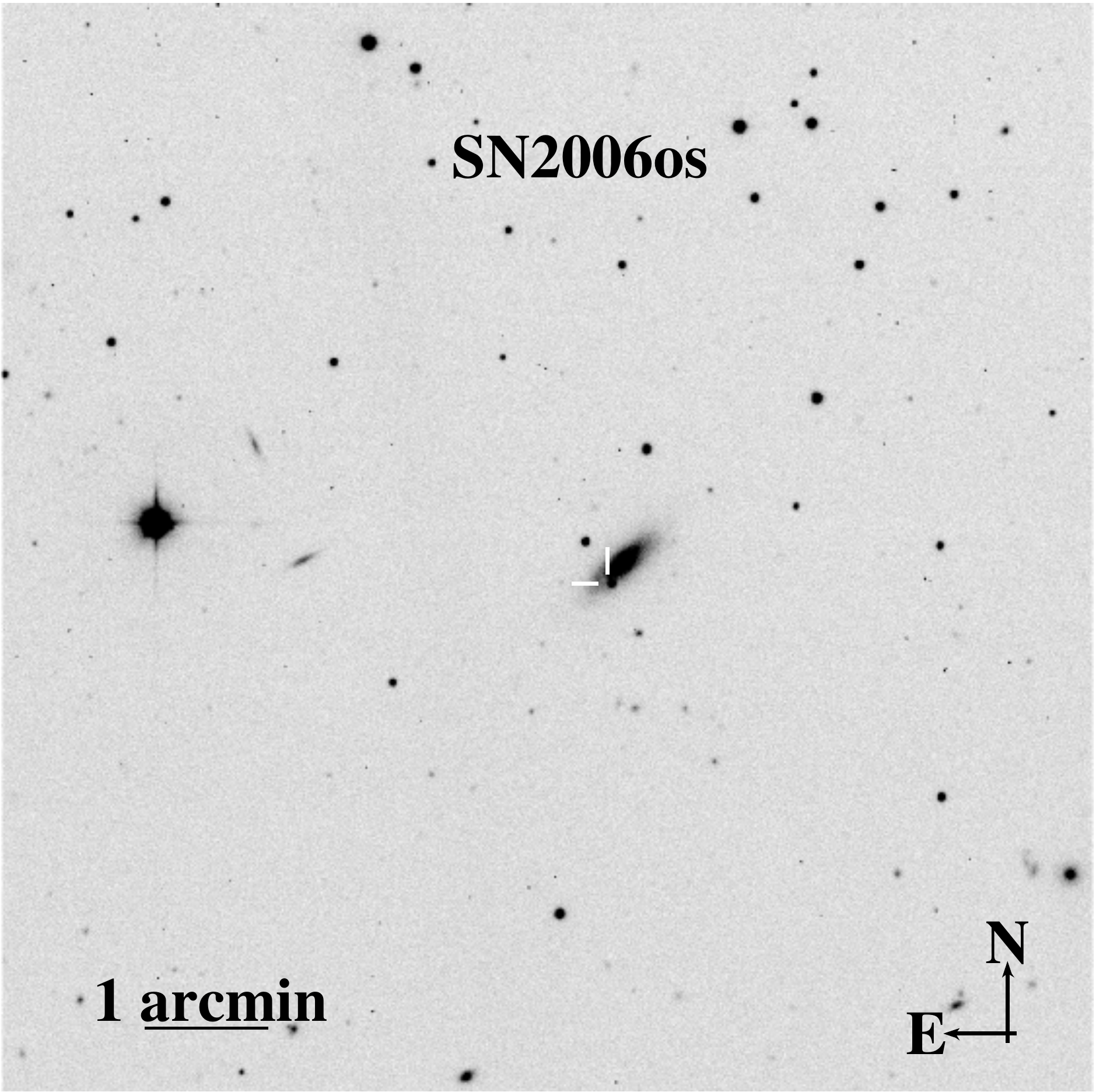}{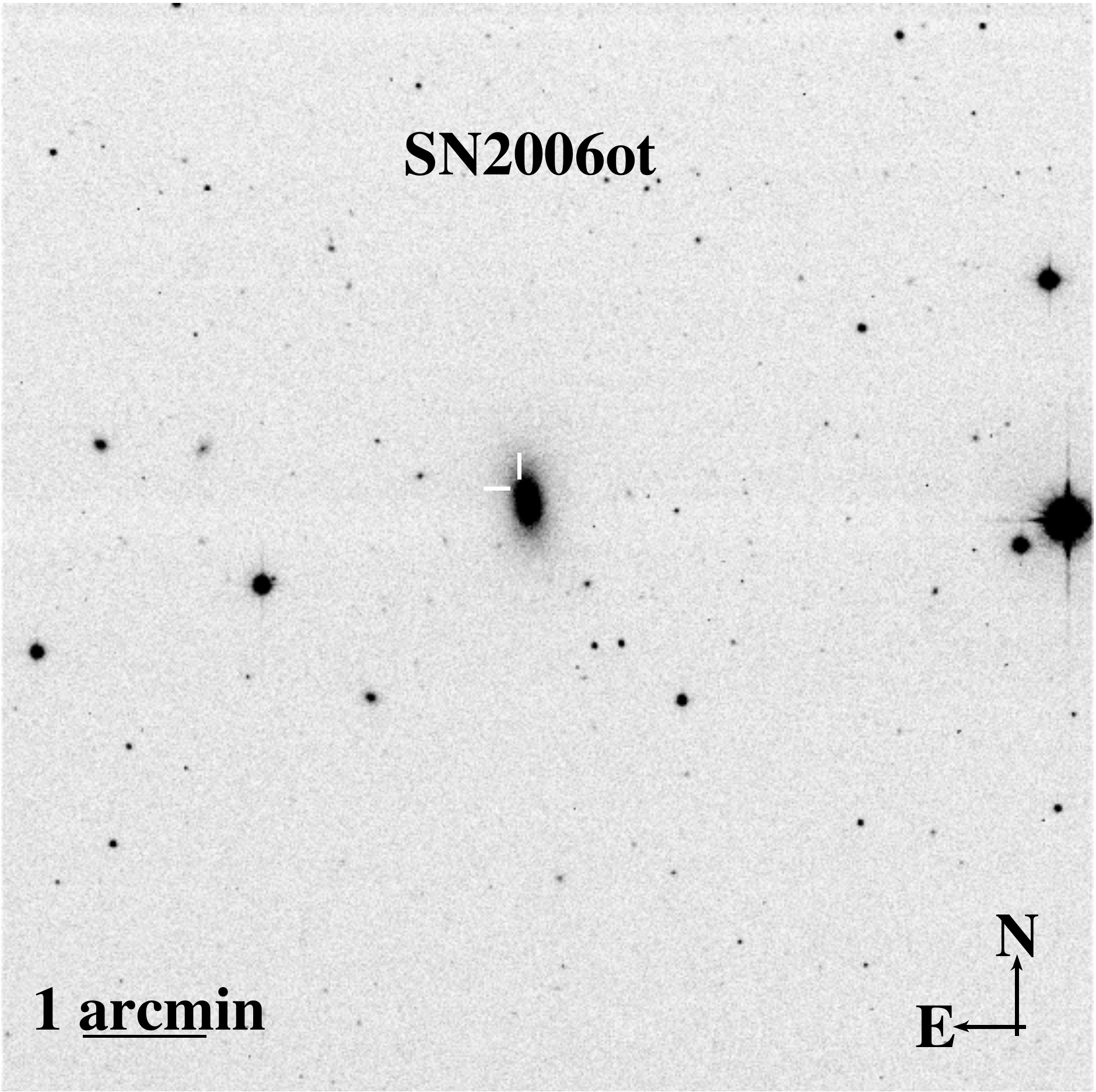}
{\center Stritzinger {\it et al.} Fig.~\ref{fig:fcharts}}
\end{figure}
\clearpage
\newpage

\begin{figure}[t]
\epsscale{.54}
\plottwo{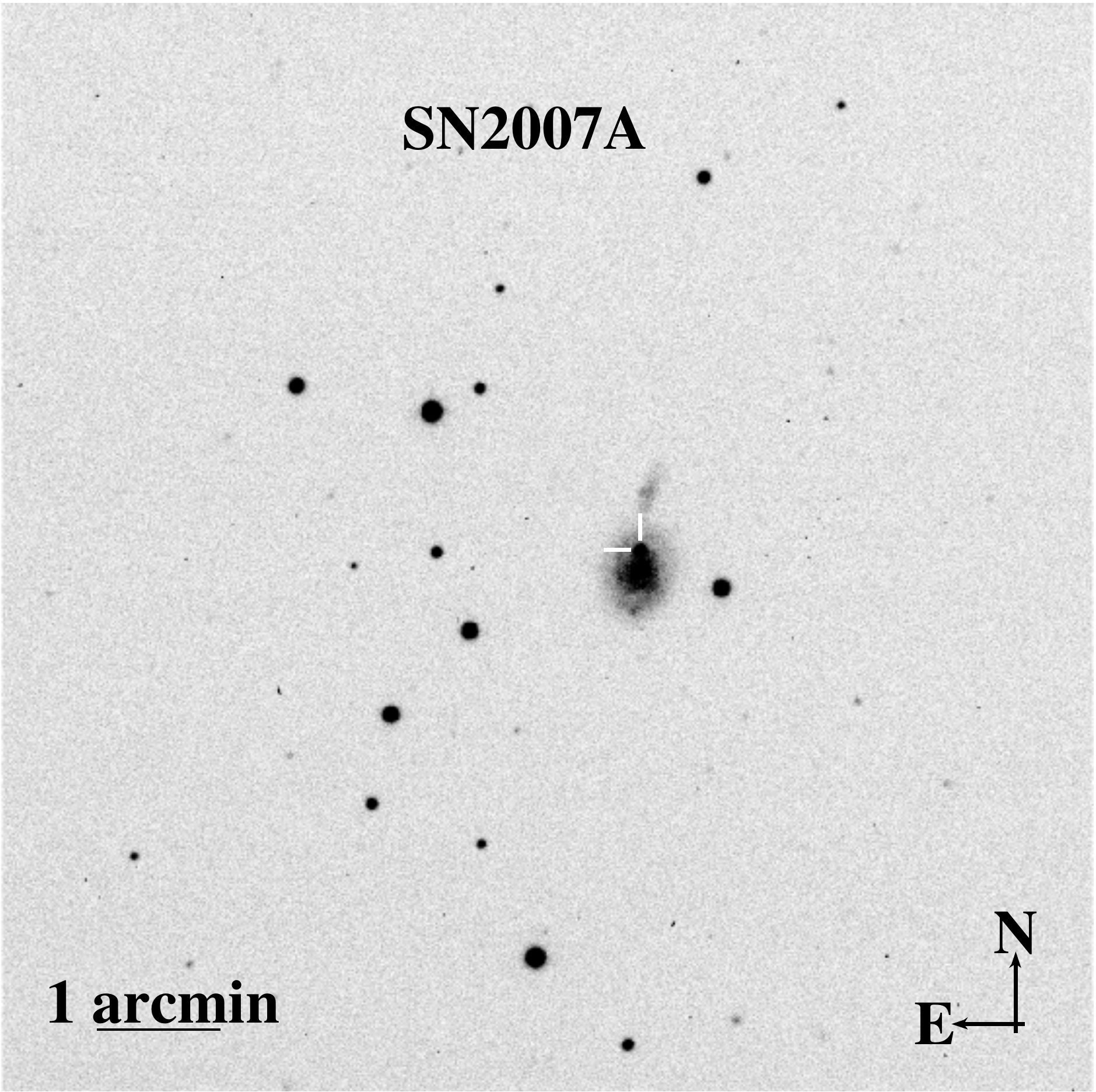}{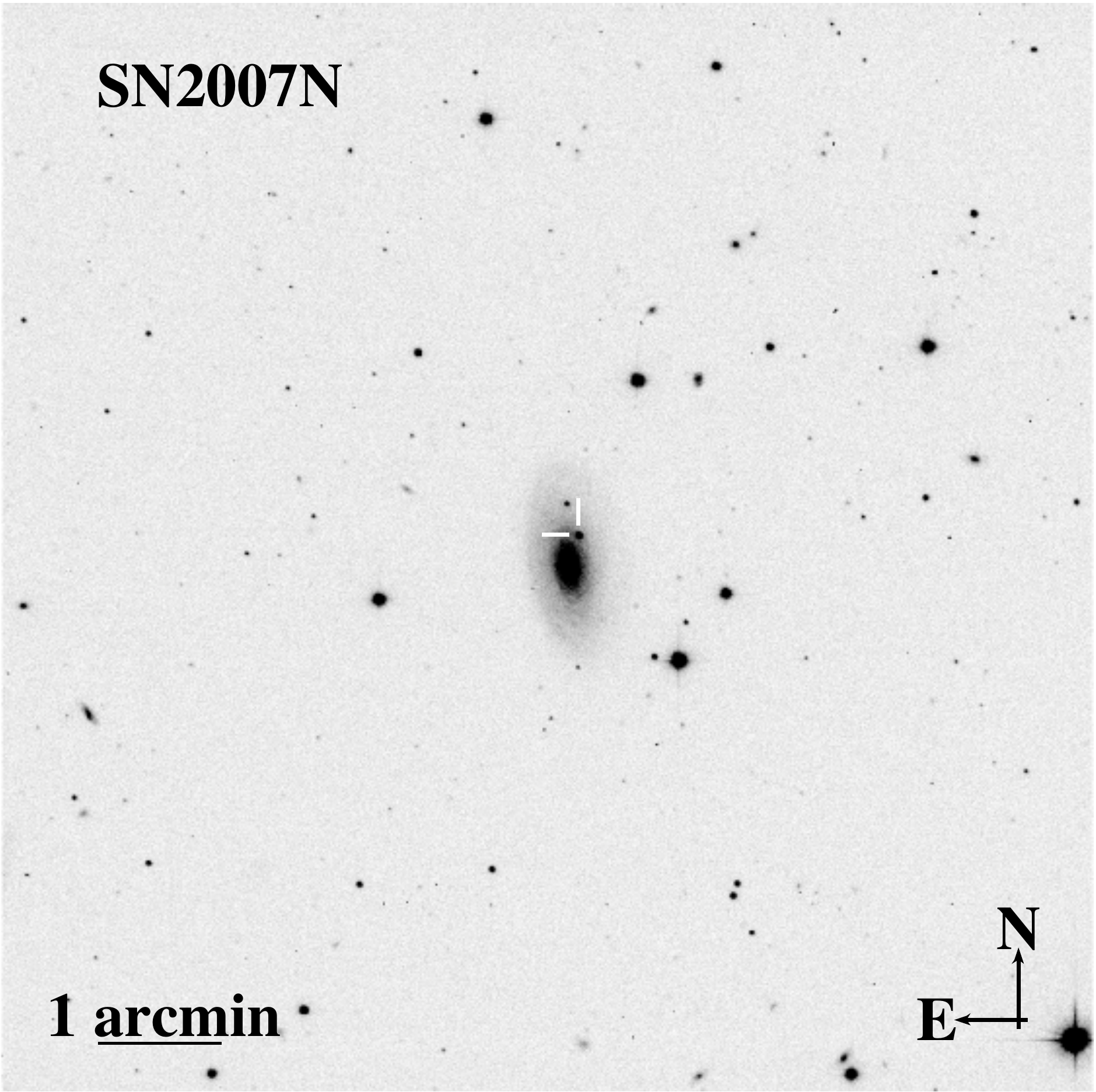}
\plottwo{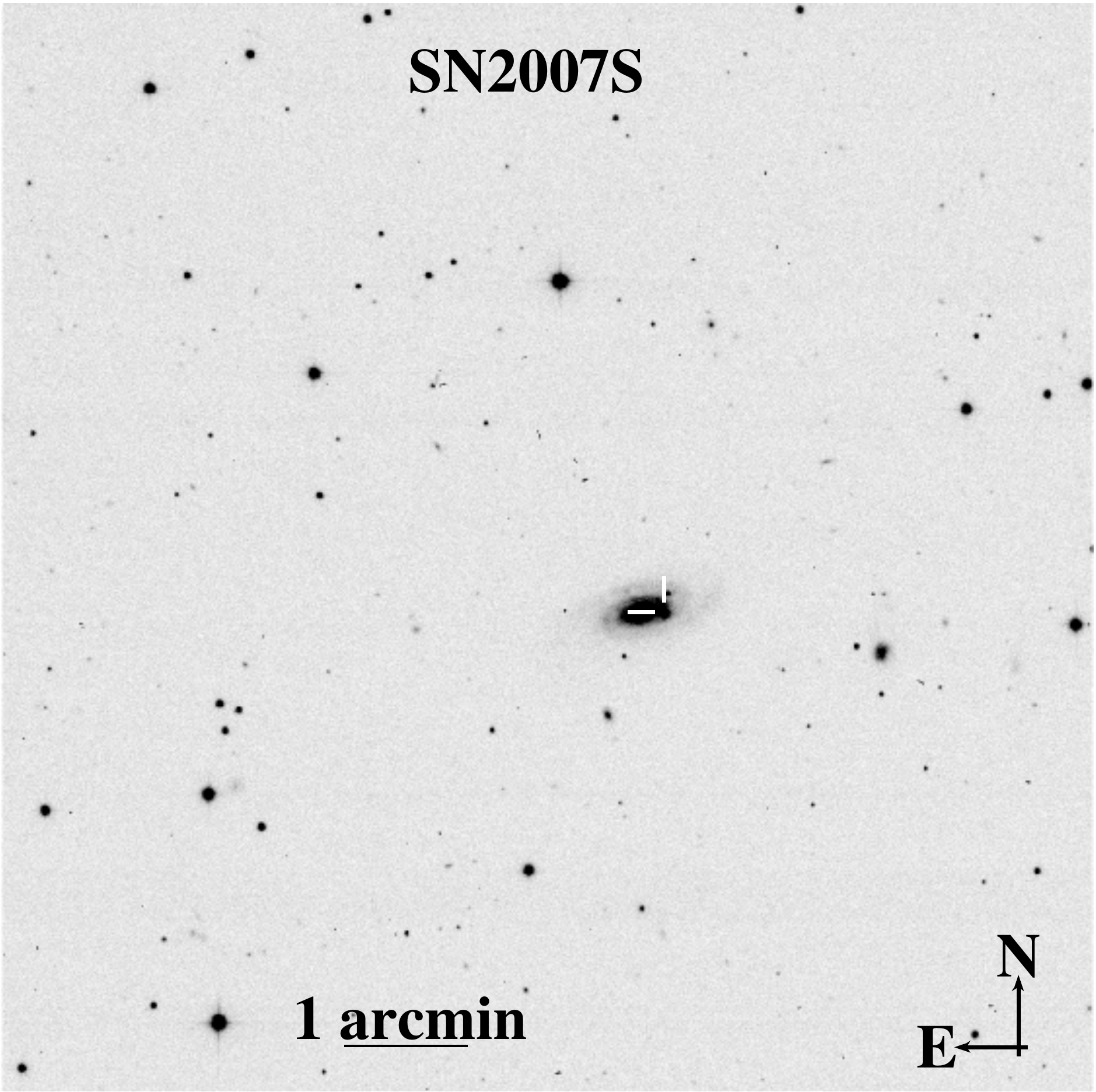}{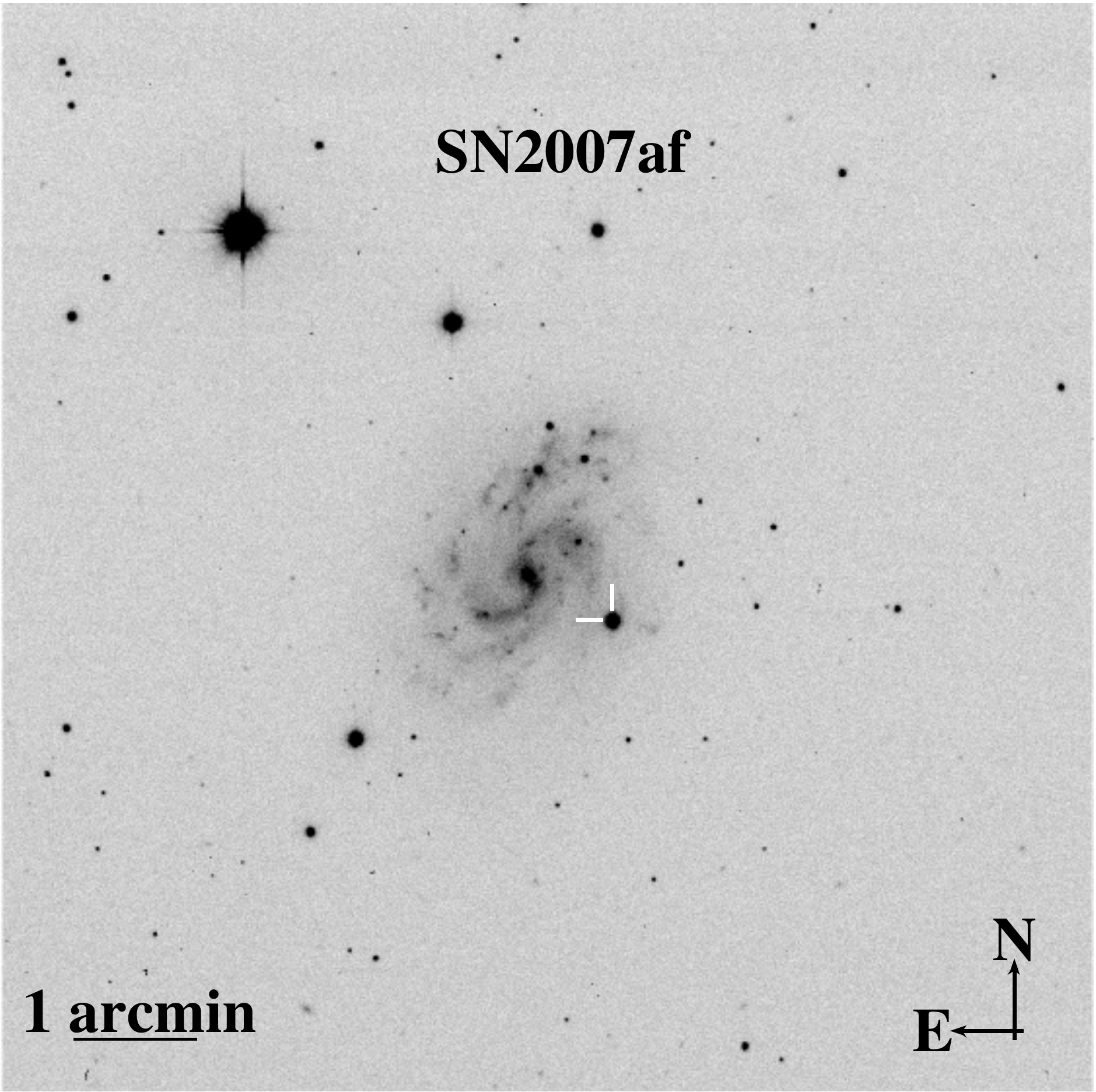}
\newline                                                                     
\plottwo{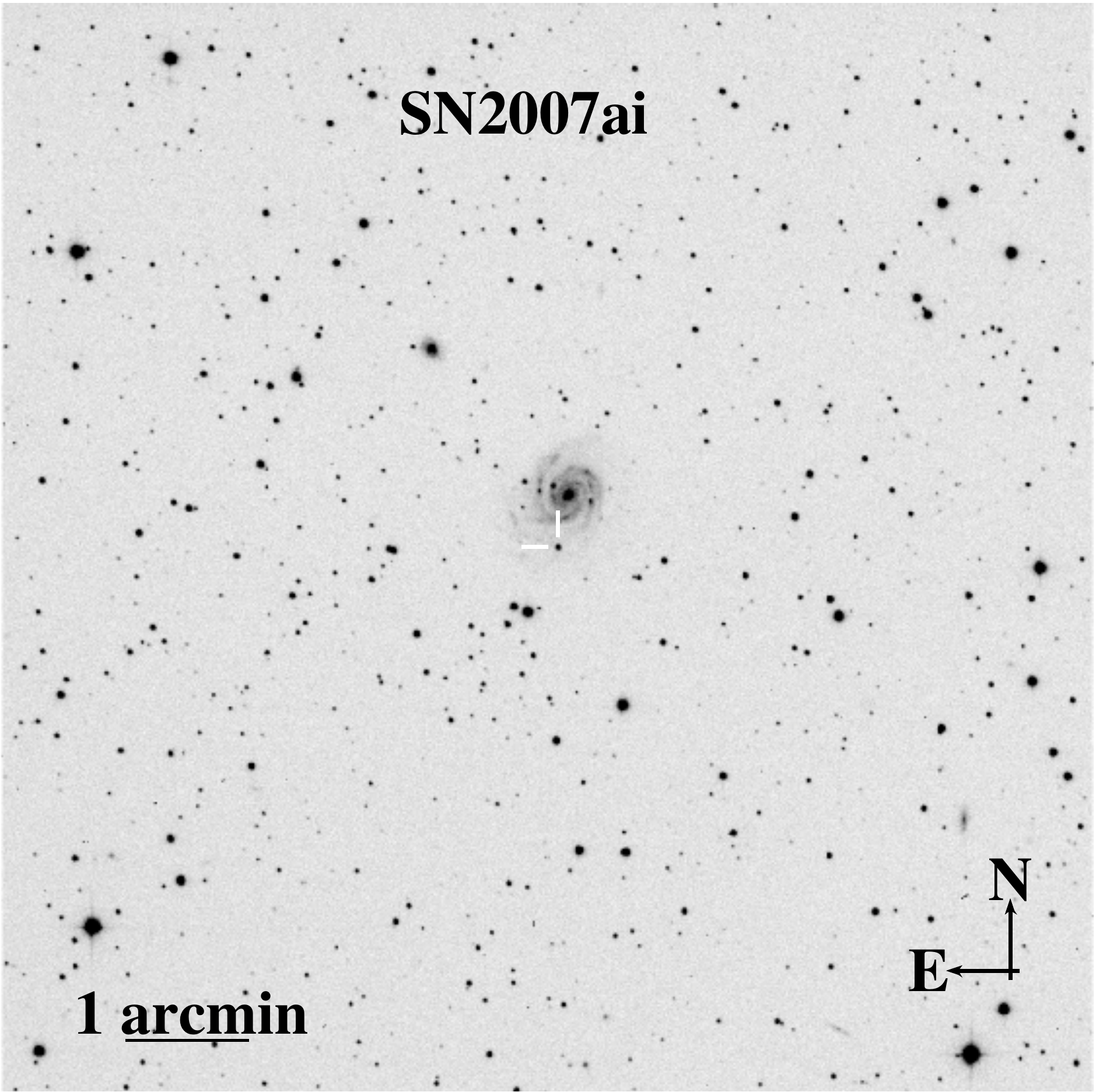}{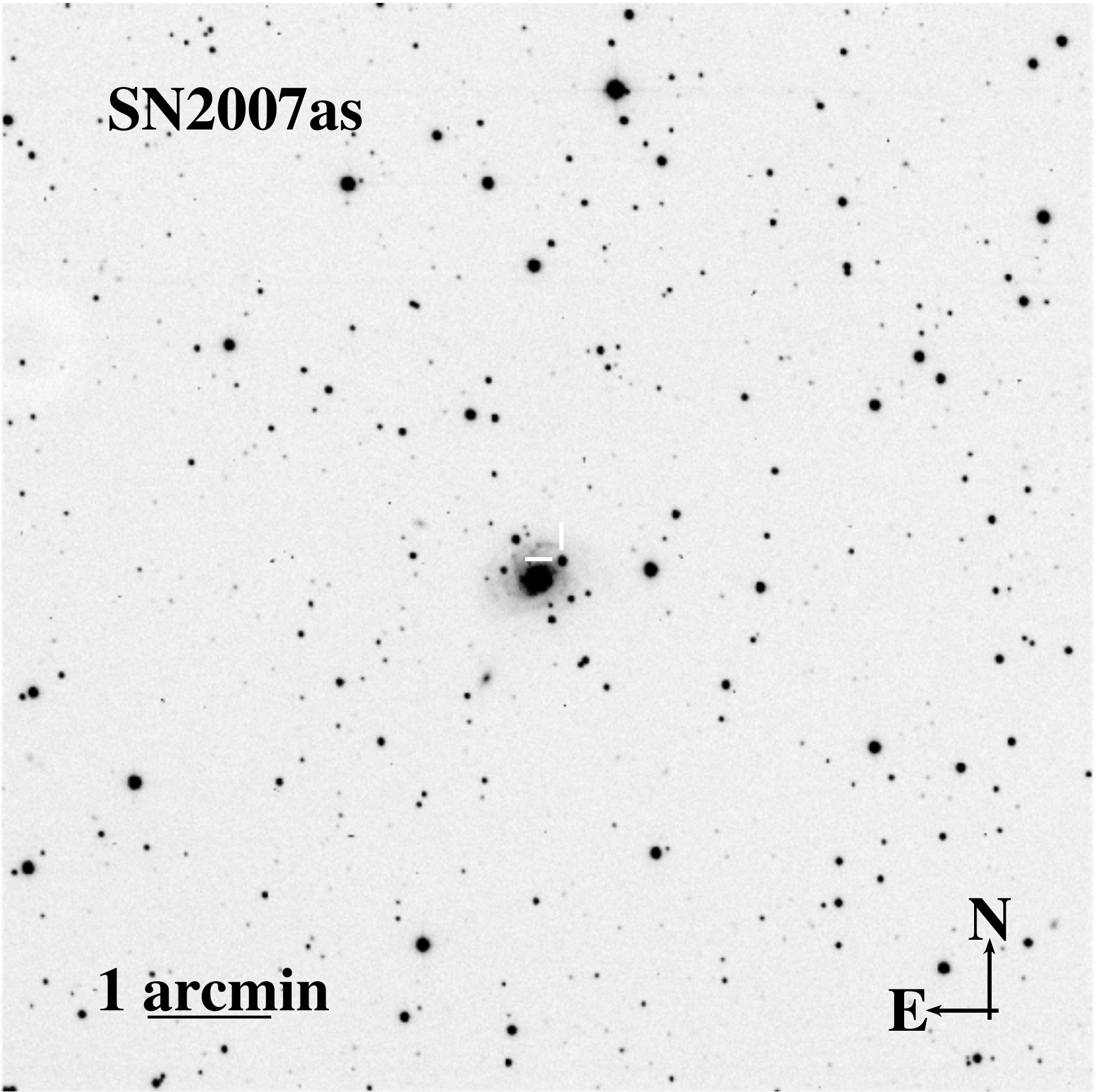}
\plottwo{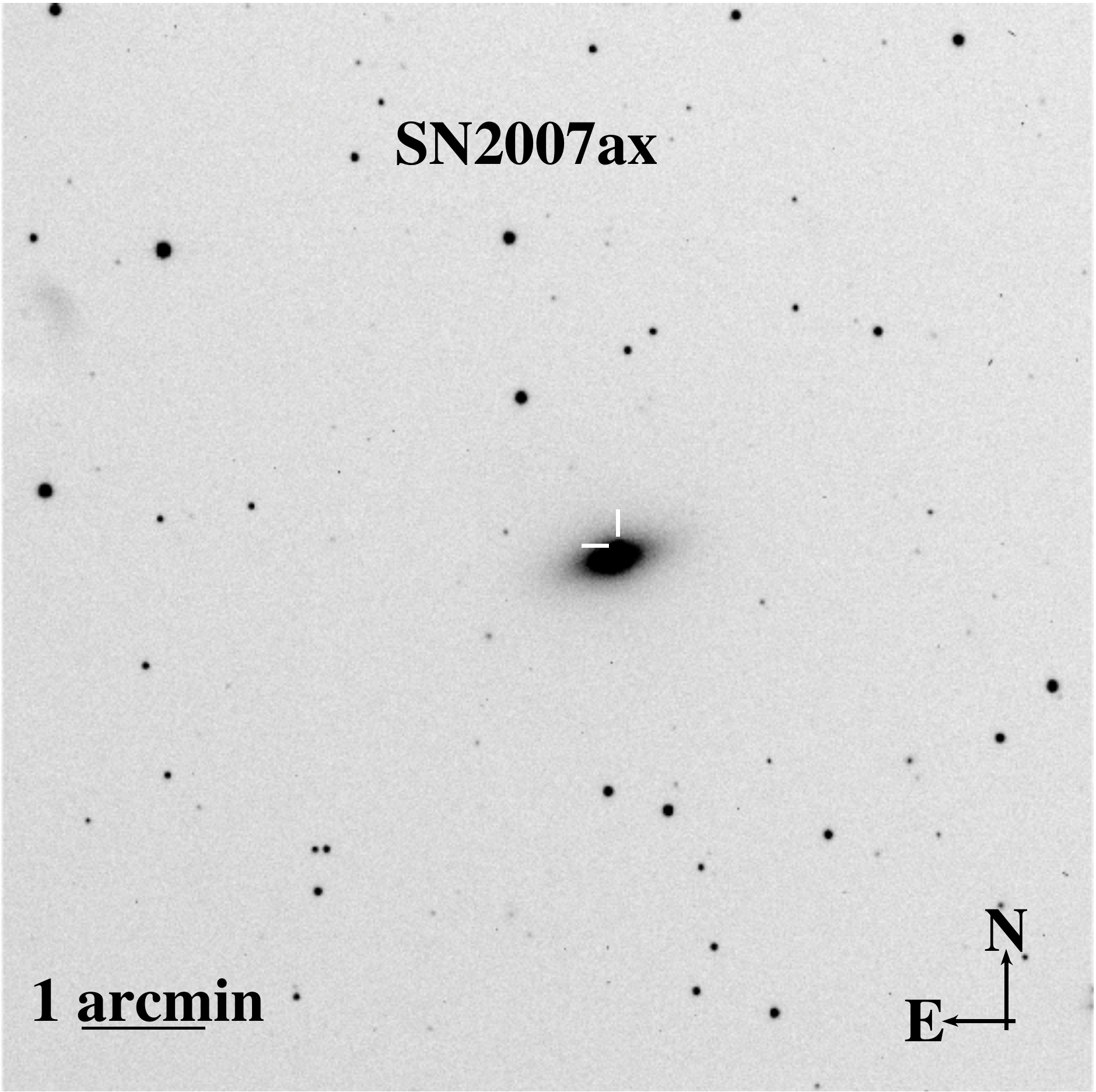}{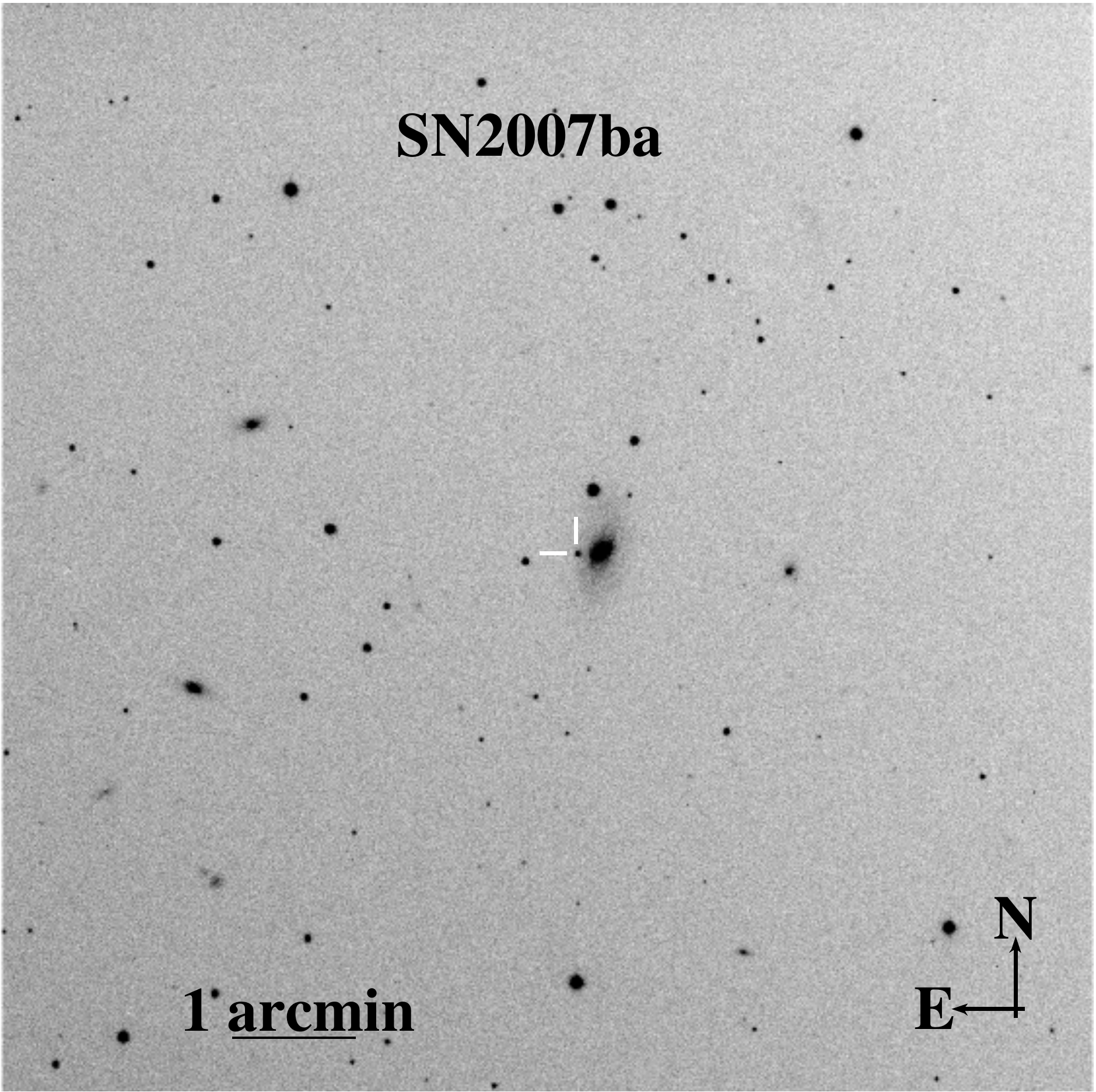}
\newline                                                                     
\plottwo{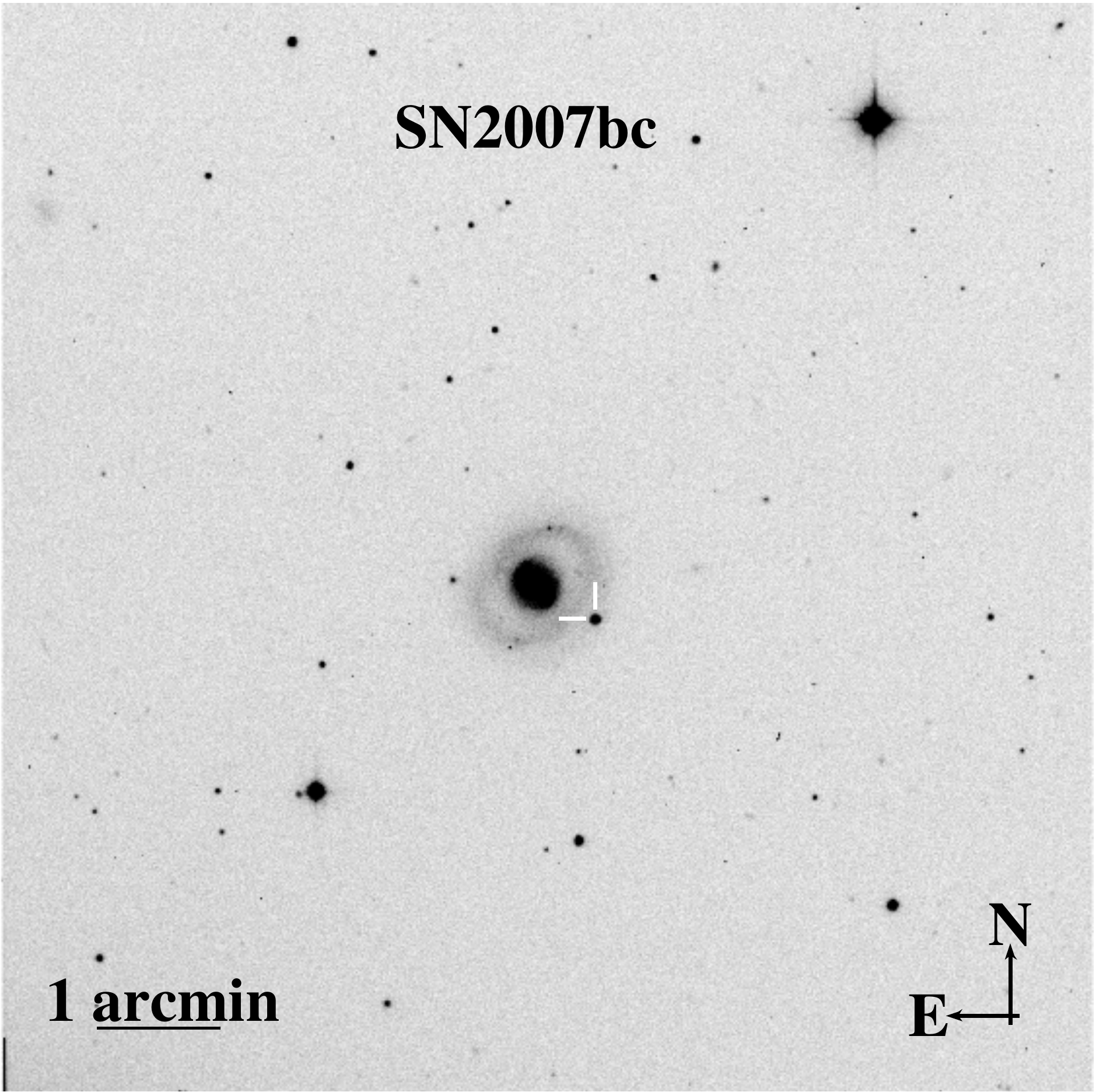}{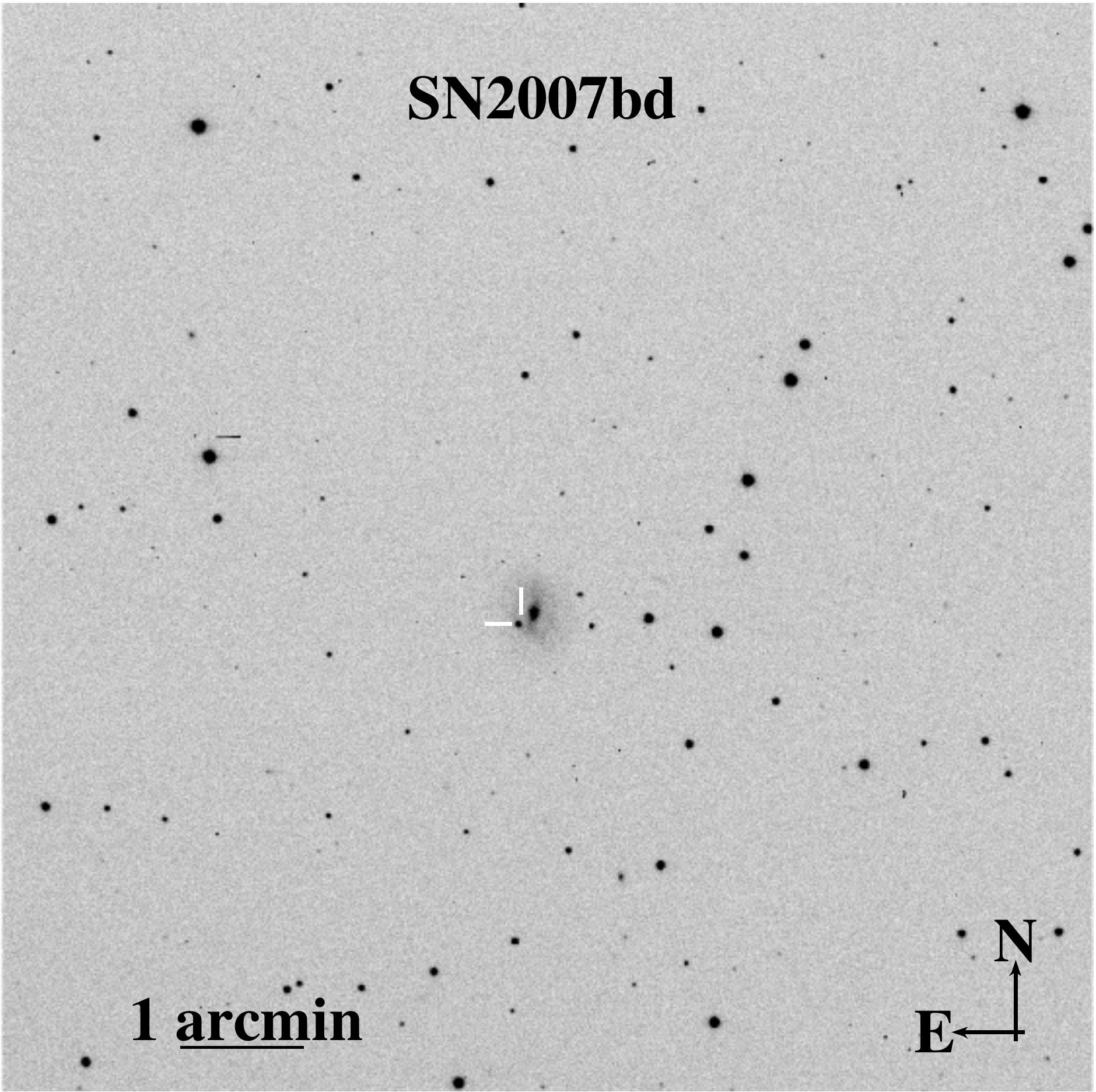}
\plottwo{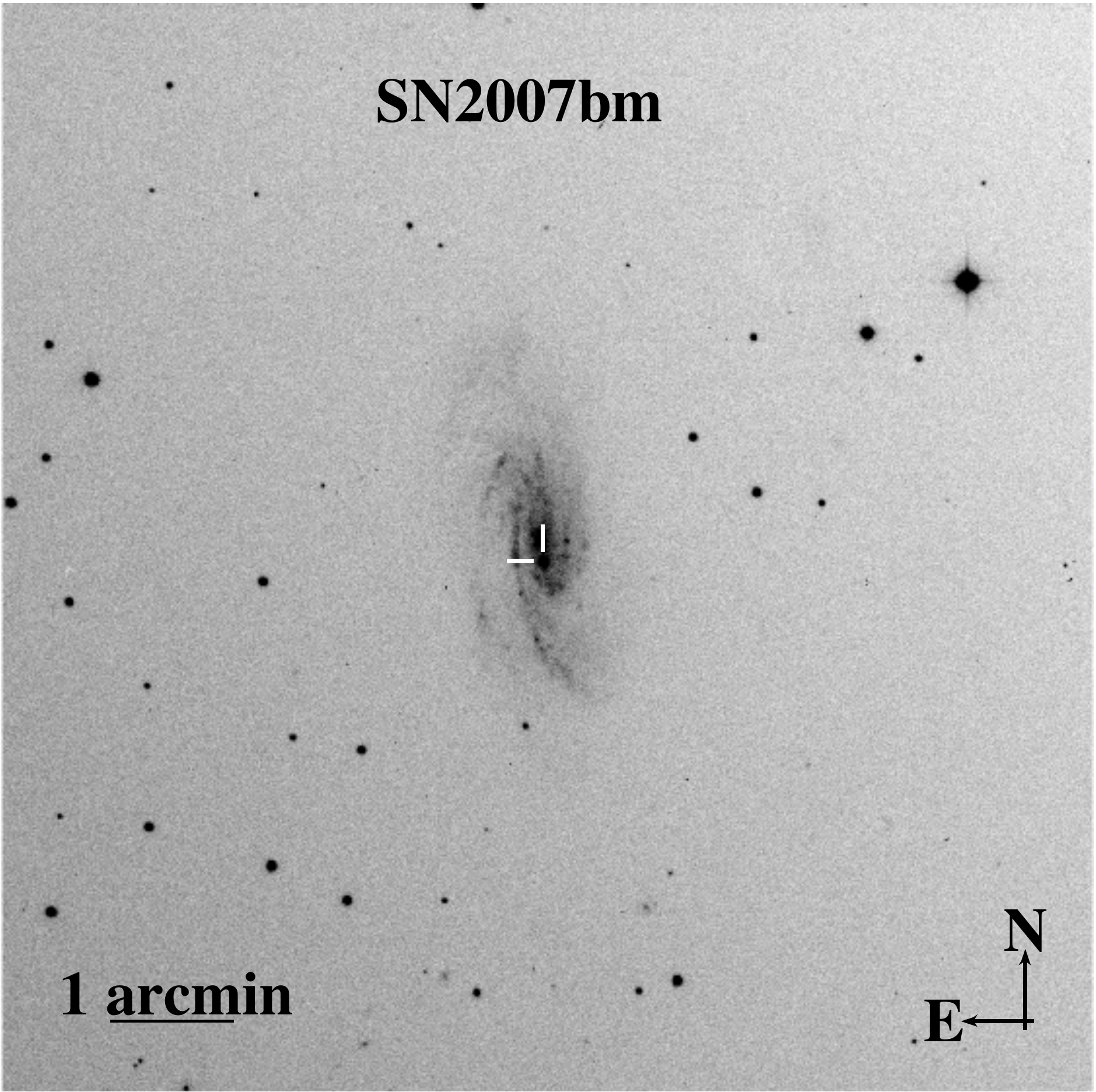}{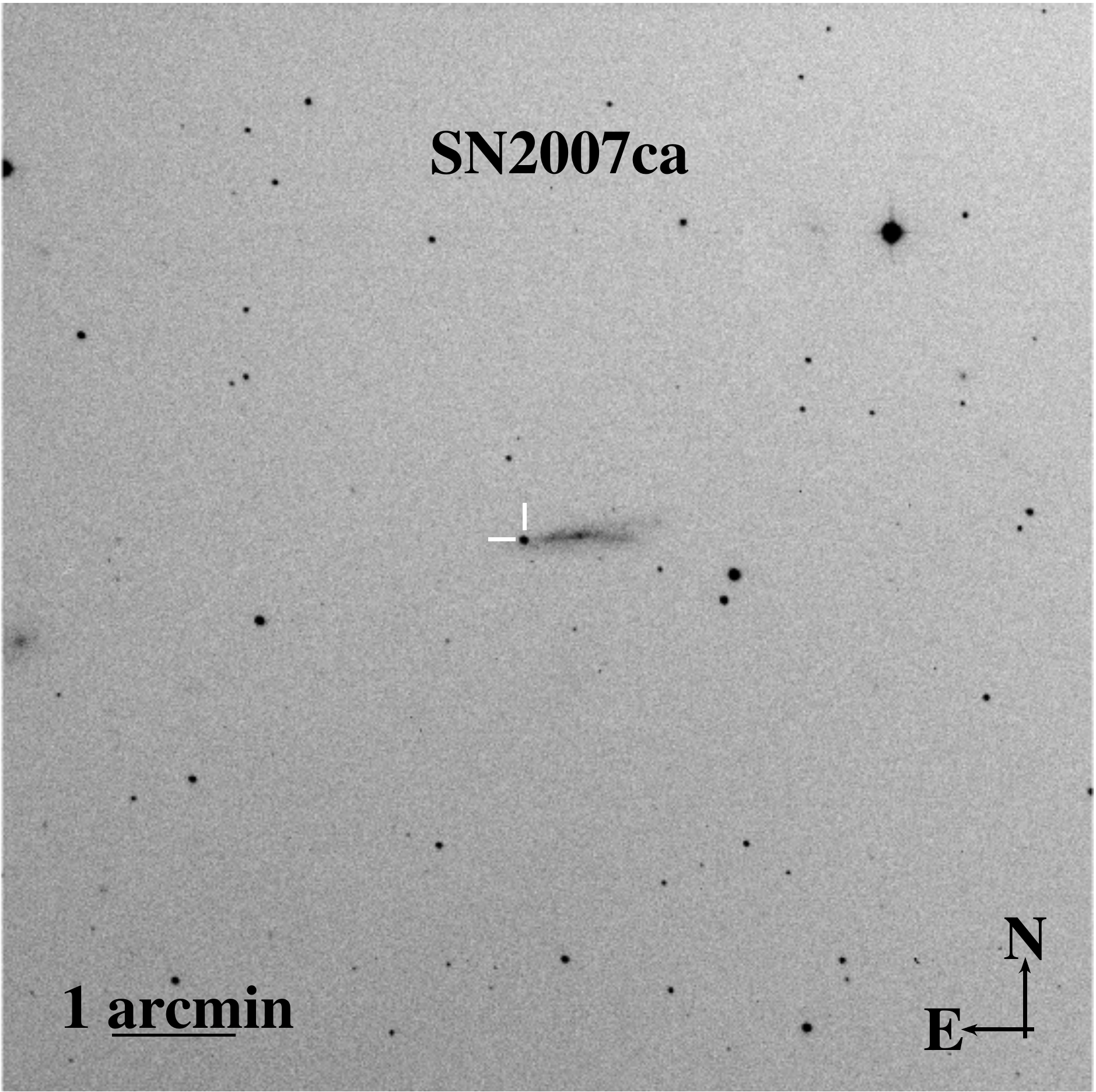}
\newline
\plottwo{charts/SN2006hx.pdf}{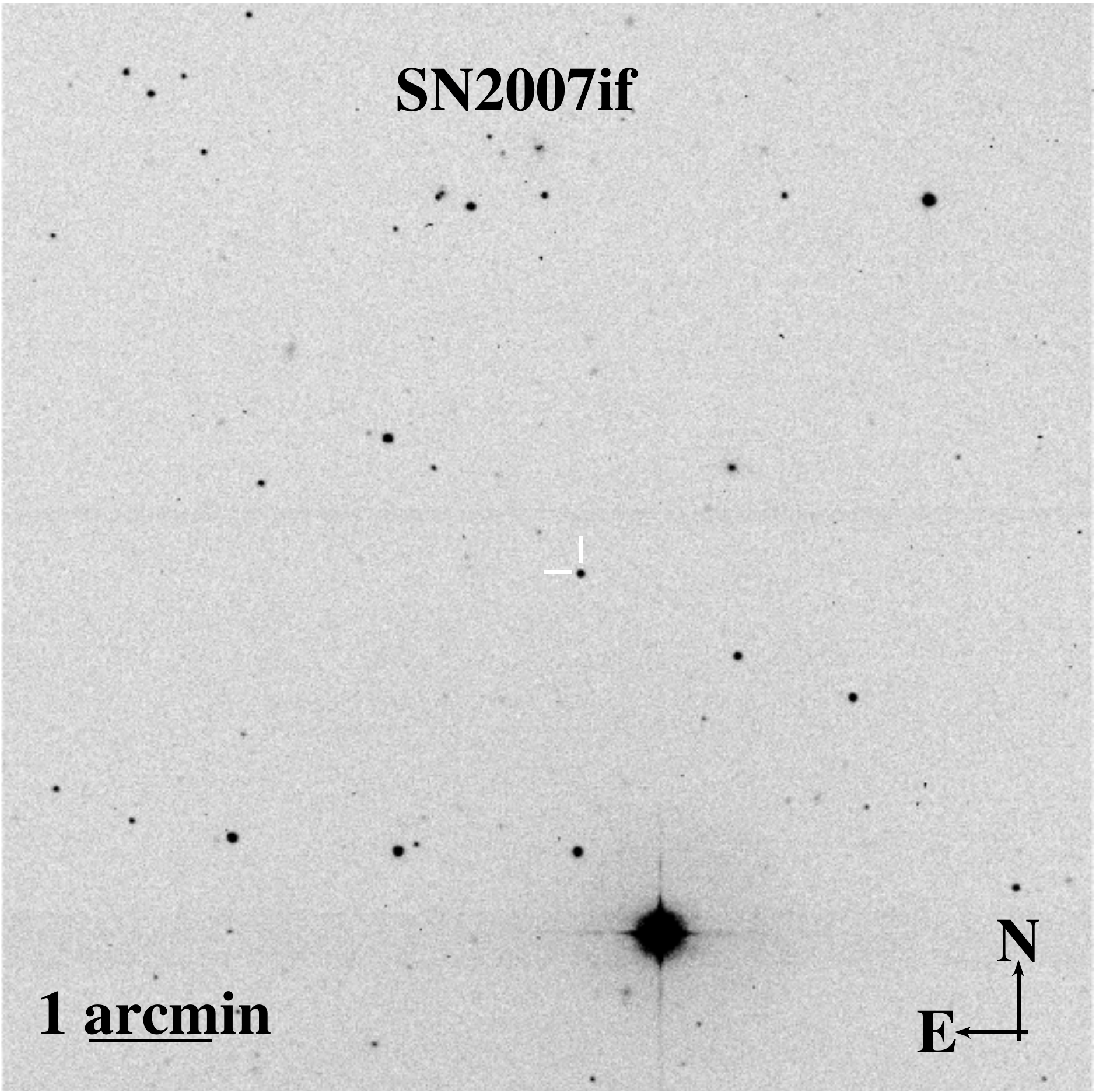}
\plottwo{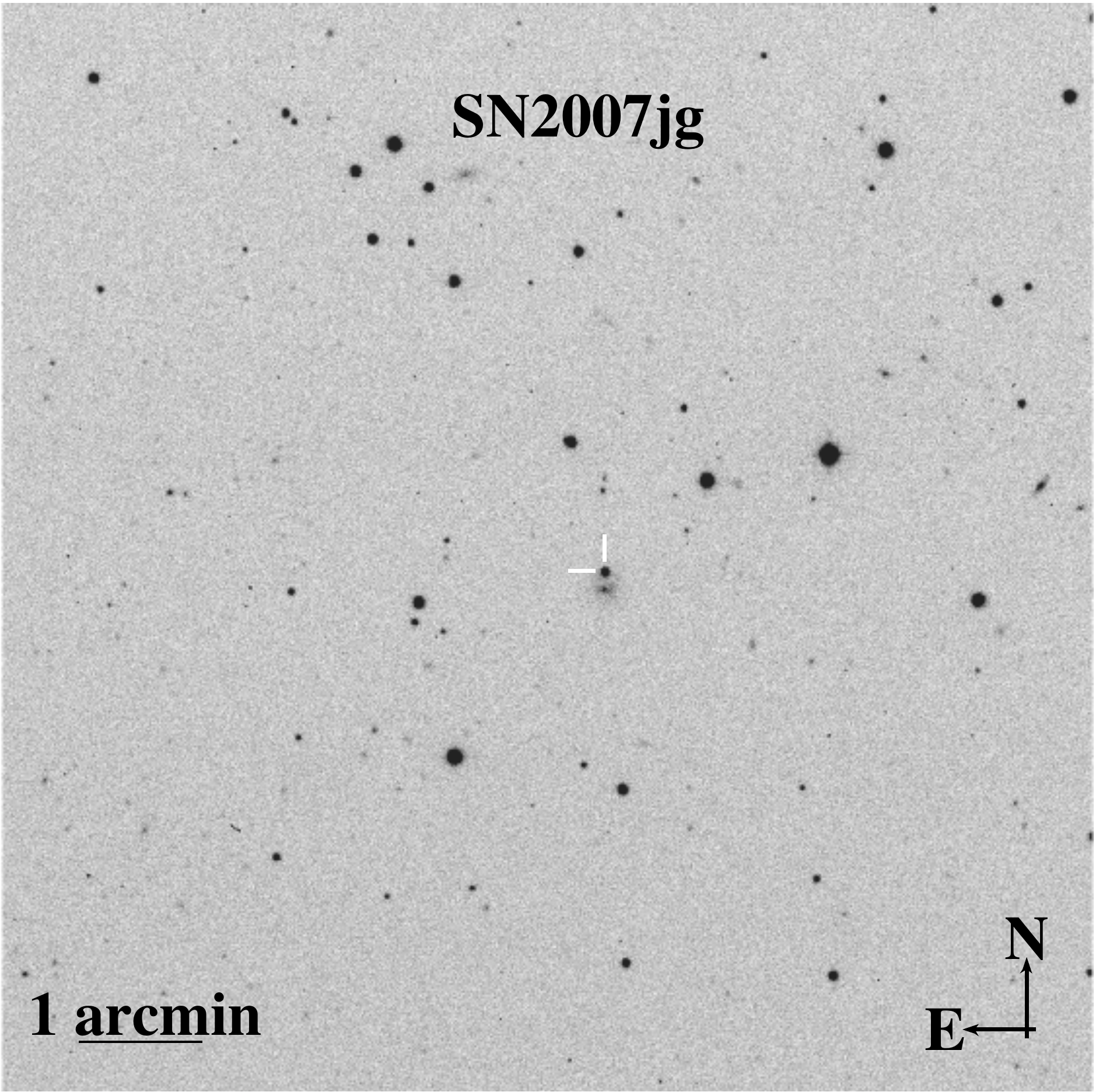}{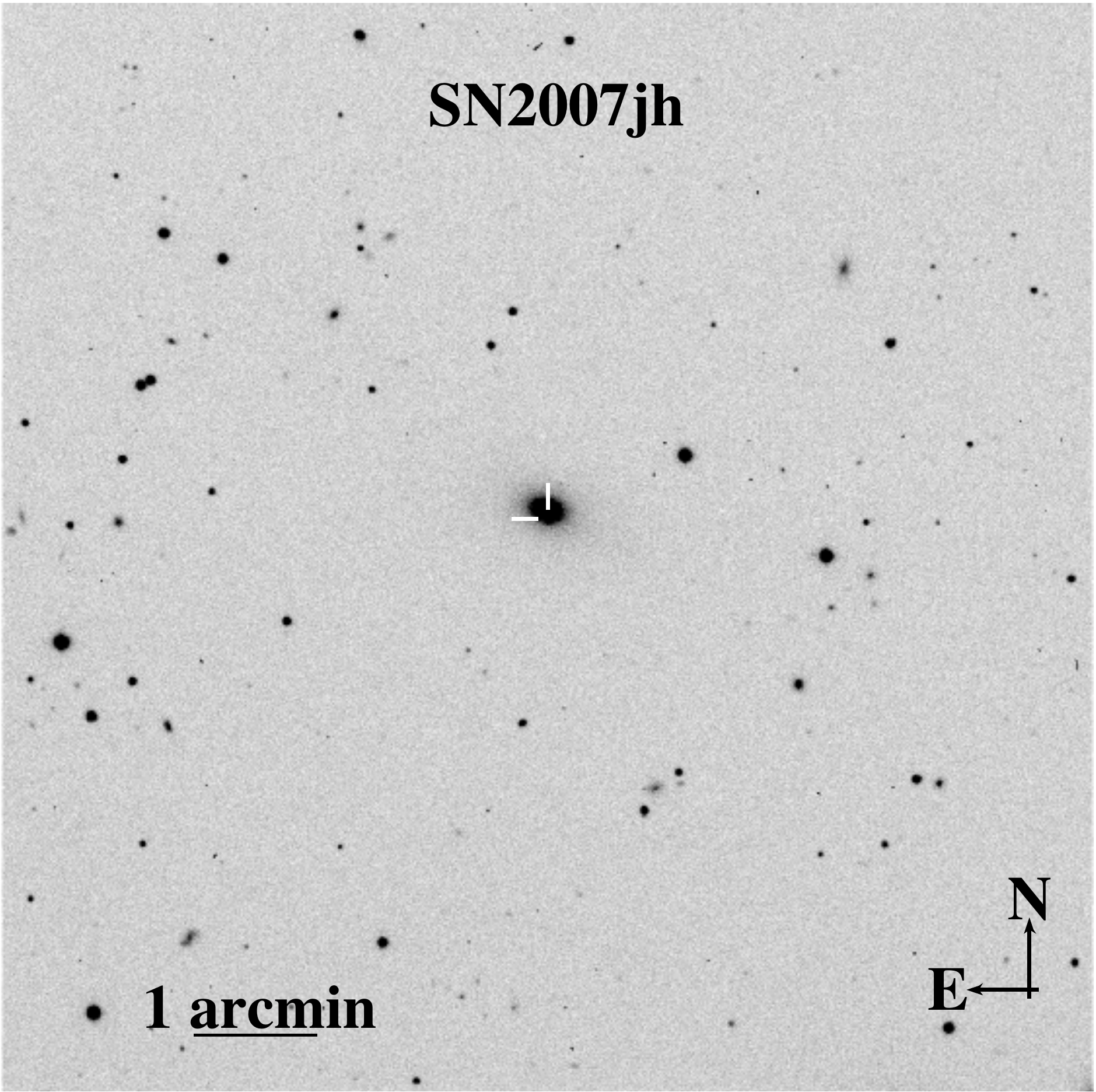}
\newline
\plottwo{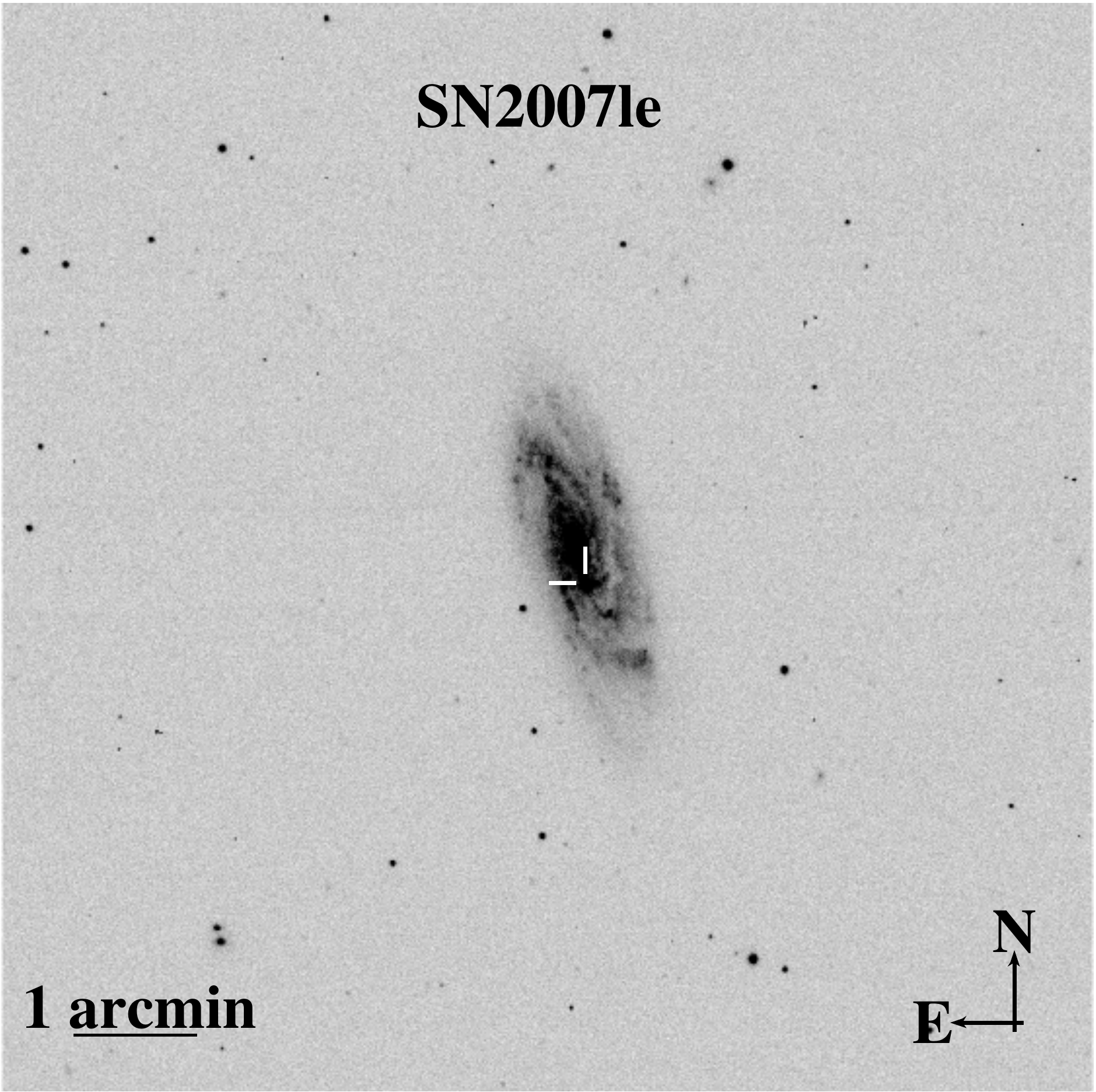}{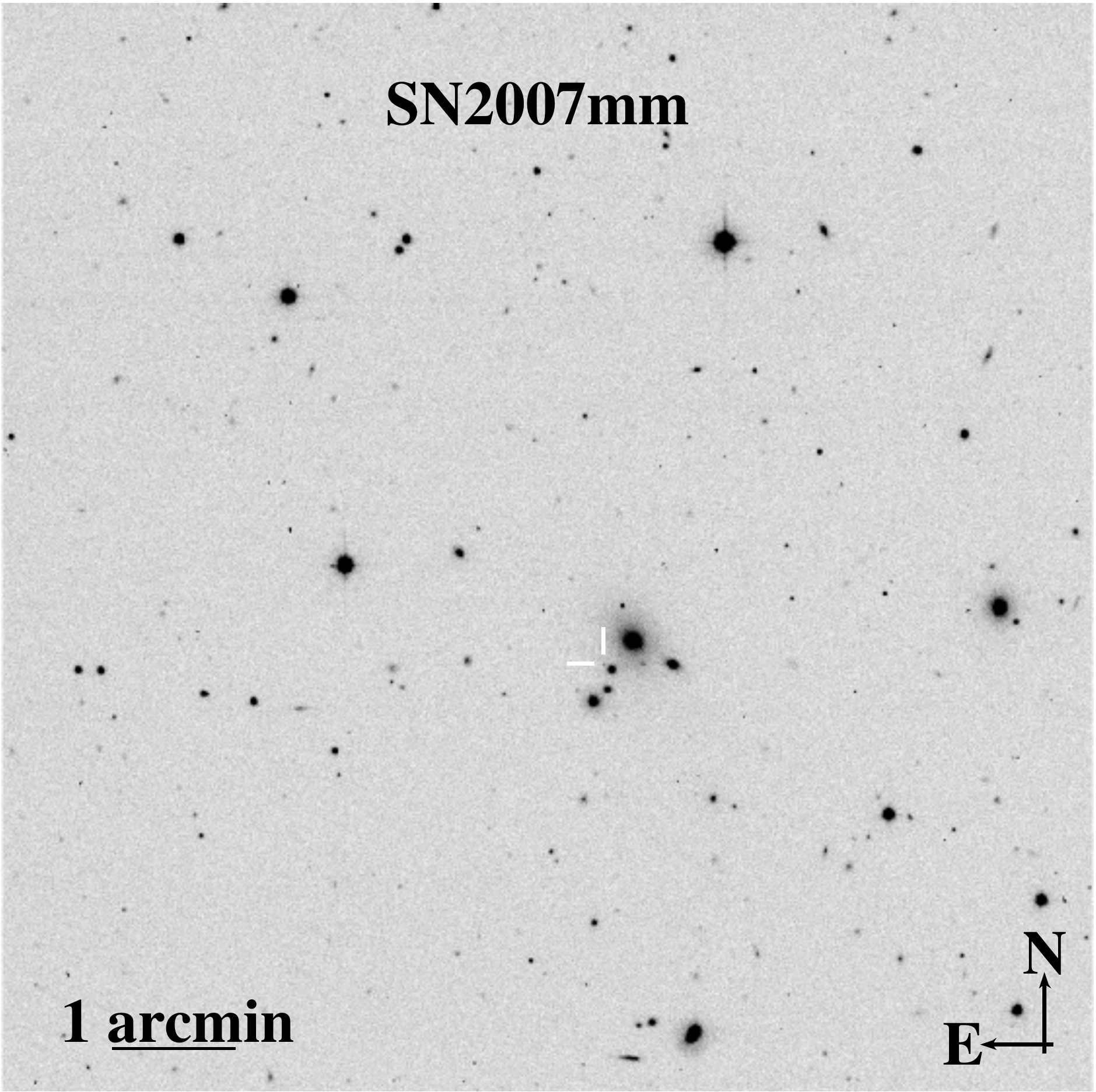}
\plottwo{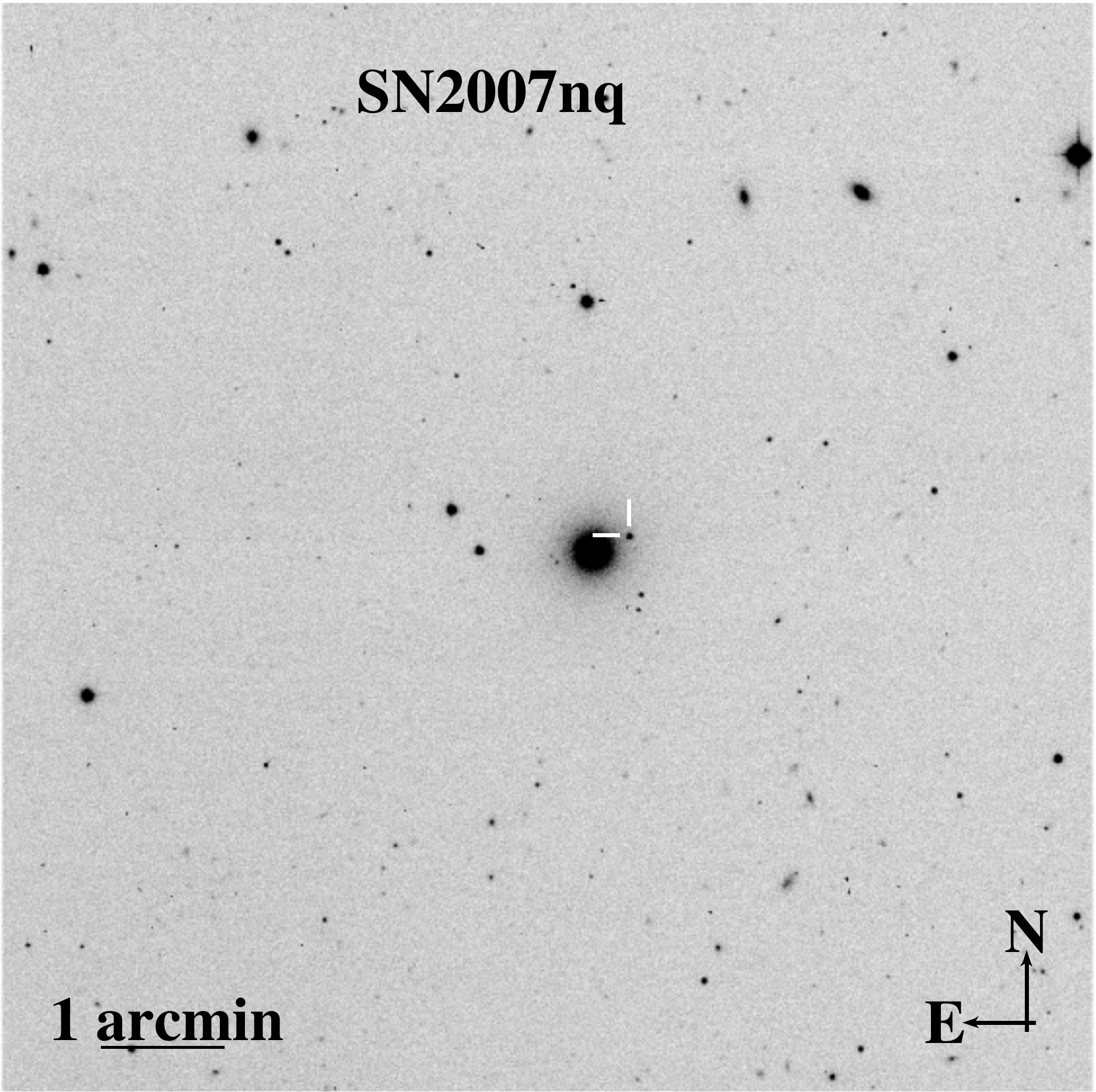}{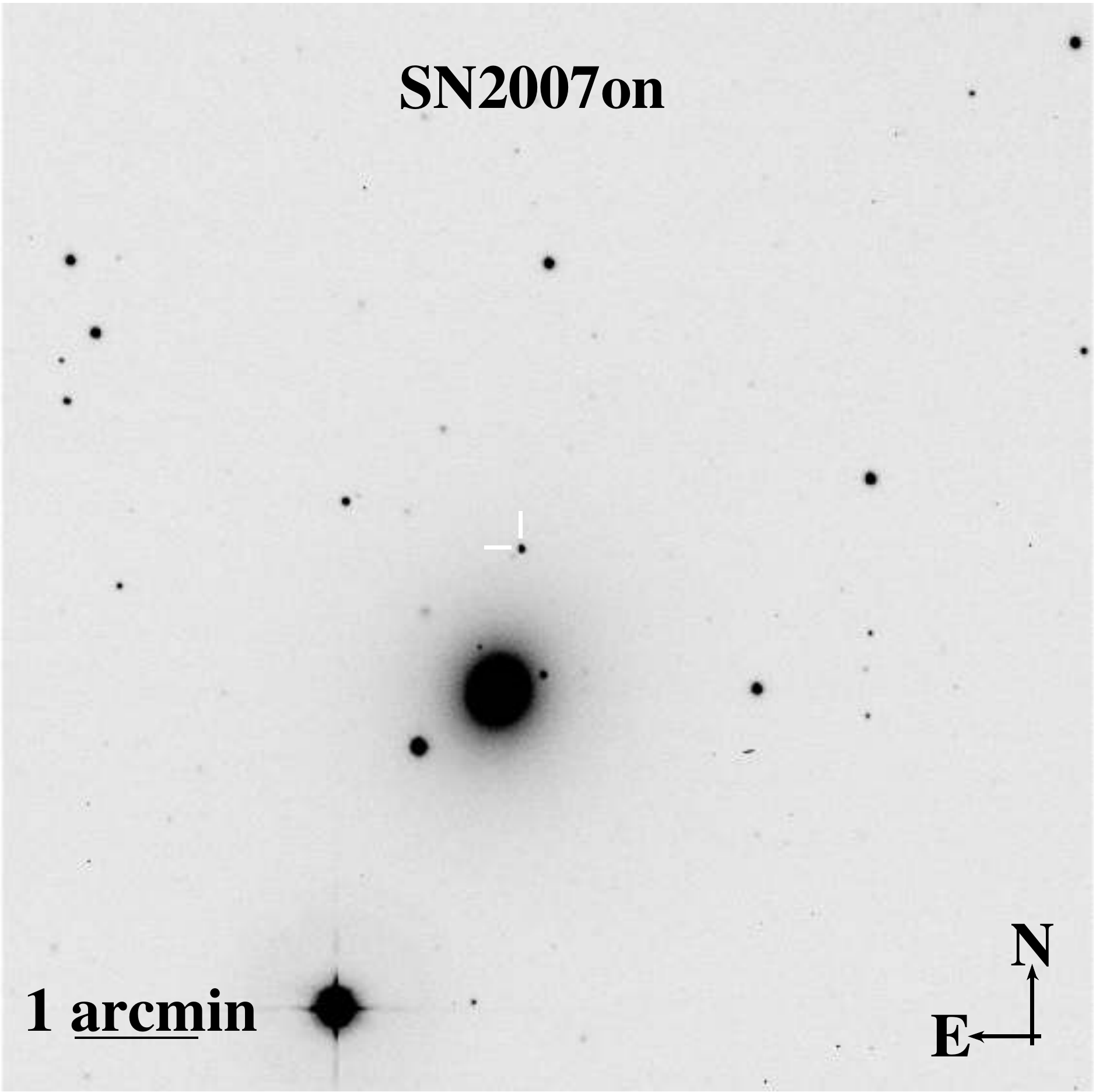}

{\center Stritzinger {\it et al.} Fig.~\ref{fig:fcharts}}
\end{figure}
\clearpage
\newpage

\begin{figure}[t]
\epsscale{.54}
\plottwo{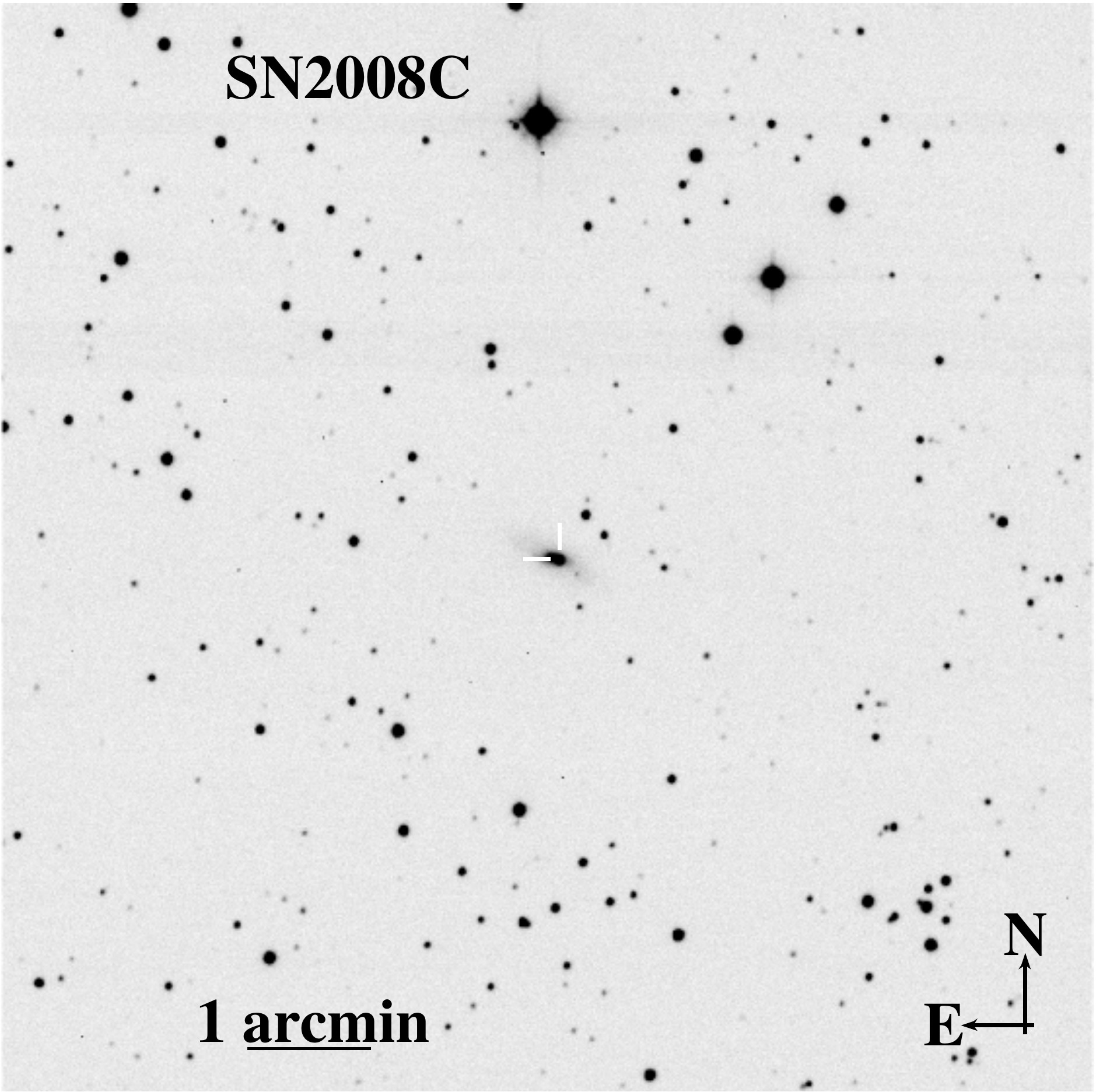}{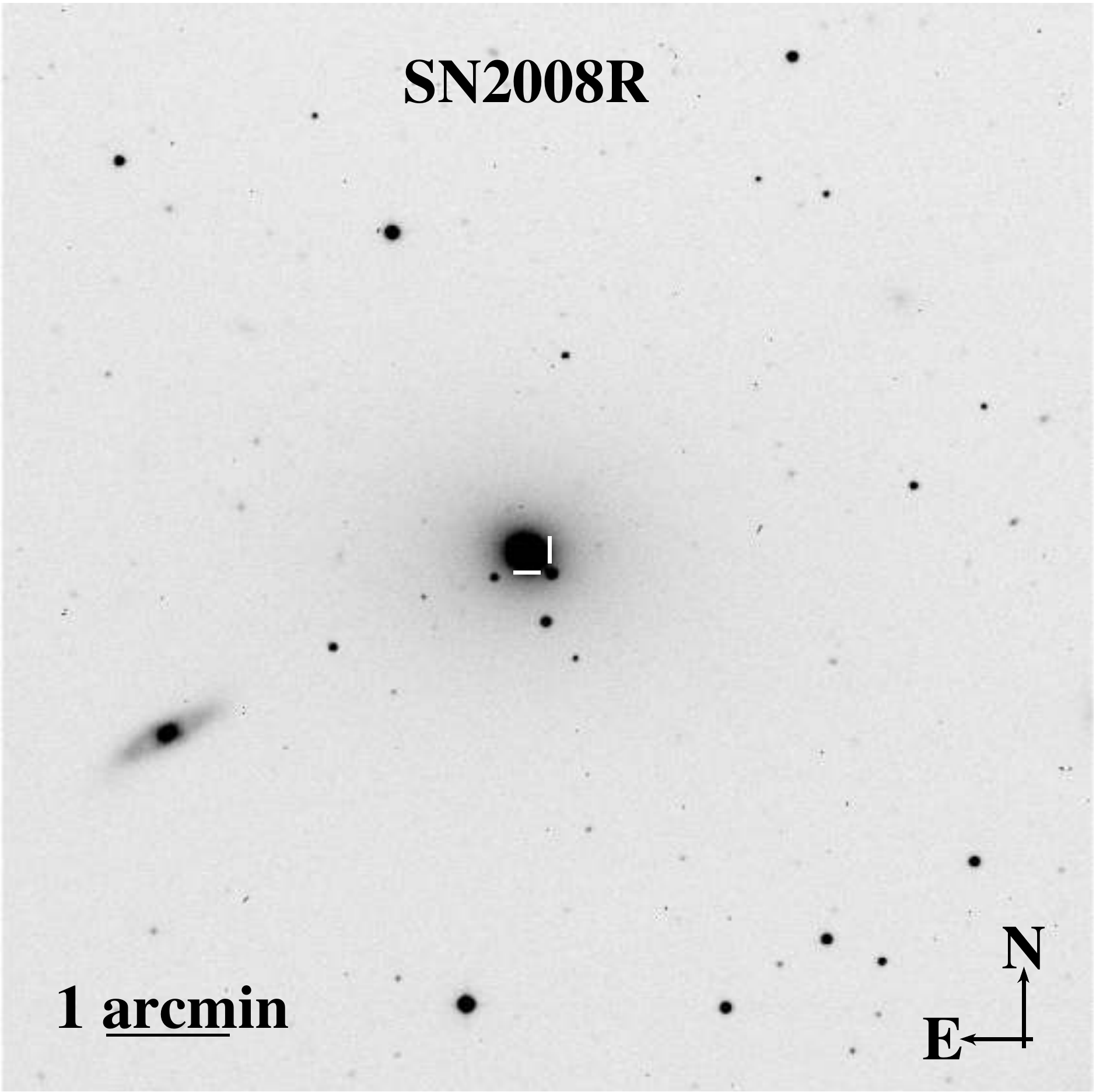}
\plottwo{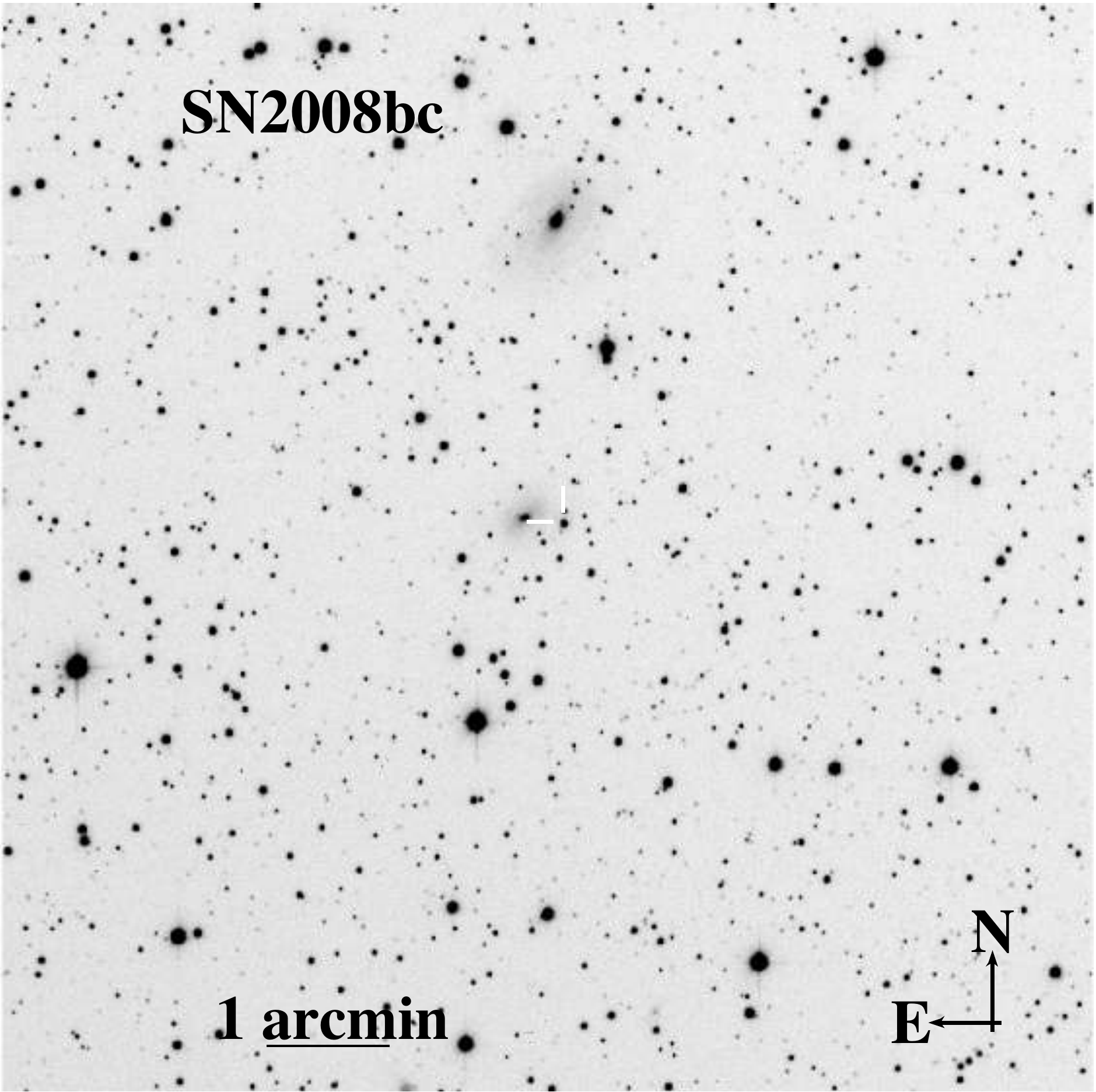}{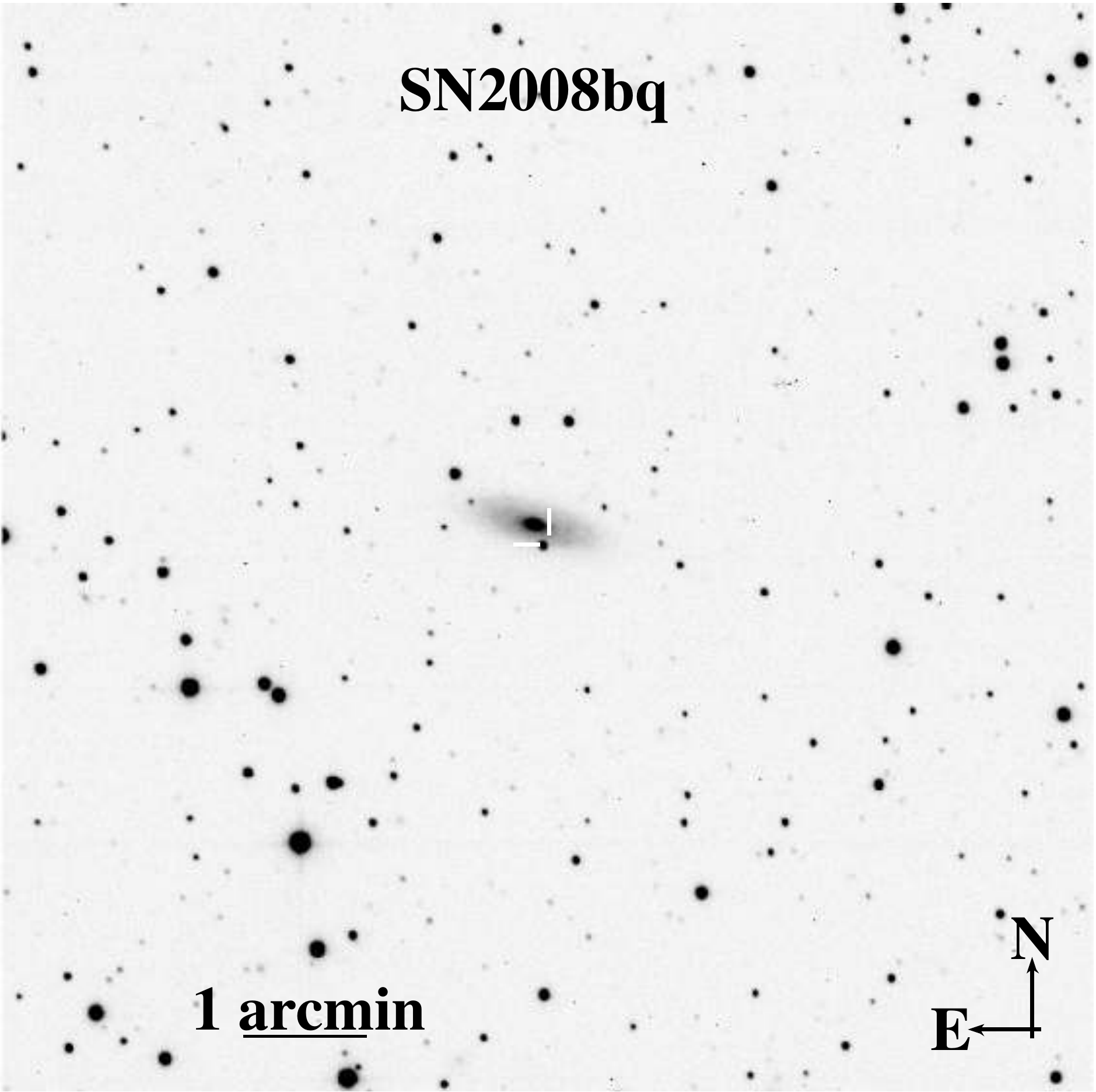}
\newline                                                                     
\plottwo{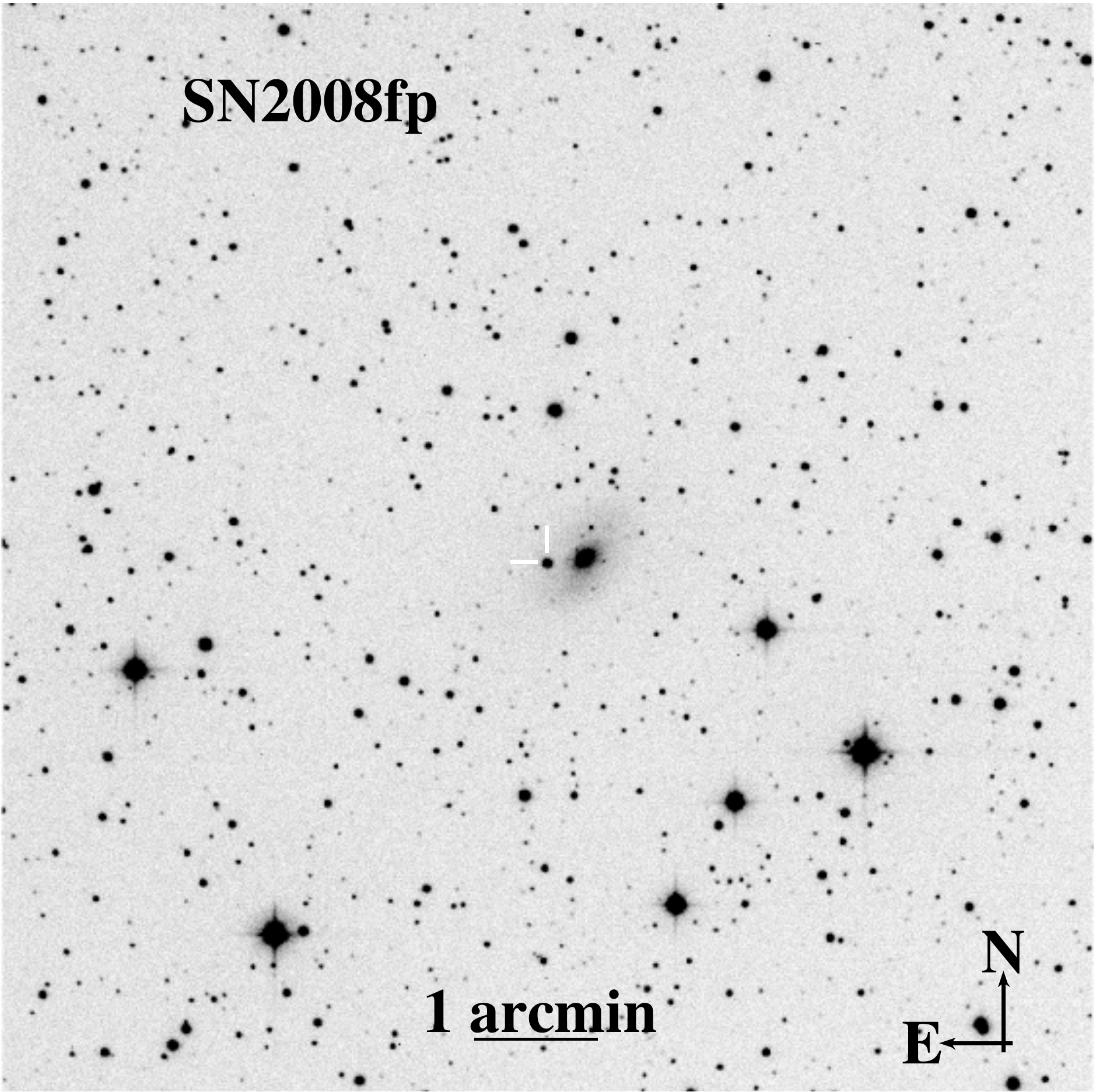}{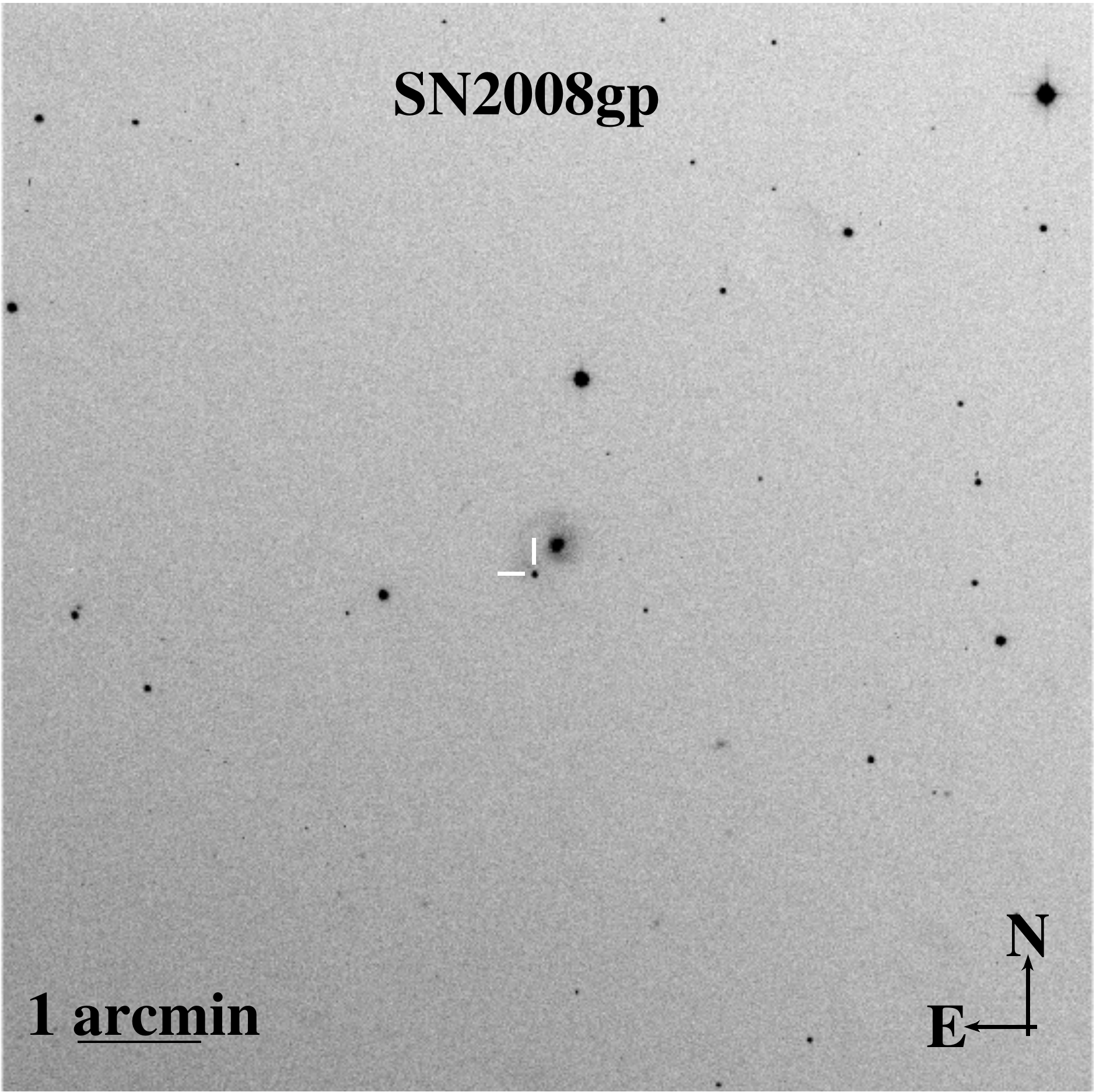}
\plottwo{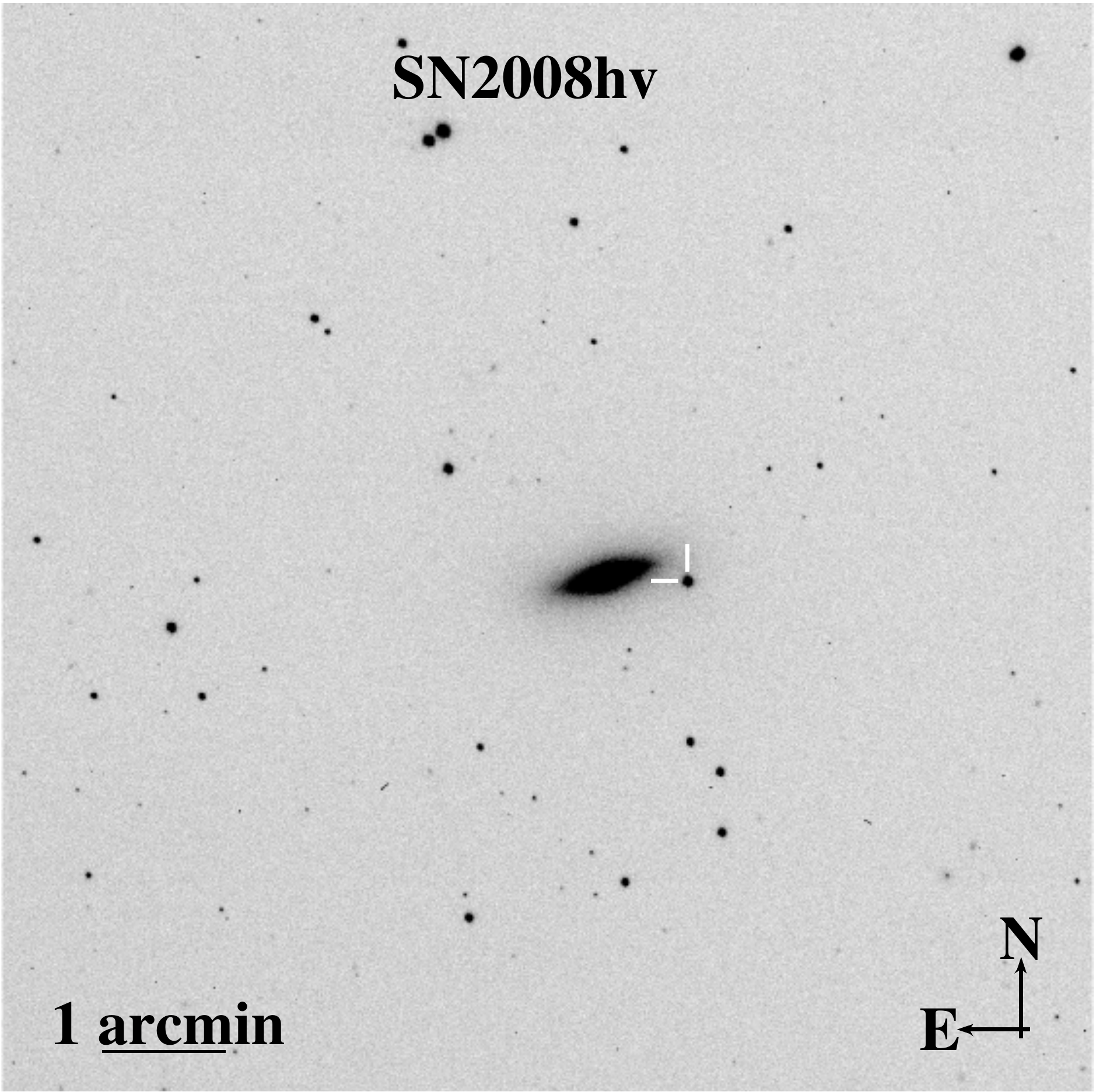}{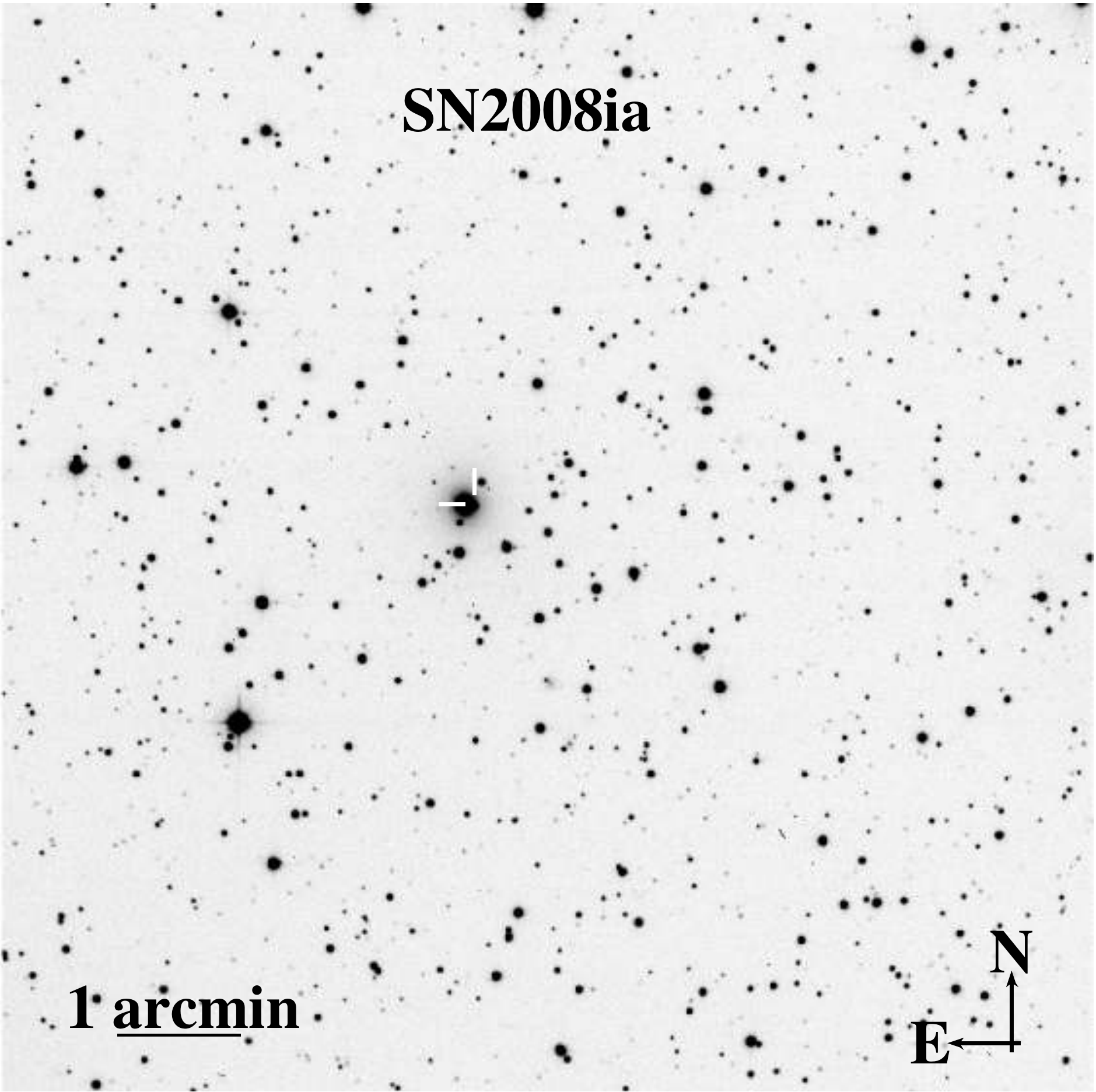}
\newline                                                                     
\plottwo{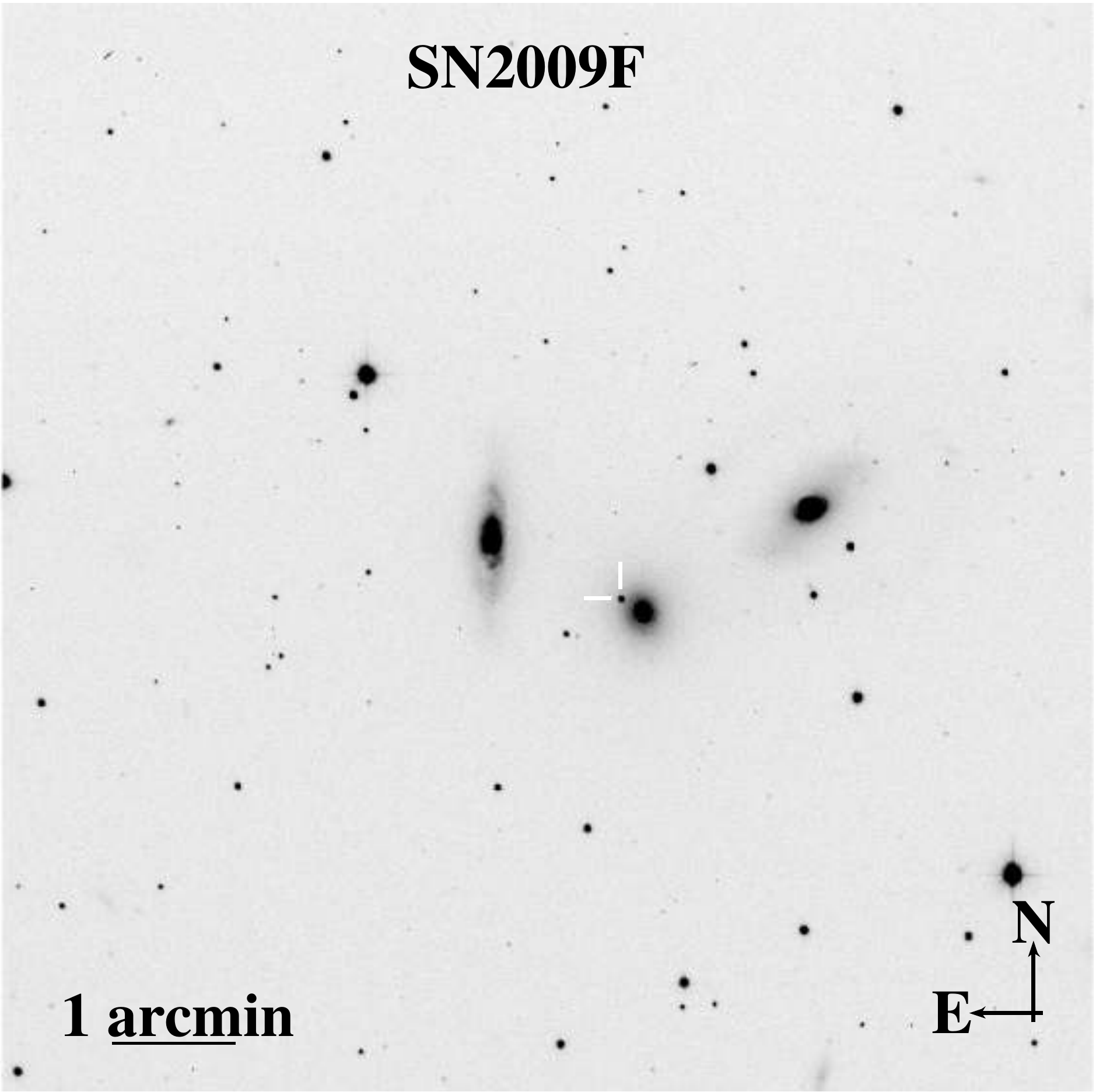}{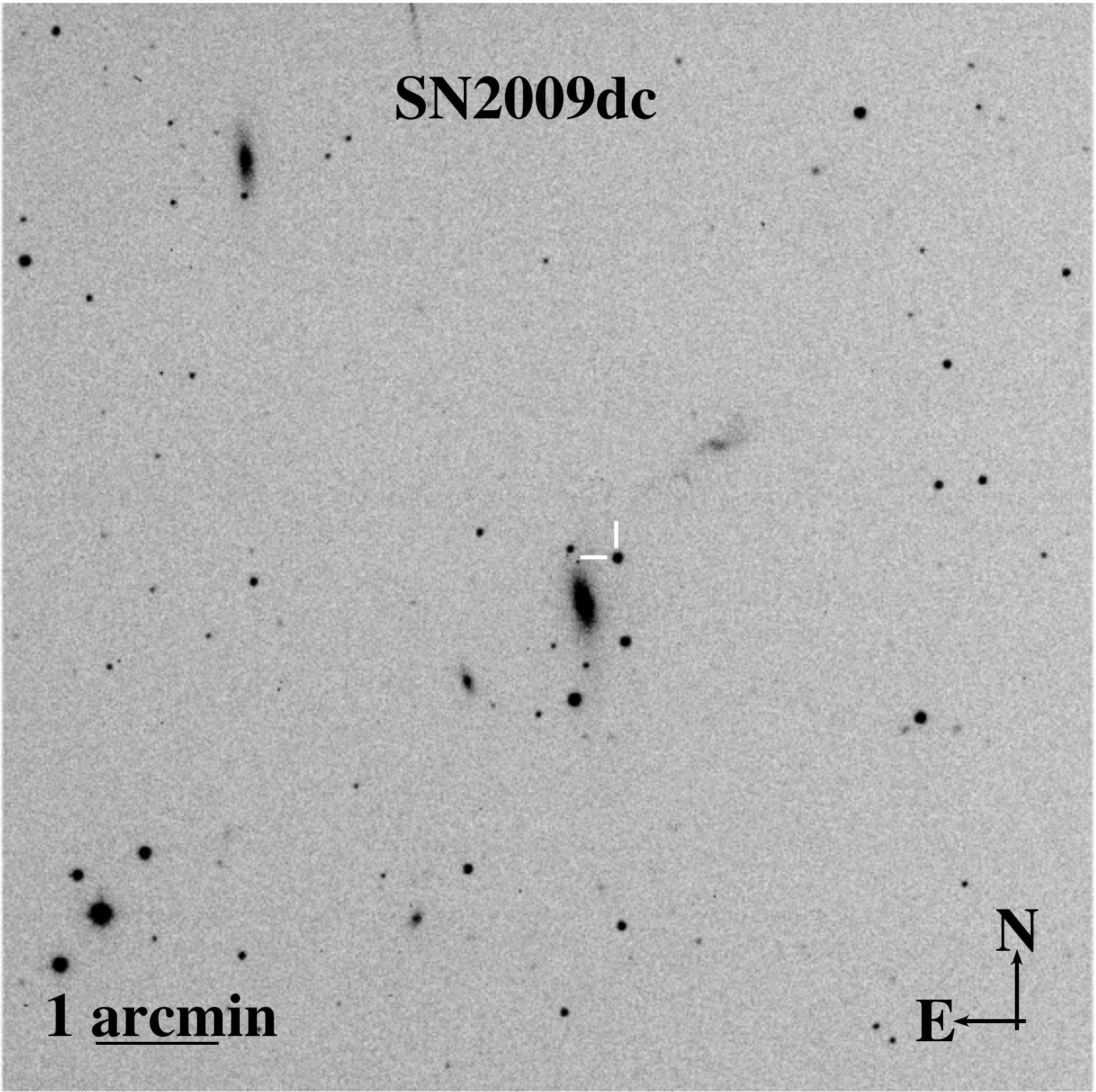}
{\center Stritzinger {\it et al.} Fig.~\ref{fig:fcharts}}
\end{figure}
\clearpage
\newpage

\clearpage
\begin{figure}[t]
\epsscale{1.0}
\plotone{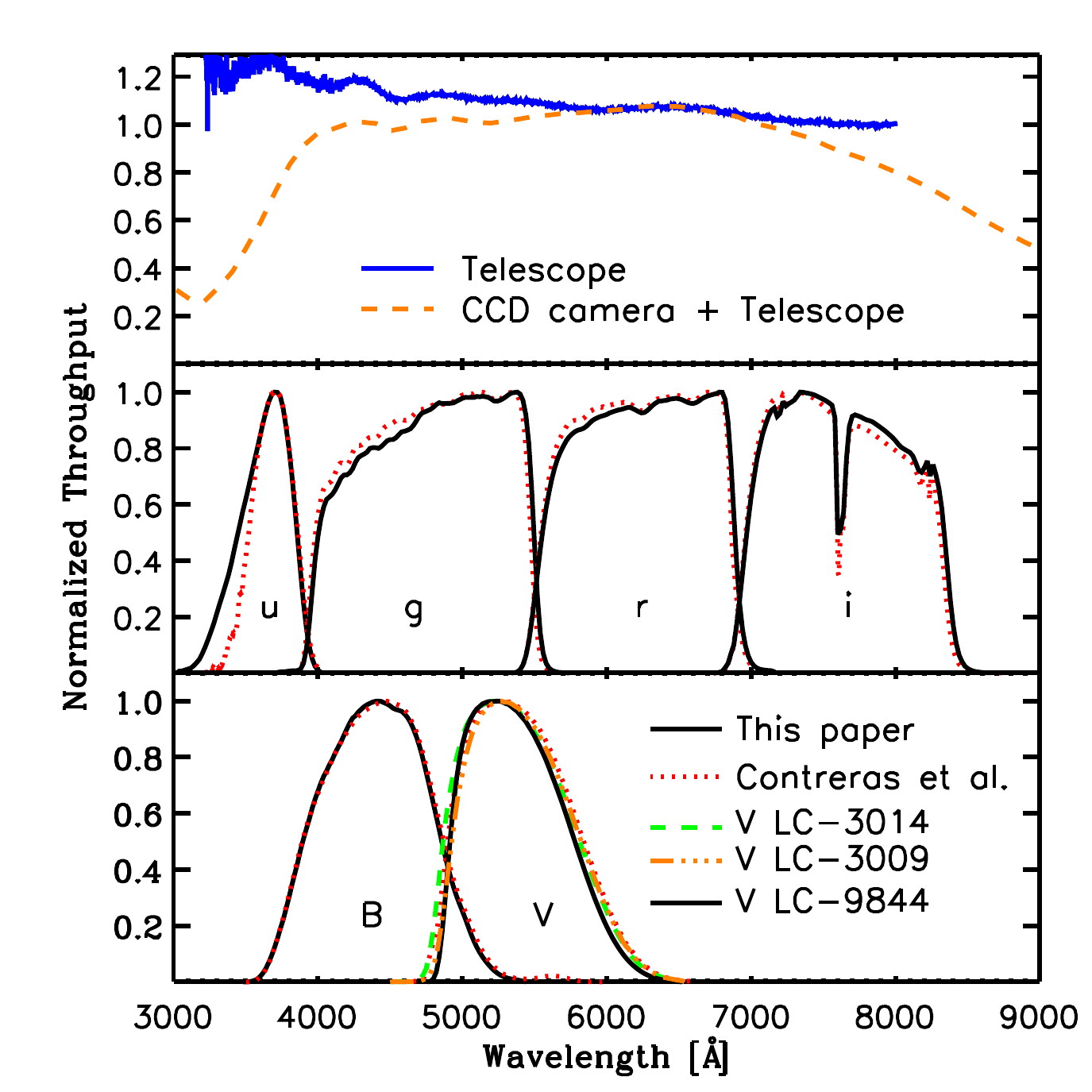}{\center Stritzinger {\it et al.} Fig.~\ref{fig:optfilters}}
\end{figure}

\clearpage
\newpage
\begin{figure}[t]
 \plottwo{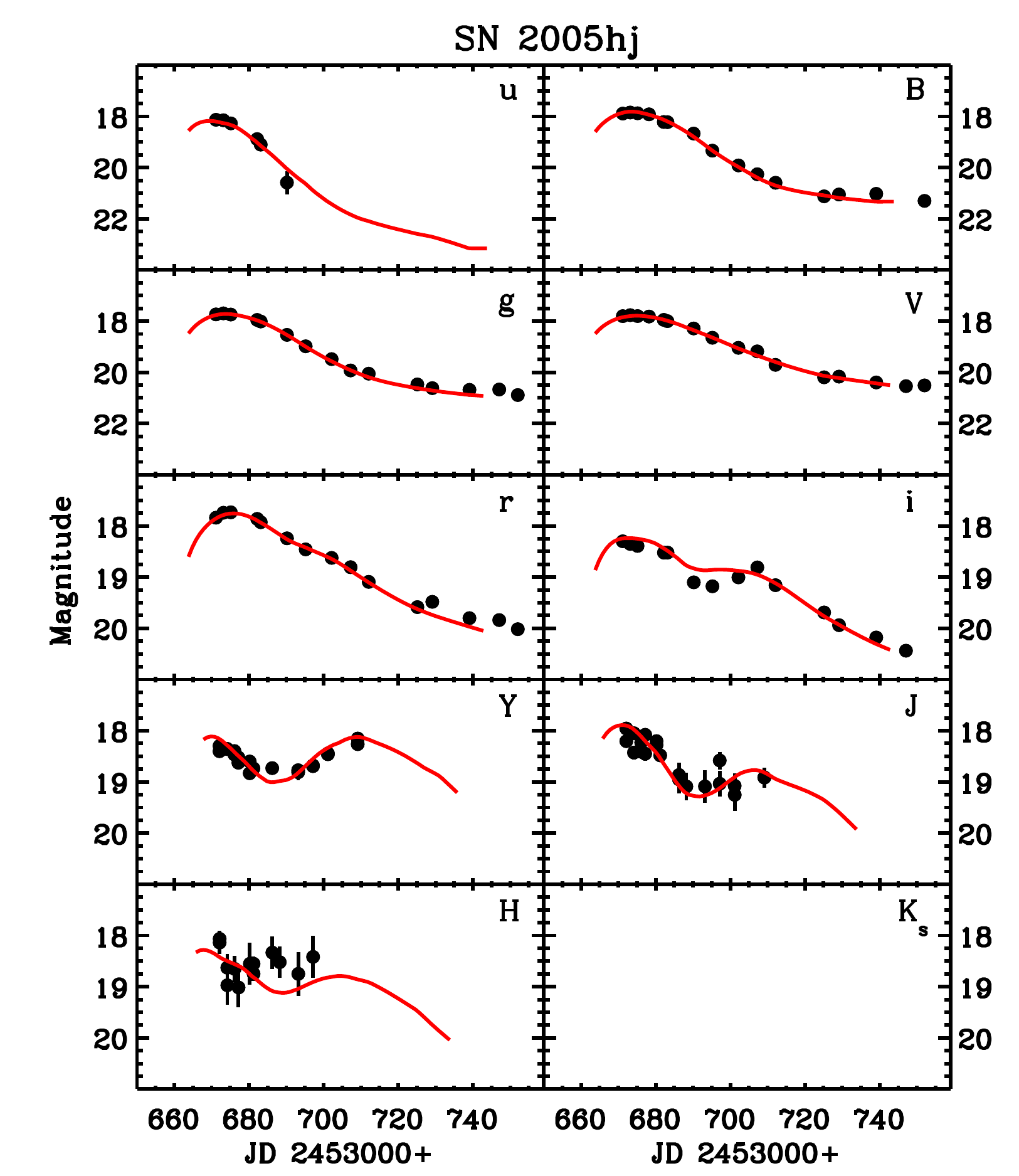}{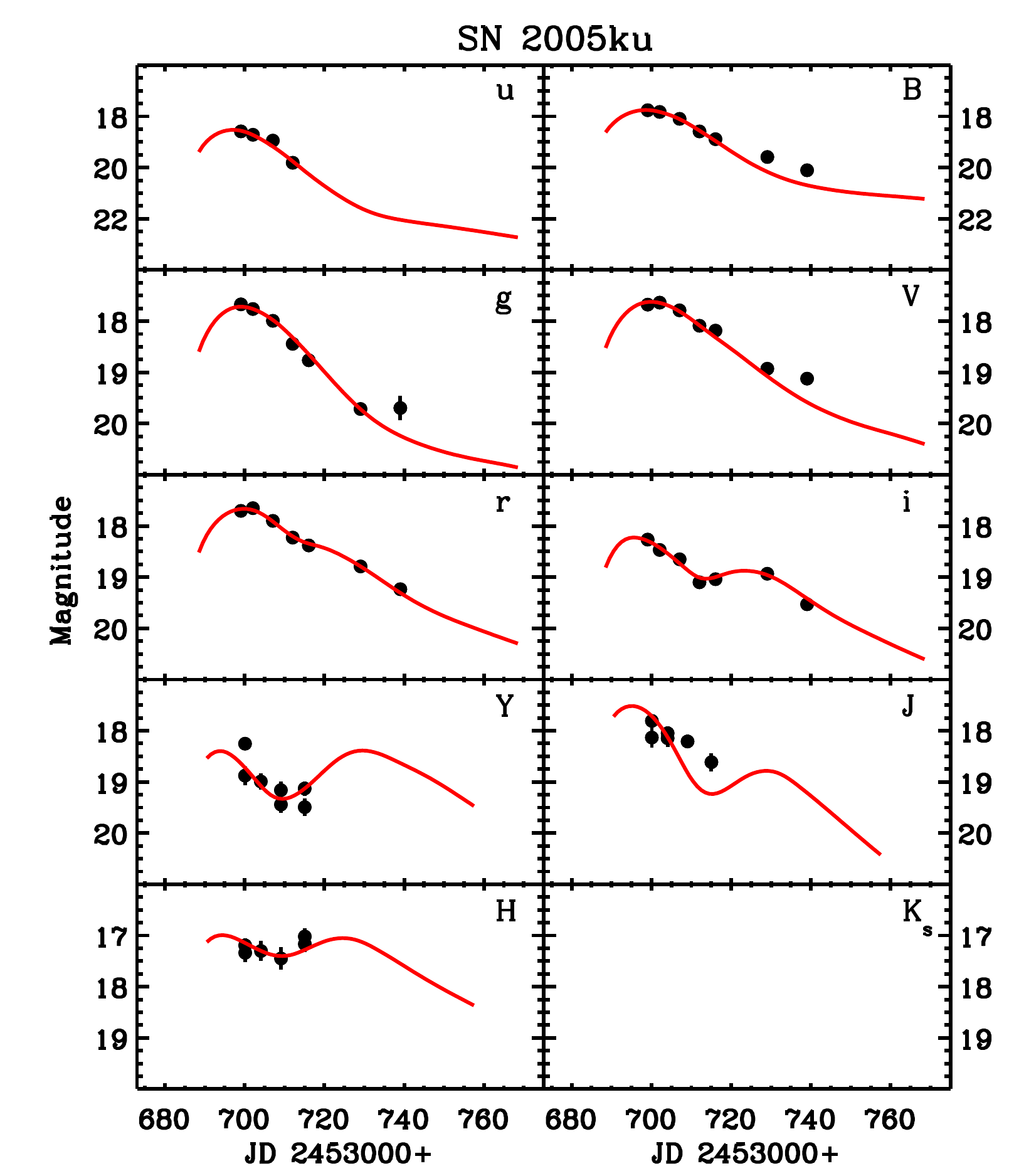}
 \newline
   \newline
\plottwo{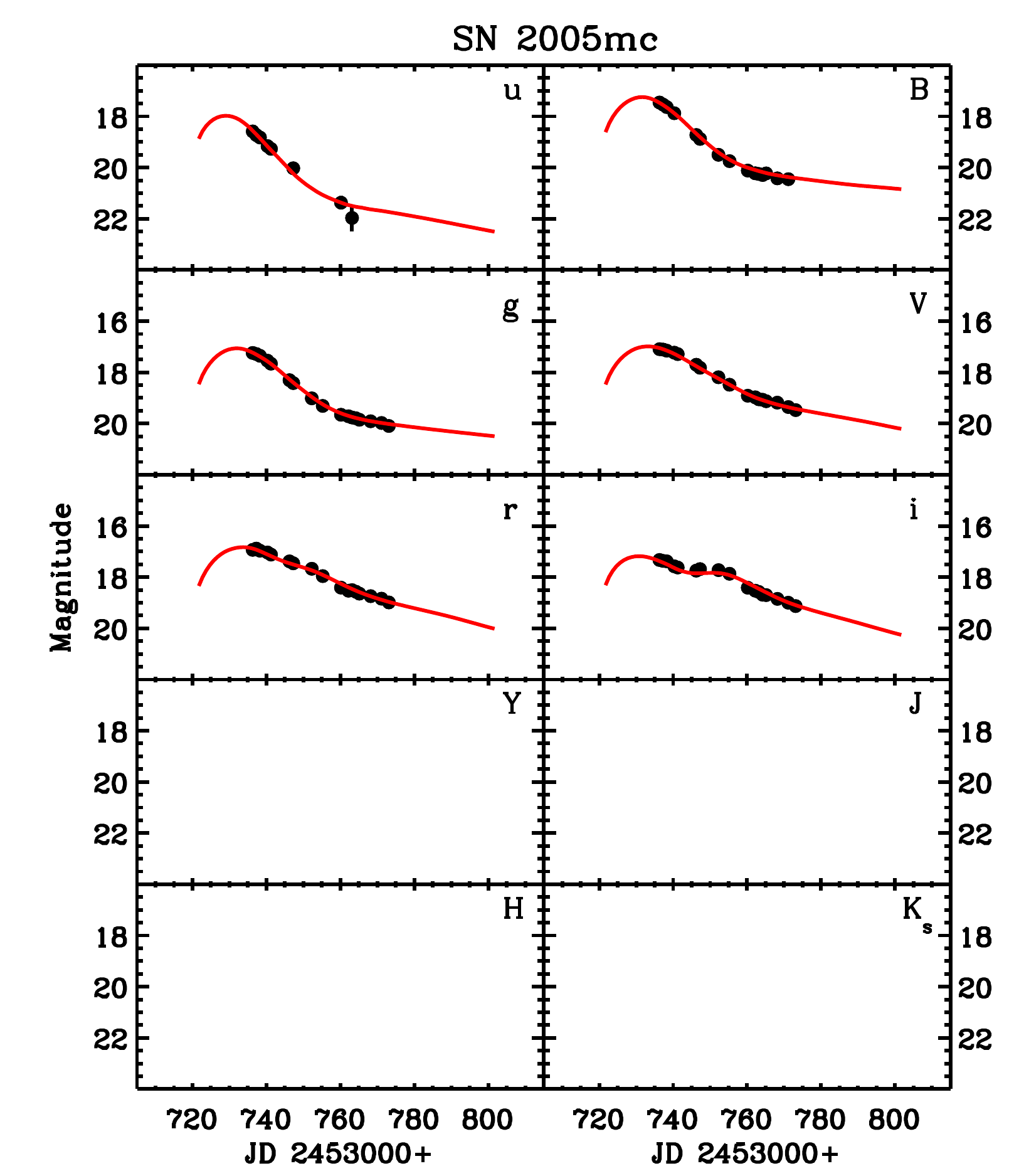}{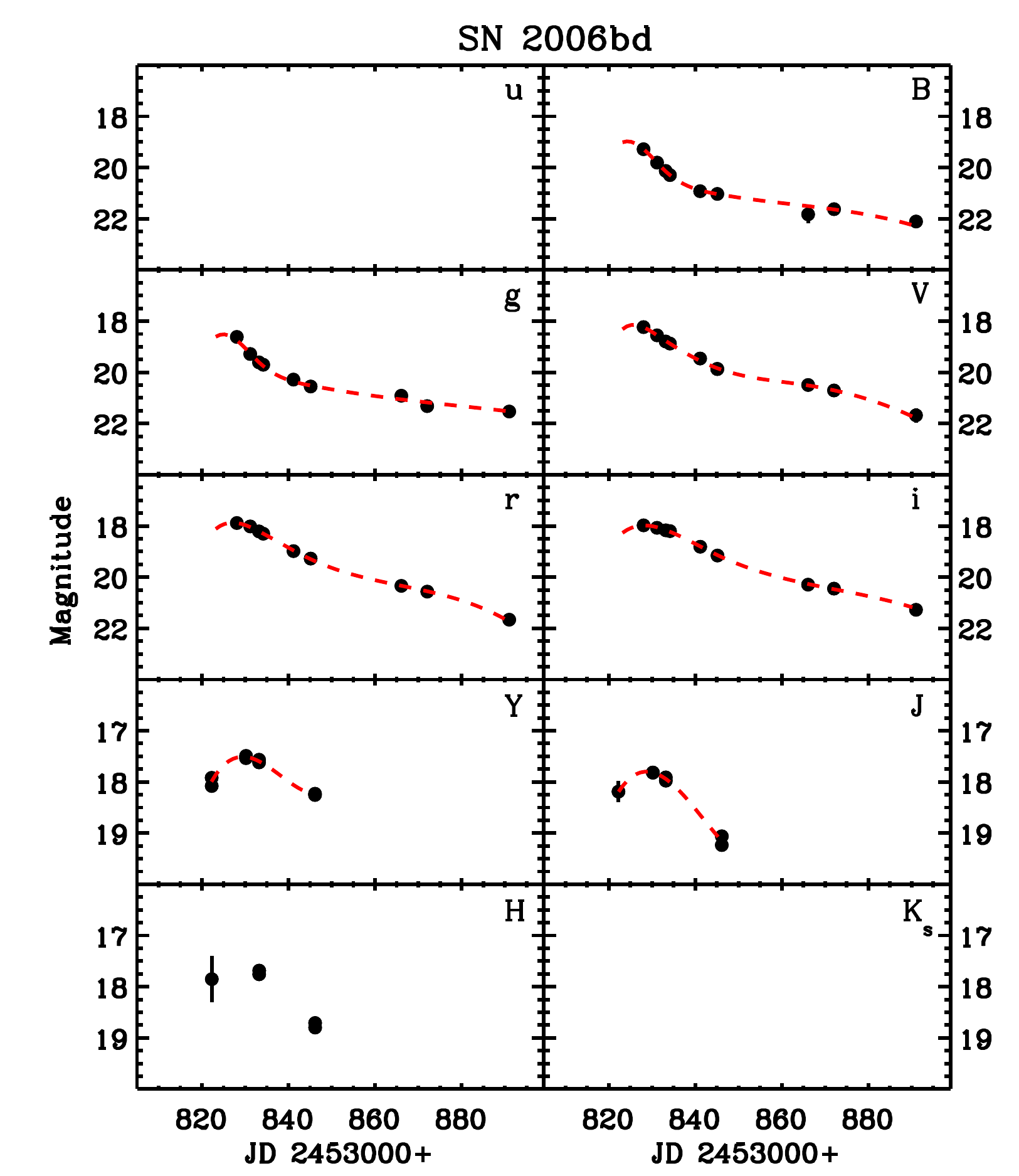}
  {\center Stritzinger {\it et al.} Fig. \ref{fig:flcurves}}
\end{figure}

\clearpage
\newpage
\begin{figure}[t]
 \plottwo{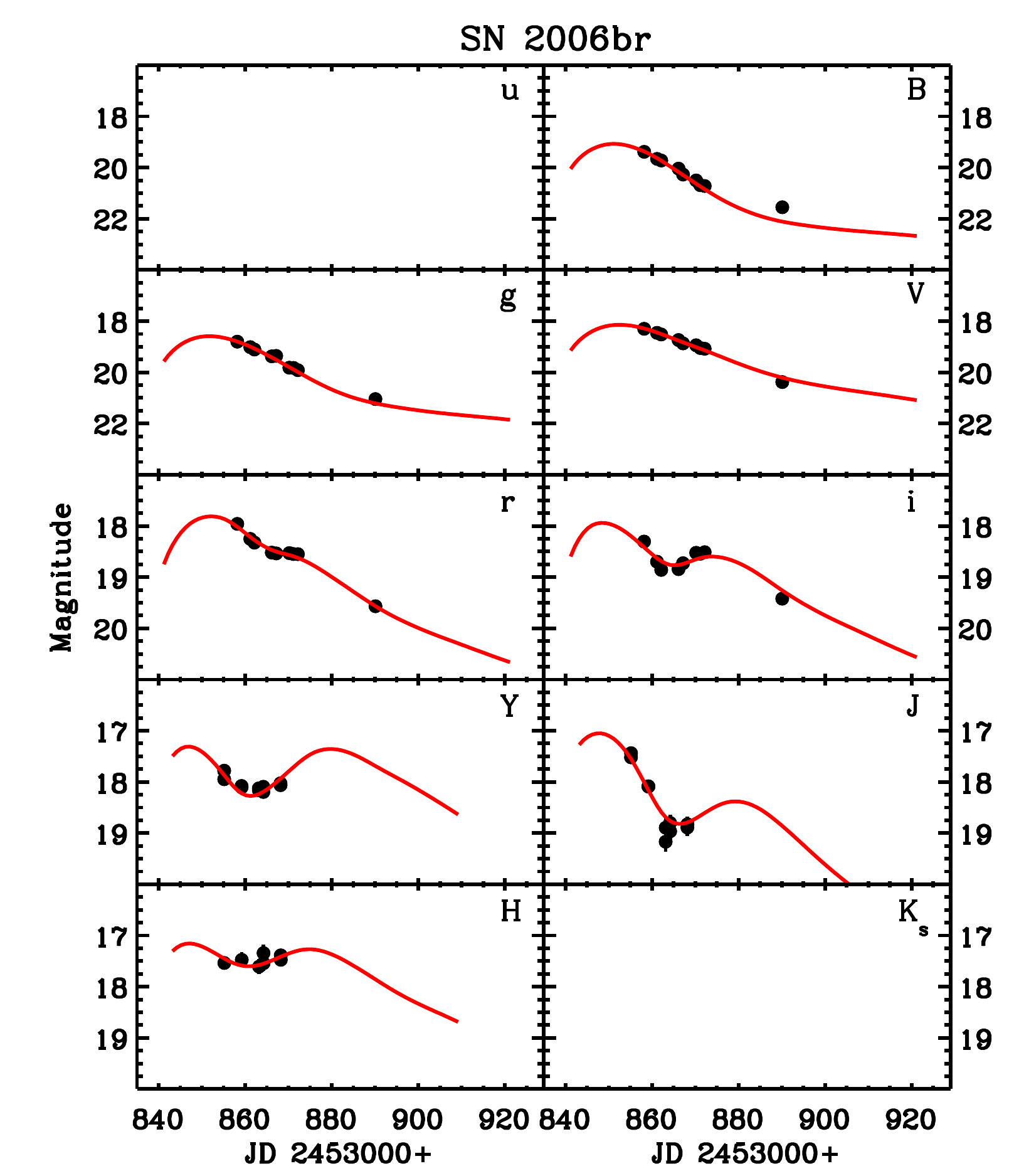}{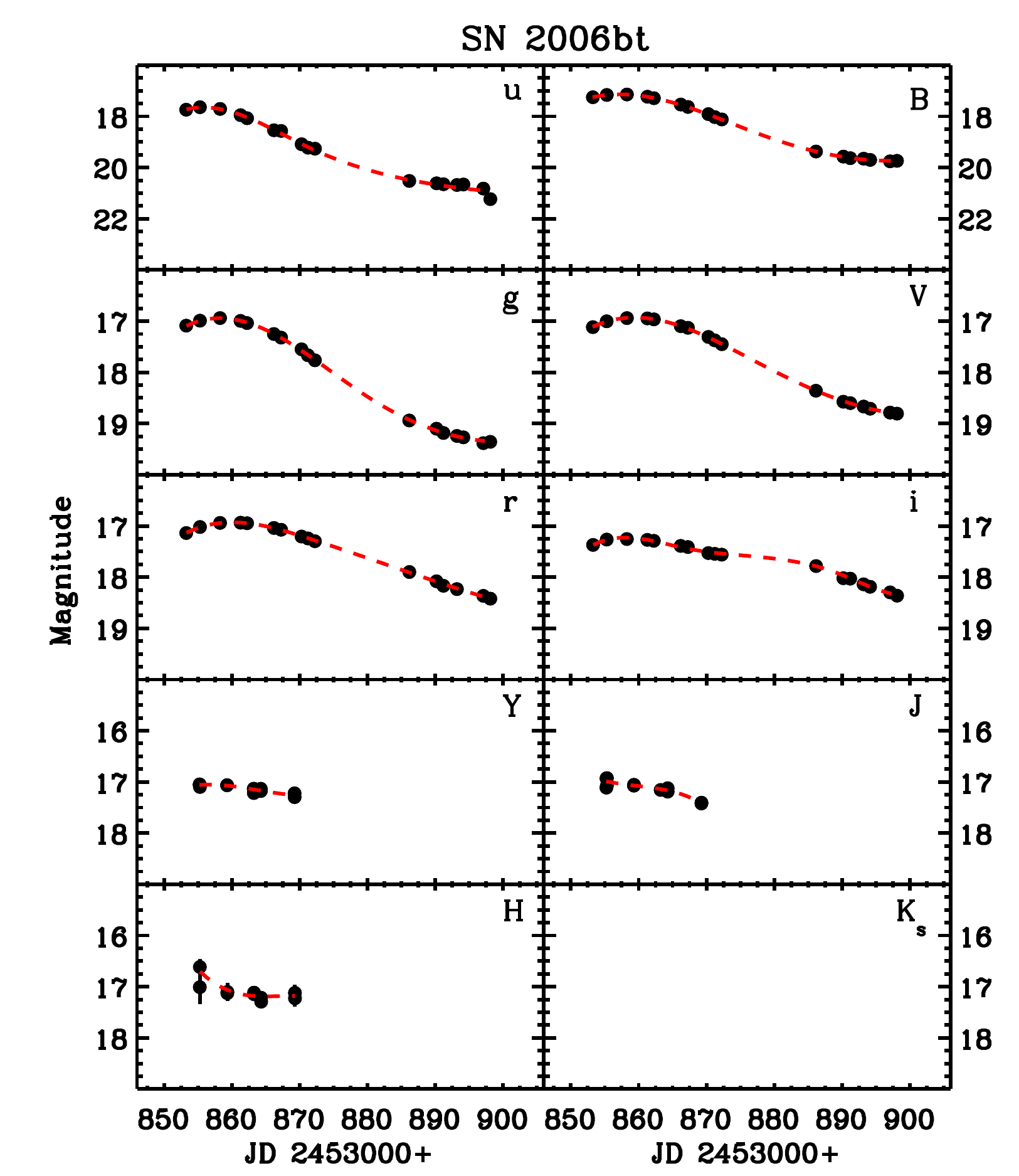}
 \newline
   \newline
\plottwo{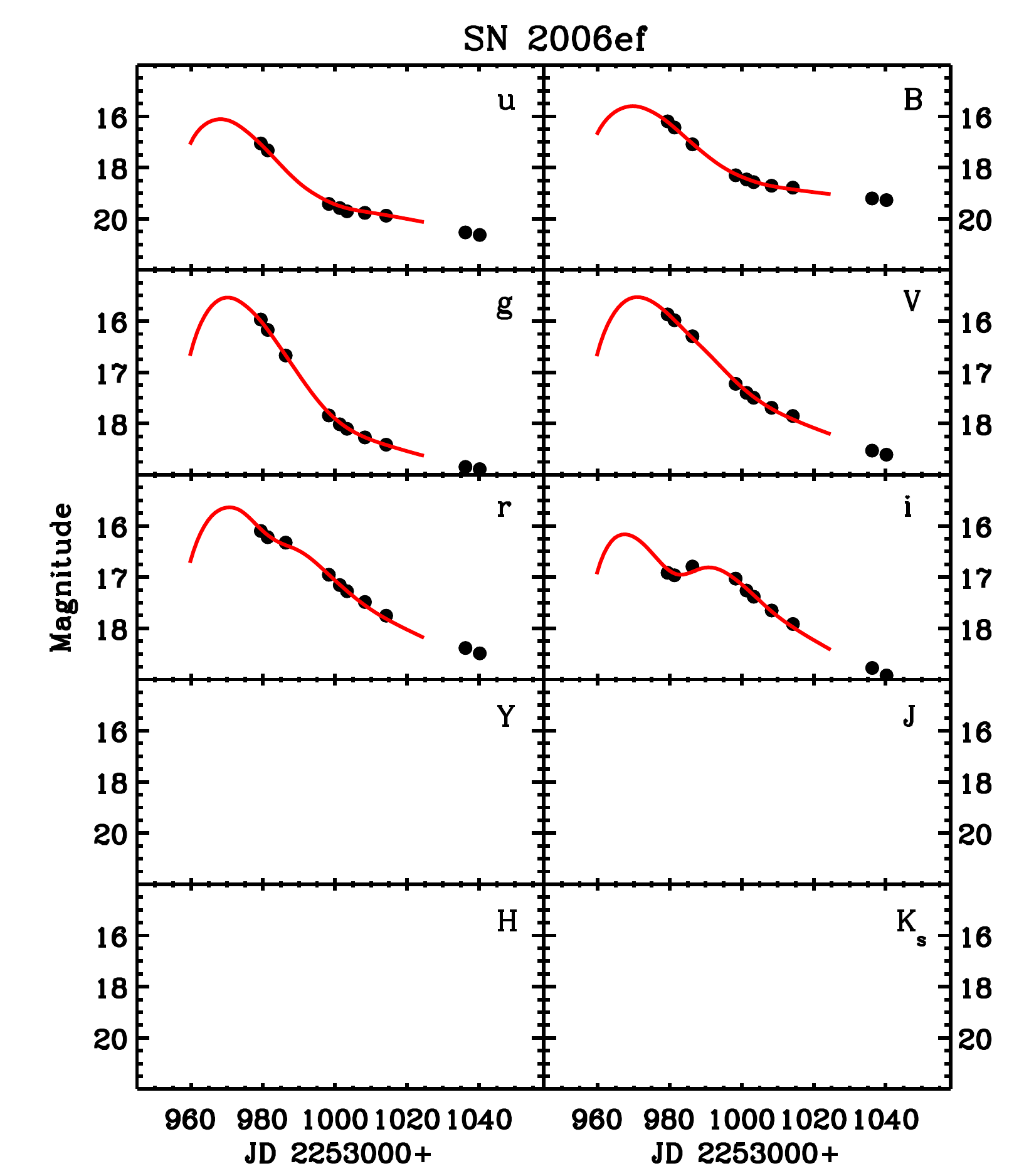}{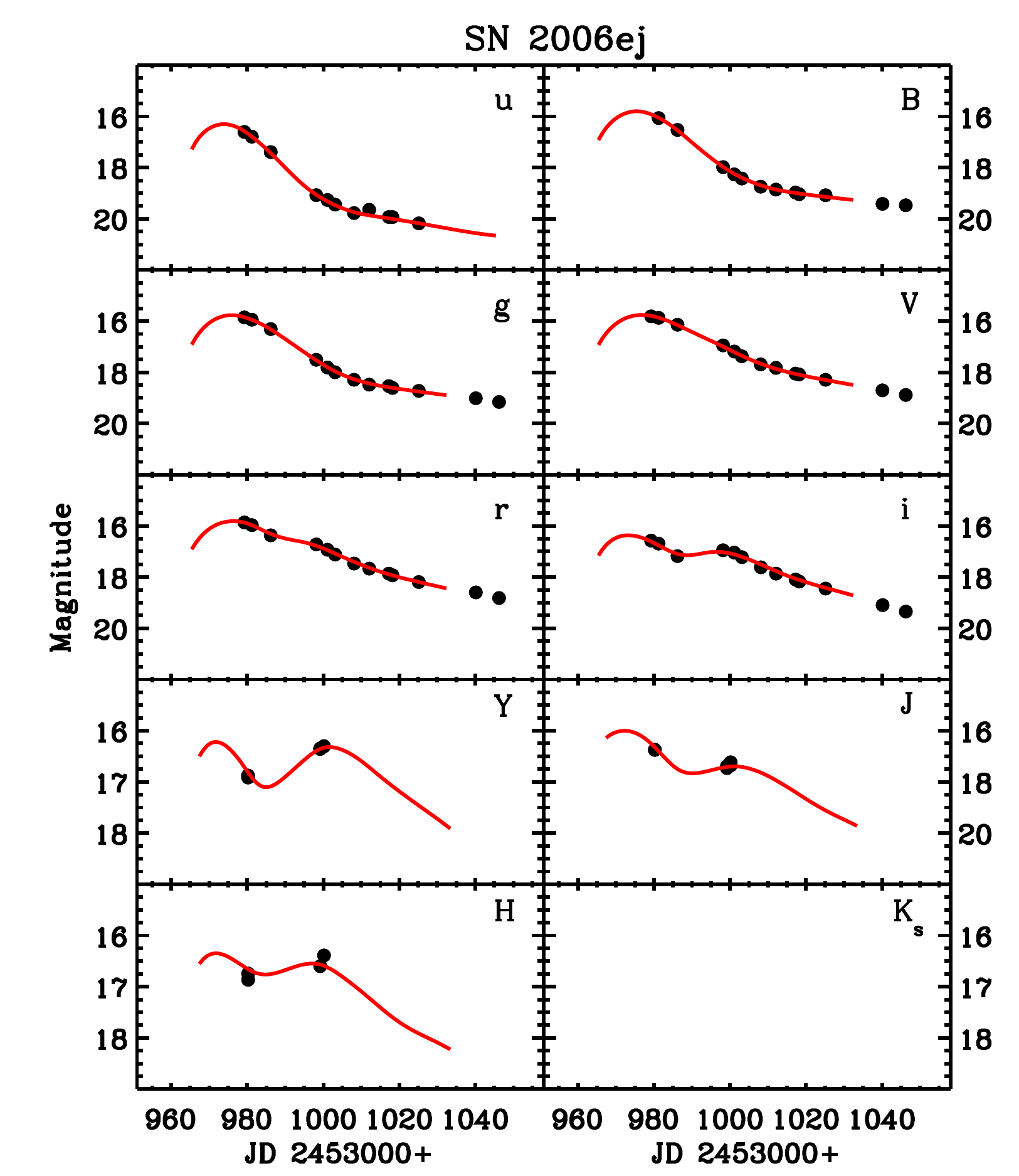}
  {\center Stritzinger {\it et al.} Fig. \ref{fig:flcurves}}
\end{figure}

\clearpage
\newpage
\begin{figure}[t]
 \plottwo{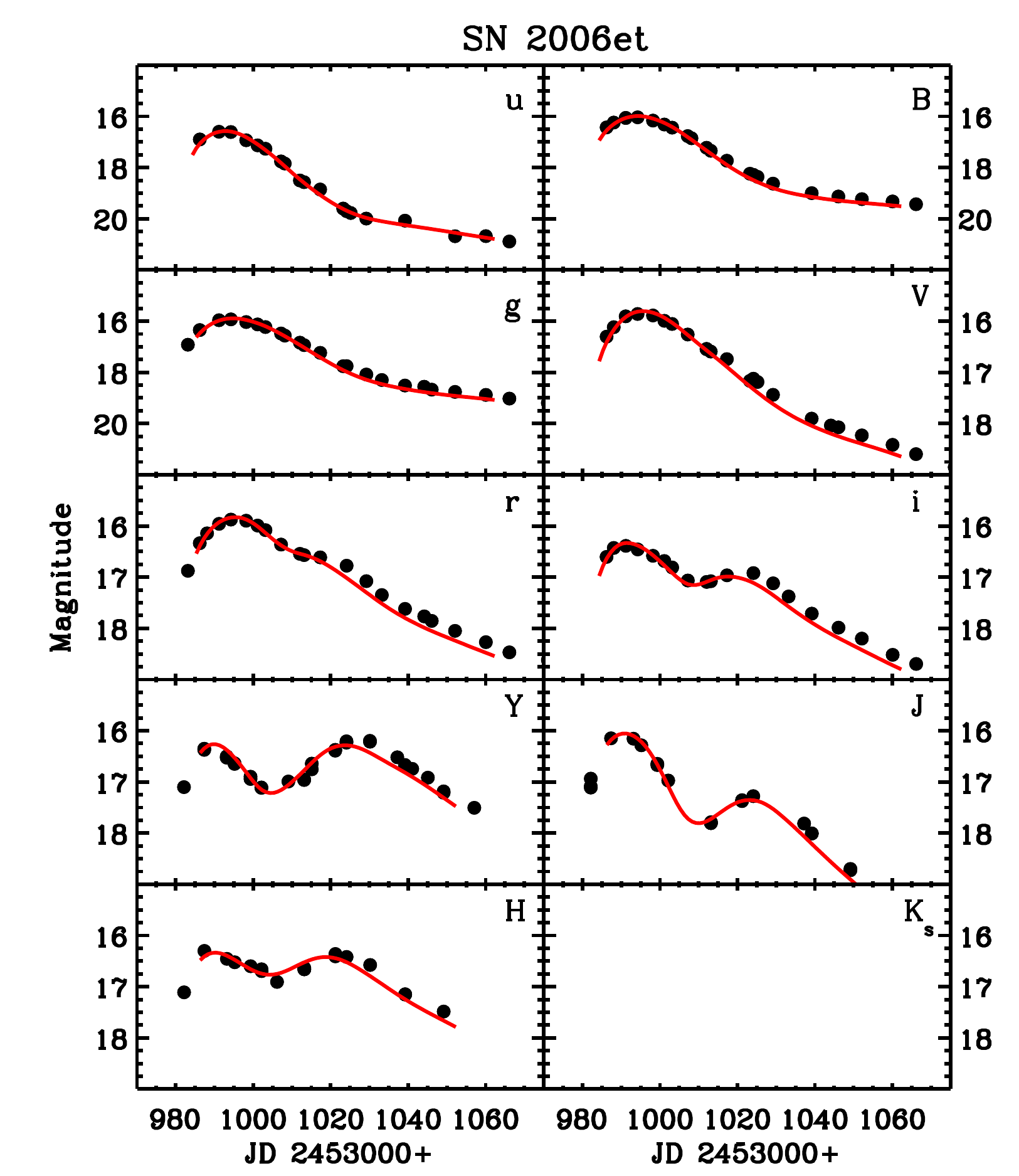}{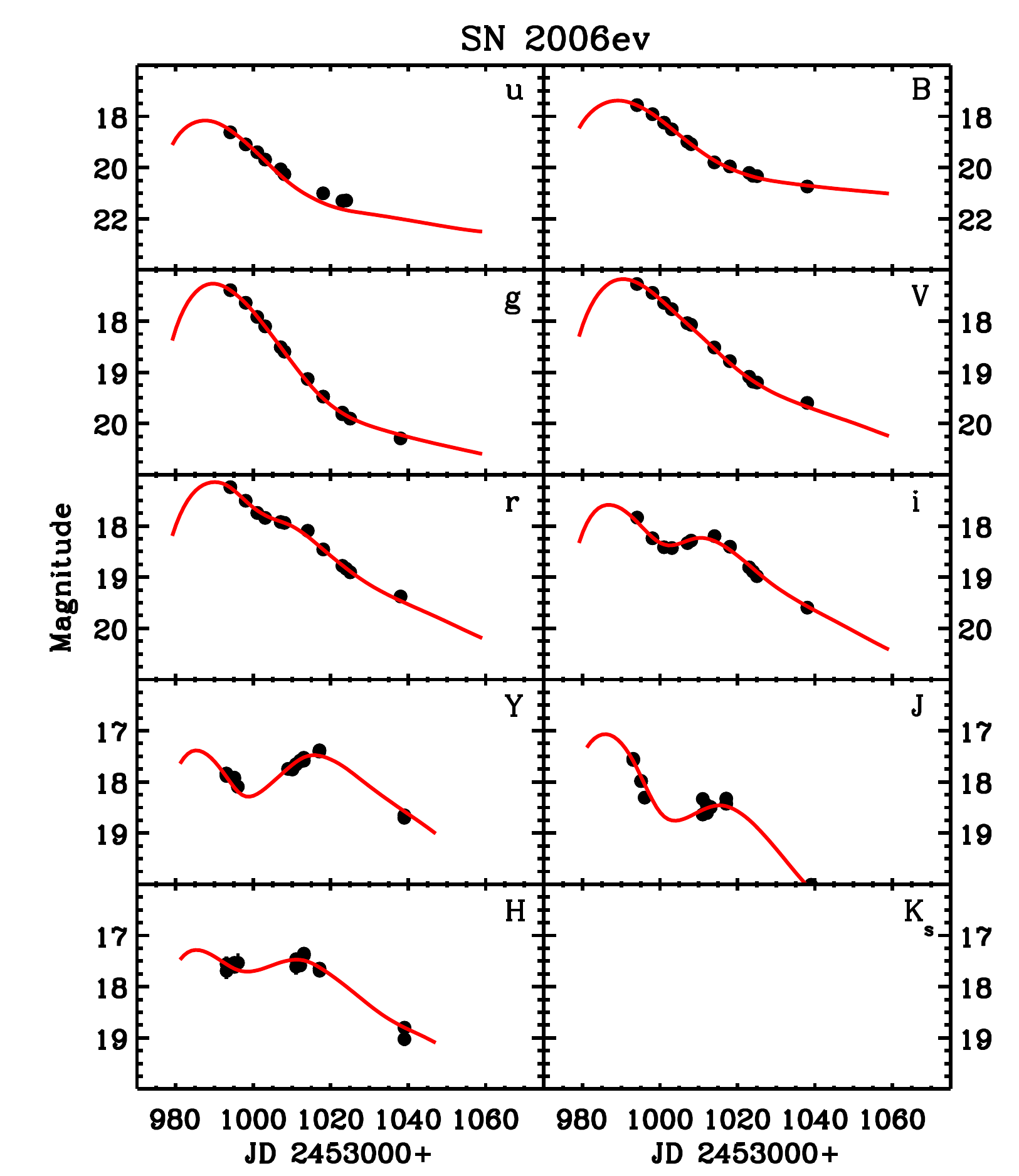}
 \newline
   \newline
\plottwo{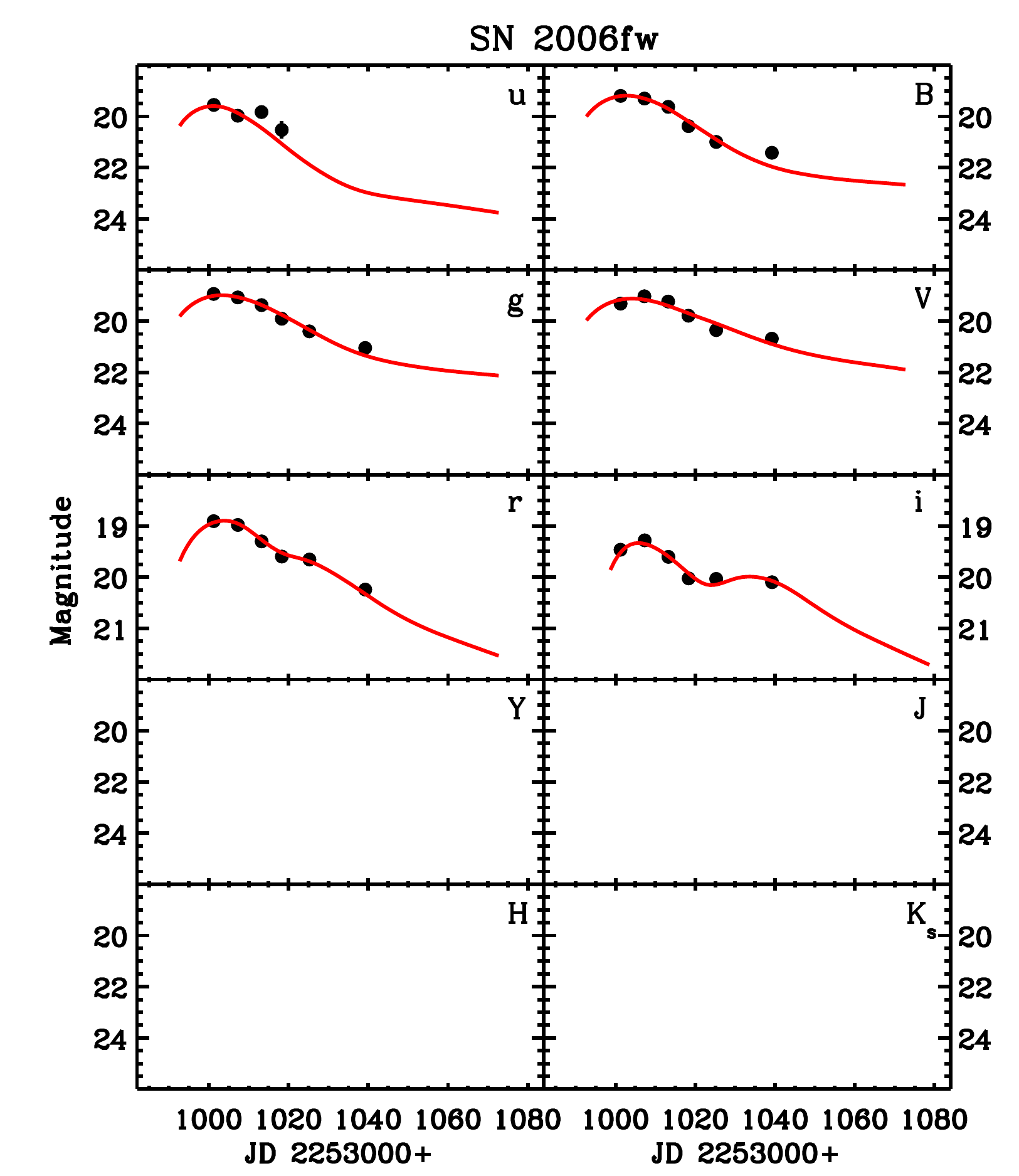}{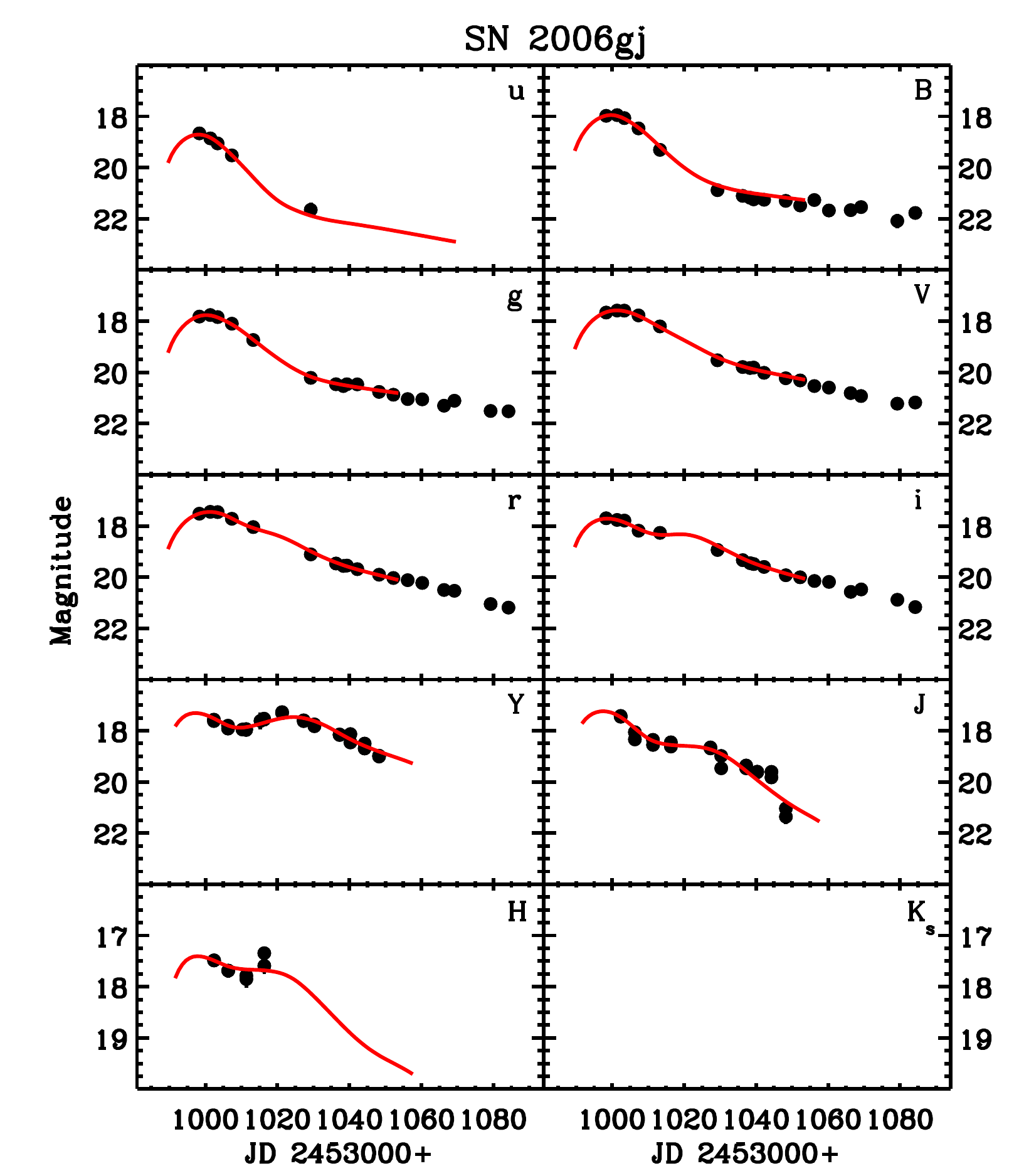}
  {\center Stritzinger {\it et al.} Fig. \ref{fig:flcurves}}
\end{figure}

\clearpage
\newpage
\begin{figure}[t]
 \plottwo{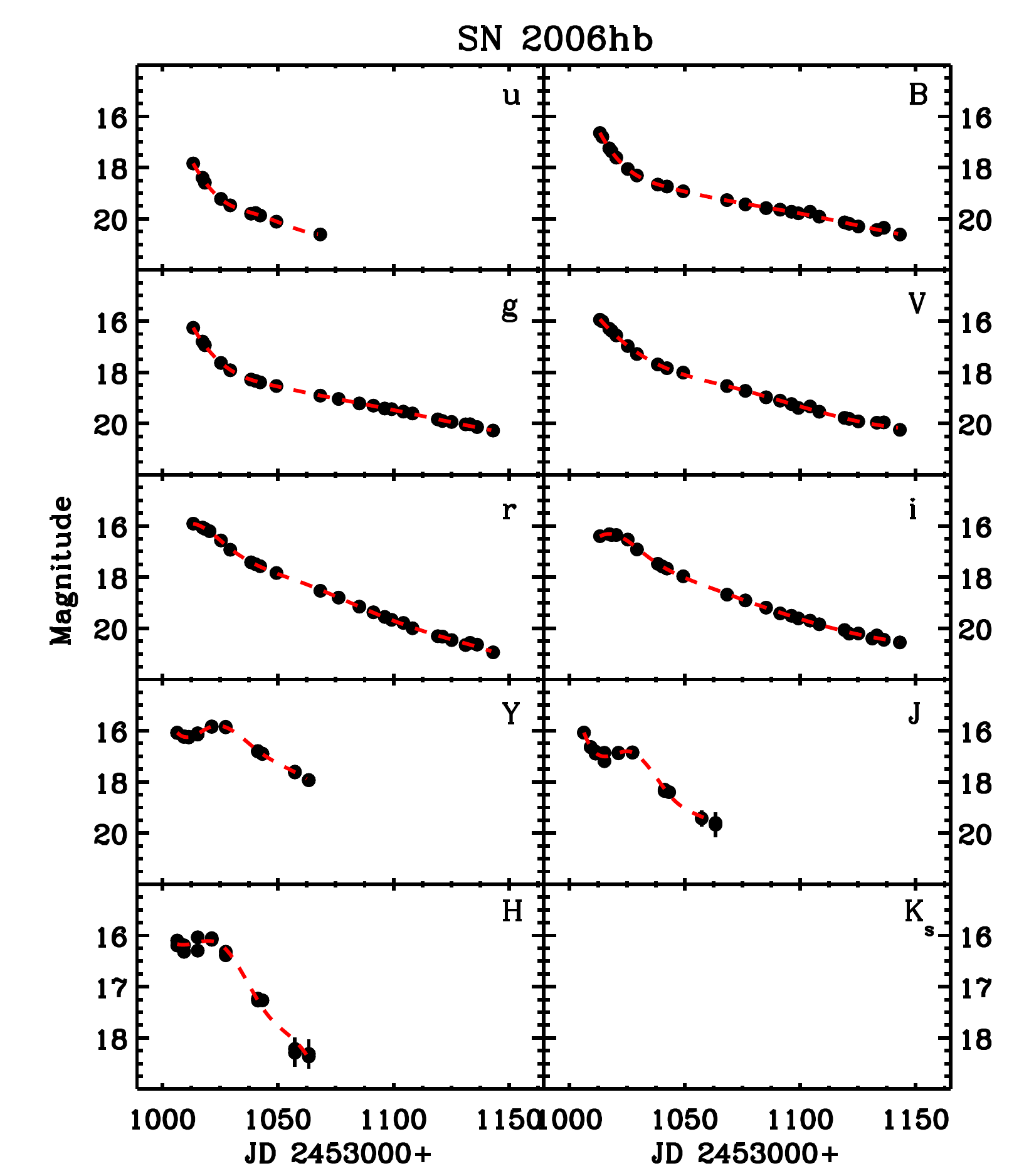}{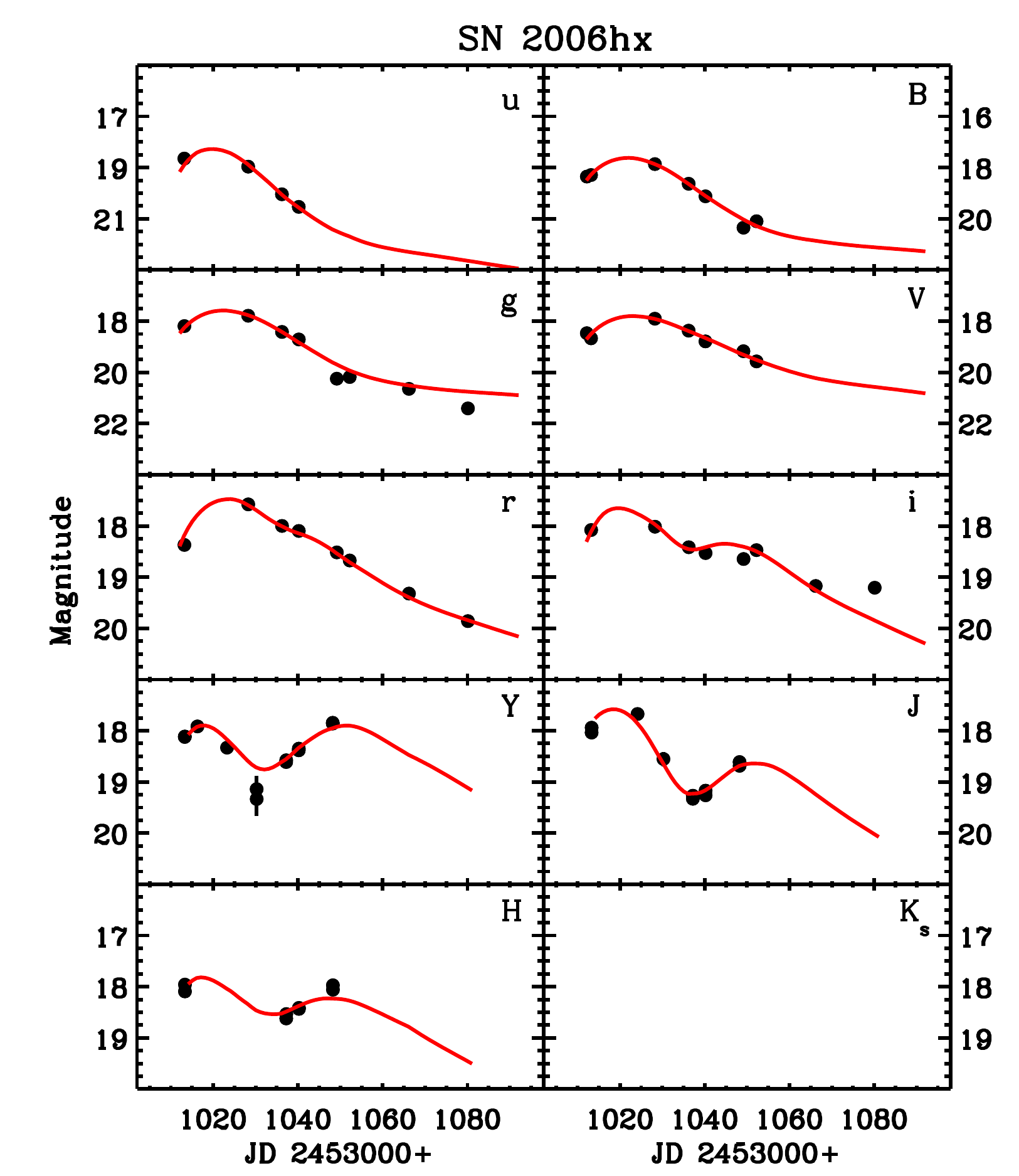}
 \newline
   \newline
\plottwo{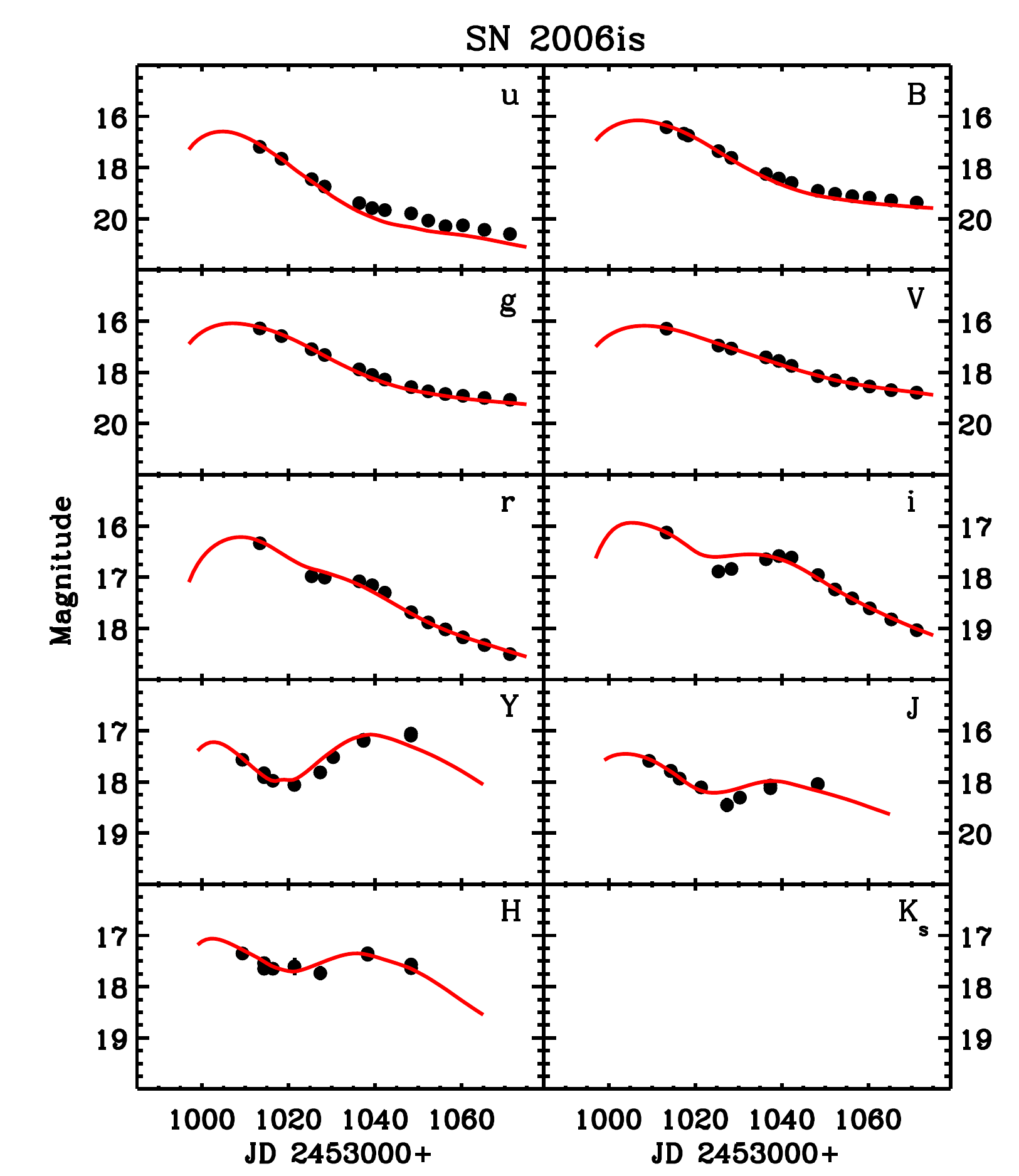}{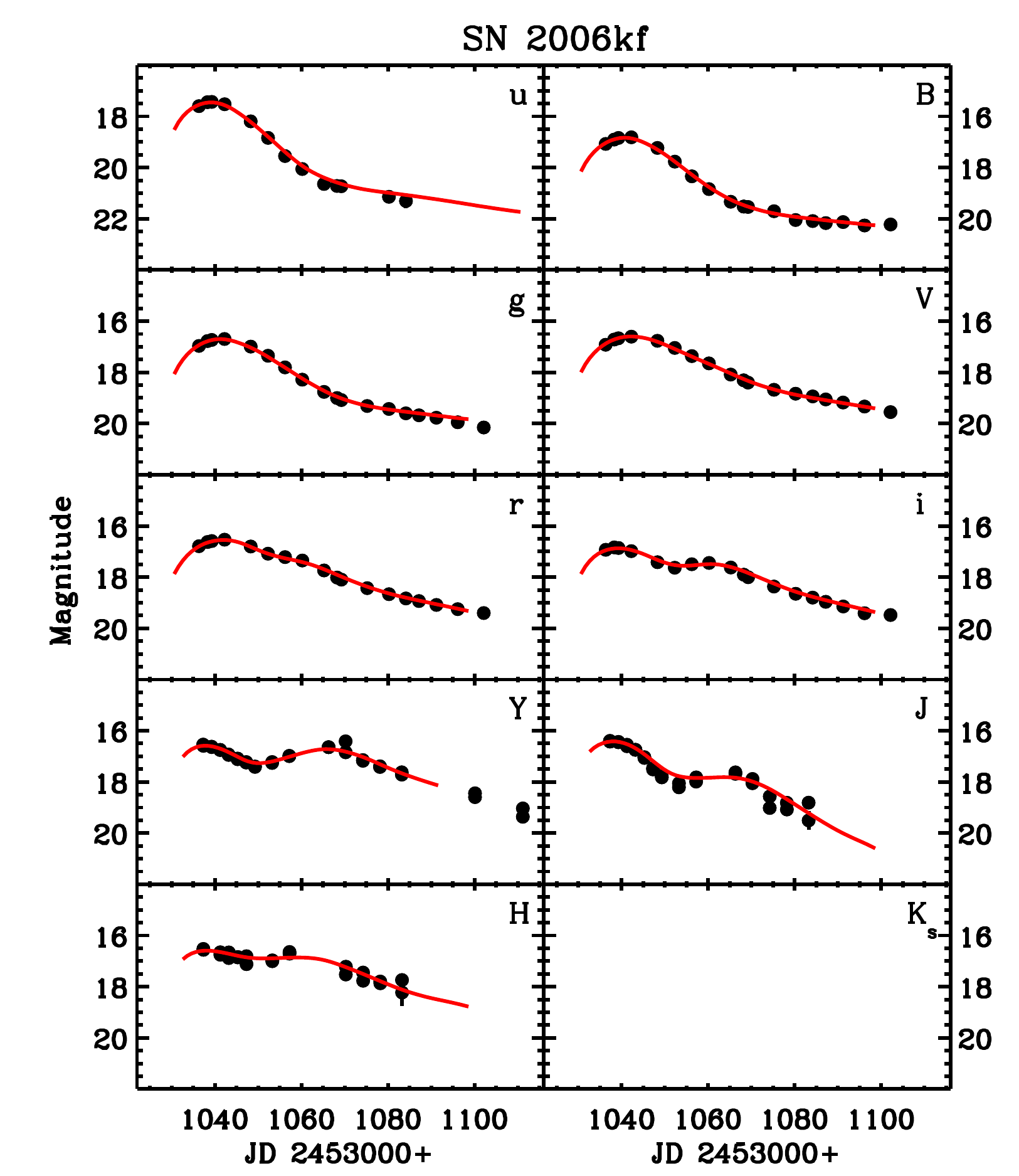}
  {\center Stritzinger {\it et al.} Fig. \ref{fig:flcurves}}
\end{figure}

\clearpage
\newpage
\begin{figure}[t]
 \plottwo{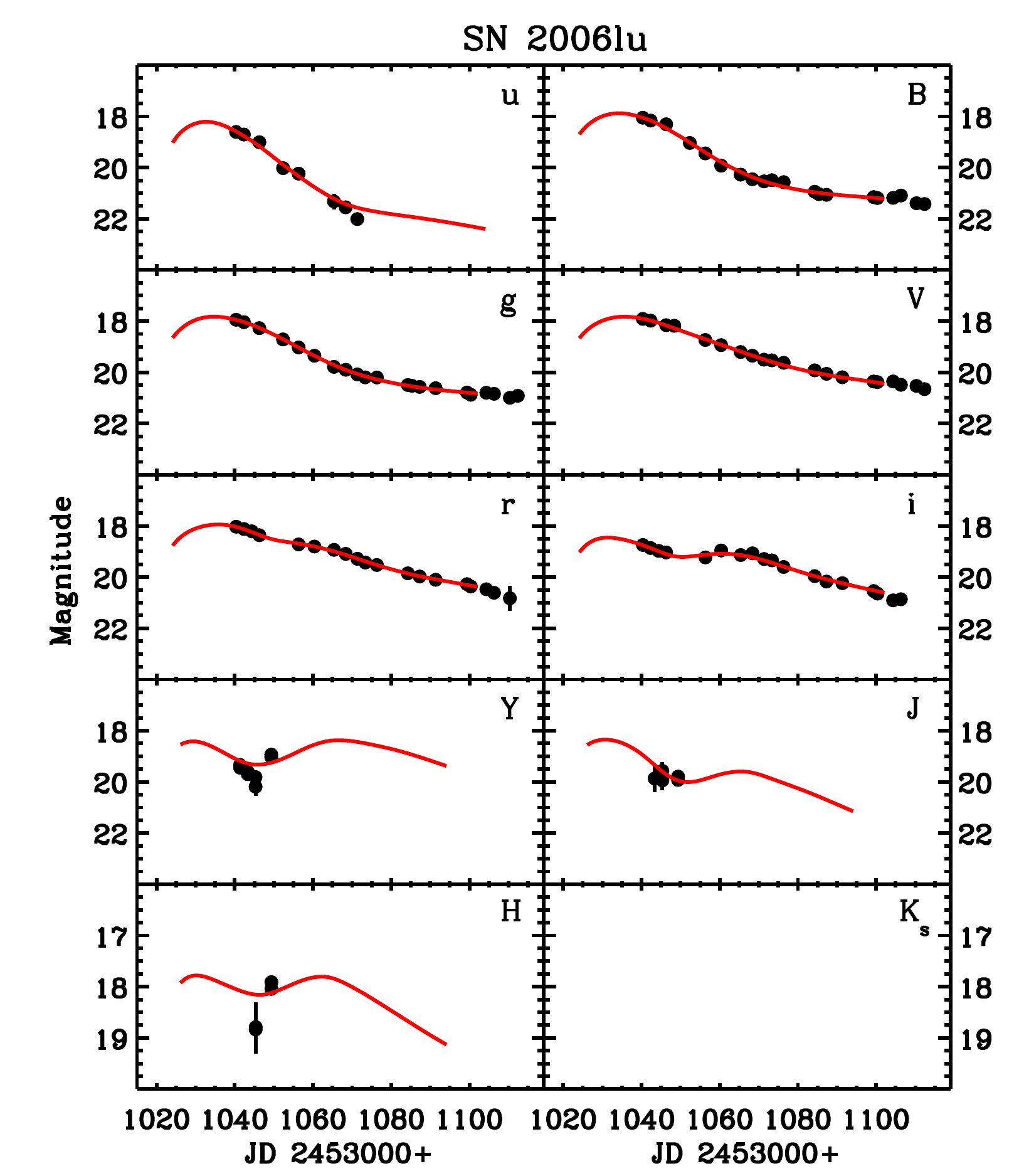}{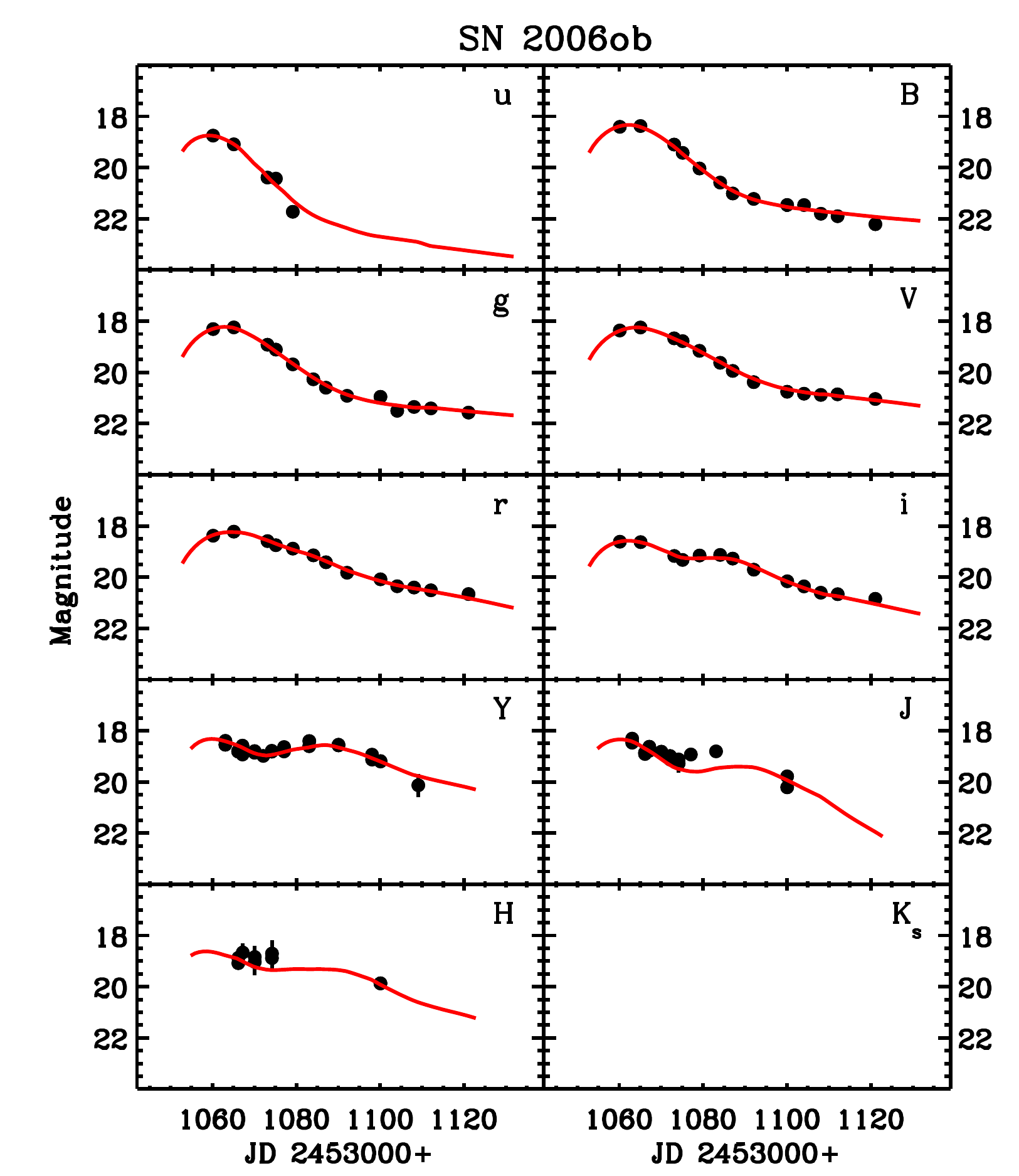}
 \newline
   \newline
\plottwo{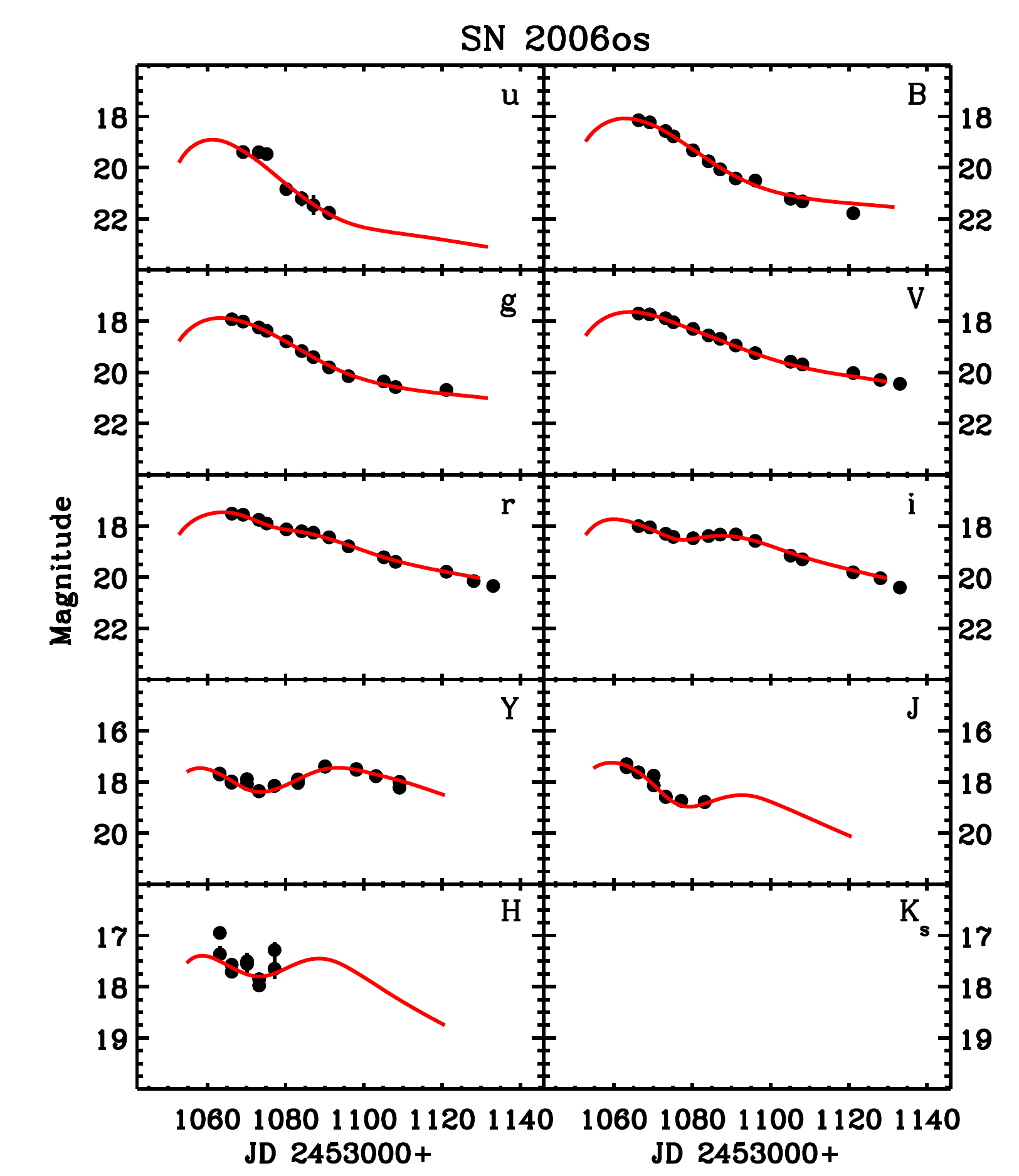}{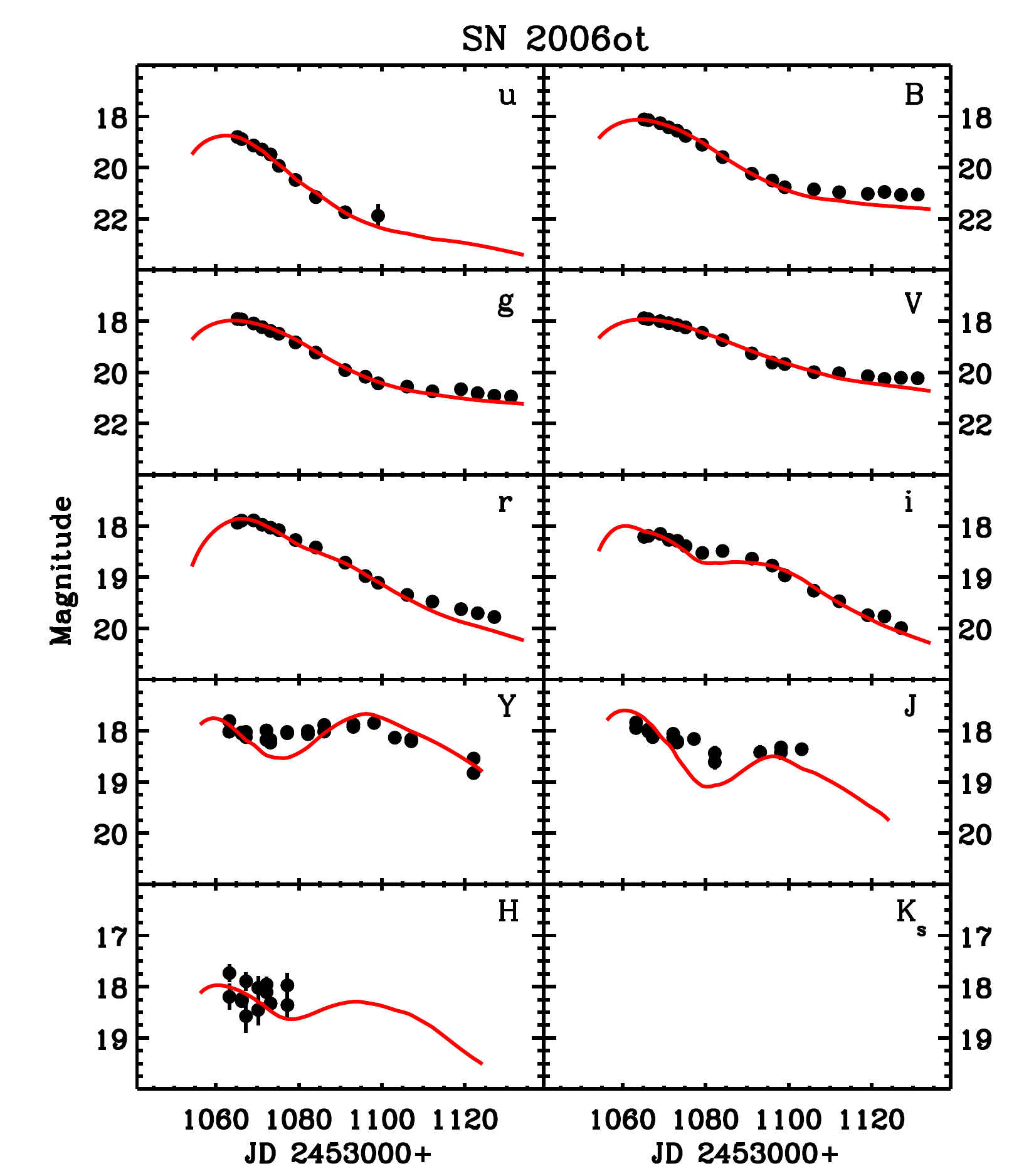}
  {\center Stritzinger {\it et al.} Fig. \ref{fig:flcurves}}
\end{figure}

\clearpage
\newpage
\begin{figure}[t]
 \plottwo{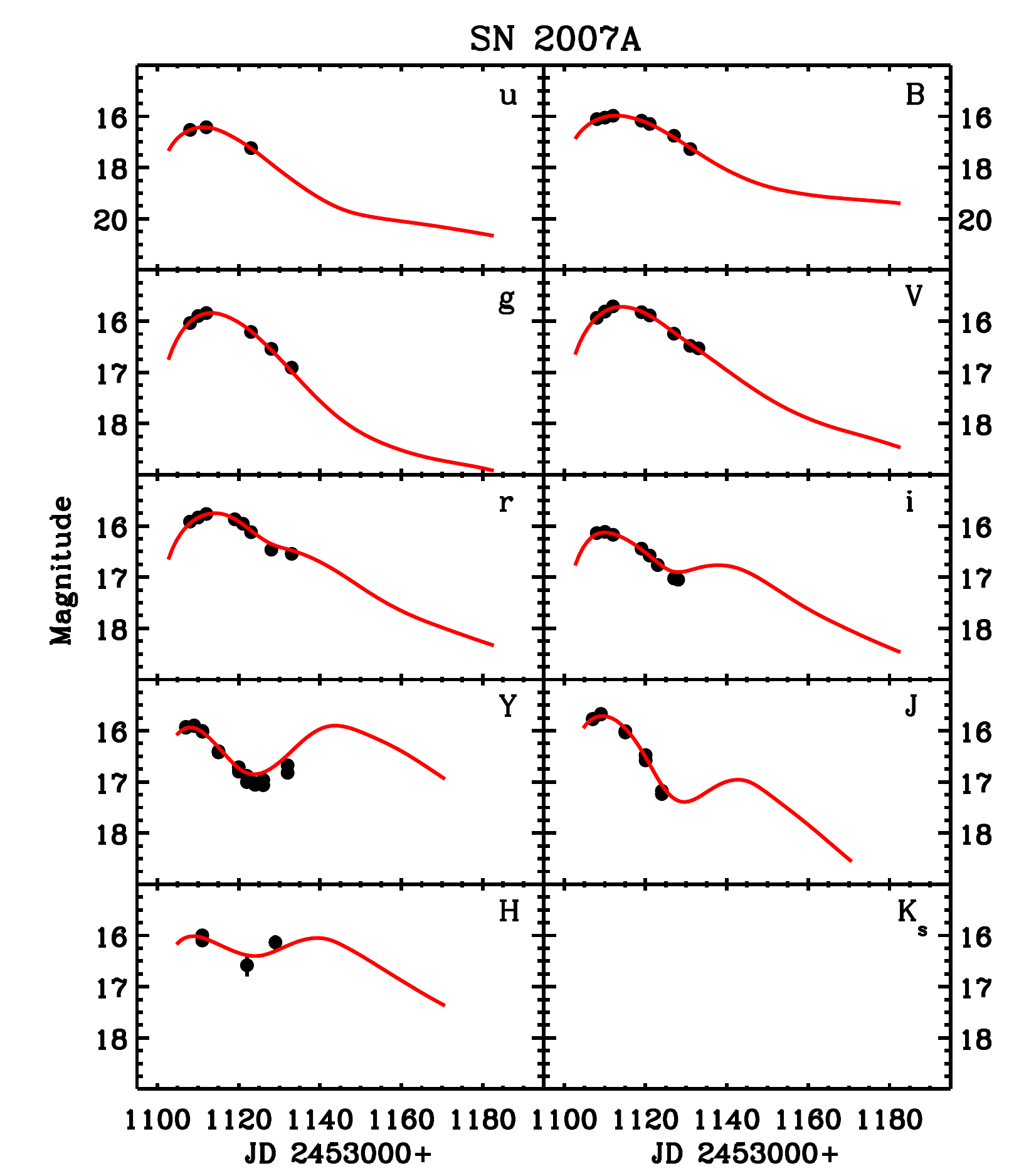}{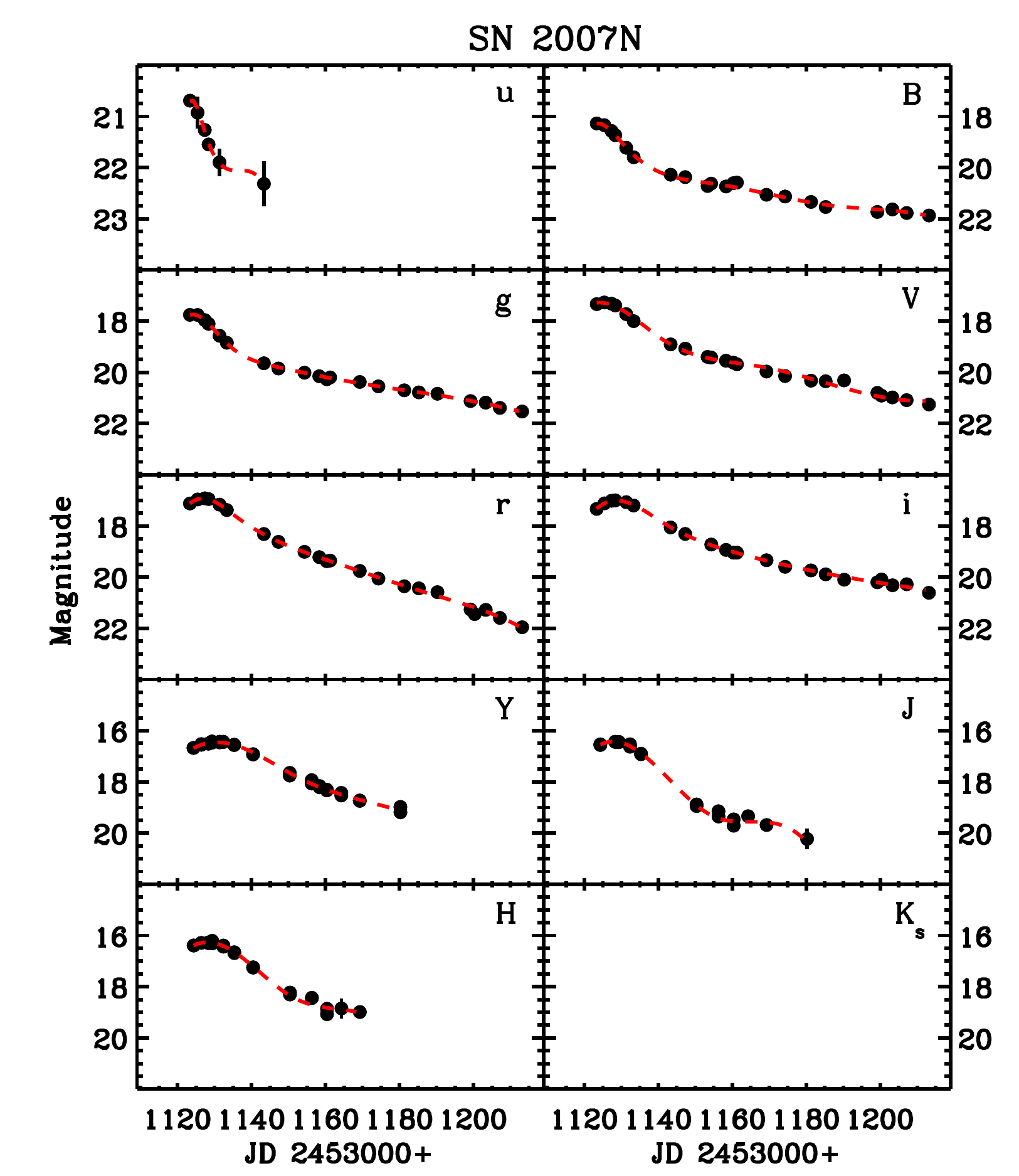}
 \newline
   \newline
\plottwo{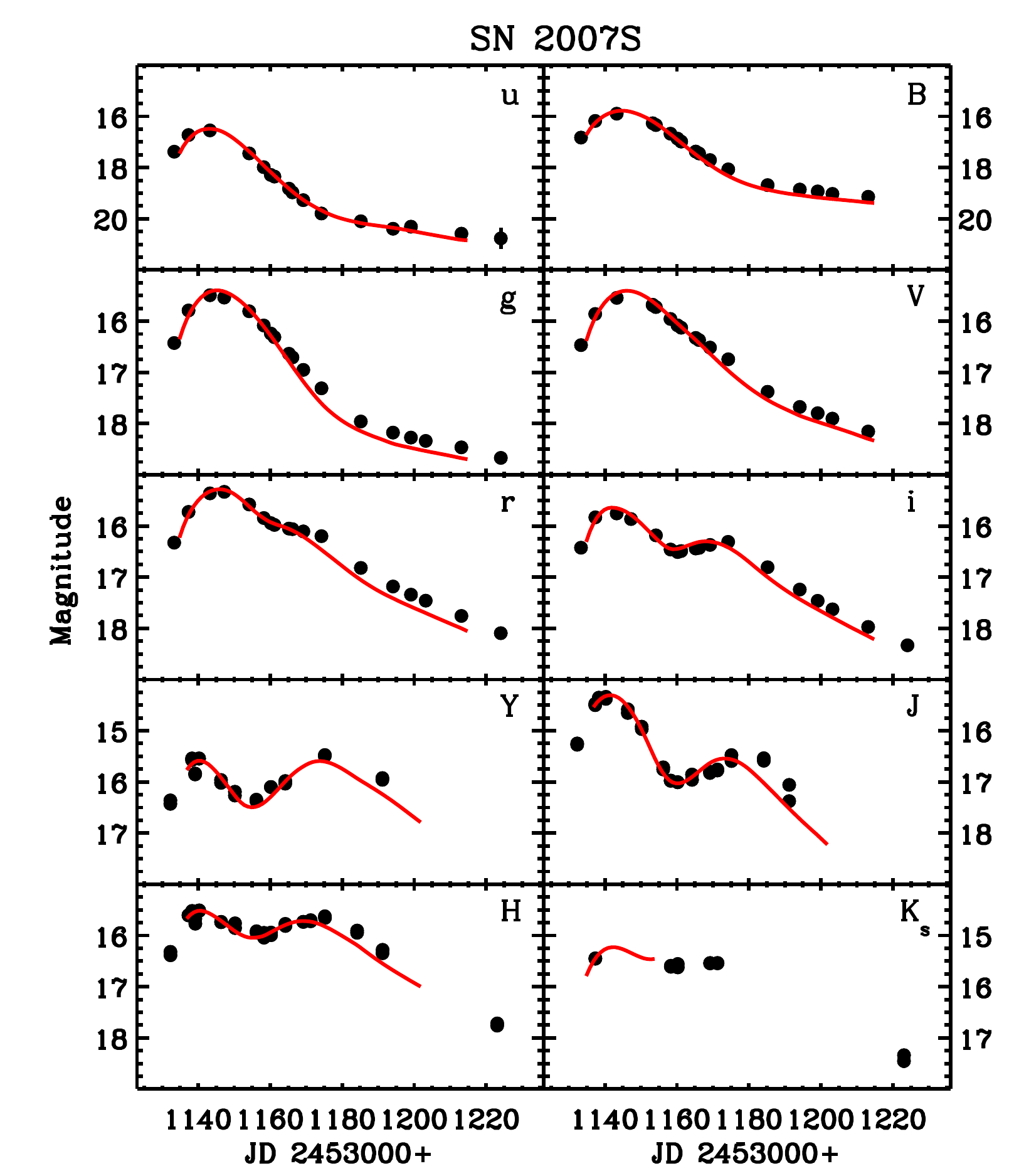}{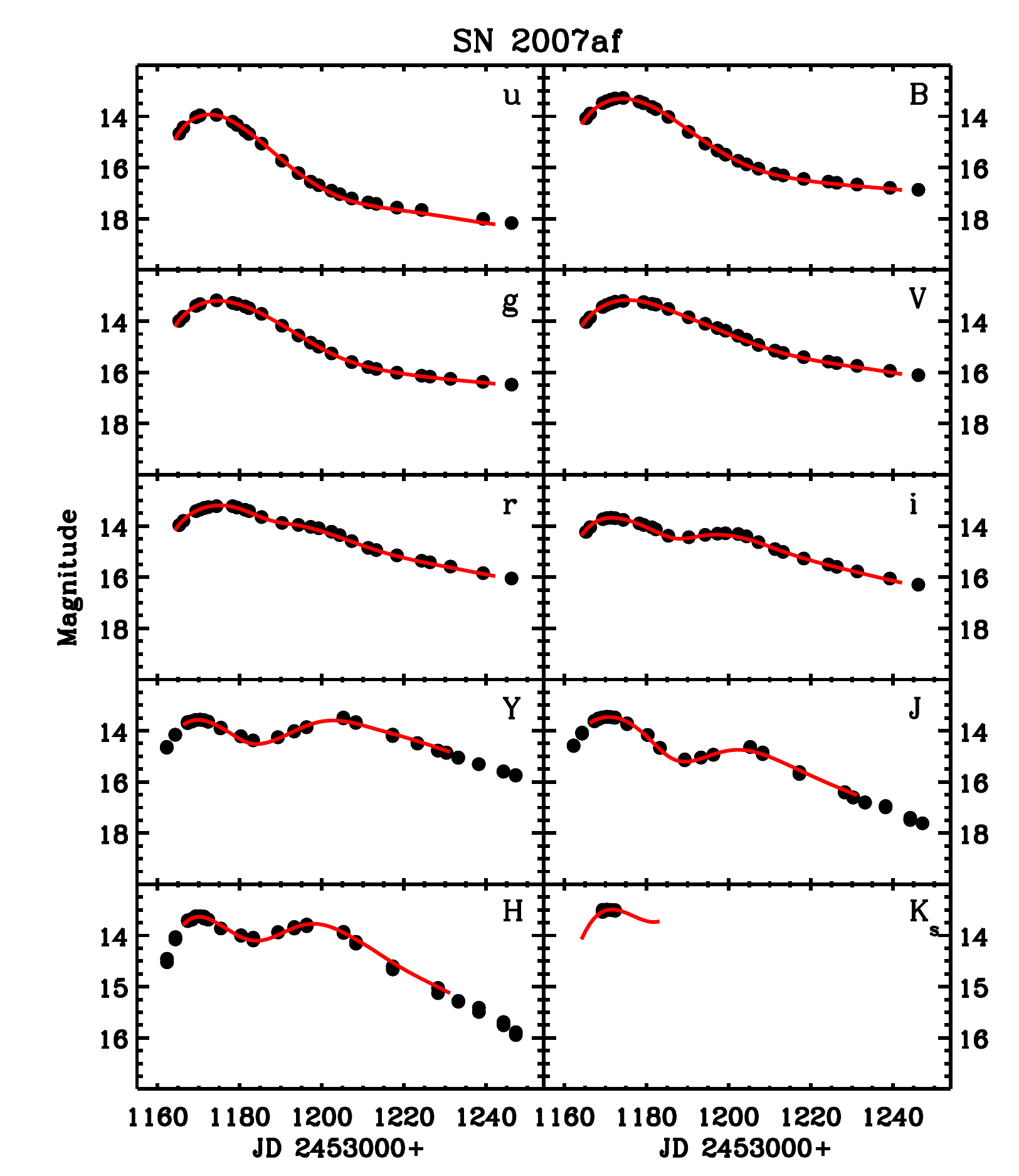}
  {\center Stritzinger {\it et al.} Fig. \ref{fig:flcurves}}
\end{figure}

\clearpage
\newpage
\begin{figure}[t]
 \plottwo{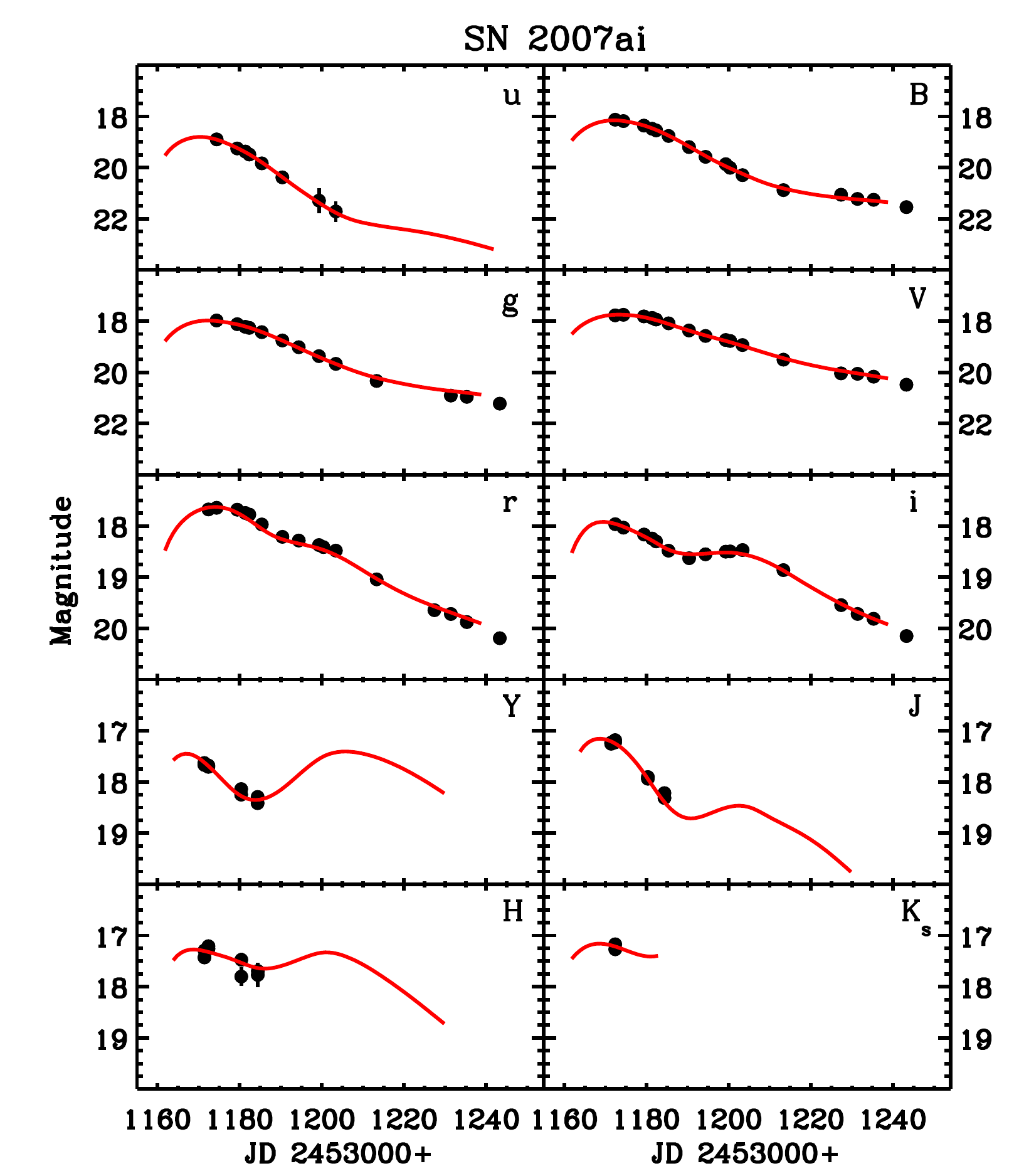}{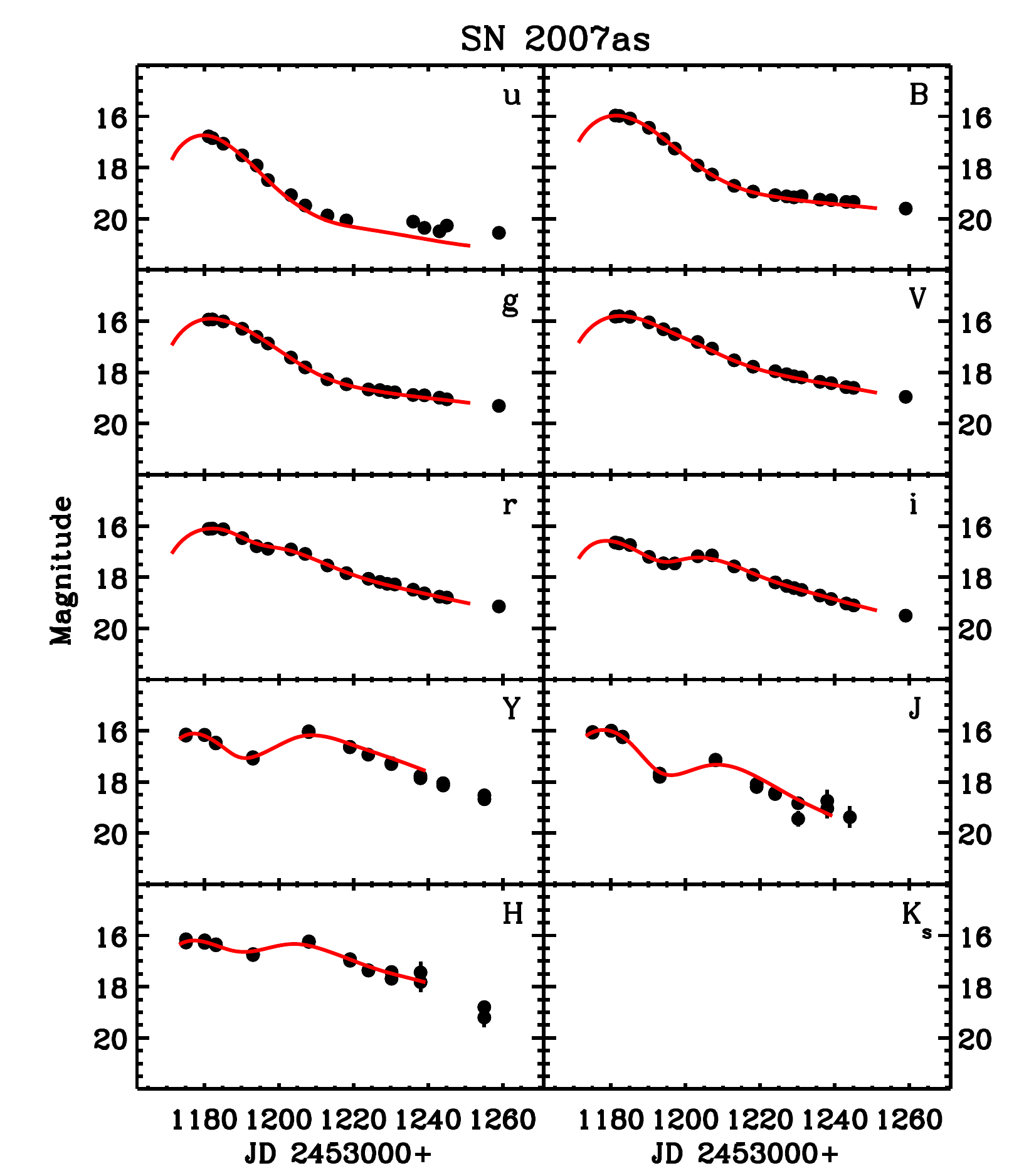}
 \newline
   \newline
\plottwo{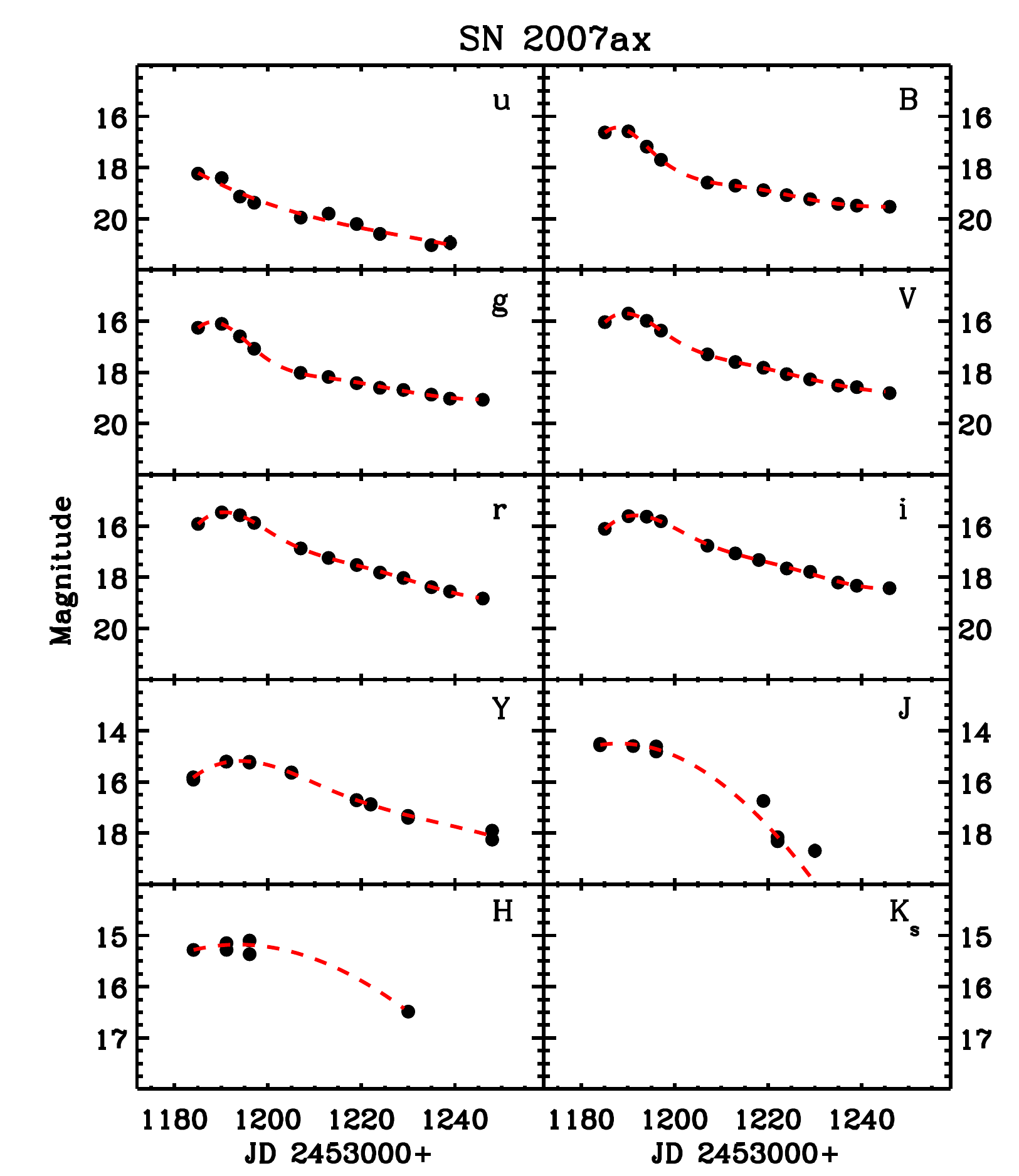}{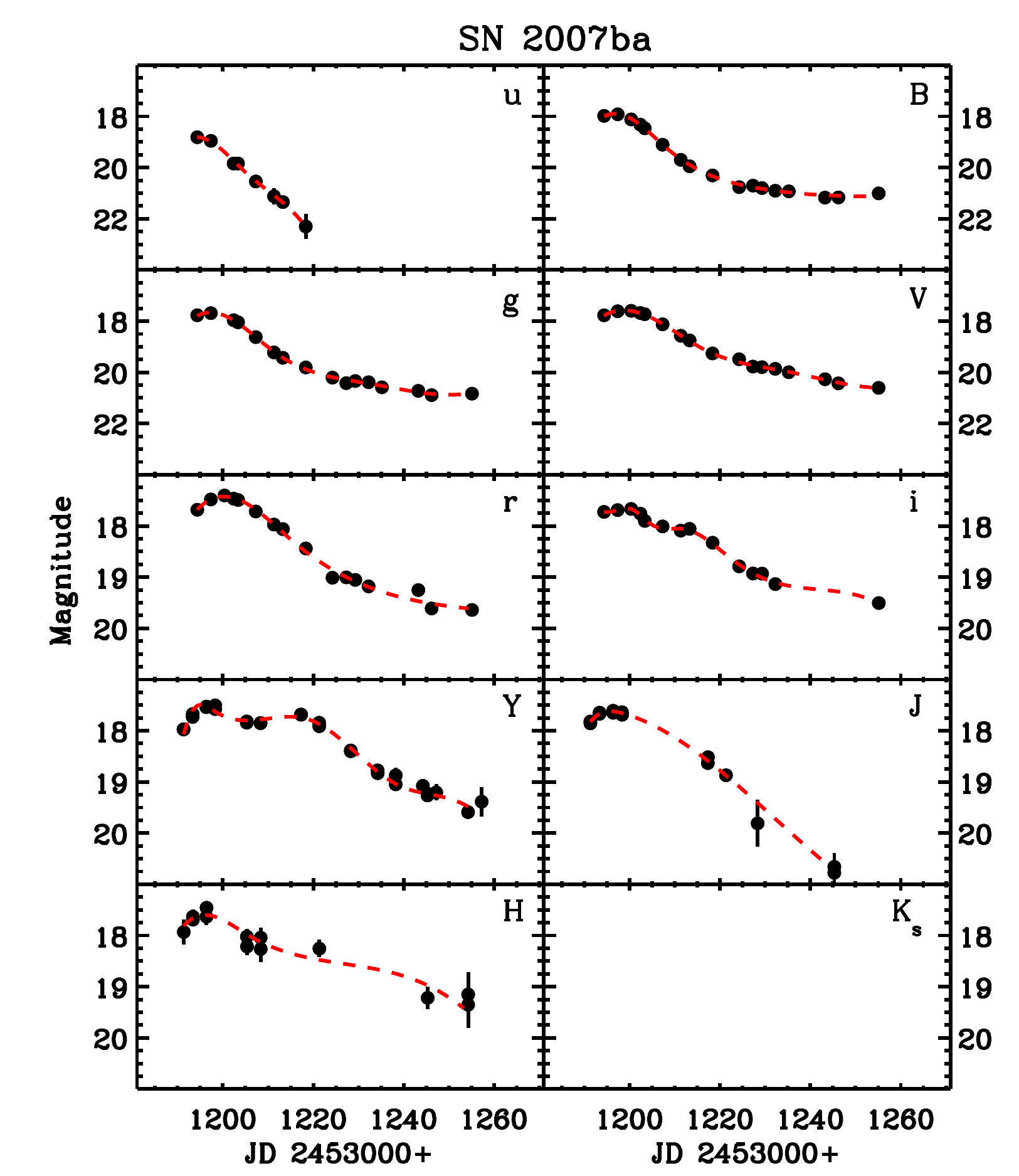}
  {\center Stritzinger {\it et al.} Fig. \ref{fig:flcurves}}
\end{figure}

\clearpage
\newpage
\begin{figure}[t]
 \plottwo{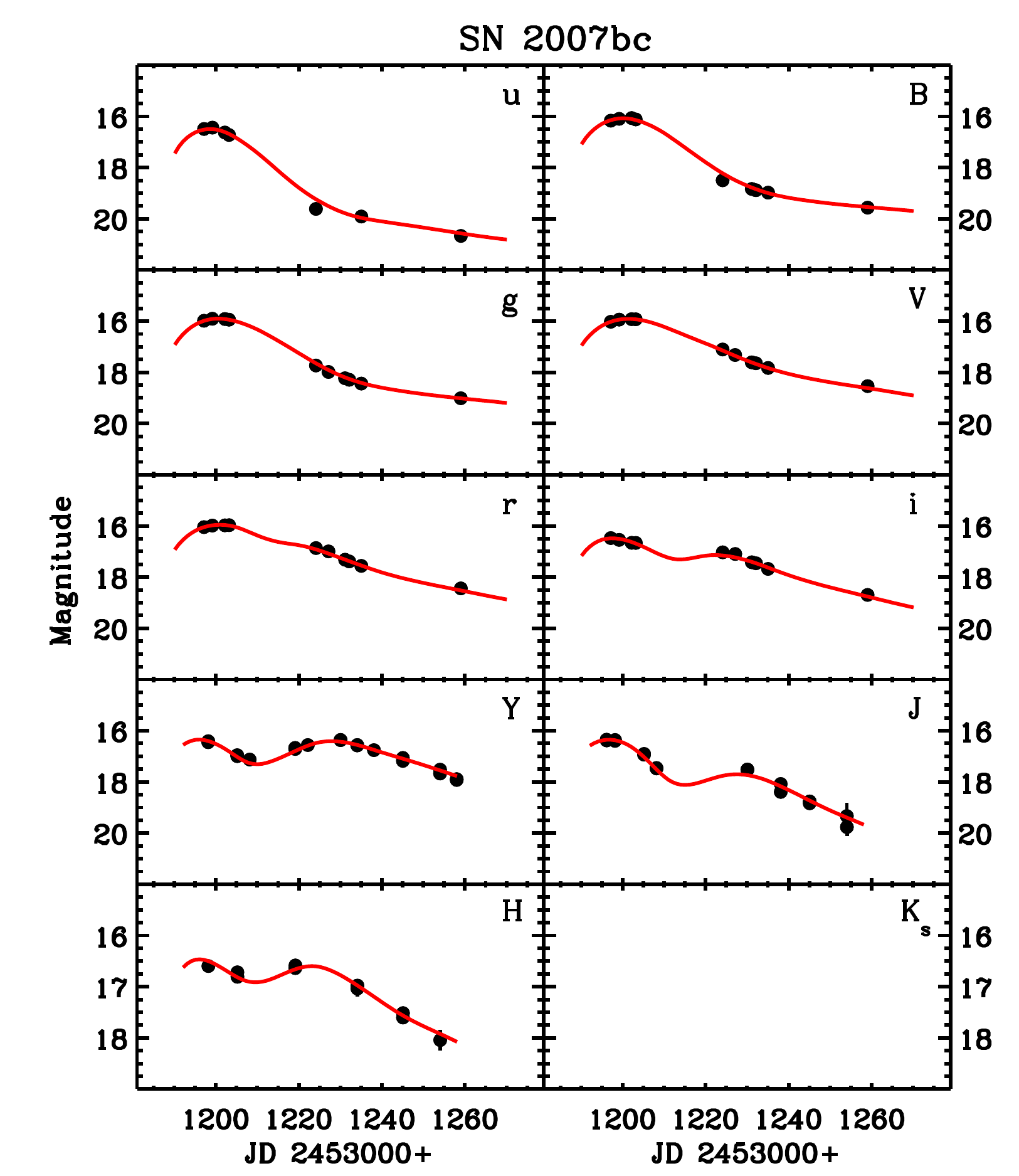}{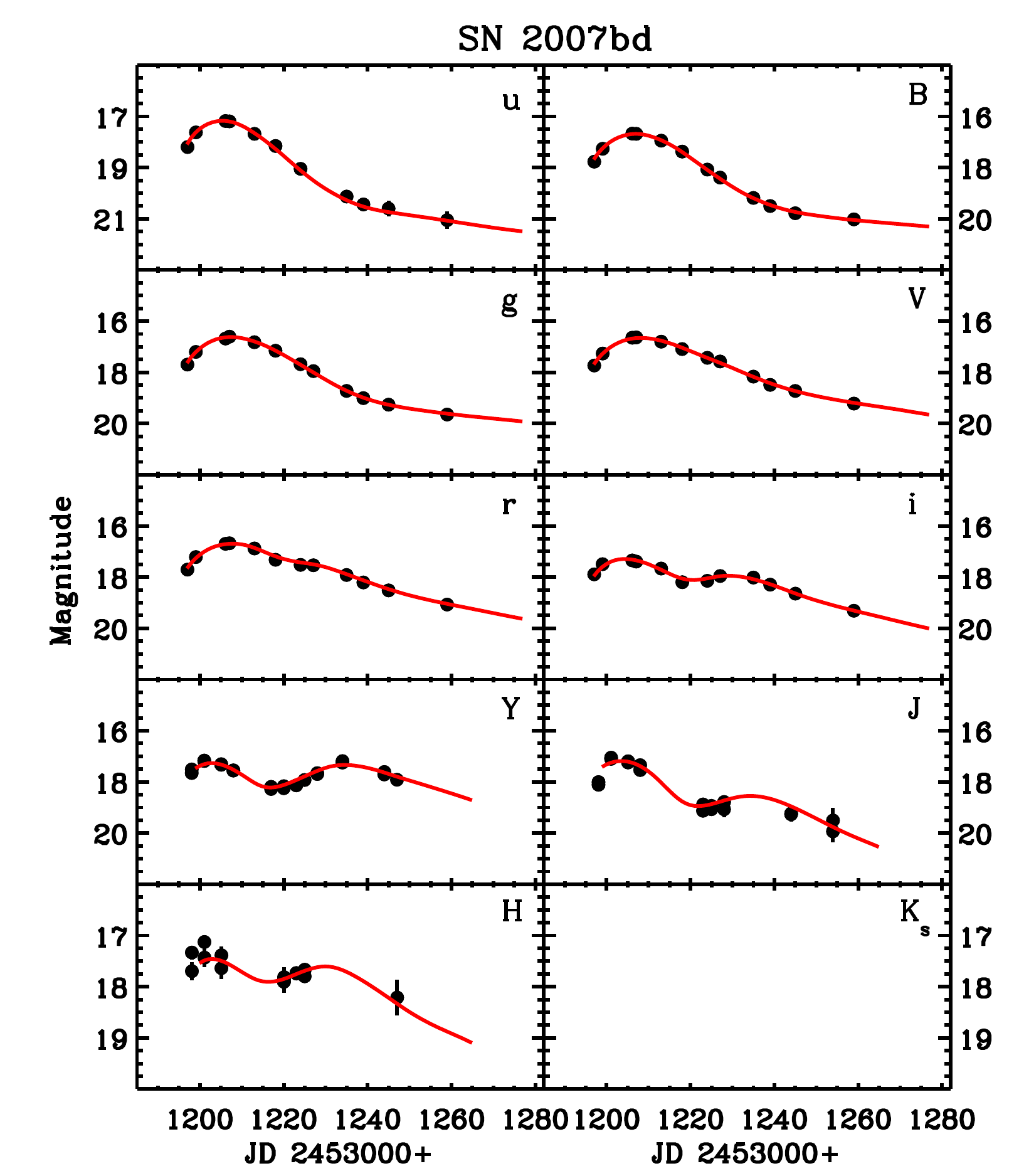}
 \newline
   \newline
\plottwo{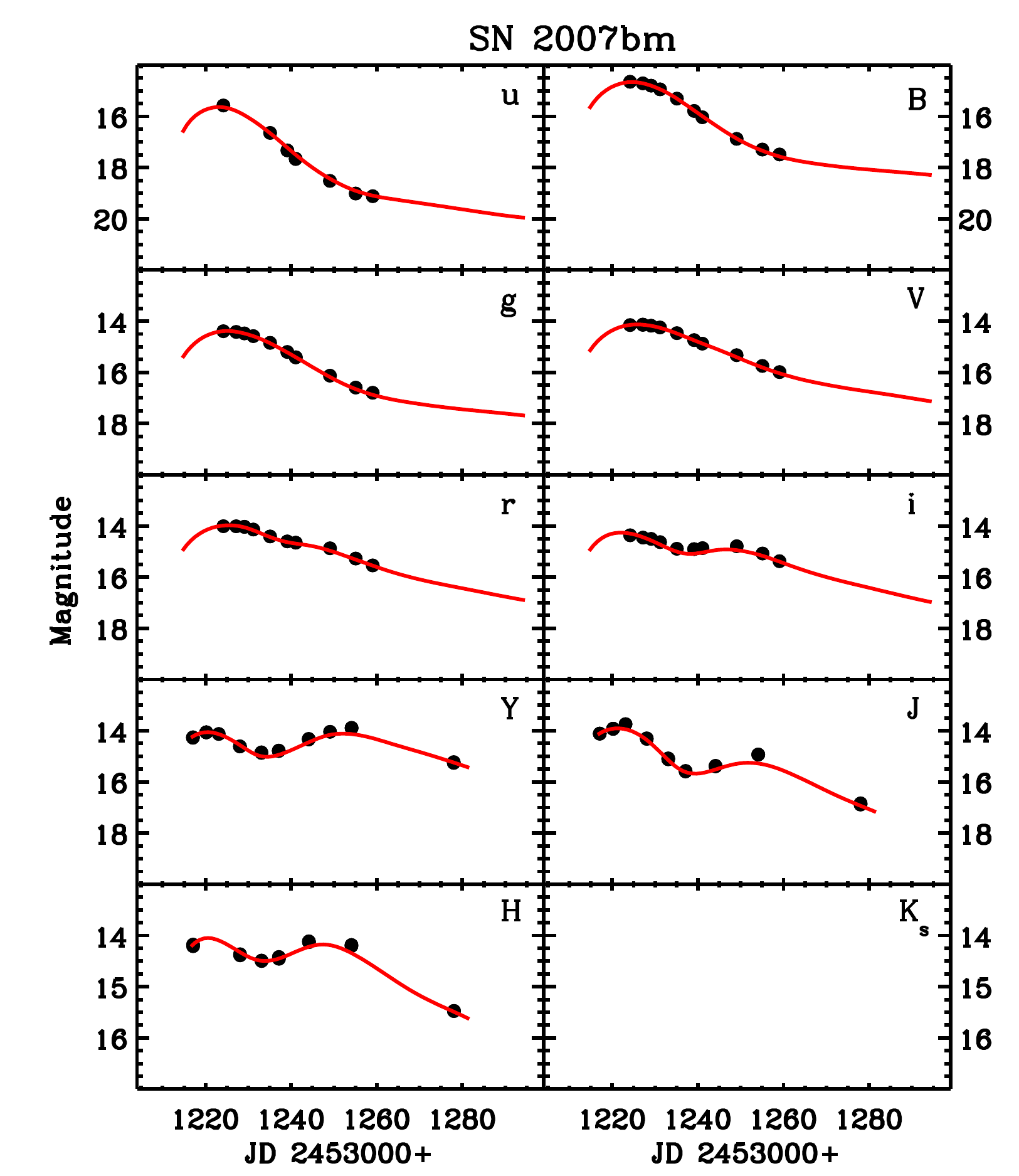}{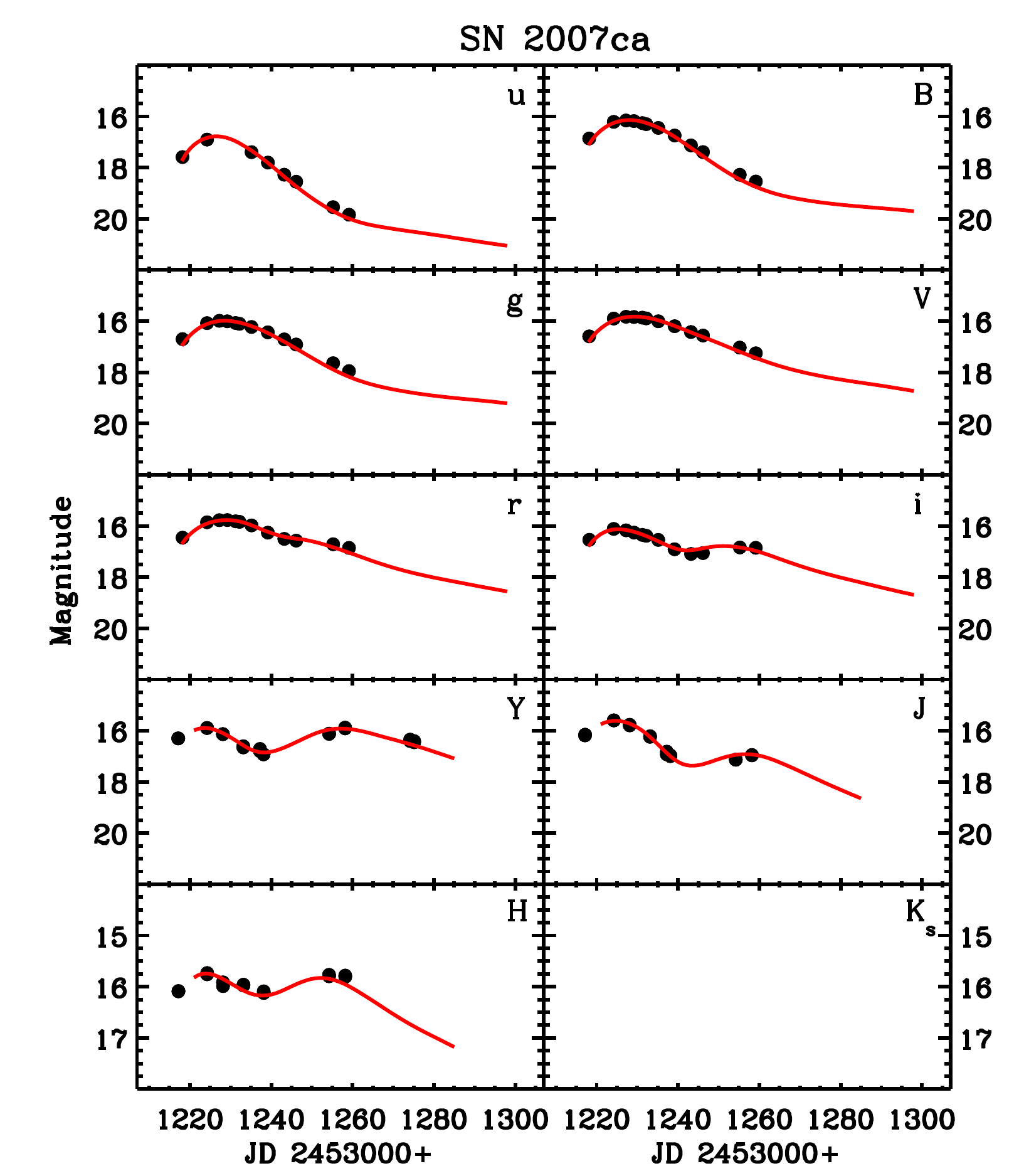}
  {\center Stritzinger {\it et al.} Fig. \ref{fig:flcurves}}
\end{figure}

\clearpage
\newpage
\begin{figure}[t]
 \plottwo{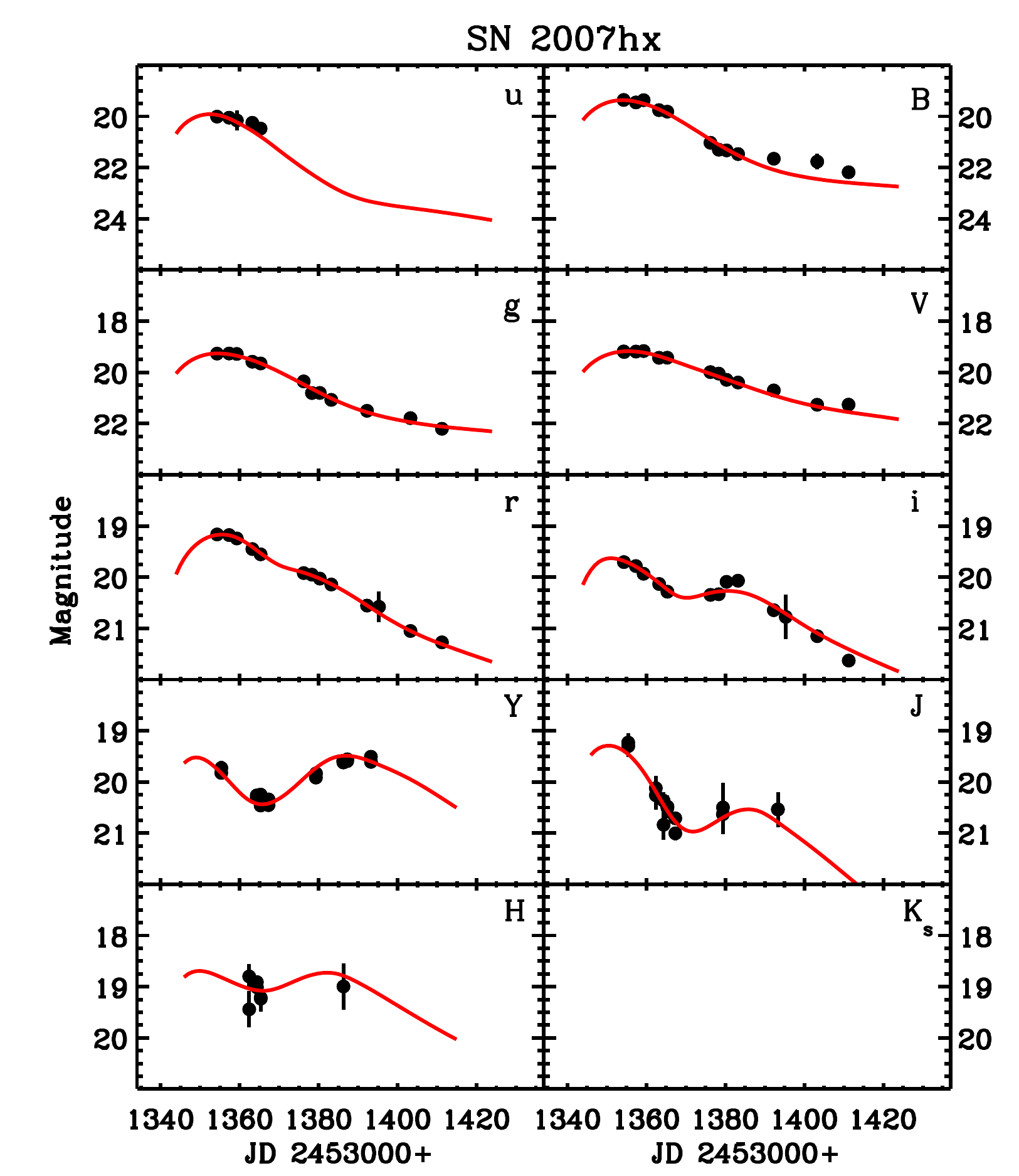}{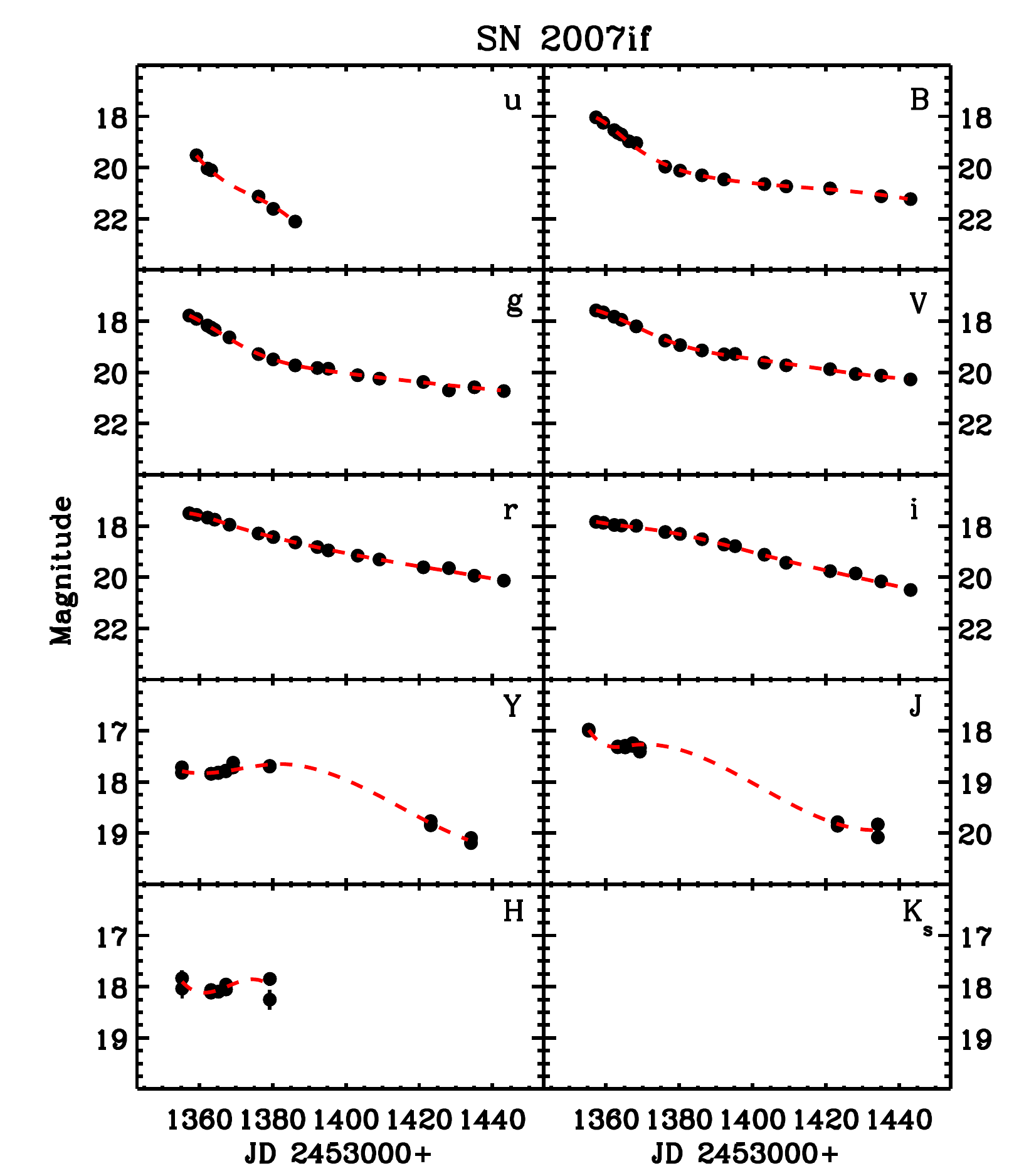}
 \newline
   \newline
 \plottwo{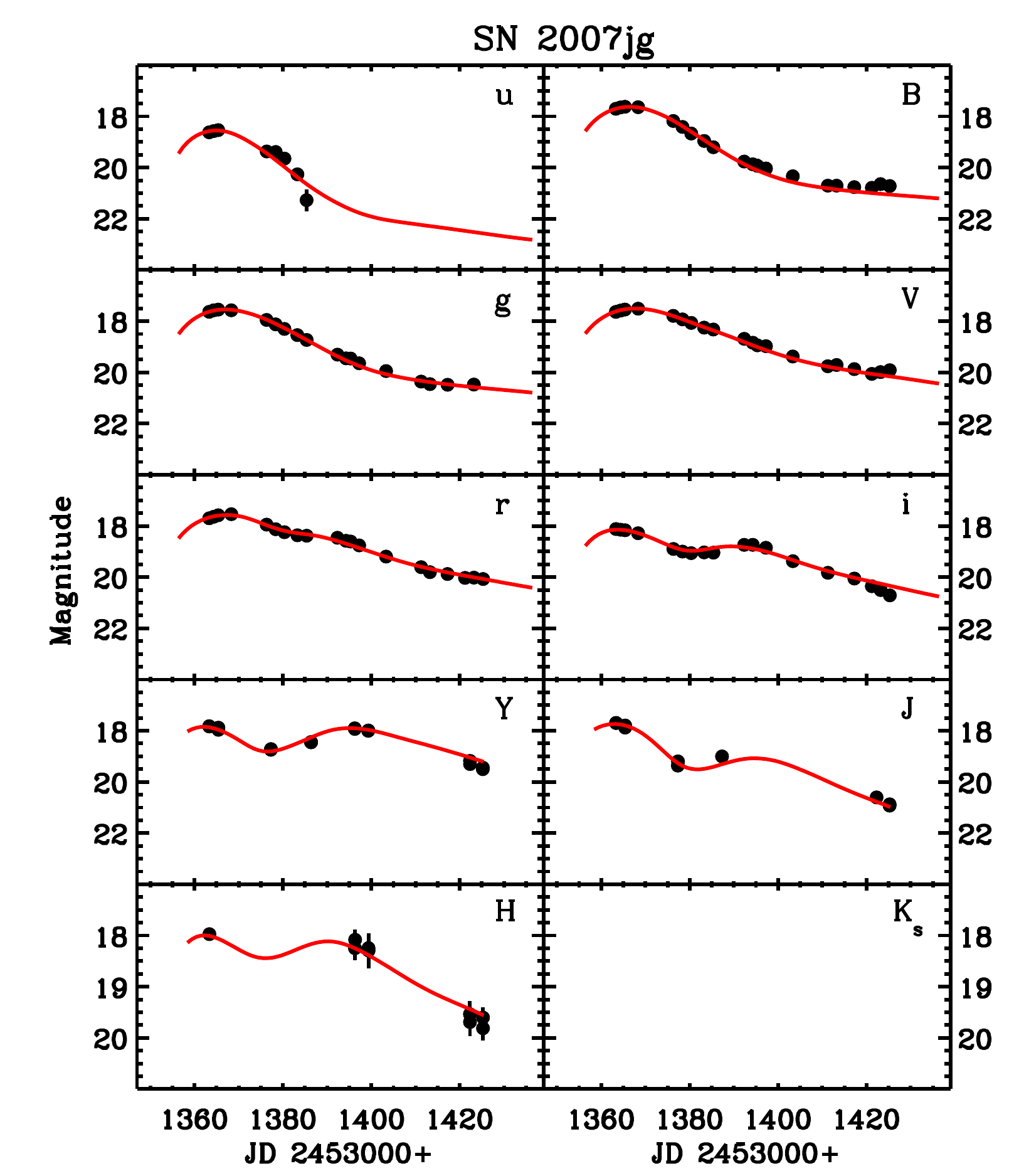}{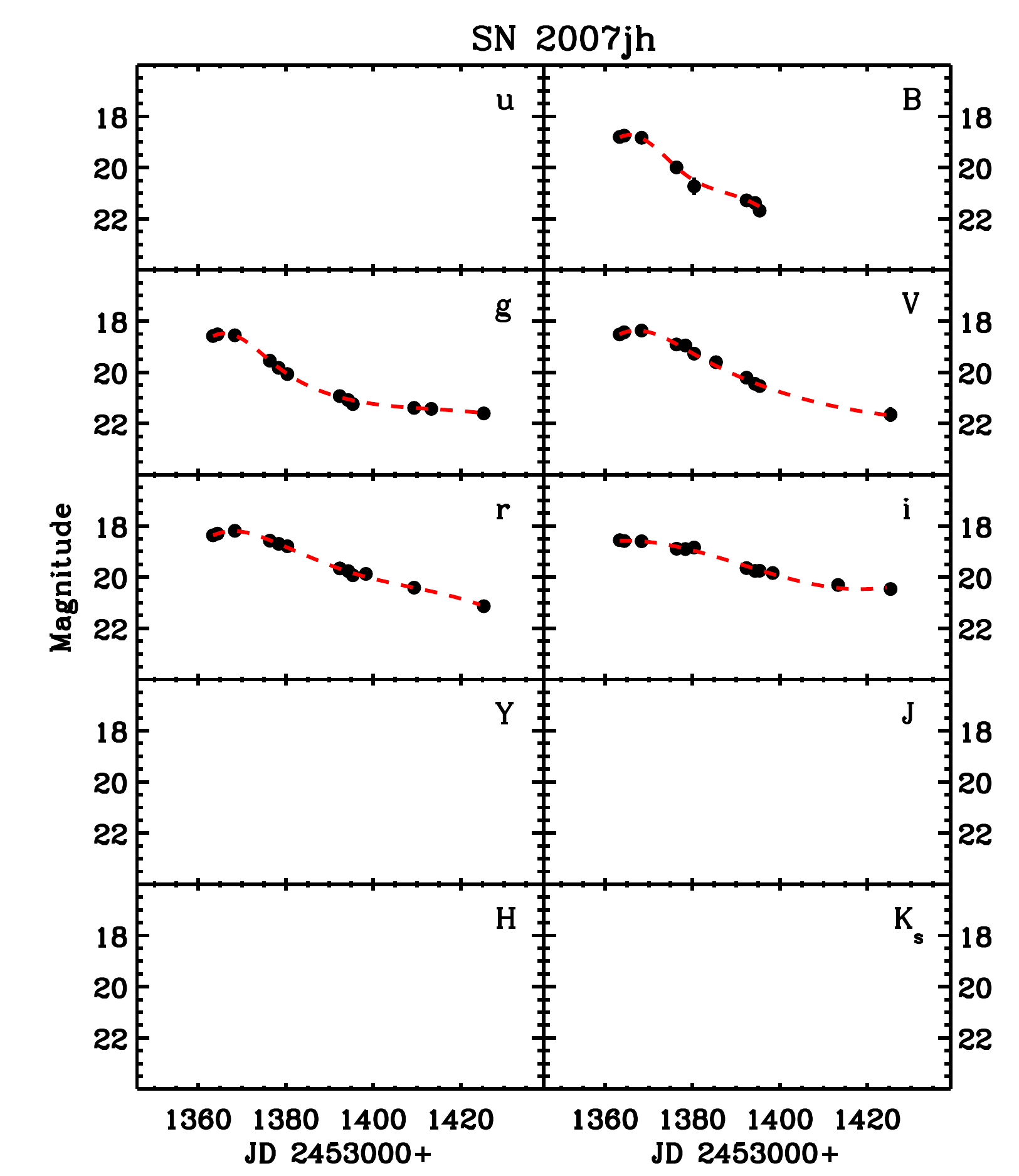}
  {\center Stritzinger {\it et al.} Fig. \ref{fig:flcurves}}
\end{figure}

\clearpage
\newpage
\begin{figure}[t]
\plottwo{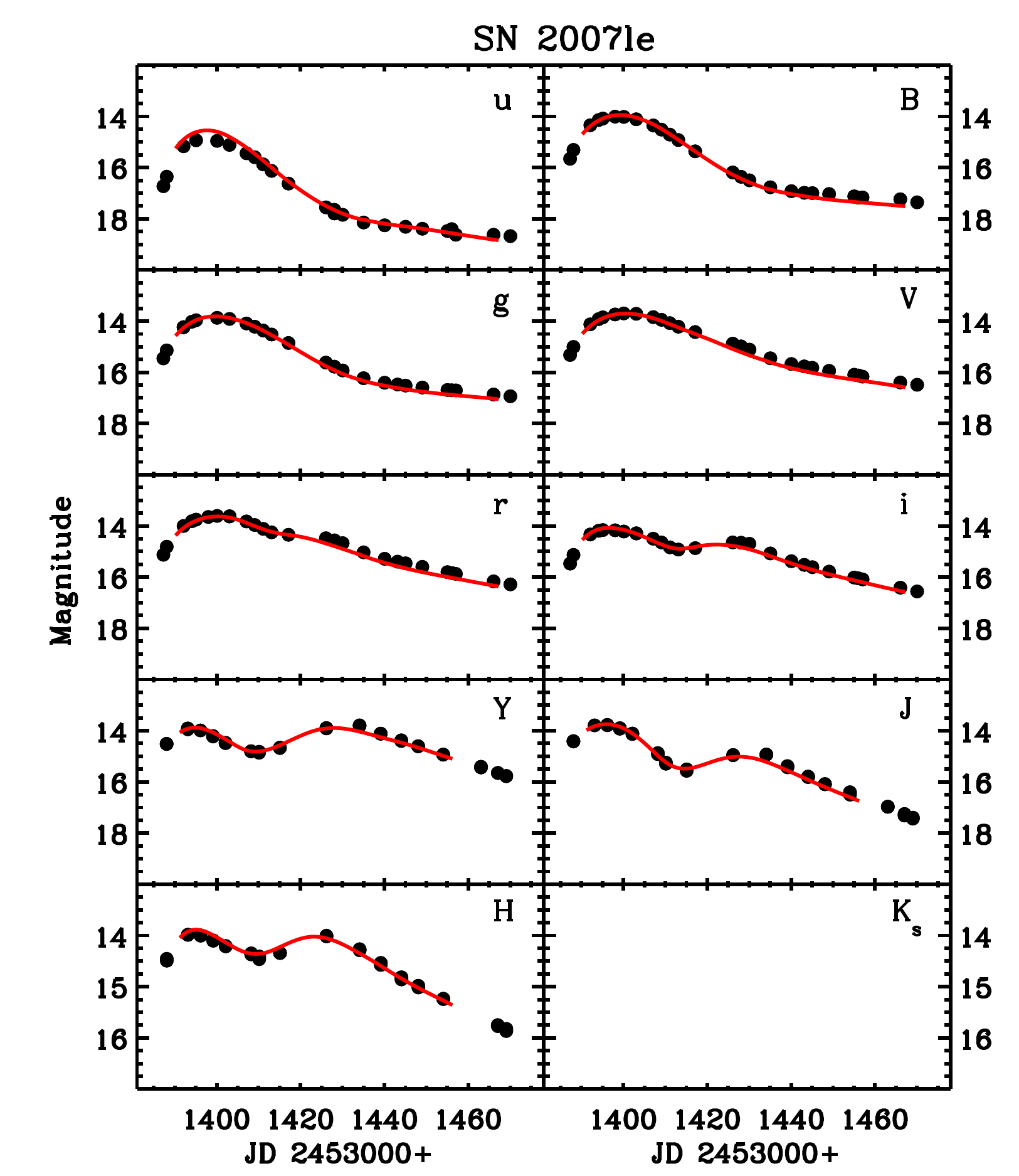}{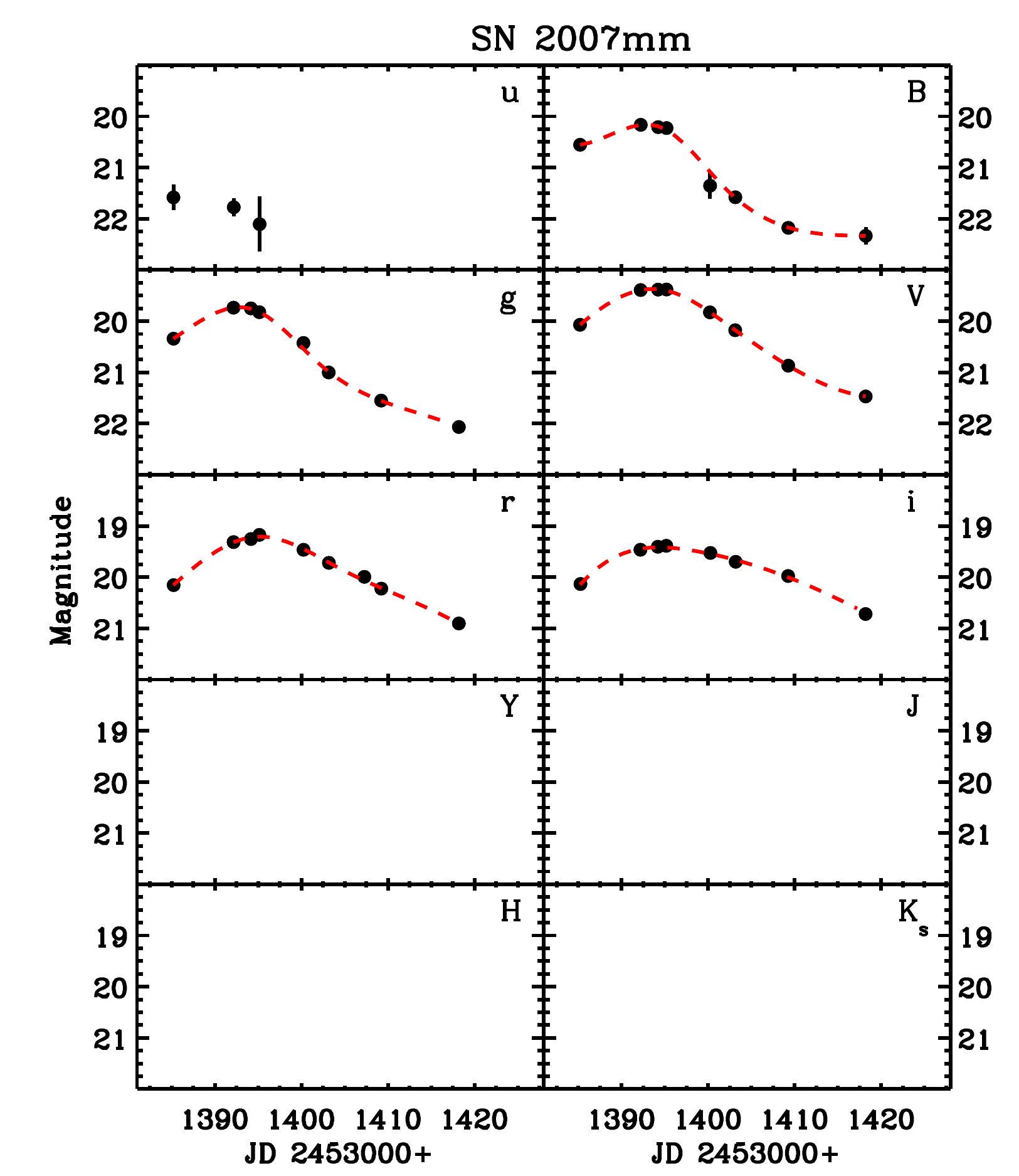}
 \newline
   \newline
 \plottwo{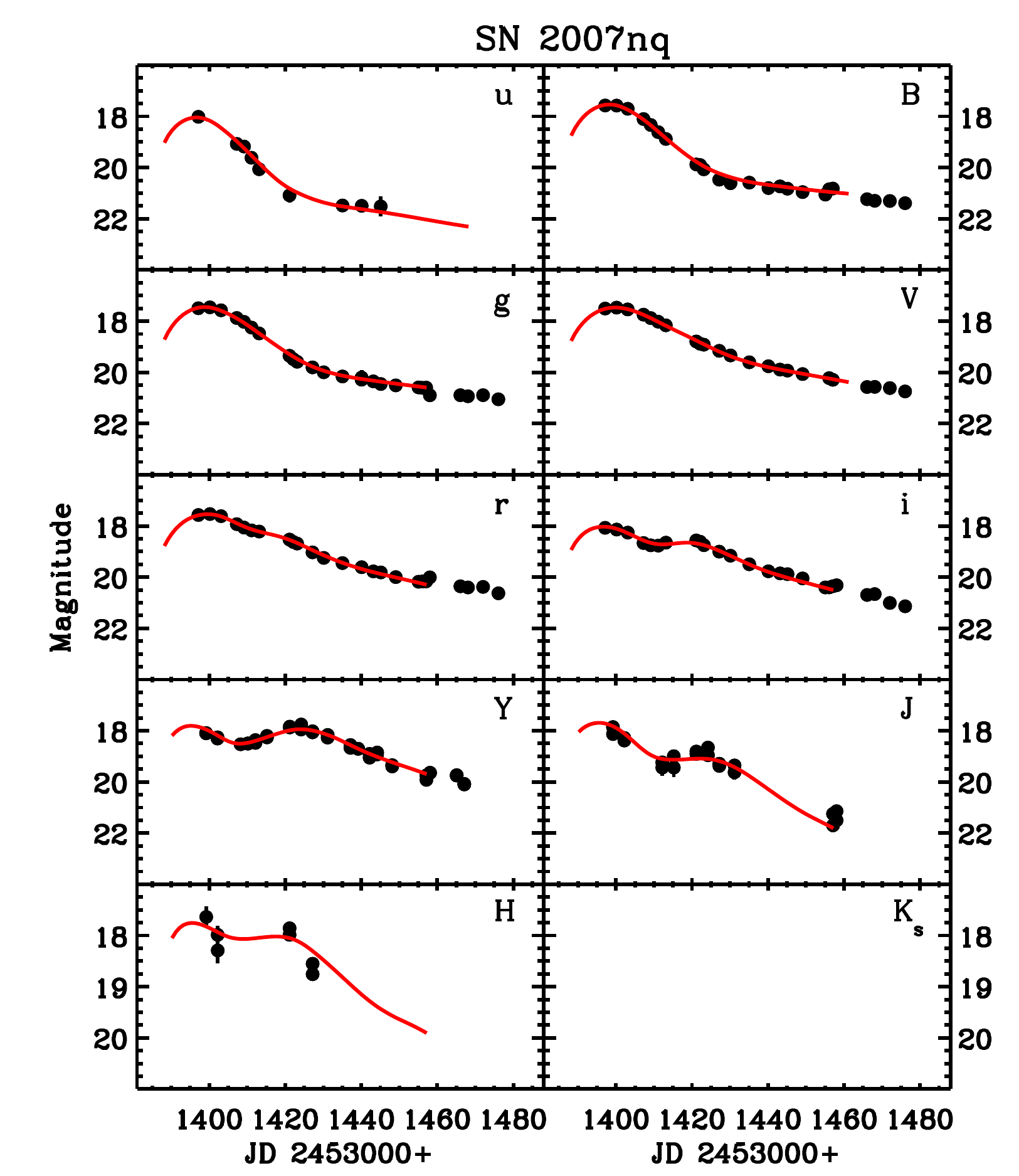}{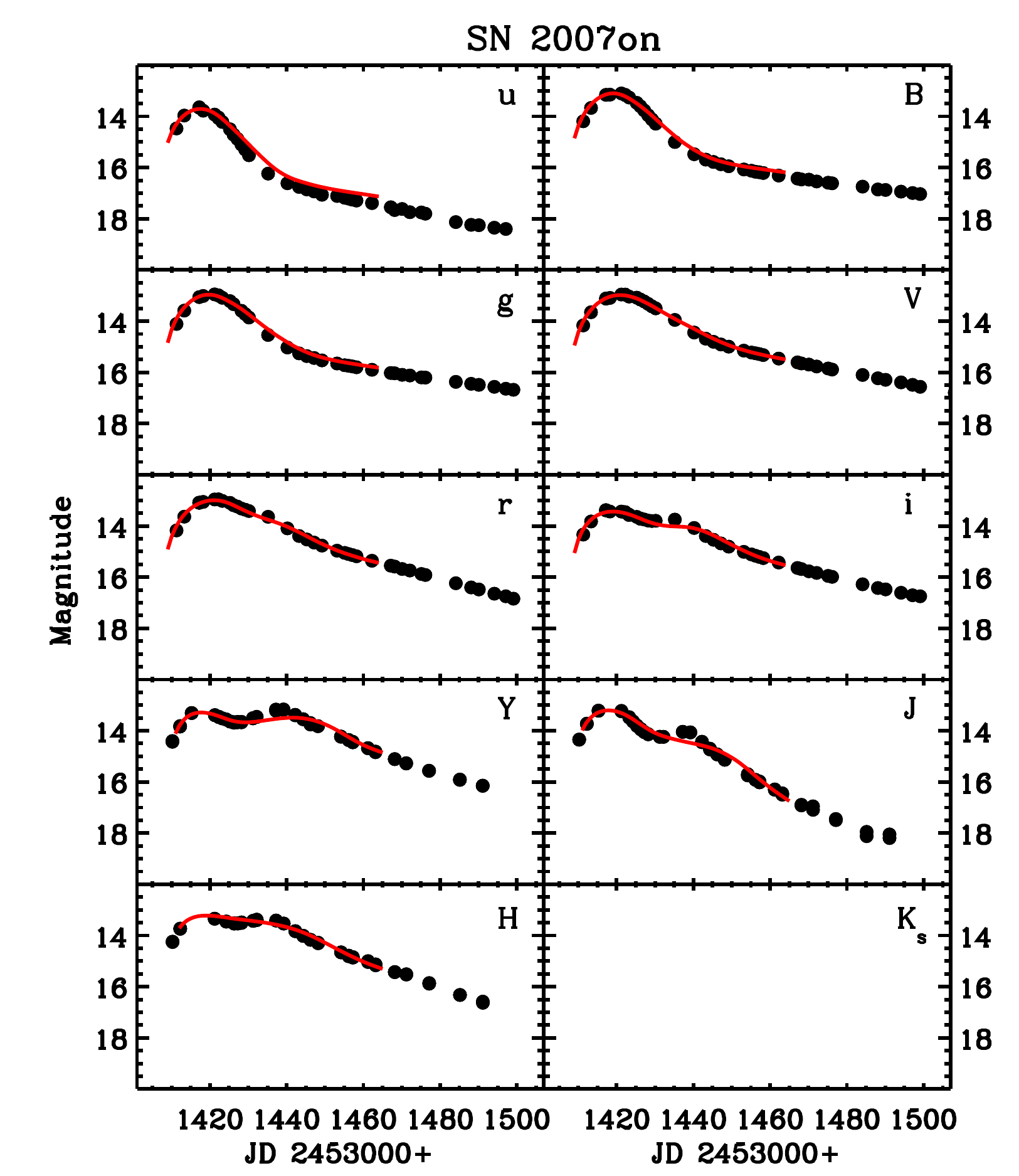}
  {\center Stritzinger {\it et al.} Fig. \ref{fig:flcurves}}
\end{figure}

\clearpage
\newpage
\begin{figure}[t]
 \plottwo{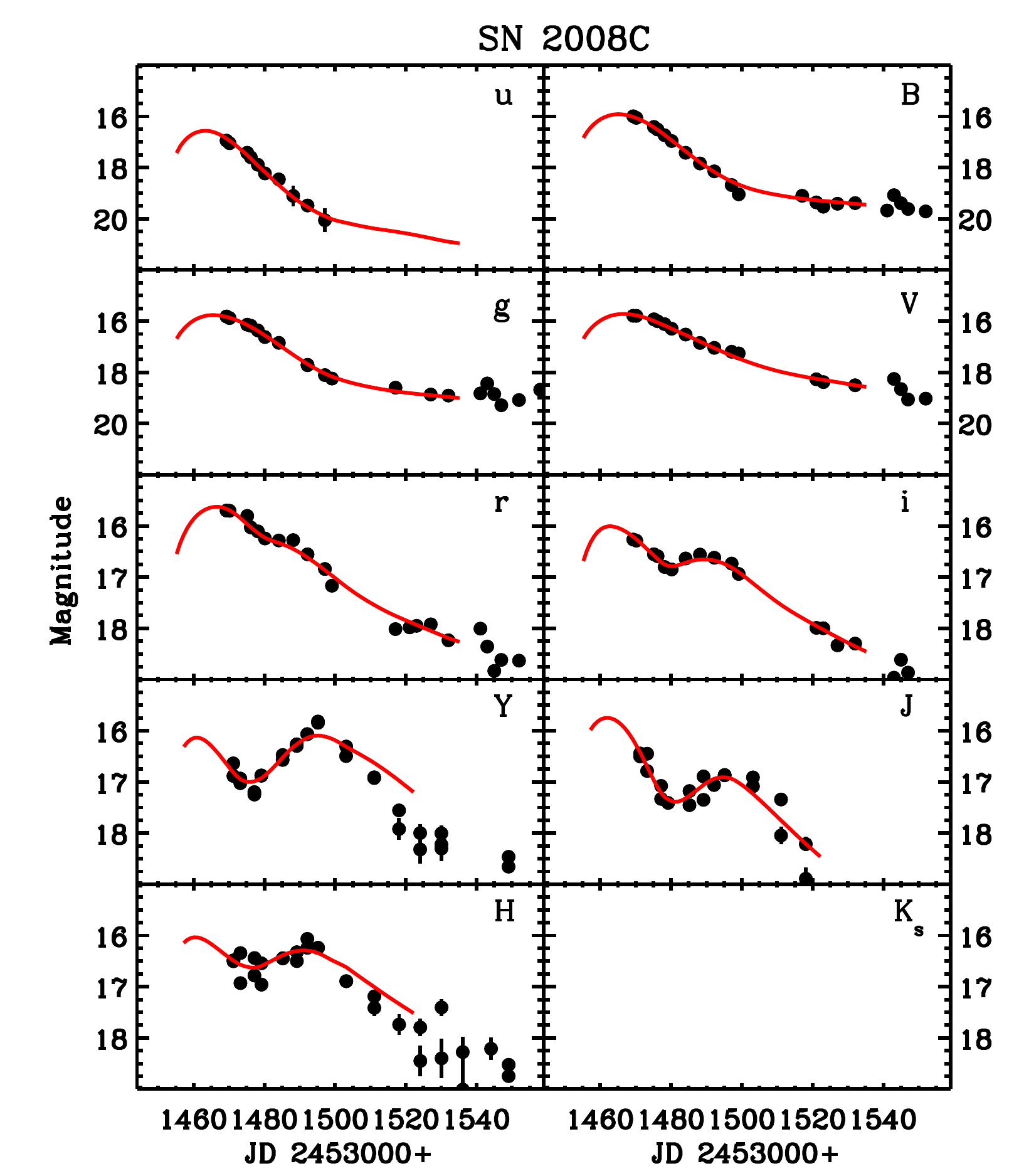}{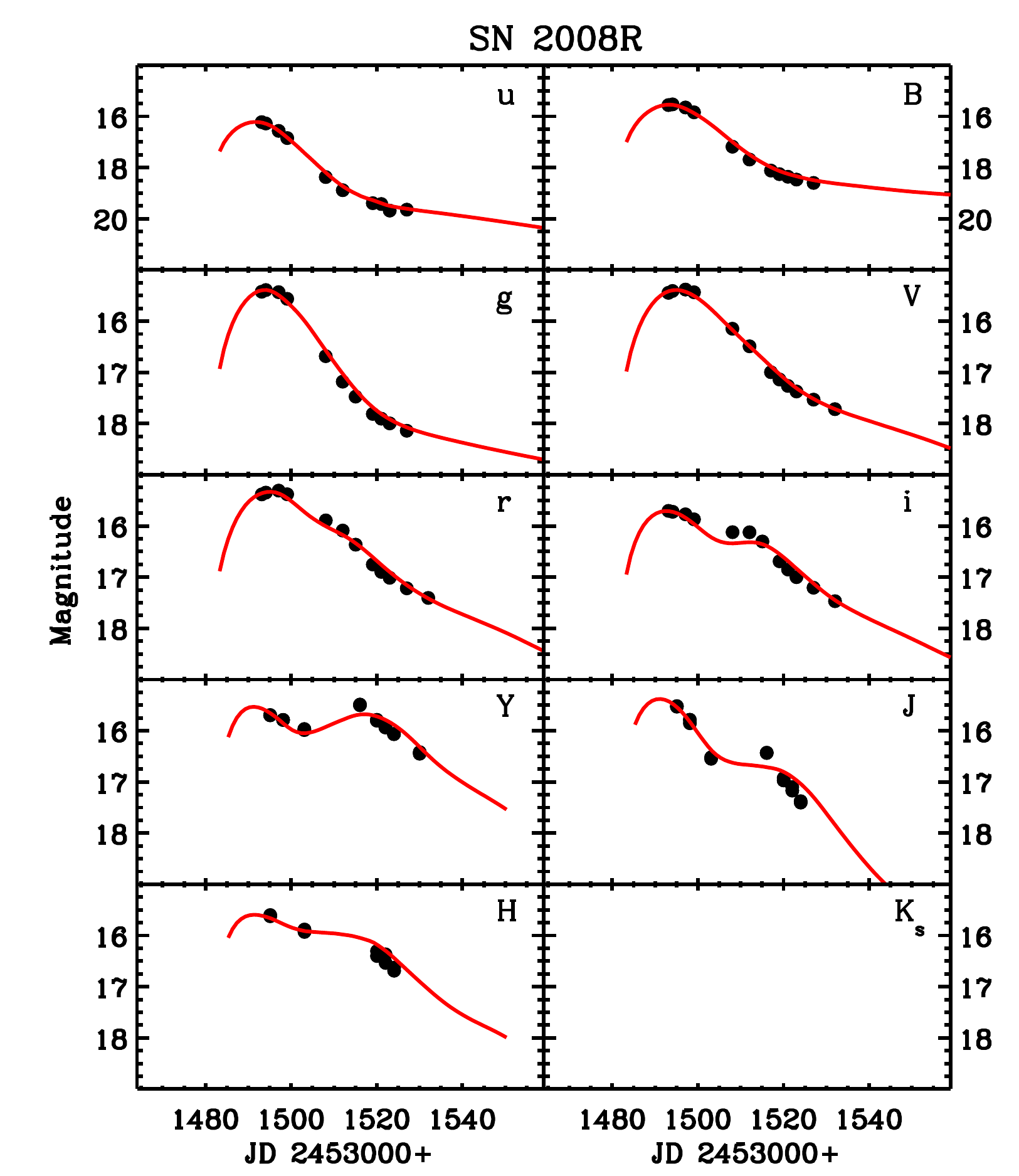}
 \newline
   \newline
\plottwo{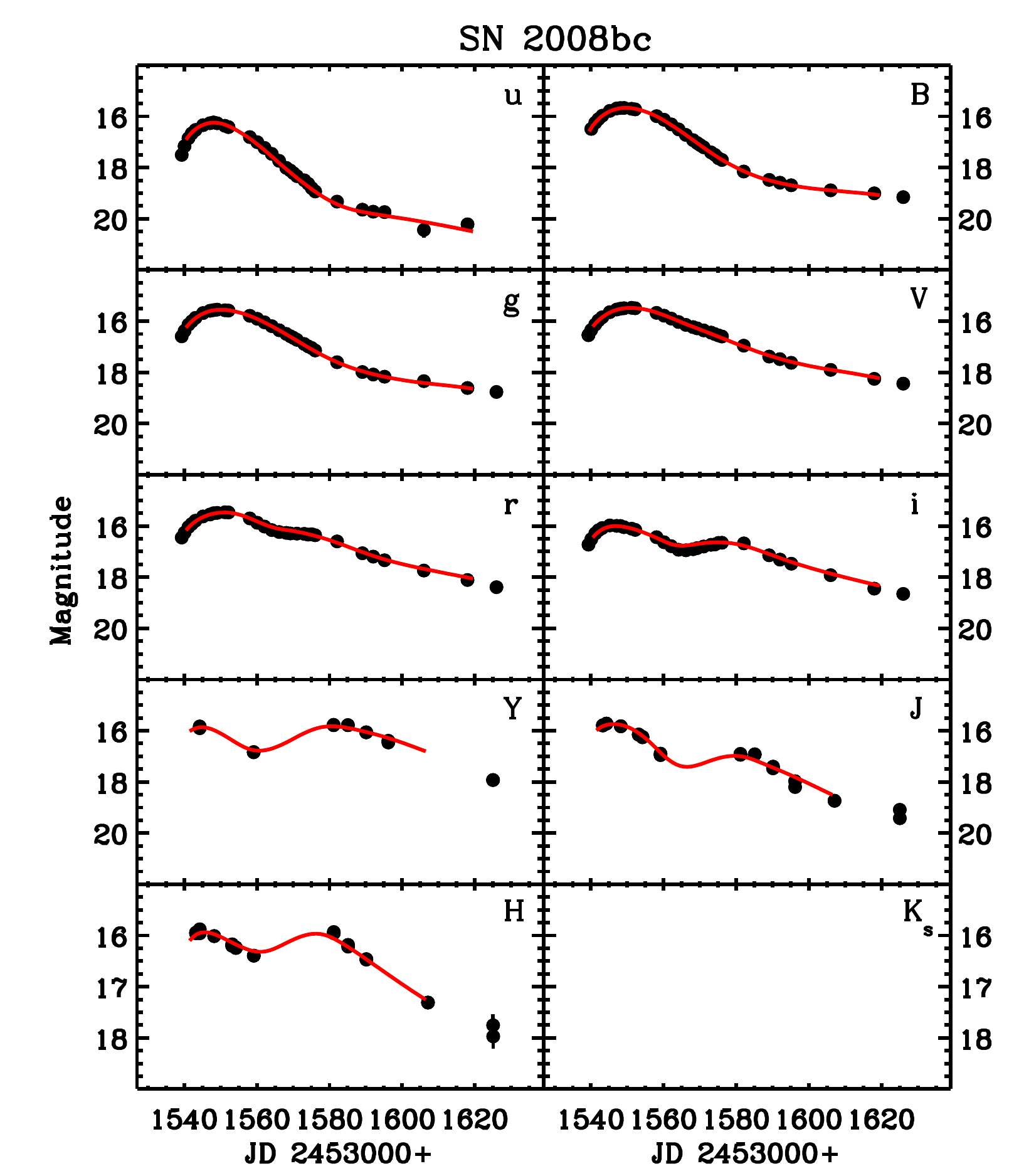}{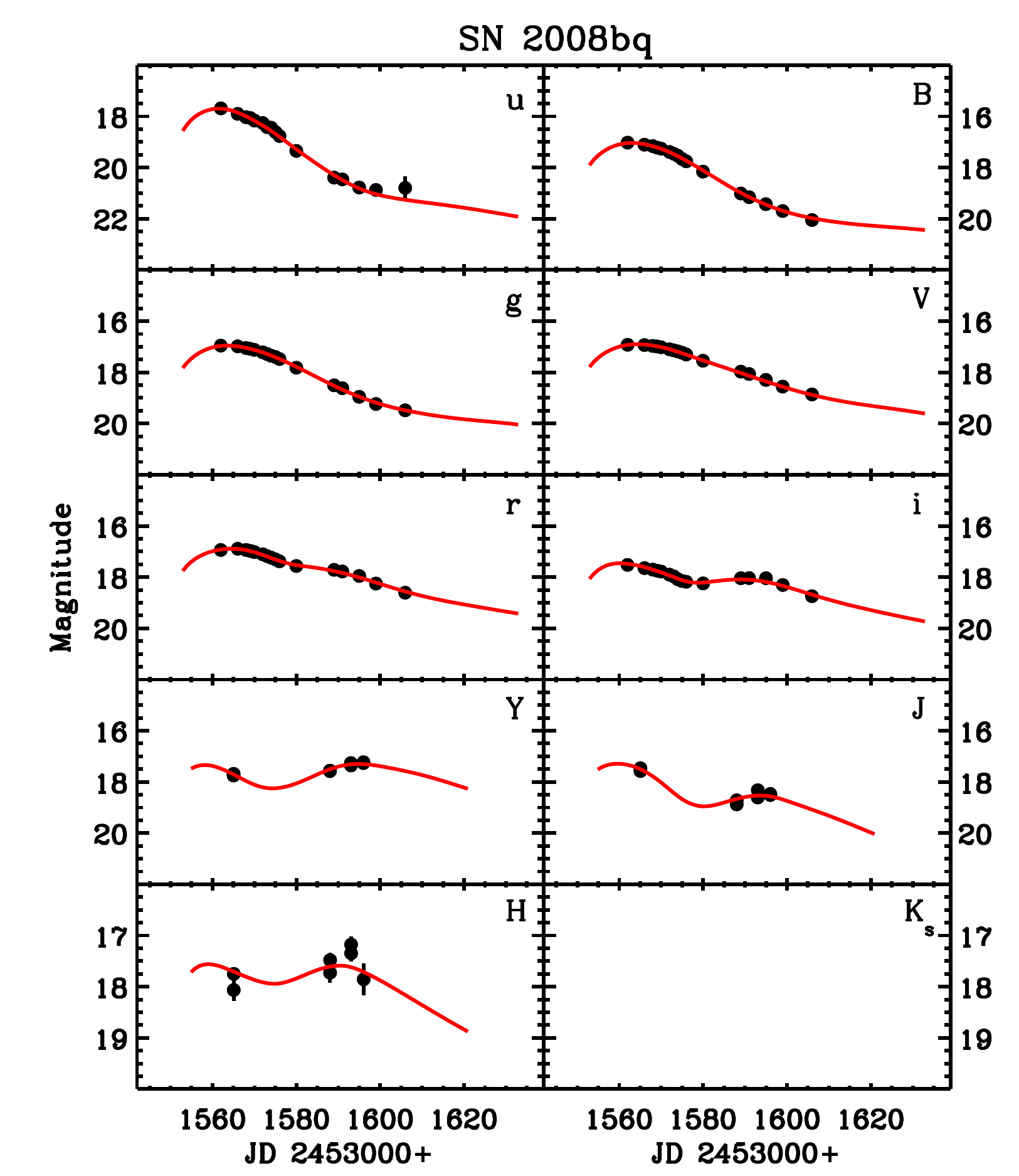}
  {\center Stritzinger {\it et al.} Fig. \ref{fig:flcurves}}
\end{figure}

\clearpage
\newpage
\begin{figure}[t]
 \plottwo{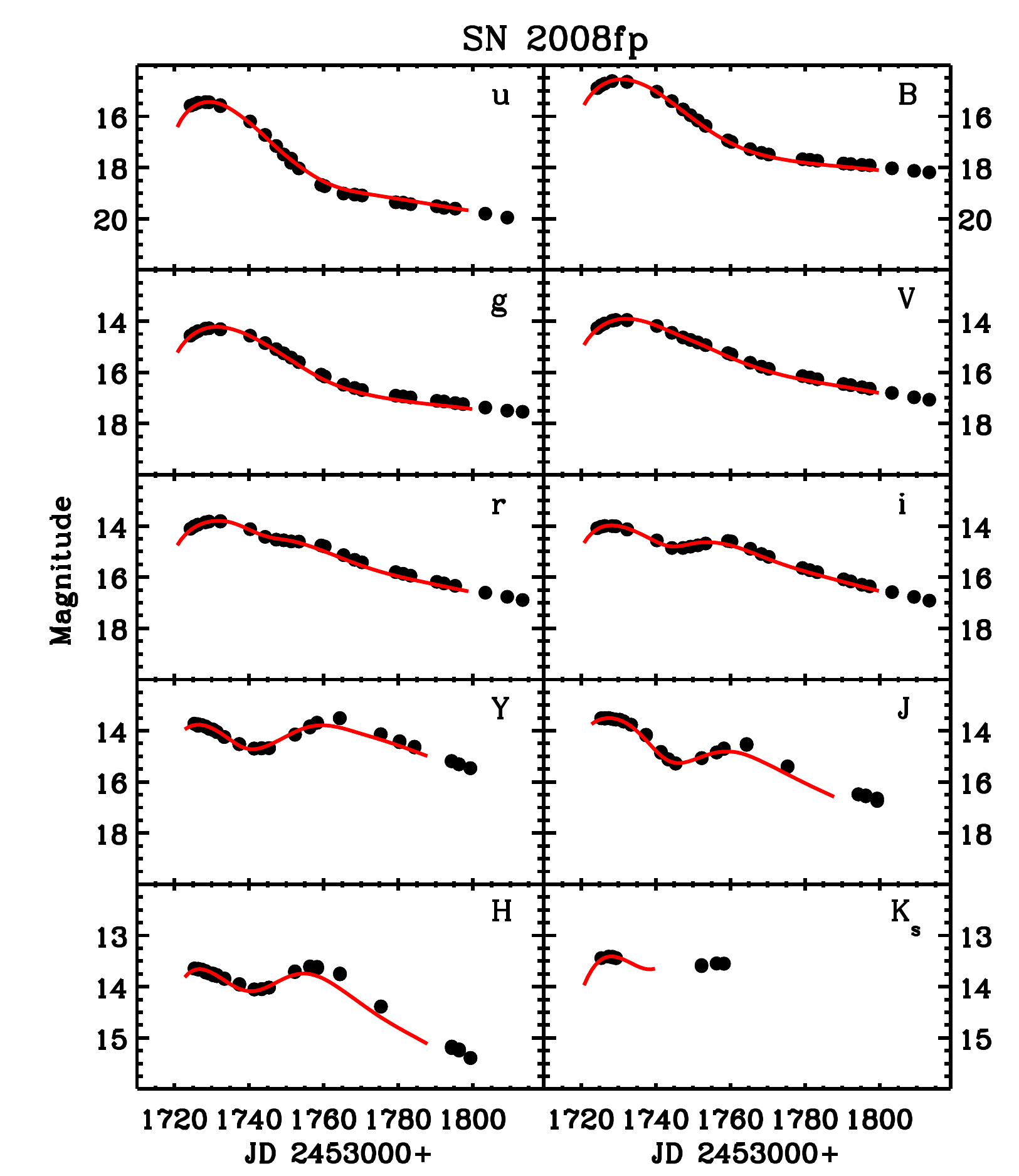}{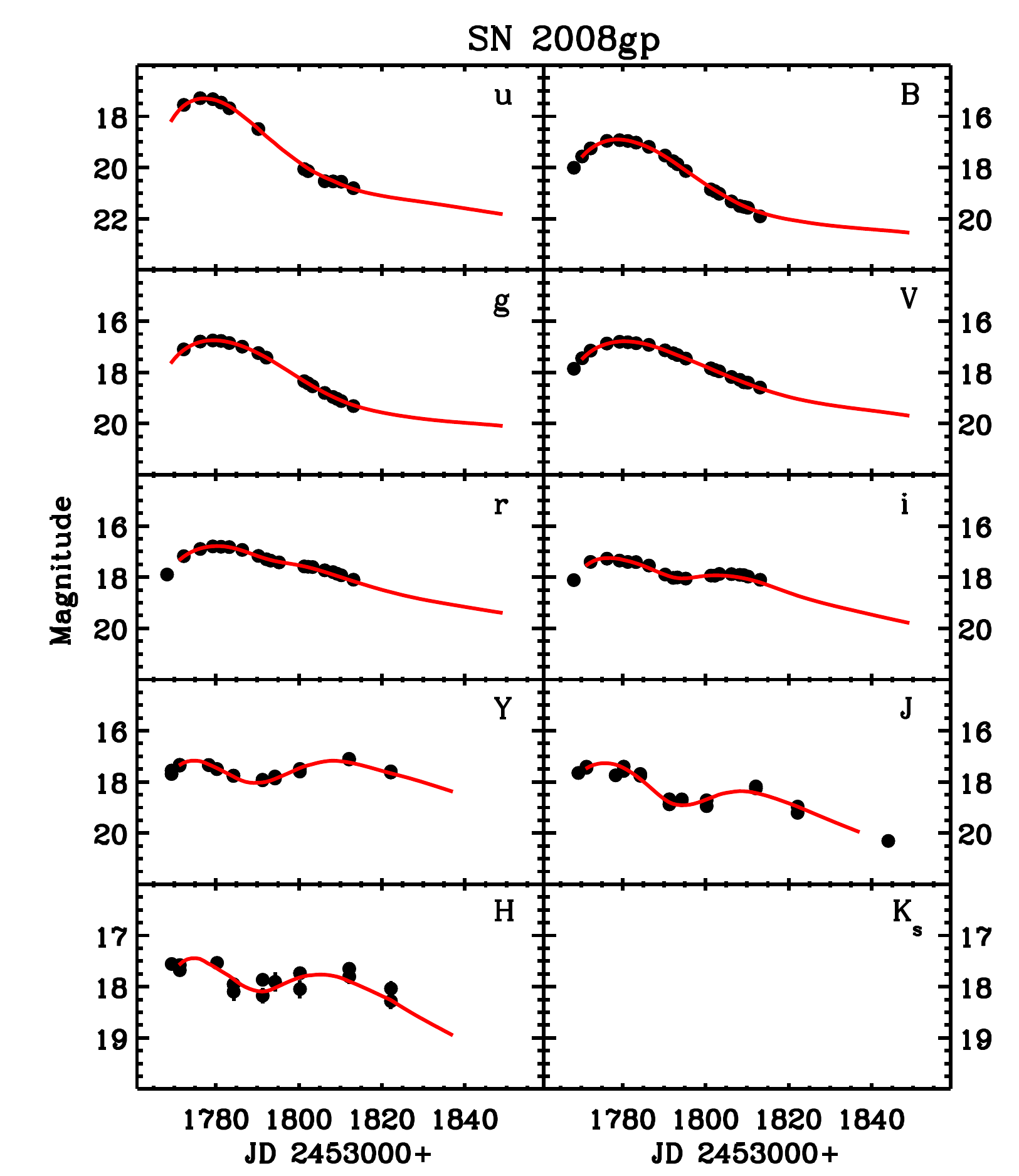}
  \newline
   \newline
 \plottwo{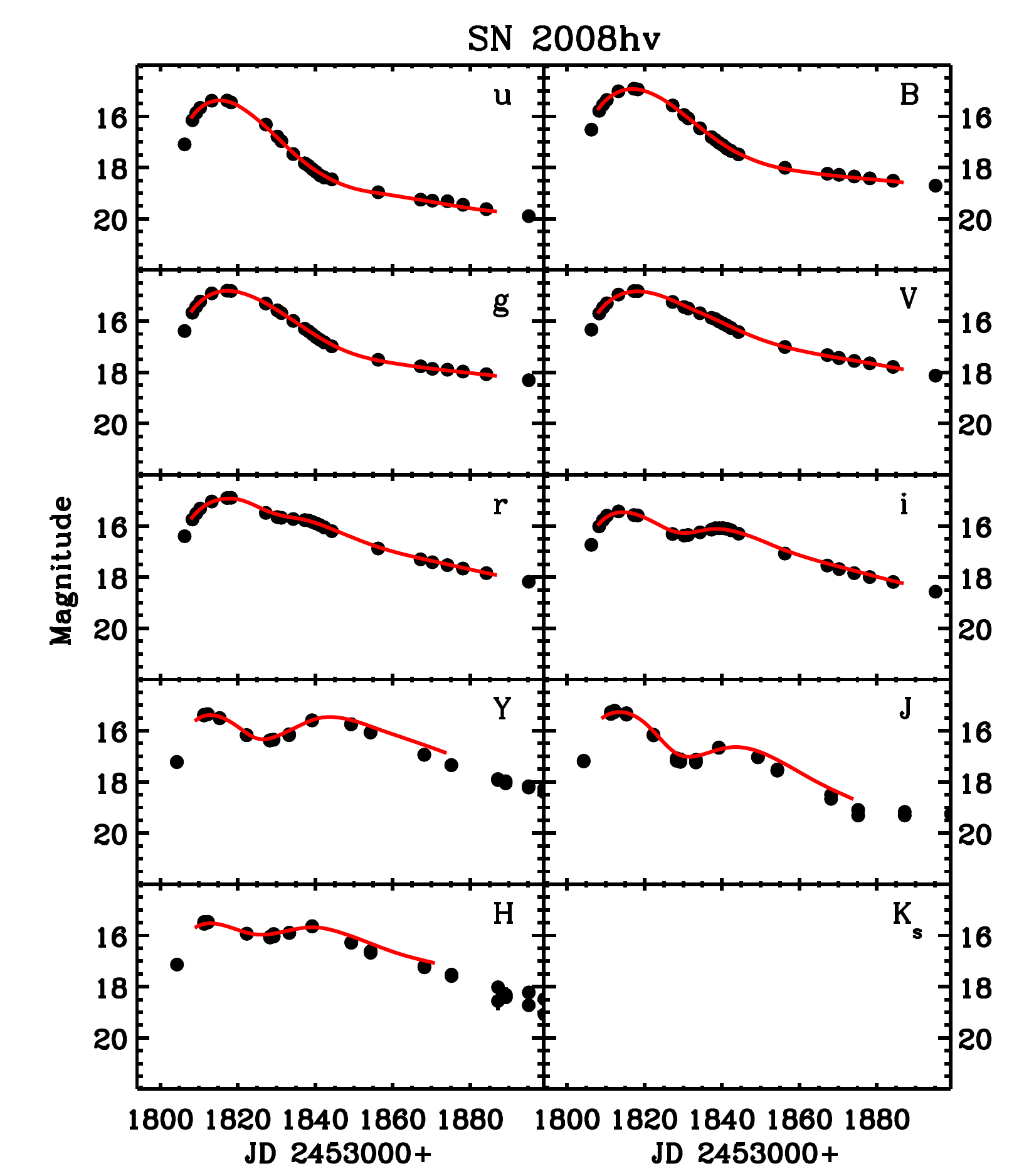}{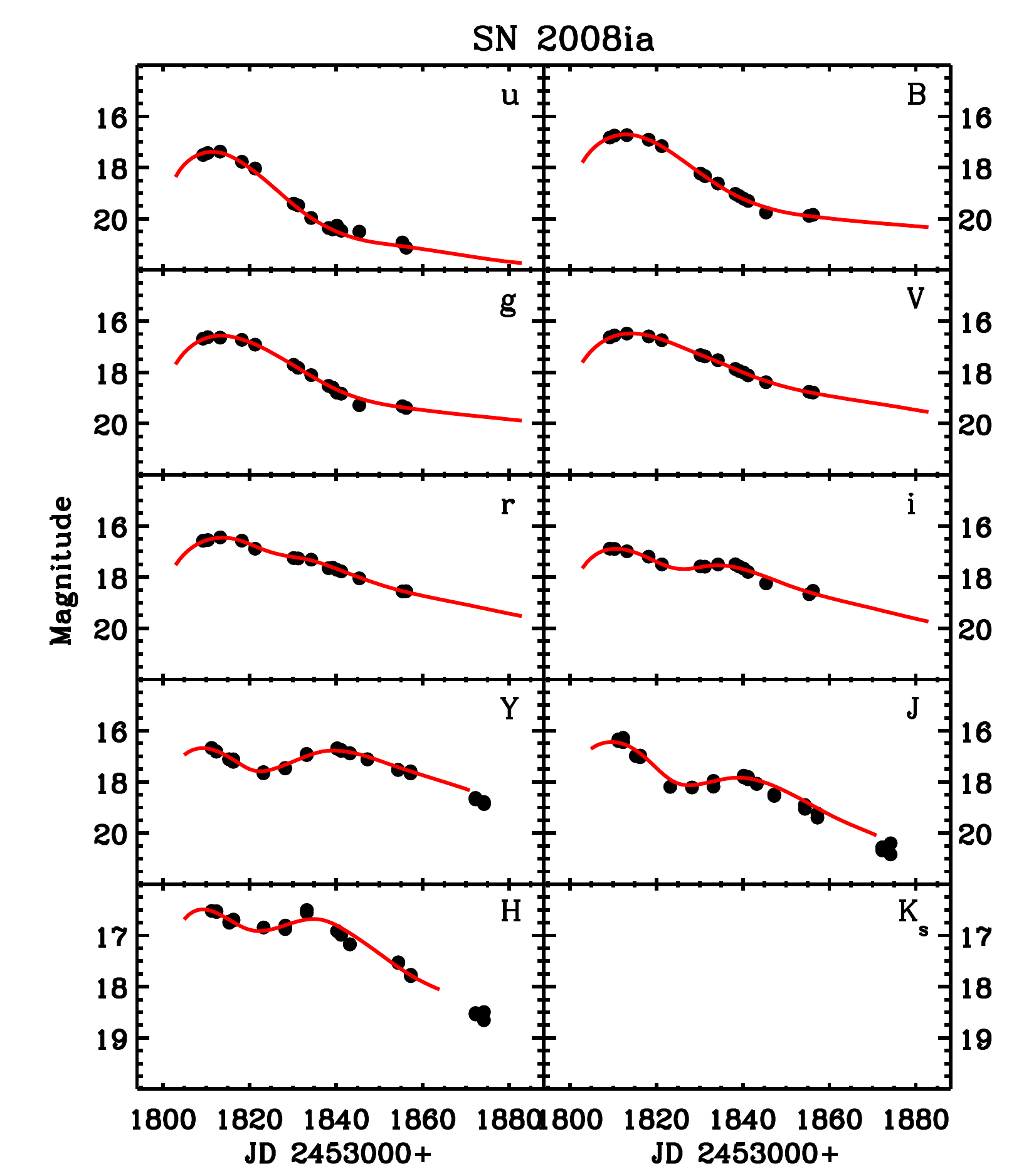}
   {\center Stritzinger {\it et al.} Fig. \ref{fig:flcurves}}
\end{figure}

\clearpage
\newpage
\begin{figure}[t]
 \plottwo{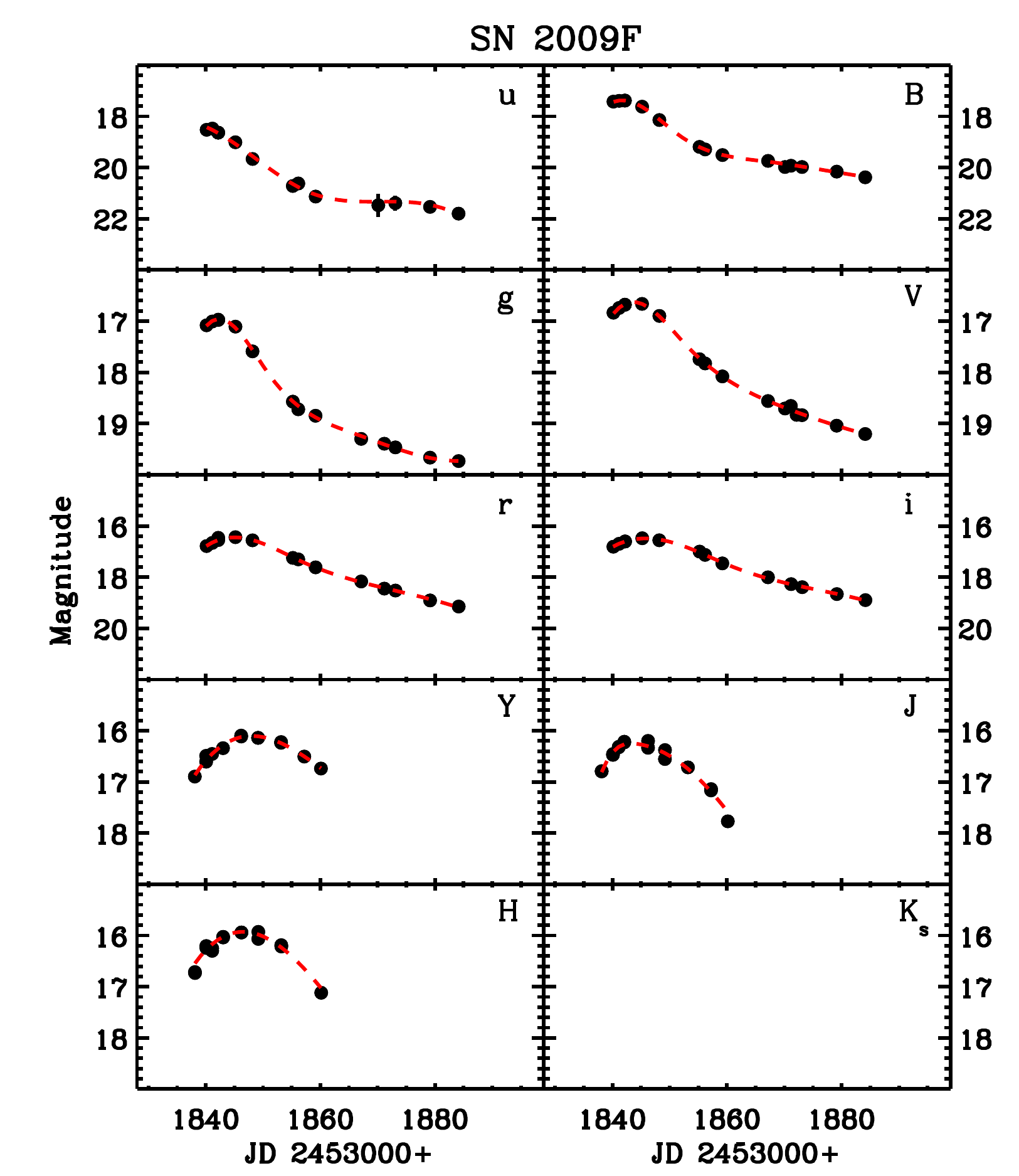}{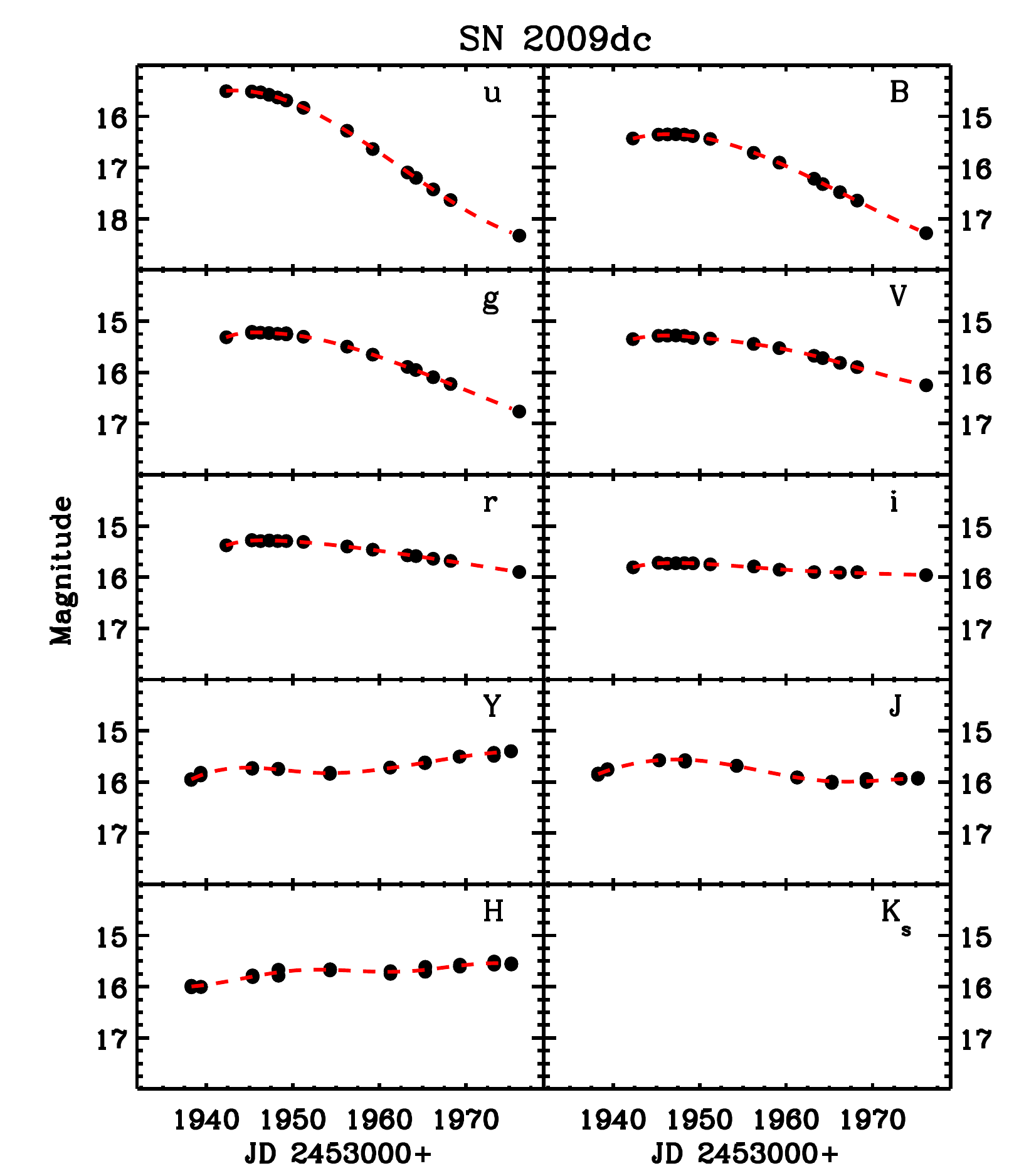}
  {\center Stritzinger {\it et al.} Fig. \ref{fig:flcurves}}
\end{figure}

\clearpage
\begin{figure}[t]
\plotone{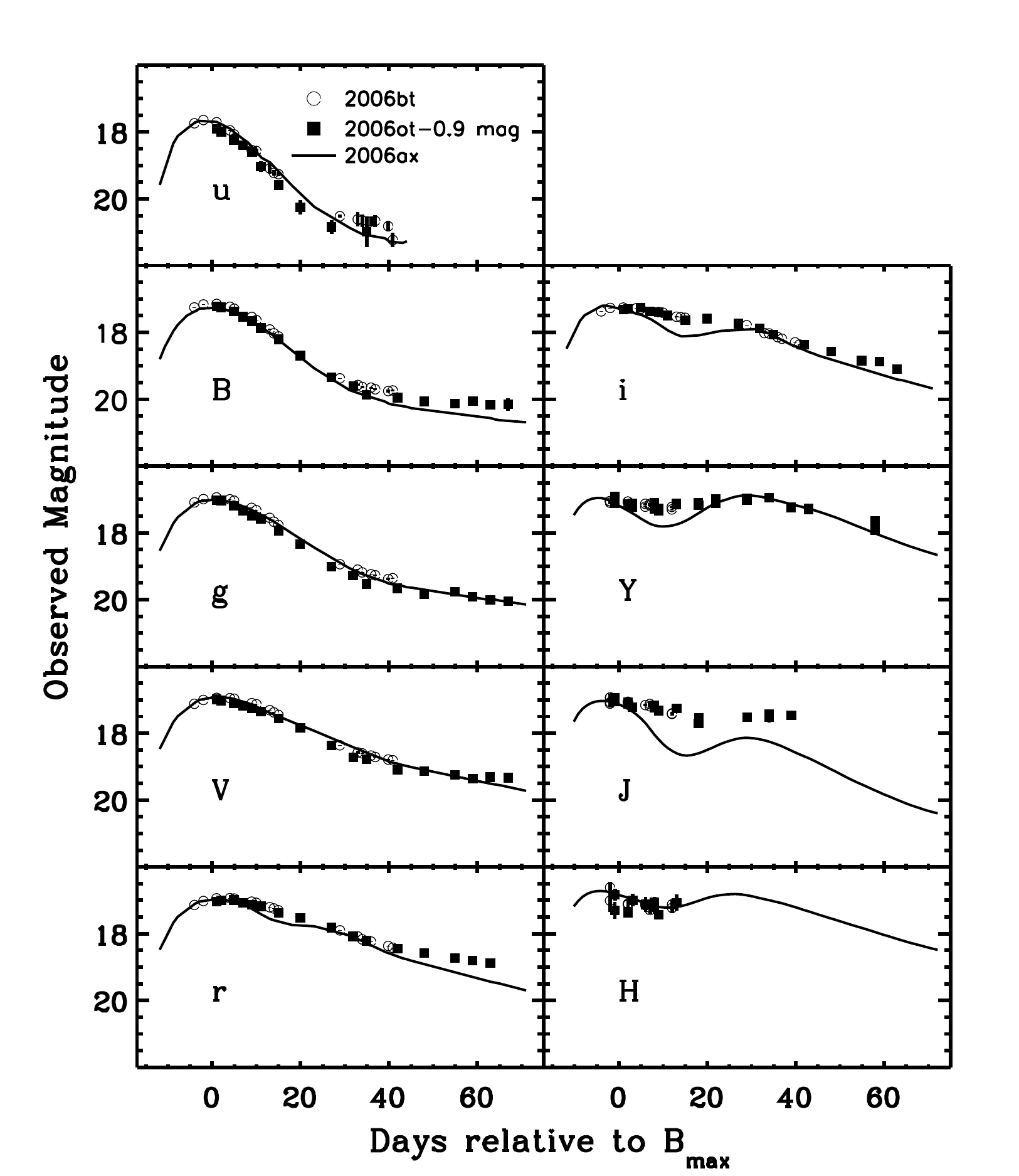}{\center Stritzinger {\it et al.} Fig.~\ref{fig:06otphot}}
\end{figure}

\clearpage
\begin{figure}[t]
\plotone{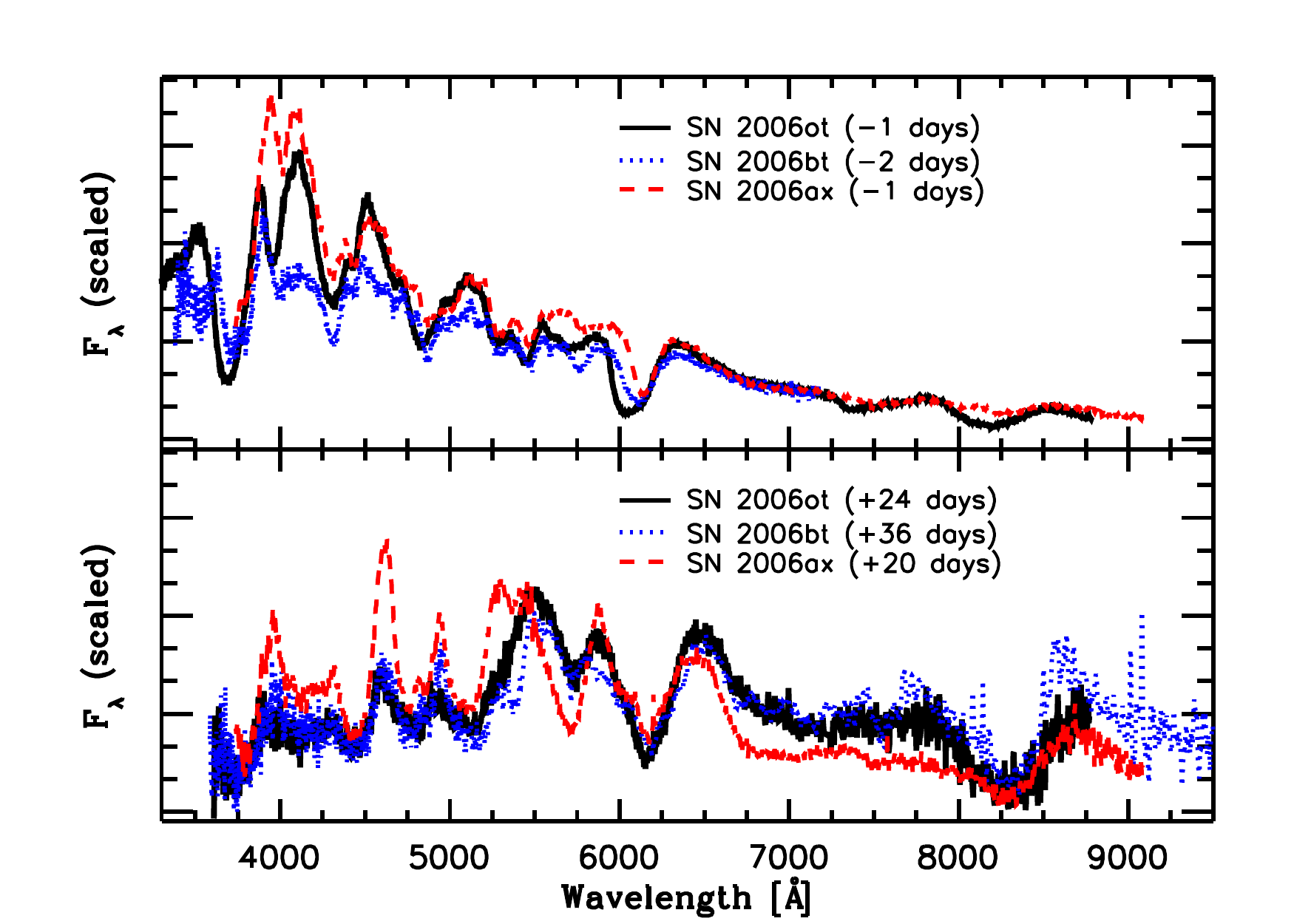}{\center Stritzinger {\it et al.} Fig.~\ref{fig:06otspec}}
\end{figure}

\clearpage
\begin{figure}[t]
\plotone{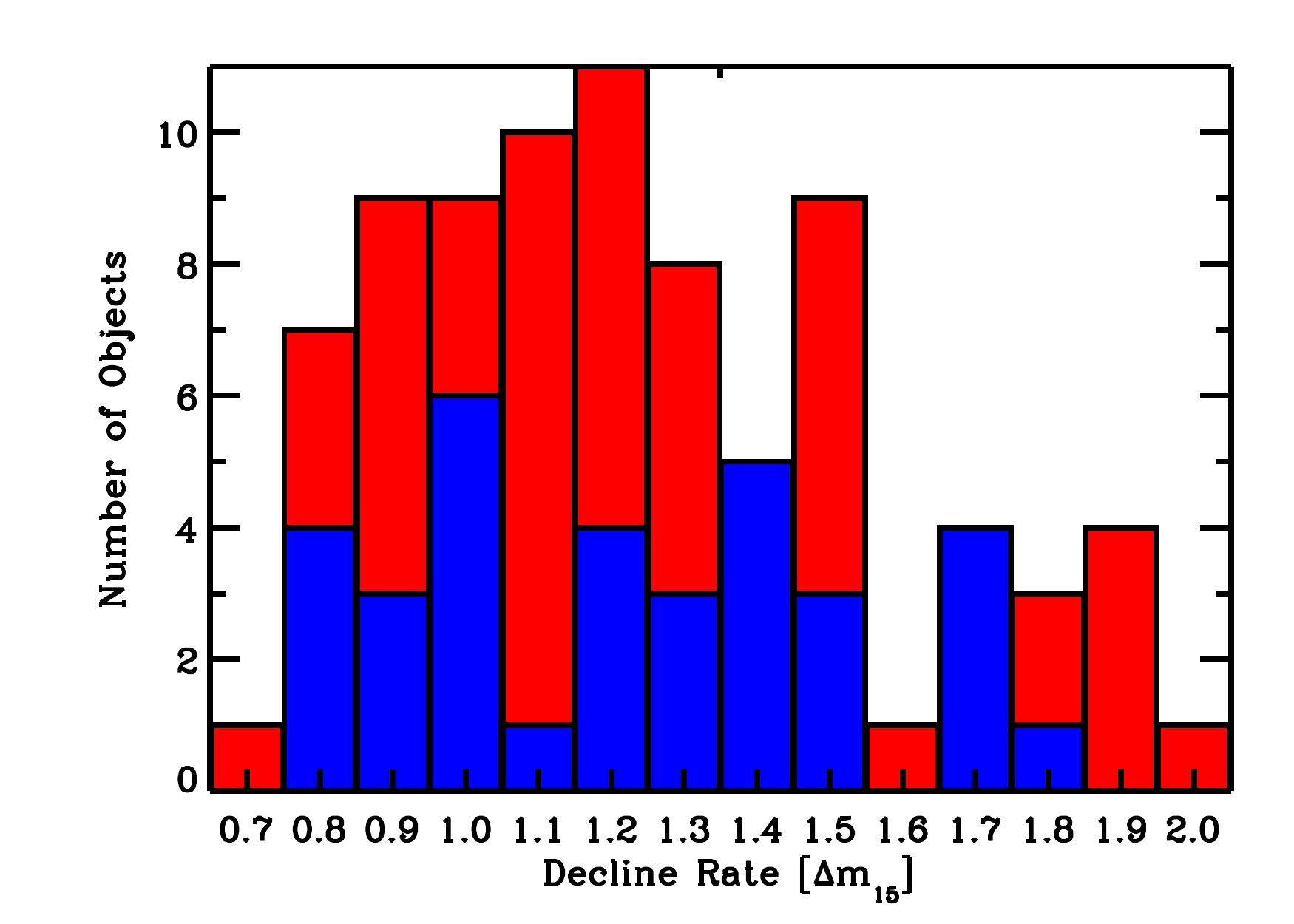}{\center Stritzinger {\it et al.} Fig.~\ref{fig:dm15}}
\end{figure}

\clearpage
\begin{figure}[t]
 \plottwo{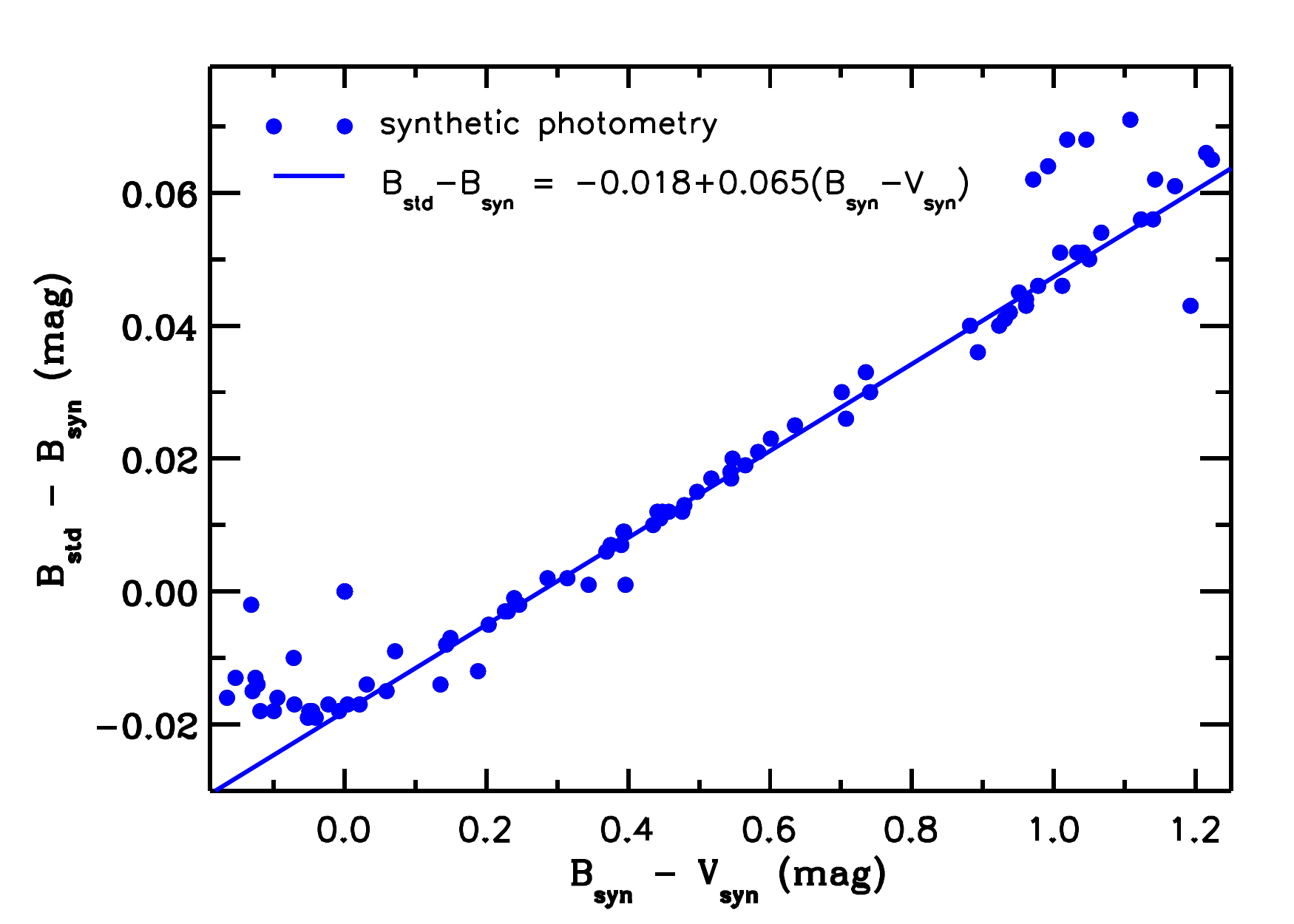}{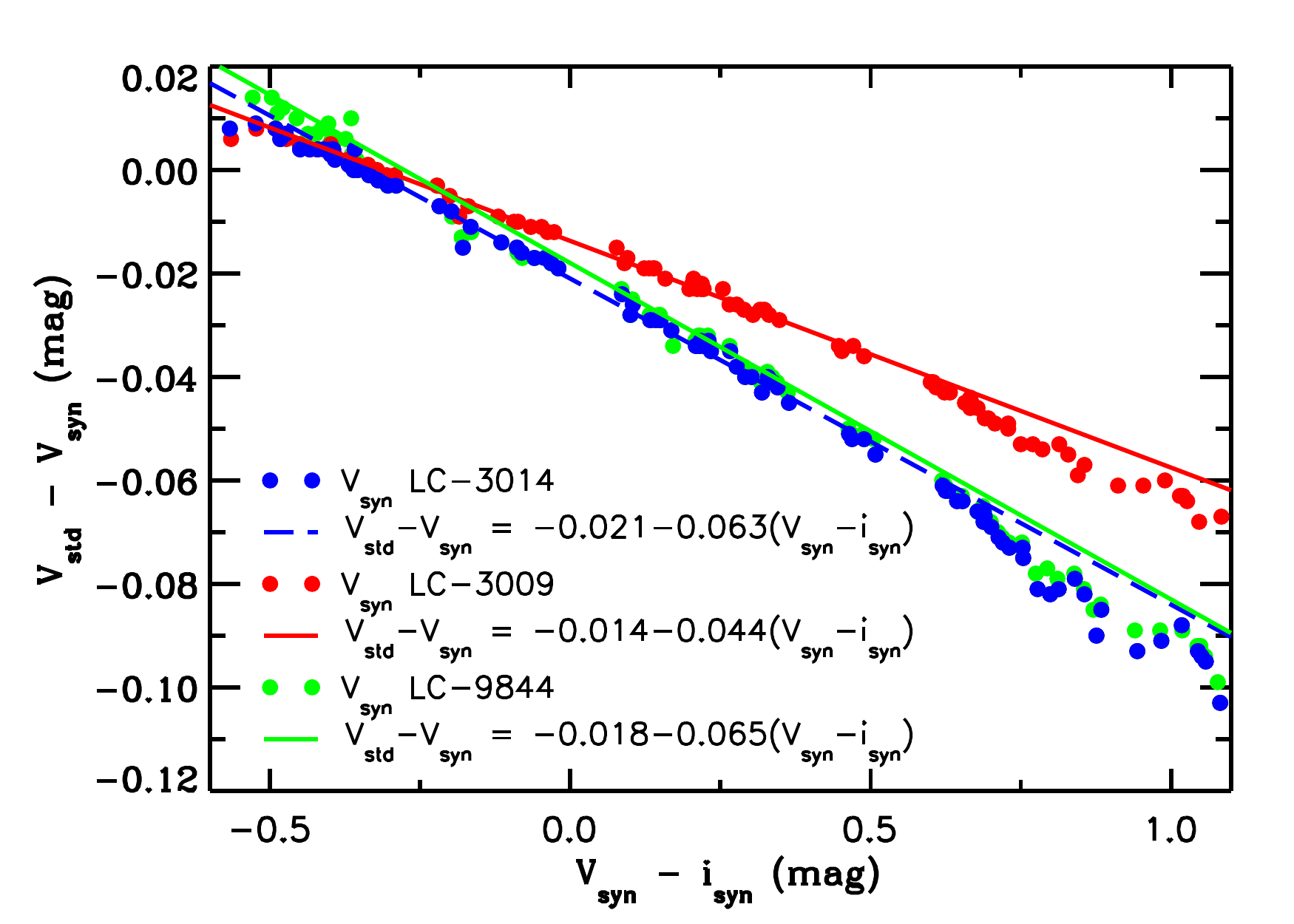}
  {\center Stritzinger {\it et al.} Fig.~\ref{fig:CTs}}
\end{figure}

\end{document}